\documentclass[12pt, letterpaper]{article} 
\usepackage[english]{babel}
\usepackage[latin1]{inputenc}
\usepackage{graphicx}       % For Ã¥ inkludere figurer
\usepackage{listings}
\usepackage{url}
\usepackage{hyperref}
\usepackage{inputenc}
\usepackage{listings}
\usepackage{epsfig}
\usepackage{amsmath,amssymb,amsfonts} 
\usepackage{relsize}
\usepackage{mathrsfs}
\usepackage{pstricks}
\usepackage{multirow}
\usepackage[all]{xypic}
\usepackage{colortbl}
\usepackage{rotating}
\usepackage{subfigure}
\usepackage{float}
\usepackage{fancyhdr}
\usepackage{ae}
\usepackage{natbib}
\usepackage{multicol}
\usepackage{enumerate}
\usepackage{wrapfig}
\usepackage[hang,flushmargin]{footmisc}
\usepackage{bold-extra}

\newtheorem{definition}{Definition}
\newtheorem{theorem}{Theorem}

\newtheorem{proof1}{Proof of Theorem}

\usepackage[explicit]{titlesec}
\usepackage[normalem]{ulem}
\usepackage{xr}
\usepackage{ccaption}
\captionnamefont{\bfseries}
\bibpunct{(}{)}{;}{a}{}{,}

\usepackage[margin=1.0in]{geometry}

\DeclareMathOperator{\card}{card}
\newcommand{\ABBB}{\scriptsize\arraycolsep=0.0pt\def\arraystretch{0.0}\begin{array}{@{}l@{}l@{}} & \\\framebox(5.0,5.0){} & \end{array}}
\newcommand{\AABB}{\scriptsize\arraycolsep=0.0pt\def\arraystretch{0.0}\begin{array}{@{}l@{}l@{}}\framebox(5.0,5.0){}  & \framebox(5.0,5.0){} \\ &\end{array}}
\newcommand{\ABAB}{\scriptsize\arraycolsep=0.0pt\def\arraystretch{0.0}\begin{array}{@{}l@{}l@{}} \framebox(5.0,5.0){}& \\ \framebox(5.0,5.0){}  & \end{array}}
\newcommand{\ABBA}{\scriptsize\arraycolsep=0.0pt\def\arraystretch{0.0}\begin{array}{@{}l@{}l@{}}\framebox(5.0,5.0){} & \\ &\framebox(5.0,5.0){}\end{array}}
\newcommand{\BAAB}{\scriptsize\arraycolsep=0.0pt\def\arraystretch{0.0}\begin{array}{@{}l@{}l@{}} &\framebox(5.0,5.0){} \\ \framebox(5.0,5.0){} & \end{array}}
\newcommand{\BAAA}{\scriptsize\arraycolsep=0.0pt\def\arraystretch{0.0}\begin{array}{@{}l@{}l@{}} &\framebox(5.0,5.0){} \\ \framebox(5.0,5.0){} & \framebox(5.0,5.0){} \end{array}}
\newcommand{\ABAA}{\scriptsize\arraycolsep=0.0pt\def\arraystretch{0.0}\begin{array}{@{}l@{}l@{}} \framebox(5.0,5.0){}& \\ \framebox(5.0,5.0){} & \framebox(5.0,5.0){} \end{array}}
\newcommand{\AABA}{\scriptsize\arraycolsep=0.0pt\def\arraystretch{0.0}\begin{array}{@{}l@{}l@{}} \framebox(5.0,5.0){} &\framebox(5.0,5.0){} \\ & \framebox(5.0,5.0){} \end{array}}
\newcommand{\AAAB}{\scriptsize\arraycolsep=0.0pt\def\arraystretch{0.0}\begin{array}{@{}l@{}l@{}} \framebox(5.0,5.0){} &\framebox(5.0,5.0){} \\ \framebox(5.0,5.0){} & \end{array}}

\newcommand{\AAAA}{\scriptsize\arraycolsep=0.0pt\def\arraystretch{0.0}\begin{array}{@{}l@{}l@{}}
      \framebox(5.0,5.0){} &\framebox(5.0,5.0){} \\
      \framebox(5.0,5.0){} & \framebox(5.0,5.0){} \end{array}}
\newcommand{\cc}{\scriptsize\arraycolsep=0.0pt\def\arraystretch{0.0}\begin{array}{@{}l@{}l@{}l@{}}
      \framebox(5.0,5.0){} && \\
       &\makebox(5.0,5.0){} & \framebox(5.0,5.0){} \end{array}}

\newcommand{\cd}{\scriptsize\arraycolsep=0.0pt\def\arraystretch{0.0}\begin{array}{@{}l@{}l@{}}
      \framebox(5.0,5.0){}&  \\
       &\makebox(5.0,5.0){} \\
& \framebox(5.0,5.0){} \end{array}}

\newcommand{\neigf}{\scriptsize\arraycolsep=0.0pt\def\arraystretch{0.0}\begin{array}{@{}l@{}l@{}l@{}l@{}l@{}}
      &\framebox(5.0,5.0){$\times$}&&& \\
     \framebox(5.0,5.0){$\times$}&&\framebox(5.0,5.0){$\times$}&& \\
&&\framebox(5.0,5.0){$\cdot$}&&\\
&&\framebox(5.0,5.0){$\times$}&&\framebox(5.0,5.0){$\times$}\\
&&&\framebox(5.0,5.0){$\times$}& \end{array}}

\newcommand{\neigs}{\scriptsize\arraycolsep=0.0pt\def\arraystretch{0.0}\begin{array}{@{}l@{}l@{}l@{}l@{}l@{}}
      &&&& \\
     &\framebox(5.0,5.0){$\times$}&\framebox(5.0,5.0){$\times$}&& \\
& \framebox(5.0,5.0){$\times$}&\framebox(5.0,5.0){$\cdot$}& \framebox(5.0,5.0){$\times$}&\\
&&\framebox(5.0,5.0){$\times$}&\framebox(5.0,5.0){$\times$}&\\
&&&& \end{array}}

\newcommand{\neigt}{\scriptsize\arraycolsep=0.0pt\def\arraystretch{0.0}\begin{array}{@{}l@{}l@{}l@{}l@{}l@{}}
      &\framebox(5.0,5.0){$\times$}&&& \\
     \framebox(5.0,5.0){$\times$}&\framebox(5.0,5.0){$\times$}&&& \\
& &\framebox(5.0,5.0){$\cdot$}&&\\
&&&\framebox(5.0,5.0){$\times$}&\framebox(5.0,5.0){$\times$}\\
&&&\framebox(5.0,5.0){$\times$}& \end{array}}

\newcommand{\neigfo}{\scriptsize\arraycolsep=0.0pt\def\arraystretch{0.0}\begin{array}{@{}l@{}l@{}l@{}l@{}l@{}}
      &\framebox(5.0,5.0){$\times$}&&& \\
     \framebox(5.0,5.0){$\times$}&&\framebox(5.0,5.0){$\times$}&& \\
&\framebox(5.0,5.0){$\times$} &\framebox(5.0,5.0){$\cdot$}&\framebox(5.0,5.0){$\times$}&\\
&&\framebox(5.0,5.0){$\times$}&&\framebox(5.0,5.0){$\times$}\\
&&&\framebox(5.0,5.0){$\times$}& \end{array}}

\newcommand{\neigfi}{\scriptsize\arraycolsep=0.0pt\def\arraystretch{0.0}\begin{array}{@{}l@{}l@{}l@{}l@{}l@{}}
      &\framebox(5.0,5.0){$\times$}&&& \\
     \framebox(5.0,5.0){$\times$}&&&& \\
&\framebox(5.0,5.0){$\times$} &\framebox(5.0,5.0){$\cdot$}&\framebox(5.0,5.0){$\times$}&\\
&&&&\framebox(5.0,5.0){$\times$}\\
&&&\framebox(5.0,5.0){$\times$}& \end{array}}

\newcommand{\neigse}{\scriptsize\arraycolsep=0.0pt\def\arraystretch{0.0}\begin{array}{@{}l@{}l@{}l@{}l@{}l@{}}
      &&&& \\
     \framebox(5.0,5.0){$\times$}&\framebox(5.0,5.0){$\times$}&\framebox(5.0,5.0){$\times$}&& \\
& &\framebox(5.0,5.0){$\cdot$}&&\\
& &\framebox(5.0,5.0){$\times$}&\framebox(5.0,5.0){$\times$}&\framebox(5.0,5.0){$\times$}\\
&&&& \end{array}}

\newcommand{\neigni}{\scriptsize\arraycolsep=0.0pt\def\arraystretch{0.0}\begin{array}{@{}l@{}l@{}l@{}l@{}l@{}}
      & \framebox(5.0,5.0){$\times$}&&& \\
    &\framebox(5.0,5.0){$\times$}&&& \\
&\framebox(5.0,5.0){$\times$} &\framebox(5.0,5.0){$\cdot$}&\framebox(5.0,5.0){$\times$}&\\
& &&\framebox(5.0,5.0){$\times$}&\\
&&&\framebox(5.0,5.0){$\times$}& \end{array}}

\newcommand{\neigte}{\scriptsize\arraycolsep=0.0pt\def\arraystretch{0.0}\begin{array}{@{}l@{}l@{}l@{}l@{}l@{}}
      &&&& \\
     \framebox(5.0,5.0){$\times$}&&\framebox(5.0,5.0){$\times$}&& \\
&\framebox(5.0,5.0){$\times$} &\framebox(5.0,5.0){$\cdot$}&\framebox(5.0,5.0){$\times$}&\\
& &\framebox(5.0,5.0){$\times$}&&\framebox(5.0,5.0){$\times$}\\
&&&& \end{array}}

\newcommand{\neigEleven}{\scriptsize\arraycolsep=0.0pt\def\arraystretch{0.0}\begin{array}{@{}l@{}l@{}l@{}l@{}l@{}}
      &\framebox(5.0,5.0){$\times$}&&& \\
     &&\framebox(5.0,5.0){$\times$}&& \\
&\framebox(5.0,5.0){$\times$} &\framebox(5.0,5.0){$\cdot$}&\framebox(5.0,5.0){$\times$}&\\
& &\framebox(5.0,5.0){$\times$}&&\\
&&&\framebox(5.0,5.0){$\times$}& \end{array}}

\usepackage{xr}
\begin{document}
\begin{center}
{\noindent{\Large \textbf{Prior specification of neighbourhood
 and interaction\\[-0.1cm] structure 
in binary Markov random fields} \vspace{0.6cm} \\} 
%{\noindent{\LARGE \textbf{Fully Bayesian prior specification of
%      dependency structure in stationary binary Markov random fields} \vspace{1cm} \\} 
{\large \textsc{Petter Arnesen}} and {\large \textsc{H\aa kon Tjelmeland}}\\{\it Department of
    Mathematical Sciences, Norwegian University of Science and
    Technology}\let\thefootnote\relax\footnote{Correponding author is
    Petter Arnesen, Department of Mathematical Sciences,
Norwegian University of Science and Technology, Trondheim 7491, Norway. E-mail: \textit{petterar@math.ntnu.no}.}
\vspace{0.4cm}
%\\{\large \textsc{H\aa kon Tjelmeland}}\\{\it Department of
%    Mathematical Sciences, Norwegian University of Science and
%    Technology}\vspace{1cm}
%\let\thefootnote\relax\footnote{Petter Arnesen is PhD student, Department of Mathematical Sciences, 
%Norwegian University of Science and Technology, Trondheim 7491, Norway 
%(E-mail: \textit{petterar@math.ntnu.no}).
%H\aa kon Tjelmeland is Professor, Department of Mathematical Sciences, 
%Norwegian University of Science and Technology, Trondheim 7491, Norway 
%(E-mail: \textit{haakont@math.ntnu.no}).}
}
\end{center}
                
\begin{abstract}
%{   \bf \noindent ABSTRACT:
In this paper we propose a prior distribution for the clique
set and dependence structure of binary Markov random fields (MRFs).
In the formulation we allow
both pairwise and higher order interactions. We construct the prior
by first defining a prior for the neighbourhood system of the MRF,
and conditioned on this we define a prior for the appearance of higher 
order interactions. Finally, for the parameter values we adopt
a prior that allows for parameter values to equal, and in this way 
we reduce the effective number of free parameters. 
To sample from a resulting posterior distribution conditioned on an 
observed scene we construct a reversible
jump Markov chain Monte Carlo (RJMCMC) algorithm.
We circumvent evaluations of the intractable normalising constant of the MRF when 
running this algorithm by adopting a previously defined approximate auxiliary 
variable algorithm. We demonstrate the usefulness of our prior in two 
simulation examples and one real data example.
\end{abstract}
 
\vspace{0.5cm}
\vspace{-0.1cm}

\renewcommand{\baselinestretch}{1.5}   \small\normalsize

\section{Introduction}
                
MRFs are a well used model class in
spatial statistics, see \cite{snell} and \cite{moller2003} for an
introduction. In this paper we consider binary MRFs which is a
subclass of discrete MRF. An MRF $x$ is usually defined by conditioning on a
parameter vector $\phi$, so that we have $p(x|\phi)$. This $\phi$
models interactions between the components  of $x$, and in applications these
interactions are typically limited only to be pairwise, see for instance \cite{Besag1974},
\cite{Besag1986}, \cite{friel2007}, \cite{Everitt2012} and \cite{art145}. 
Mainly three arguments are made for constraining
these models to include only pairwise interactions. 
Firstly, the interpretation of the parameters of higher order
interaction can be difficult, for instance assigning values to the parameters
in order to give the MRF some desired property can be a hard
task. Secondly, if one
allows for higher order interactions, the number of parameters grows 
exponentially as a function of the number of variables in a clique. 
And thirdly, because of the intractable normalising constant, binary
and discrete MRFs are a computationally intensive
class of models, especially if parameter inference is required for
large dependency structures with higher order interactions. 
It has however been shown, see 
\citet{DescombesEtAl} and \citet{TjelmelandBesag}, that for many data sets higher order
interaction models outperforms pairwise models, for instance when an
MRF is used as a prior distribution when recovering a noisy image. In
those papers
reasonable interpretation of higher order interactions are in fact
given, and computations are carried out in reasonable time. In both
the above papers, the number of parameters are reduced by manually
setting the same value to groups of parameters. Only the second of the
two papers are however performing parameter inference, which is made
possible by approximating the maximal likelihood estimator obtained by applying
a Markov chain Monte Carlo (MCMC) 
technique to estimate the normalising constant \citep{geyer1992}.  
Other, and more recent approximation techniques to circumvent the
computational problems associated with discrete MRFs have also been proposed, 
see for instance \cite{phd6},
\cite{Everitt2012} and \cite{art145}. Note however that \cite{phd6} is the
only work of the three just mentioned which performs parameter
inference for MRFs with higher order cliques. We discuss options for handling the
normalising constant later in the paper.    
 
When assigning a prior to the interaction structure of an MRF this can 
be done on three different levels. First one can assume the 
neighbourhoods and the parametric form of the MRF to be given, and assign a prior
distribution for the values of the parameters only. This is for instance done in
\citet{art109} and \citet{ryden1998}. A second level of prior formulation is
adopted in \cite{arnesen2013}. The neighbourhoods are still fixed, 
but here priors are formulated for both the parametric form of the MRF and 
for the associated parameter values. An interpretational parametrisation inspired by
\citet{TjelmelandBesag} is used, and to reduce the effective number of free
parameters without reducing the flexibility of the MRF one adopts a prior
where groups of parameters are allowed to have exactly the same value.  
In this article we go to the third level of prior formulation
by assigning a prior also for the neighbourhood system of the MRF,
and to our knowledge this is the first attempt to learn the neighbourhood
system of an MRF from observed data by such a fully Bayesian approach.
As in \citet{arnesen2013} we keep the prior property that
groups of parameters are allowed to be assigned the same parameter value, but we
need to adopt a new parametrisation of the MRF. To obtain posterior
samples we develop an RJMCMC sampling algorithm
\citep{green1995}, and to cope with the intractable normalising
constant we adopt the approximate exchange algorithm
\citep{pro20} of \citet{Everitt2012}.   

%Taking the three issues with higher order MRFs into account,
%\cite{arnesen2013} propose a prior distribution for the parametric structure
%in such models. Firstly an interpretational parametrisation inspired by
%\citet{TjelmelandBesag} is adopted. Secondly their prior distribution do not pu%t any absolute
%restrictions on the parameter values, but allows for groups of
%parameters to have exactly the same value in order to reduce the
%number of parameters. And thirdly, the approximation
%technique presented in \cite{phd6} is used to jointly sample both the
%number of parameters and the parameter values from an approximation of
%the posterior
%distribution.

% Such an approach
%is taken in \cite{arnesen2013}, and in this paper we address the
%third option, assigning a prior distribution also to the neighbourhood
%of the MRF.  

%A limitation of \citet{arnesen2013} is the need to set the size of the
%cliques in the MRF prior to sampling. In this paper we address this
%issue by assigning a prior distribution also to the appearance of
%different cliques in the MRF, and this is to our knowledge the first attempt to% do
%so for these type of models. We still keep the prior property that
%groups of parameters are allowed to be assigned the same parameter value, but w%e
%need to adopt a new parametrisation of MRFs. To obtain posterior
%samples we develop a RJMCMC sampling algorithm
%\citep{green1995}, and to cope with the intractable normalising
%constant we adopt in this paper the approximate exchange algorithm
%\citep{pro20} of \citet{Everitt2012}.   

Assigning a prior distribution to the
dependence structure of $x$ is a much used approach in the
theory of graphical modelling for categorical data, see
\cite{spiegelhalter1993}, \cite{Madigan1995}, \cite{Lauritzen1996},
\cite{Forster1999} and \cite{massam2009}, but the problems
considered there are often small enough so that the normalising
constant is easily calculated. Also, the variables in such problems
often represent features of different nature, whereas we assume translational
invariance of $x$ similar to all the other work on MRFs mentioned
above. For graphical models this lack of invariance
is compensated for by multiple observations of $x$, whereas we consider
situations where only one scene $x$ is observed. For more details regarding
the link between these two type of models the reader is 
referred to \cite{arnesen2013}.   

The paper is organised as follows. In the next section, Section
\ref{sec:preliminary}, we introduce notation and define some properties
for the nodes in a rectangular lattice, while in Section
\ref{sec:representation} we assign binary random variables to these nodes and
discuss two possible parametrisation of their joint probability
distribution. In Section \ref{sec:binaryMRF} we let this probability
distribution be a stationary MRF, and using both parametrisations we
construct in Section \ref{sec:prior} a prior distribution for the dependence 
structure and the parameter values
of an MRF. In Section \ref{sec:MCMC} we
present the proposal distributions we use when sampling from the
posterior distribution along with a discussion on how to handle the
normalising constant. We investigate three examples in Section
\ref{sec:examples}, and lastly we give some closing remarks in Section \ref{sec:remarks}.

\section{Preliminaries}
\label{sec:preliminary}

In this section we consider a rectangular lattice with periodic boundary
conditions and introduce the notation we need in later sections to 
define a prior for stationary MRFs on such a lattice.

Consider a rectangular $m\times n$ lattice and let $(i,j)$, where
$i\in\{ 0,1,\ldots,m-1\}$ and $j\in\{ 0,1,\ldots,n-1\}$,
denote an arbitrary node in this lattice. Here we adopt the 
convention that $i$ and $j$ specify the vertical and horizontal
positions of the node in the lattice, respectively,
that $i=0$ is at the 
top of the lattice and $i=m-1$ at the bottom, and that 
$j=0$ and $j=n-1$ are at the left and right ends of the lattice,
respectively. To denote sets of nodes we use lower case Greek letters, 
and in particular we denote the set of all 
nodes in the lattice by
$\chi=\{(i,j);i=0,...,m-1,j=0,...,n-1\}$ and
use $\lambda$ and $\lambda^\star$ to denote arbitrary subsets
of $\chi$, i.e. $\lambda,\lambda^\star\subseteq \chi$. 
In much of what follows we assume the lattice to have
{\it torus boundary conditions}.
\begin{definition}\label{def:torus}
If, for a rectangular lattice
$\chi = \{ (i,j);i=0,\ldots,m-1,
j=0,\ldots,n-1\}$, the translation of a node
$(i,j)\in \chi$ with an amount $(t,u)\in \chi$
is defined to be
\begin{equation*}
(i,j) \oplus (t,u) = (i+t \mod m, j+u \mod n),
\end{equation*}
we say that the lattice has torus boundary conditions.
\end{definition}
The translation of each node in a set $\lambda\subseteq \chi$ by an
amount $(t,u)\in \chi$ we denote by 
$\lambda \oplus (t,u) = \{ (i,j)\oplus (t,u) ; (i,j)\in \lambda\}$.

To denote sets of subsets of $\chi$ we use upper case Greek letters, 
and at the next level we denote sets of sets of subsets of $\chi$ by
upper case Roman letters. In particular we let
$\Omega(\chi) = \{\lambda; \lambda\subseteq \chi\}$
denote the set of all subsets of $\chi$
and use $\Lambda,\Lambda^\star\subseteq\Omega(\chi)$ to denote arbitrary 
subsets of $\Omega(\chi)$. The $\Omega(\chi)$ is
often called the power set of $\chi$, and one should note that in 
particular it includes the empty set and $\chi$ itself.
We let $L$ denote the partition
of $\Omega(\chi)$ we get by identifying all subsets of $\chi$ that 
are translations of each other.
\begin{definition}
\label{def:partition}
Let $L$ be the partition of 
$\Omega(\chi)$ where, for any two distinct subsets of nodes $\lambda,\lambda^\star\subseteq \chi$, 
there exists a $\Lambda\in L$ so that
 $\{\lambda,\lambda^\star\} \subseteq
\Lambda$ if and only if 
there exists a $(t,u)\in \chi$ so that
$\lambda^\star=\lambda\oplus (t,u)$.
\end{definition} 
The elements in the partition we call {\it cells}. One can note that two of the cells in 
$L$ consist of only one element
each, namely $\{\emptyset\}$ and $\{ \chi\}$, whereas all other
cells have $mn$ elements. One example of a cell with $mn$ elements is 
$\{\{(i,j)\};(i,j)\in \chi\}$, and another example is
$\{\{(i,j),(i,j)\oplus (1,0)\};(i,j)\in \chi\}$. One can also note 
that all elements in a cell $\Lambda\in L$ contains
the same number of nodes, and we denote this number by
$\tau(\Lambda)$. For instance we have $\tau(\{\{(i,j)\};(i,j)\in \chi\})=1$
and $\tau(\{\{(i,j),(i,j)\oplus (1,0)\};(i,j)\in \chi\})=2$. 

As all 
elements in a cell $\Lambda\in L$ are translations of 
each other we may specify a cell by specifying the relative positions
of the nodes in the elements in that cell. 
This suggests a more intuitive notation
for the cells, where one represents each node in an element of a cell by a box and
arranges the boxes according to the relative positions of these nodes.
For instance
we get $\ABBB=\{\{(i,j)\};(i,j)\in \chi\}$, 
$\ABAB=\{\{(i,j),(i,j)\oplus
(1,0)\};(i,j)\in \chi\}$, and $\BAAA=\{\{(i,j),(i,j)\oplus
(1,-1),(i,j)\oplus (1,0) \};(i,j)\in \chi\}$.
In the following we use this 
box representation to refer to specific cells in $L$,
and use the $\Lambda\in L$ notation whenever we need to 
refer to a generic cell.

For two distinct cells $\Lambda,\Lambda^\star\in L$ we let $N(\Lambda,\Lambda^\star)$
denote the number of elements 
$\lambda^\star\in\Lambda^\star$ that is a strict subset of an (arbitrary) 
element $\lambda\in\Lambda$, i.e. 
\begin{equation*}
N(\Lambda,\Lambda^\star) = \card( \{ \lambda^\star\in\Lambda;\lambda^\star
\subset \lambda \mbox{~for some~}\lambda\in\Lambda\}),
\end{equation*}
where $\card(A)$ denotes the cardinality, or the number of elements, in the set $A$.
That this number is the same for all 
$\lambda\in\Lambda$ follows from the fact that $\lambda\in\Lambda \Leftrightarrow
\lambda\oplus (t,u)\in \Lambda$ for any $(t,u)\in \chi$. 
Using the box
representation of the cells in $L$ it is easy to see
what $N(\Lambda,\Lambda^\star)$ becomes. We have for 
instance $N(\ABAB,\ABBB)=2$, $N(\BAAA,\ABBB)=3$,
$N(\BAAA,\ABAB)=1$ and $N(\ABBB,\ABAB)=0$.
%Next we define the function $N(\lambda,\Lambda^\star)=
%|\{\lambda^\star\in \Lambda^\star ; \lambda^\star\subset \lambda\}|$, 
%as the number of elements from $\Lambda^\star$ that is an element of 
%$\lambda \in \Lambda$. By translation by some amount $(t,u)\in S$ 
%of $\lambda$ and all elements of $\Lambda^\star$, 
%$\Lambda^\star\oplus(t,u)=\{\lambda^\star\oplus(t,u);
%\lambda^\star\in \Lambda^\star\}$ we may 
%write $N(\lambda,\Lambda^\star)=N(\lambda\oplus(t,u),\Lambda^\star\oplus (t,u))=
%|\{\lambda^\star\in \Lambda^\star \oplus (t,u);
%\lambda^\star\subset \lambda \oplus (t,u)\}|$, and since we 
%obviously have $\Lambda^\star\oplus (t,u)=\Lambda^\star$ 
%from Definition \ref{def:partition} we get $N(\lambda,\Lambda^\star)=
%N(\lambda \oplus (t,u),\Lambda^\star)\equiv N(\Lambda,\Lambda^\star)$, 
%as the function $N(\lambda,\Lambda^\star)$ is invariante for all 
%$\lambda\in \Lambda$. 
We use whether or not some elements in a cell $\Lambda^\star\in L$ 
is a subset of an element in another (arbitrary) cell in $\Lambda\in L$ 
to define a partial ordering of the cells in $L$.
\begin{definition}
\label{def:partial}
For any $\Lambda,\Lambda^\star\in L$ we define
\begin{equation*}
\Lambda^\star \prec \Lambda \Leftrightarrow N(\Lambda,\Lambda^\star)>0.
\end{equation*}
\end{definition}
For instance, we have $\ABBB\ \prec \ABAB\ $, $\ABBB\ \prec\BAAA\ $ and $\ABAB\ \prec\BAAA\ $, 
but $\AABB\not\prec \ABAB$ and
$\ABAB\ \not\prec\ABBB\ $. Clearly we also have $\{\emptyset\}\prec\Lambda$ for any
$\Lambda\in L\setminus\{ \emptyset\}$.

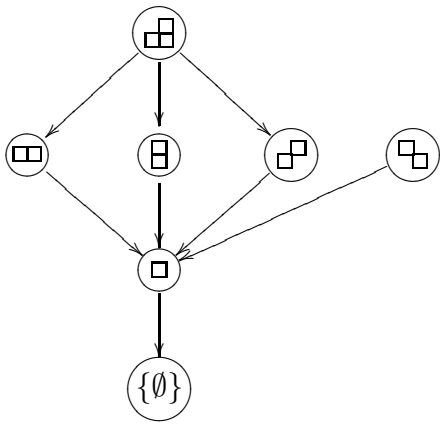
\begin{figure}
  \centering
\centerline{
\xymatrix{
  \\
& *++[o][F]{\BAAA}\ar[d]\ar[dl]\ar[dr]& &\\
*++[o][F]{\AABB}\ar[dr]& *++[o][F]{\ABAB}\ar[d]&*++[o][F]{\BAAB}\ar[dl] &*++[o][F]{\ABBA}\ar[dll]\\
&*++[o][F]{\ABBB}\ar[d]&&\\
&*++[o][F]{\{\emptyset\}}&&
  }}

\vspace*{0.5cm}

  \caption{DAG representation of a dense set $M\subset L$.}
  \label{fig:exampleDAG}
\end{figure}

In Section \ref{sec:binaryMRF} we consider MRFs defined on the lattice $\chi$
and use a subset of the cells in $L$ to specify the 
set of cliques for the MRF. We denote such a subset by 
$M\subseteq L$, and use the term {\it clique types} for the elements 
in $M$. The cliques of the MRF will thereby be all $\lambda\in \Lambda$ for all
$\Lambda \in M$, and will say that $\lambda\in \Lambda$ is of clique type
$\Lambda$ for $\Lambda \in M$. We also say that a 
$\Lambda\in L$ is {\em on} if $\Lambda\in M$, and otherwise %
$\Lambda$ is {\em off}. We also define a prior for $M$ and 
then restrict the attention to sets $M\subseteq L$
that are {\it dense}.
\begin{definition}
\label{def:dense}
A set of cells $M\subseteq L$ is said to be dense
if $\{ \Lambda^\star\in L; \Lambda^\star\prec\Lambda\}\subseteq M$
for all $\Lambda \in M$.
\end{definition} 
We visualise the elements in a dense set $M\subseteq L$ as
a directed acyclic graph (DAG), where the DAG has one vertex for each 
cell $\Lambda\in M$ and an edge from the vertex 
representing a $\Lambda^\star\in M$ to the vertex representing 
a $\Lambda\in M$ whenever $\Lambda^\star\prec\Lambda$ and 
$\tau(\Lambda^\star) = \tau(\Lambda) + 1$. For instance, 
Figure \ref{fig:exampleDAG}
shows such a DAG representation for the dense set of cells
$\{\{\emptyset\},\ABBB\ ,\AABB\ ,\ABAB\ ,\BAAB\ ,\ABBA\ ,\BAAA\ \}$.
In the next section we present two parametrisations of binary stationary
distributions defined on $\chi$.

\section{Binary stationary distributions on a torus}
\label{sec:representation}

Consider a rectangular lattice $\chi=\{(i,j);i=0,\ldots,m-1,j=0,\ldots,n-1\}$ with torus 
boundary conditions as in Section
\ref{sec:preliminary}.  Now, to each node $(i,j)\in \chi$ we
associate a binary variable $x_{i,j}\in \{0,1\}$, and let 
$x=(x_{i,j};(i,j)\in \chi)$ denote a vector of all these binary variables. Associating the value
$0$ with white and the value $1$ with black we can think of an 
$x$ as a colouring of the lattice $\chi$. A colouring where the nodes in a set $\lambda\subseteq \chi$ are 
black and all other nodes are white we denote by $\bold 1^\lambda$, i.e. 
$\bold 1^\lambda = ( x_{i,j}=I( (i,j)\in \lambda); (i,j)\in \chi)$,
where $I(\cdot )$ is the indicator function that equals unity whenever the argument is true and zero 
otherwise.

We denote a distribution for $x$ by $p(x)$ and assume the positivity 
condition $p(x)>0$ for all $x\in \{ 0,1\}^{mn}$.
Thus, we can write
\begin{equation}
\label{eq:binaryDist}
p(x)=Z\exp\{U(x)\},
\end{equation}
where $Z$ is a (typically computationally intractable) normalising constant and
$-U(x)$ is often referred to as the energy function of $x$. In the following our focus is
on {\it stationary} distributions $p(x)$. 
\begin{definition}\label{def:stationarity}
A random vector $x$ defined on a rectangular lattice
$\chi$ with torus boundary
conditions is said to be stationary if and only if
$p(\bold 1^\lambda) = p(\bold 1^{\lambda \oplus (t,u)})$ for all
$\lambda\in \Omega(\chi)$ and $(t,u)\in \chi$.
\end{definition}
In the remainder of this section we consider two representations of $U(x)$ and 
in particular discuss the consequences of imposing the stationarity assumption.
In Section \ref{sec:binaryMRF} we use each of these two representations as a 
basis to define a corresponding parametrisation for stationary binary MRFs.
The first representation is defined using
pseudo-Boolean function theory, while the second representation is defined
directly on $U(x)$. 
%Combining these two
%representations will in Section \ref{sec:prior} allow us to define a reasonable prior distribution for both the
%dependence structure and the parameter values of an MRF.

Following \cite{tjelmeland2012}, we note that
$U(x)$ is a pseudo-Boolean function and
thereby can be expressed as
\begin{equation}
\label{eq:pseudoParam}
U(x)=\sum_{\lambda \in
    \Omega(\chi)}\beta^{\lambda}\prod_{(i,j)\in \lambda}x_{i,j}.
\end{equation}
We refer to the $\beta$ parameters as interaction
parameters. More
details on pseudo-Boolean functions and their properties can be found
in \cite{Hammer1992} and \cite{Grabisch2000}. 
%By introducing for instance the constraint $\beta^\emptyset=0$, the linear independence of the products
%of $x$ makes this an identifiable parametrisation of the probability
%distribution $p(x)$.  
Using Definition
\ref{def:stationarity} above, we state a theorem that identifies what 
restrictions the set $\{\beta^\lambda;\lambda \in \Omega(\chi)\}$ must fulfil for $p(x)$ to be stationary. 
\begin{theorem}\label{th:betaStationarity}
A random vector $x$ defined on a rectangular lattice
$\chi=\{ (i,j);i=0,\ldots,m-1,j=0,\ldots,n-1\}$ with 
torus boundary conditions is stationary if and only 
if $\beta^{\lambda}=\beta^{\lambda\oplus (t,u)}$ for all 
$\lambda \in \Omega(\chi)$ and $(t,u)\in \chi$. We then say
that $\beta^\lambda$ is translational invariant.
\end{theorem}
\stepcounter{proof1}
The result in this theorem is a special case of Theorem 2 of \cite{arnesen2013}. 
The theorem above states that two interaction parameters
$\beta^{\lambda}$ and $\beta^{\lambda^\star}$ must have the same value whenever
$\lambda$ and $\lambda^\star$ belong to the same cell $\Lambda\in L$
in the partition of $\Omega(\chi)$ defined in Section \ref{sec:preliminary}.
Thus, for $x$ to be stationary $\beta^\lambda$ must take the same value for all $\lambda\in\Lambda$
for each $\Lambda\in L$, and we denote the common value of
$\beta^\lambda$, for all $\lambda\in\Lambda$, 
by $\beta^\Lambda$. Using the box notation introduced above we then in particular have that 
$\beta^{\ABBB}$, $\beta^{\ABAB}$ and $\beta^{\BAAA}$ are the common values for all 
$\beta^\lambda$ for $\lambda\in\ABBB$, $\lambda\in\ABAB$ and $\lambda\in\BAAA$, 
respectively.

%In Section \ref{sec:prior} we will construct a prior distribution for the parameters presented
%in this section. However, the interpretation of the
%$\beta$ parameters is rather complex, making it difficult to specify a
%reasonable prior for these. In particular we think that higher order
%interaction parameters apriori should tend to take smaller values
%than lower order interaction parameters. Therefore we present an
%alternative parametrisation where all the parameters are on the same
%scale, and thereafter combining these two parameterisations through a
%one-to-one relation. 

The second representation we consider for $U(x)$ is the quite naive one obtained 
by just representing the value of $U(x)$ for each possible vector $x$. 
For each $\lambda\in\Omega(\chi)$ we introduce a parameter $\phi^\lambda$ and 
write
\begin{equation}
U(\bold 1^\lambda)=\phi^\lambda, \ \lambda\in\Omega(\chi).
\label{eq:phiParam}
\end{equation}
%This parametrisation is of course extremely impractical
%to work with in addition to being grossly overparametrised. We will
%however see in the two next theorems that the stationarity assumption puts
%natural constraints on the $\phi$ parameters such that a one-to-one
%relation with the $\beta$ parameters can be established. 
%%In addition we will see next that assuming $\beta$ parameters to
%%be equal to zero, or correspondingly setting  $\mathcal{C}=\mathcal{C}_L$ for some
%%$L\subset \mathcal{L}$, introduce further constrains so that the $\phi$
%%parametrisation becomes a tractable alternative. 
The stationarity assumption puts similar restrictions on the $\phi$ parameters
as it did for the $\beta$ parameters.
\begin{theorem}
A random vector $x$ defined on a rectangular lattice
$\chi=\{ (i,j);i=0,\ldots,m-1,j=0,\ldots,n-1\}$ with 
torus boundary conditions is stationary if and only 
if $\phi^\lambda=\phi^{\lambda \oplus (t,u)}$ for all 
$\lambda \in \Omega(\chi)$  and $(t,u)\in \chi$. We then say
that $\phi^\lambda$ is translational invariant.
\end{theorem}
\begin{proof1}
Combining (\ref{eq:binaryDist}) and (\ref{eq:phiParam}) we get
\begin{equation*}
p({\bold 1^\lambda}) = Z \exp\left\{ \phi^\lambda\right\},
\end{equation*}
for all $\lambda\in\Omega(\chi)$. Thus we have that
\begin{equation*}
\phi^\lambda=\phi^{\lambda\oplus (t,u)} \Leftrightarrow
p(\bold 1^\lambda) = p(\bold 1^{\lambda\oplus (t,u)})
\end{equation*}
for any $\lambda\in\Omega(\chi)$ and $(t,u)\in \chi$, and the result follows.
%If we assume stationarity we have from Definition
%\ref{def:stationarity} that $p(\bold 1^\lambda)=p(\bold 1^{\lambda\oplus (t,u)})$,
%and translation invariance for $\phi^{\bold 1^\lambda}$ follows directly
%from $p(\bold 1^\lambda)=\exp(U(\bold 1^\lambda))$. If we assume $\phi^{\bold
%  1^\lambda}$ to be translation invariant, stationarity follows directly
%from the same.  
\end{proof1}
Corresponding to what we have for the $\beta$ parameters, we thus have that
for $x$ to be stationary we must have that $\phi^\lambda$ must take the same 
value for all $\lambda\in\Lambda$ for each $\Lambda\in L$.
We therefore also introduce a notation corresponding to what we did for the 
$\beta$ parameters, and denote the 
common value of $\phi^\lambda$ for all $\lambda\in\Lambda$ by $\phi^\Lambda$.
In particular we then have that $\phi^{\ABBB}$ is the common value for 
all $\lambda\in\ABBB$ and correspondingly $\phi^{\ABAB}$
and $\phi^{\BAAA}$ for all $\lambda \in \ABAB$ and all $\lambda\in \BAAA$, respectively. It should be noted that even though we use a
similar notation for the $\beta$ and $\phi$ parameters, the 
correspondence is not as strong as our notation may suggest.
For a given colouring $x$ of the nodes in $\chi$, the associated 
$\phi$ parameter is a function of all the components of $x$, 
whereas the associated $\beta$ parameter is only a function of
a subset of the nodes.
%In the following we refer to all colourings $x$ that is associated 
%with the same $\phi$ parameter, as a configuration set, denoted 
%by $\bold 1 (\Lambda) = \{ 1^\lambda; \lambda\in\Lambda\}$ for
%$\Lambda\in L$, and the $\phi^\Lambda$ parameters as configuration set
%parameters.
The next theorem states a one-to-one correspondence between
$\{\beta^\Lambda;\Lambda\in L\}$ and
$\{\phi^{\Lambda};\Lambda\in L\}$.
\begin{theorem}\label{th:oneToOne}
Consider a stationary random vector $x$ defined on a rectangular lattice
$\chi=\{ (i,j);i=0,\ldots,m-1,j=0,\ldots,n-1\}$ with torus boundary
conditions. Then there is a one-to-one
correspondence between $\{\beta^\Lambda;\Lambda\in L\}$ and
$\{\phi^{\Lambda};\Lambda\in L\}$. 
\end{theorem}
\begin{proof1}
From \eqref{eq:pseudoParam} and \eqref{eq:phiParam} we have that
\begin{equation}\label{eq:eq1}
\phi^\lambda=\sum_{\lambda^\star\in \Omega(\chi)}\beta^{\lambda^\star}\prod_{(i,j)\in
  \lambda^\star}\bold 1^{\lambda^\star}_{i,j}=\beta^\lambda + \sum_{\lambda^\star\subset
  \lambda}\beta^{\lambda^\star} \text{ for all }  \lambda\in \Omega(\chi).
\end{equation}
By using Moebious inversion \citep{Lauritzen1996} it follows from this that
\begin{equation}\label{eq:eq2}
\beta^\lambda=\phi^\lambda-\sum_{\lambda^\star\subset
  \lambda}(-1)^{\card(\lambda)-\card(\lambda^\star)-1}\phi^{\lambda^\star}.
\end{equation}
If we invoke the translation invariance for both representations
we get from \eqref{eq:eq1} that
\begin{equation}\label{eq:phiFrombeta}
\phi^\Lambda=\beta^\Lambda + \sum_{\Lambda^\star\prec \Lambda}N(\Lambda,\Lambda^\star) \beta^{\Lambda^\star}.
\end{equation}
We see that we can use this expression to compute all
$\{\phi^\Lambda;\Lambda\in L\}$ from
$\{\beta^\Lambda;\Lambda\in L\}$ recursively. Firstly we can compute the
$\phi^\Lambda$ for which $\tau(\Lambda)=0$,
i.e. $\phi^{\{\emptyset\}}=\beta^{\{\emptyset\}}$. Next we
can calculate the $\phi^\Lambda$ for
which $\tau(\Lambda)=1$, thereafter all $\phi^\Lambda$ for which
$\tau(\Lambda)=2$, and so forth until $\phi^\Lambda$ have been calculated
for all $\Lambda\in L$. Thus, $\{\phi^\Lambda;\Lambda\in
L\}$ is uniquely specified by $\{\beta^\Lambda;\Lambda\in
L\}$.

If we again invoke the translation invariance for both
representations we get from \eqref{eq:eq2}
\begin{equation}\label{eq:betaFromPhi}
\beta^\Lambda=\phi^\Lambda -\sum_{\Lambda^\star \prec
  \Lambda}
(-1)^{\card(\Lambda)-\card(\Lambda^\star)-1}N(\Lambda,\Lambda^\star)\phi^{\Lambda^\star},
\end{equation}
where we correspondingly get that $\{\beta^\Lambda;\Lambda\in
L\}$ can be recursively computed from $\{\phi^\Lambda;\Lambda\in
L\}$. Again one must first compute $\beta^\Lambda$ for 
$\tau(\Lambda)=0$, i.e. $\beta^{\{\emptyset\}}$, then $\beta^\Lambda$ for which
$\tau(\Lambda)=1$, thereafter all $\beta^\Lambda$
for which $\tau(\Lambda)=2$,
and so forth. Thereby $\{\beta^\Lambda;\Lambda\in
L\}$ is also uniquely specified by $\{\phi^\Lambda;\Lambda\in
L\}$ and the proof is complete.
\end{proof1}

In the next section we present stationary binary MRFs, and use 
the two representations introduced above to establish a parametrisation of MRFs 
that is suitable for a fully Bayesian modelling.

\section{Binary MRFs}\label{sec:binaryMRF}
In this section we first give a brief introduction to binary MRFs
\citep{cressie1993,moller2003}. Thereafter we assume the MRF to 
be stationary and defined on a lattice with torus 
boundary conditions, and discuss what 
consequences the Markov assumption then has for the $\beta$ and 
$\phi$ representations introduced above. Finally we 
use each of the two representations to define corresponding 
parametrisations of an MRF.

Consider a rectangular lattice $\chi=\{(i,j);i=0,\ldots,m-1,j=0,\ldots,n-1\}$ as 
in Section \ref{sec:preliminary} and a corresponding random vector 
$x=(x_{i,j};(i,j)\in \chi)$ where $x_{i,j}\in\{0,1\}$.
In addition to the notation introduced above we let, for
$\lambda \subseteq \chi$, $x_{\lambda}=(x_{i,j};(i,j)\in \lambda)$
denote the collection of random variables associated with the nodes in $\lambda$,
and let $x_{-(i,j)}=x_{\chi\setminus \{(i,j)\}}$. A binary MRF is defined with 
respect to a so-called neighbourhood system 
$\Psi=\{\psi_{(i,j)};(i,j)\in \chi\}$ where
$\psi_{(i,j)}\subseteq \chi\setminus\{ (i,j)\}$ and
$(i,j)\in\psi_{(t,u)}\Leftrightarrow (t,u)\in\psi_{(i,j)}$. The 
random vector $x$ is then said to be an MRF with respect to $\Psi$ if 
$p(x)>0$ for all $x\in\{0,1\}^{mn}$ and it fulfils the Markov property
\begin{equation*}
p(x_{i,j}|x_{-(i,j)})=p(x_{i,j}|x_{\psi_{(i,j)}})
\end{equation*}
for all $(i,j)\in \chi$ and $x\in\{0,1\}^{nm}$. Given a neighbourhood system
$\Psi$, a set of nodes $\lambda\subseteq \chi$ is said to be a clique
if $(i,j)\in \psi_{(t,u)}$ for all distinct pair of nodes
$(i,j),(t,u)\in \lambda$. In particular one can note that the empty set
$\emptyset$ is always a clique, and so is $\{(i,j)\}$ for all $(i,j)\in \chi$. 
We denote the set of all cliques
by $\Upsilon$, and let $\Upsilon_{max}$ denote the set of all 
maximal cliques, where a clique is said to be maximal if it is not a 
subset of another clique. Clearly
$\Upsilon\subseteq\Omega(\chi)$, $\Upsilon_{max}\subset\Omega(\chi)$
and $\Upsilon_{max}\subset\Upsilon$.
The Hammersley-Clifford theorem
\citep{Clifford} then states that $x$ is an MRF with respect to $\Psi$
if and only if $U(x)$ can be expressed as
\begin{equation}
U(x)=\sum_{\lambda \in \Upsilon_{max}}
    V_{\lambda}(x_{\lambda}),
\label{eq:cliqueRep}
\end{equation}
where $V_{\lambda}(x_{\lambda})$ is referred to as the
potential function for the maximal clique
$\lambda\in\Upsilon_{max}$. Combining \eqref{eq:pseudoParam} and \eqref{eq:cliqueRep}
it follows that $\beta^\lambda=0$ for all $\lambda\not\in\Upsilon$. Thus, 
$U(x)$ for an MRF is given by 
\begin{equation}
\label{eq:interactionParam}
U(x)=\sum_{\lambda \in
    \Upsilon}\beta^{\lambda}\prod_{(i,j)\in \lambda}x_{i,j},
\end{equation}
see \cite{tjelmeland2012} for a formal proof.

In the following we assume $x$ to be defined on a lattice with torus
boundary conditions, to be stationary, and 
to be an MRF with respect to a neighbourhood system $\Psi$. 
Theorem \ref{th:betaStationarity} then gives that $\beta^\lambda$ must be 
translational invariant, and in particular we must then have 
$\beta^\lambda=0 \Leftrightarrow \beta^{\lambda\oplus(t,u)}=0$ for 
$\lambda\in\Omega(\chi),(t,u)\in \chi$. Without loss of 
generality we can thereby assume also the clique system to be 
translational invariant in that 
\begin{equation}
\label{eq:cliqueTransInvariant}
\lambda\in\Upsilon\Leftrightarrow
\lambda\oplus (t,u)\in\Upsilon \mbox{ for all } \lambda\in\Omega(\chi),
(t,u)\in \chi.
\end{equation}

In turn this implies that the neighbourhoods are 
translational invariant in that $\psi_{(i,j)\oplus (t,u)} = 
\psi_{(i,j)}\oplus (t,u)$ for all $(i,j),(t,u)\in \chi$.
From (\ref{eq:cliqueTransInvariant}) it follows that we can find 
an $M\subseteq L$ so that $\Upsilon=\bigcup_{\Lambda\in M}\Lambda$,
and again using that $\beta^\lambda$ is translational invariant we get that 
(\ref{eq:interactionParam}) can be reformulated as
\begin{equation}
\label{eq:paramModelBeta}
U(x) = \sum_{\Lambda\in M}\left[ \beta^\Lambda \sum_{\lambda\in\Lambda}
\prod_{(i,j)\in \lambda} x_{i,j}\right].
\end{equation}
Thus, a stationary MRF defined on a lattice with torus boundary conditions 
is specified by the set $M$
and the values of the parameters $\{ \beta^\Lambda,\Lambda\in M\}$. 
A natural and frequently used assumption in statistics is to 
allow higher-order interactions in a model only if corresponding
lower-order interactions already are present. In the 
formulation of the MRF discussed above this means that if 
$\Lambda^\star\prec\Lambda$ we can allow $\Lambda\in M$ only if
$\Lambda^\star\in M$, i.e. that $M$ is dense. In the following 
we therefore restrict $M$ to be dense. In addition we restrict,
without loss of generality, that $\{\emptyset\}$ always is contained
in $M$.

Introducing for instance the constraint $\beta^{\{\emptyset\}}=0$, 
the linear independence of the products
of $x$ in (\ref{eq:paramModelBeta}) makes the corresponding 
$p(x)$ an identifiable parametric model. In a fully Bayesian 
model formulation one would then need to put a prior on $M$
and the parameter set $\{\beta^\Lambda,\Lambda\in M\setminus\{\{\emptyset\}\}\}$.
However, the $\beta$ parameters relate to cliques of 
different sizes and this makes the interpretation difficult.
In particular we think the values of $\beta$ parameters associated with
large cliques apriori should be expected to be closer to zero than 
values of parameters associated with smaller cliques, but we are not able to 
make this statement more precise than this.
In the following we find an alternative
parametrisation based on the $\phi$ representation discussed 
above and argue that this makes it easier to formulate a 
reasonable prior.

Given a stationary MRF defined on a lattice with torus boundary conditions
specified by $M$ and $\{\beta^\Lambda;\Lambda\in M\}$ and recalling
that $\beta^\Lambda=0$ for $\Lambda\in L\setminus M$, we can use 
(\ref{eq:phiFrombeta}) to find the $\phi$ representation 
$\{\phi^\Lambda;\Lambda\in L\}$ of the model. In contrast to 
the situation for the $\beta$ parameters, the Markov assumption 
does not induce any particular zero structure for the $\phi$ parameters.
However, recalling that we have assumed $M$ to be dense, 
we easily see from (\ref{eq:phiFrombeta}) and (\ref{eq:betaFromPhi})
that there is a one-to-one correspondence between 
$\{\beta^\Lambda;\Lambda\in M\}$ and $\{\phi^\Lambda;\Lambda\in M\}$. 
An alternative to the $\beta$ parametrisation discussed above is therefore
to parametrise the model by $M$ and $\{\phi^\Lambda;\Lambda\in M\}$. One could 
note that then the remaining $\phi$ parameters, $\{\phi^\Lambda;
\Lambda\in L\setminus M\}$, are given by (\ref{eq:phiFrombeta}) and 
(\ref{eq:betaFromPhi}) as linear functions of $\{\phi^\Lambda;
\Lambda\in M\}$. From the definition of the $\phi$ parameters in 
(\ref{eq:phiParam}) we see that they essentially just specify the probabilities
for different realisations of $x$. If no specific prior information 
is available and suggests otherwise it therefore seems reasonable to
assign a prior to $\{\phi^\Lambda;\Lambda\in M\}$ where the various
parameters are exchangeable.

%For a stationary MRF defined on a lattice with torus boundary
%conditions we now have the parameters $\beta=\{\beta^\Lambda;\Lambda\in
%L\}$. If $L\subset \mathcal{L}$ one can still use
%\eqref{eq:phiFrombeta} to calculate all $\phi^{\bold 1(\Lambda)}$ for $\Lambda\in
%\mathcal{L}$ just by setting $\beta^\Lambda=0$ for all $\Lambda \not
%\in L$. This makes $\phi=\{\phi^{\bold
%  1(\Lambda)};\Lambda\in L\}$ a tractable parameterisations, as the
%parameters $\{\phi^{\bold 1(\Lambda)},\Lambda \not \in L\}$ is reduced
%to be linear functions of $\{\phi^{\bold
%  1(\Lambda)};\Lambda\in L\}$. In particular, using obvious notation,
%we see from \eqref{eq:betaFromPhi} that we can write 
%\begin{equation}\label{eq:link}
%\beta^\Lambda=\phi^{\bold 1(\Lambda)} - \mathcal{D}(\phi,\Lambda),
%\end{equation}
%recalling that we assume all our DAGs to be dense. We may now view the $\phi$
%parameters as a reparametrisation of
%the non-zero interaction parameters, or correspondingly the interaction
%parameters for the included clique types $L$. As already mentioned the $\phi$
%parameters scale with each other and hence is more natural to define
%a common prior distribution to. One of the focus points in the next section is to assign a positive prior probability
%of $\beta^{\Lambda}$ parameters to be zero, or equivalent $\Lambda$ to
%be off, which in the $\phi$ parametrisation
%simply corresponds to assigning positive prior probability to the event 
%\begin{equation}
%\phi^{\bold{1}(\Lambda)}=\mathcal{D}(\phi,\Lambda),
%\label{eq:phiConstrain}
%\end{equation}
%by Equation \eqref{eq:link}.

\section{Prior distribution}\label{sec:prior}

When constructing our prior distribution for the parameters in an MRF we
have three properties in mind. Firstly, we want to assign a positive prior
probability to the event that $\Lambda \in L$ is on or
off. This is naturally done in the $\beta$ parametrisation, but
Theorem \ref{th:oneToOne} still enables us to work with the $\phi$
parameters. Secondly, we want a positive prior probability for the event that
groups of clique types have exactly the same parameter
value, and thirdly we want to assign a prior distribution to the
parameter values of the MRF. As discussed in the last two paragraphs
in the previous section,
% and also in \citet{arnesen2013}, 
the two last aspects are most naturally formulated with the $\phi$ 
parametrisation.
Thus, the main focus will be on the $\phi$ parametrisation for
the reminder of this paper, and in the following we describe how we
define a prior so that all the three properties discussed above
are obtained. 

For the first property discussed above we note that 
setting a $\Lambda\in L$ to be off is equivalent to setting
\begin{equation*}
\phi^\Lambda =\sum_{\Lambda^\star \prec
  \Lambda}
(-1)^{\card(\Lambda)-\card(\Lambda^\star)-1}N(\Lambda,\Lambda^\star)\phi^{\Lambda^\star},
\end{equation*}
by \eqref{eq:betaFromPhi}. 
Recalling that $L$ is required to be dense, we have that if
$\Lambda$ is on, all $\Lambda^\star\in L$
where $\Lambda^\star \prec \Lambda$ must
be on as well. As in the previous section we let $M\subseteq L$ denote 
the set of all clique types in $L$ that is on. Moreover,
for integers $k\geq 0$ we let $M_k=\{\Lambda\in M;\tau(\Lambda)=k\}$ be the
set of all $k$'th order clique types in $M$.
As mentioned above we restrict $M$ always to contain $\{\emptyset\}$, 
so clearly
$M_0=\{\{\emptyset\}\}$ and one must have $M_1=\emptyset$ or $M_1=\{\ABBB\}$.
Moreover, one should note
that $M_2$ is one-to-one with the neighbourhood system $\Psi$,
and because of symmetry in the neighbourhoods
$\card(\psi_{(i,j)})=2\cdot\card(M_2)$ for all $(i,j)\in \chi$.
We assume the prior distribution for $M$, $p(M)$, to have the form
\begin{equation*}
p(M)=p(M_2)p(M_1|M_2)p(\{M_k,k\geq 3\}|M_1,M_2)
\end{equation*}
where we in the following carefully explain the three
factors in this expression in turn. 

\begin{figure}
 \centering
  \includegraphics[scale=0.3]{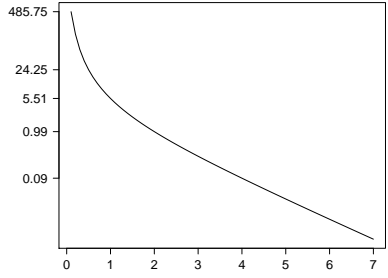} 
\caption{\label{fig:Eeta}For a $100\times 100$ lattice, the expected
number of neighbours to a node, for given value of $\eta$, $2\cdot\mbox{E}[\card(M_2) | \eta]$,
as a function of $\eta$.}
\end{figure}

The $p(M_2)$ is essentially a prior distribution for the neighbourhood system $\Psi$,
and for an arbitrary node $(i,j)\in \chi$ we assume apriori 
the various $(t,u)\in \chi\setminus \{(i,j)\}$ to be a neighbour of $(i,j)$
independently of each other, and with probabilities so that nodes close
to $(i,j)$ have the highest probabilities of being a neighbour. More 
precisely, we define $p(M_2)$ to have the form
\begin{equation*}
p(M_2)= \prod_{\Lambda\in L:\tau(\Lambda)=2}\left [ f_\eta(\Lambda)^{I(\Lambda\in
M_2)}(1-f_\eta(\Lambda))^{I(\Lambda \not \in M_2)}\right
],
\end{equation*}
%or [[alternative]]
%\begin{equation*}
%p(L_2)= \prod_{\Lambda\in
%  L_2}f_\eta(\Lambda)\prod_{\Lambda\not \in L_2}(1-f_\eta(\Lambda)),
%\end{equation*}
where $f_\eta(\Lambda)\in [0,1]$ is the probability for a 
specific node to be a neighbour, and $\eta$ is a parameter controlling
this probability. In this paper we use
\begin{equation*}
f_\eta(\Lambda)=e^{-\eta d(\Lambda)},
\end{equation*}  
where $d(\Lambda)$ is the Euclidean distance between the two nodes in 
(an arbitrary) $\lambda \in \Lambda$ when $\tau(\Lambda)=2$. 
%%%%%%%%%% Vi trenger ikke si det under, det er allerede spesifisert implesitt i ligningene.
%We constrain the possible second order cliques to be included in
%the model by the size of the lattice. 
We consider $\eta$ as random and want to assign a prior distribution to it.
For a given value of $\eta$ we can easily 
compute the expected number of neighbours of a node,
see Figure \ref{fig:Eeta}.
As we can also understand from the definition of $p(M_2)$,
the expected number of neighbours is very high for small values of $\eta$ and 
decreases rapidly as $\eta$ increases. Apriori we do not want high probability for 
the nodes to 
have a lot of neighbours, so the apriori density for $\eta$ should be 
small for values of $\eta$ close to zero, but otherwise we want the
prior to be vague. We adopt a gamma prior for $\eta$ with mean 
$3$ and standard deviation $\sqrt{3}\approx 1.73$. Close to 
zero this prior density is proportional with $\eta^2$, and our
experience is that this is sufficiently small to avoid a very large 
number of neighbours when simulating from a posterior distribution.
%%%%%ASK!!!!
%[[by Monte Carlo
%integration of $\eta$ it is easy to find that the expected
%number of second order clique types is approximately equal to 2.48 for
%this choice of hyper-parameters.]]

When defining the factor $p(M_1|M_2)$ one should recall that one has only two 
possibilities for $M_1$, either $M_1=\emptyset$ or $M_1=\{\ABBB\}$. Moreover,
as we have assumed $M$ to be dense one must have $M_1=\{\ABBB\}$ if
$\card(M_2)>0$. Thereby we only need to specify prior probabilities 
$p(M_1=\emptyset| M_2=\emptyset)$ and 
$p(M_1=\{\ABBB\}|M_2=\emptyset)$. We assume each of these probabilities to
be a half, and get
\begin{equation*}
p(M_1|M_2) = \left\{\begin{array}{ll}
\frac{1}{2^{I(M_2=\emptyset)}} & \mbox{~~if $M_1=\{\ABBB\}$ or $M_2=\emptyset$},\\
0 & \mbox{~~otherwise.}\\\end{array}\right.
\end{equation*}

\begin{figure}
  \centering
\centerline{
\xymatrix{
  \\
&*++[o][F]{\AAAA}\ar[dr]\ar[drr]\ar[d]\ar[dl]&&\\
 *++[o][F]{\ABAA}\ar[dr]\ar[drr]\ar[d] & *++[o][F]{\AABA}\ar[d]\ar[dl]\ar[dr]&*++[o][F]{\BAAA}\ar[dl]\ar[d]\ar[dr] & *++[o][F]{\AAAB}\ar[d]\ar[dl]\ar[dll]\\
*++[o][F]{\ABBA}\ar[dr] & *++[o][F]{\ABAB}\ar[d]&*++[o][F]{\AABB}\ar[dl]&*++[o][F]{\BAAB}\ar[dll]\\
&*++[o][F]{\ABBB}\ar[d]&&\\
&*++[o][F]{\{\emptyset\}}&&
  }}

\vspace*{0.5cm}

  \caption[]{Maximal possible DAG when $M_2=\{\ABBA,
    \ABAB, \AABB,\BAAB\}$.}
  \label{fig:exampleDAGmax}
\end{figure}
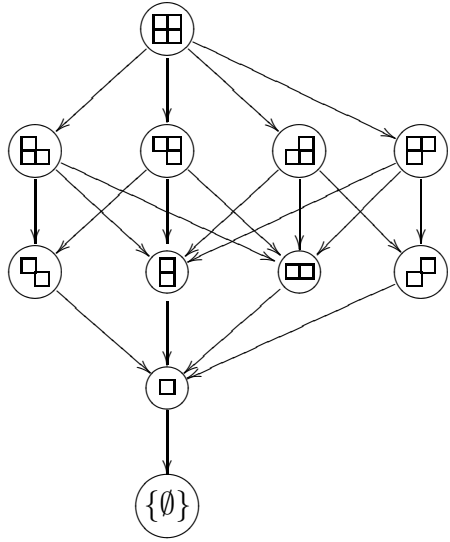

The conditional prior distribution $p(\{M_k,k\geq 3\}|M_1,M_2)$ is the probability
for including clique types of orders larger than two, given $M_1$ and $M_2$
and under the restriction that $M$ should be
dense. In particular we construct the rest of $M$ by adding
sequentially the possible interactions of orders $3, 4$ and so forth, each with a probability
$p_\star$. Given the second order interactions $M_2$ we first 
independently add the possible third order interactions 
with the probability $p_\star$. Then, given the set of third order interactions $M_3$
we independently add the possible forth order interactions with 
the same probability $p_\star$, and so forth. For
instance, there are four possible third order interactions to
add when the second order interactions are as given in Figure
\ref{fig:exampleDAG}, namely $\BAAA$ ,$\ABAA$, $\AABA$, and $\AAAB$.
If, in this example, all of the third order cliques are included it is also possible to
  include the forth order clique $\AAAA$, which gives us the maximal possible DAG for
  the $M_2$ in this example, see the illustration in
  Figure \ref{fig:exampleDAGmax}. 
For the $M$ illustrated in 
  Figure \ref{fig:exampleDAG}, we see that only one of the third
  order cliques is included in $M$, and then clearly no forth order
  clique can be included without violating the restriction that $M$ 
should be dense. 
Defining $\upsilon_k(M_{k-1}) = \{ \Lambda\in L; \tau(\Lambda)=k \mbox{ and } 
\Lambda^\star\prec \Lambda \mbox{ for all } \Lambda^\star\in M_{k-1}\}$ 
to be the set of all clique types of order $k$
that is possible to include in $M$ given the clique types of order
$k-1$ we get that
\begin{equation*}
p(\{M_k,k\geq 3\}|M_1,M_2)  = \prod_{k\geq 3} \left\{ \prod_{\Lambda\in \upsilon_k(M_{k-1})}
\left[ p_\star^{I(\Lambda\in M_k)} (1-p_\star)^{I(\Lambda\not\in M_k)}\right]\right\}
\end{equation*}
when $M$ is dense. As a prior for $p_{\star}$ we assing a uniform
distribution on the interval $(0,1)$.

Having defined the prior for $M$ as above, the next step in the prior specification 
is to define a prior that specify what sets of clique types that should have 
the same parameter value. 
%When the set $M$ is given, each element $\Lambda\in M$
%has an associated configuration set ${\bold 1}(\Lambda)$. 
We start by defining  a partition of $M$, denoted $\mathscr{S}$, and 
use $S\in\mathscr{S}$ to denote a cell in that partition. We restrict  
%These cells will in the following
%be referd to as groups, 
all clique types in a cell $S$ to have the same parameter value, $\varphi_{S}$ say.
By setting $z=\{(S,\varphi_{S});S\in\mathscr{S}\}$ we thus have
\begin{equation*}
\phi^{\Lambda}=\sum_{(S,\varphi_{S})\in
z}\varphi_{S} I(\Lambda\in S)
\end{equation*}
for $\Lambda\in M$, and we denote the vector of model parameters by
$\varphi=(\varphi_{S};S\in \mathscr{S})$. In the following we define a
prior for $\{ \phi^\Lambda,\Lambda\in M\}$ given $M$,
by specifying a prior for $z$ given $M$. We specify this conditional prior 
for $z$ as
\begin{equation*}
p(z|M)=p(\mathscr{S}|M)p(\varphi|\mathscr{S}),
\end{equation*} 
where $p(\mathscr{S}|M)$ is a
distribution over all partitions of $M$,
and $p(\varphi|\mathscr{S})$ is a
distribution for the parameters values given a partition $\mathscr{S}$.

For $p(\mathscr{S}|M)$ we first want to have a distribution that
does not favour partitions with a high number of cells, 
as this would give a corresponding high number of model parameters.
% i.e. we want $p(r|\mathcal{L})$. 
For instance, assigning a
uniform distribution directly on all the possible partitions would make the
marginal probability for approximately $\card(M)/2$ cells much higher than a
state with only one cell. Second, we want the
prior probability for any particular partition with $\card(\mathscr{S})=r$ cells to be 
smaller than any particular partition with $r+1$ cells.
To see how one can construct a prior fulfilling these requirements
one should note that it is possible to count the number of
possible partitions of a set with $\card(M)$ elements, noting that the
definition of a partition require no cells to be empty. This is known
as the Stirling numbers of the second kind \citep{Graham1988}, and is for $\card(\mathscr{S})=r$ given by
 \begin{equation*}
g(\card(M),r)=\frac{1}{r!}\sum_{k=0}^r\binom{r}{k}(-1)^{r-k}k^{\card(M)}.
\end{equation*}
The function $g(\card(M),r)$ is strictly increasing 
as a function of $r$ from $r=1$ to $r=r_{\max}=\arg\underset{r}{\max} \ g(\card(M),r)\approx \card(M)/2$, 
and strictly decreasing from $r=r_{\max}$ to $r=\card(M)$. 
We define
$p(\mathscr{S}|M)$ by a marginal distribution for the number of 
cells in the partition, $p(\card(\mathscr{S})|M)$, and 
given the number of cells in $\mathscr{S}$
we assume a uniform distribution over all partitions with the specified 
number of cells. Thus, we have 
\begin{equation*}
p(\mathscr{S}|M)=\frac{p(\card(\mathscr{S})|M)}{g(\card(M),\card(\mathscr{S}))},
\end{equation*}
and use $p(\card(\mathscr{S})|M)$ to ensure that the prior fulfils the two properties discussed above.
We define
\begin{equation*}
p(\card(\mathscr{S})=r|M) \propto \left\{ \begin{array}{ll} 1 & \mbox{~if $r < r_{\max}$},\\
\frac{g(\card(M),r)}{g(\card(M),r_{\max})2^{r-r_{\max}}} & \mbox{~otherwise},
\end{array}\right.
\end{equation*}
where the factor $2^r$ in the denominator grows faster than $g(\card(M),r)$ as a function of $r$.
One should also note that the expression for $r\geq r_{\max}$ is chosen so that it is equal to $1$
for $r=r_{\max}$.
Finally we specify $p(\varphi|\mathscr{S})$
by assuming the components of $\varphi$ to be independent and normally distributed
with zero mean and with some common variance $\sigma_{\varphi}^2$,
but to make the model identifiable we add the restriction 
\begin{equation*}
\sum_{(S,\varphi_{S})\in z}\varphi_{S}=0.
\end{equation*} 
To complete our fully Bayesian setup we assign a vague gamma
distribution for $\sigma_{\varphi}^2$ with mean $10$ and variance $10^2$.

\section{Posterior simulation}
\label{sec:MCMC}
In this section we briefly present how we simulate from the posterior
distribution
\begin{equation}\label{eq:posterior}
p(z,M,\theta|x)\propto p(x|z,M)p(z|M,\theta)p(M|\theta)
p(\theta),
\end{equation}
where $p(x|z,M)$ is a likelihood MRF for an assumed observed scene $x$,
$p(z|M,\theta)$ and $p(M|\theta)$ are the priors defined in the
previous section, and $p(\theta)$ is the hyper-prior for 
the vector of prior parameters $\theta=(\eta,p_\star,\sigma_\varphi^2)$.
When wanting to simulate from this distribution, two challenges arises. 
Firstly, as also briefly discussed in the introduction, this is
a doubly intractable distribution \citep{pro20}. So in a Metropolis--Hastings
algorithm the normalising constant for the MRF does not cancel in the 
expression for the acceptance probability. Secondly, we need to specify appropriate
proposal distributions for $z$, $M$ and the prior parameter vector 
$\theta=(\eta,p_\star,\sigma_{\varphi}^2)$. In the following two sections we
describe how we tackle these two problems.  

\subsection{Handling the normalising constant}\label{sec:constant}
As in our case a MRF $x$ is usually defined conditioned on a
parameter vector. The posterior
distribution for such models are particularly computer intensive
because the normalising constant becomes intractable as the dimension of $x$
increases, see \eqref{eq:binaryDist}. This is in the literature known as
a doubly intractable distribution \citep{pro20}, and great efforts have
been made in order to circumvent exact evaluations of the normalising
constants. These techniques can roughly be divided into three
approaches. The first is to introduce
auxiliary variables in the Bayesian model in such a way that the
resulting acceptance probabilities would be without any evaluation of
the normalising constant. These approaches are discussed in \citet{art120} and
\citet{pro20}. However to reach the true
limiting distribution in such MCMC algorithms, we are required to simulate
perfect samples \citep{propp1996} from the likelihood, which in turn can be very computationally
demanding or even practically infeasible. 
If perfect simulation from the likelihood is
infeasible \citet{CaimoFriel2011} and \citet{Everitt2012} propose to 
replace the perfect sample with a sample generated by MCMC.
Such an approach will not
exactly target the correct posterior distribution, but the experience is that in many 
situations such a procedure provides a very good approximation to what we want.
It should also be noted that the value of the
parameter vector will typically restrict this approach as well, 
as the MCMC sampler for the auxiliary variable
needs to converge for each sample. 

A second approach for coping with the computationally intractable normalising constant
is to replace it by an
estimate obtained prior to posterior simulation. The estimate is usually
obtained by an MCMC algorithm, which is also the approach taken in
\cite{geyer1992} and \citet{art144}. This approach will however not give the correct
limiting distribution, and for the approach to be practical it usually requires 
the dimension of the parameter vector to be low.
The third option for handling the intractable normalising constant is
to replace the likelihood by an
approximation, and thereby instead simulate from an approximation
to the posterior distribution. Several approximation schemes have been
developed, including approximation by pseudo-likelihood \citep{art109},
reduced dependency approximation \citep{art117, art145},
and approximation by pseudo-Boolean functions
\citep{phd6,tjelmeland2012}. As with the second approach we will not
get the true limiting distribution using approximations, however
there are typically no computational restrictions on the dimension
and value of the parameter vector, although the quality of the approximation may
depend on these quantities. 

In our situation we have an MRF with a parameter vector of possibly a
somewhat high dimension, and in fact we have also put a prior
on its dimension, so the second
approach is not a viable alternative. We have tried both the first approach, 
with an MCMC sample to replace the perfect sample, and the third
approach using pseudo-Boolean approximations. For our examples we found the
first approach to be computationally cheaper and sufficiently accurate,
so in our simulation examples we present results when adopting this approach.

\subsection{Proposal distributions}\label{sec:proposal}
In this section we only give a brief description of the proposal
distributions we have used when simulating from the posterior distribution. 
A more detailed description is given
in the supplementary material to this paper, see Section
S.1. In our algorithm we combine six proposal distributions 
making changes in $z$ or $M$, and three proposal distributions 
modifying the hyper-parameter vector $\theta$.
Each of the proposals for $z$ or $M$ is made
according to a proposal probability vector
$p_o=(p_o^1,p_o^2,p_o^3,p_o^4,p_o^5,p_o^6)$, where each of these probabilities
correspond to the order in which we discuss the proposals here.
Our first proposal is a random walk proposal
for the $\varphi$ parameters, adjusting for the sum-to-zero
constraint. Our second proposal is to take one clique type in
a given cell and propose to change its cell membership. In the third
proposal we propose to decrease or increase the number of cells by
one, changing the dimension of the parameter vector $\varphi$
correspondingly. As a forth proposal we propose to replace a
clique type that is on by
a different clique type, copying the parameter value and cell membership
from the old clique type to the new one. In the fifth proposal we
propose to turn on a new clique type or turn off an already existing 
clique type. When
proposing to turn on a new clique type we propose to insert the new clique type into an
already existing cell, and when proposing to turn off an existing
clique type we choose to delete a clique type from a cell with size
larger than one, preserving the dimension of the parameter vector in
both cases. In
the sixth and last of the proposals for $z$ and $M$, we propose to turn on a clique
type and assigning it to it a new cell, or to
turn off a clique type that is the only clique type in its cell. 
This proposal will also result in a dimension change in
the parameter vector $\varphi$. Note that we allow some of these
proposals to lead to states of zero prior probability. However for such
proposals we do not need to evaluate the likelihood, so such proposals
are very cheap to evaluate.
The first three of the proposals described above make
changes to already existing parameter values or to the
cells of the partition. The last three proposals also makes
changes to $M$, i.e. to what clique types that are on or
off. In the supplementary materials we give the proposals we use when updating 
the elements of $\theta$. These proposals are cheap to evaluate
and we do one of each of these for each time we perform one of the six
proposals described above.

\section{Simulation examples}\label{sec:examples}

In this section we investigate our prior and proposal distributions on
three examples. Firstly we consider a simulation example where data is
generated from an Ising model, secondly we generate data from a model with
$2\times 2$ cliques, and thirdly we look at a real data example
consisting of presence and absence of red deer. In all our simulations
we evaluate the acceptance 
probabilities by using the exchange algorithm of \cite{pro20} with a Gibbs sampler to
obtain the auxiliary variables \citep{CaimoFriel2011}. The length of
the burn in period of the Gibbs sampler are in all simulations set to 200 iterations, although experiments suggest that only about half of is
sufficient in all our examples. We set
$p_o=(0.1,0.15,0.15,0.15.0.20,0.25)$ for the two first examples and a
slightly different $p_o$ for the last example, see Section \ref{sec:deer}. For
the tuning parameters described in the supplementary materials,
see Section S.1, we set $\sigma_w=0.2$, $\sigma_g=\sigma_c=0.3$,
$\sigma_\eta=2.0$, $\sigma_{\sigma_\varphi}=0.7$
and $\rho=0.1$, based on some preliminary runs. 

\subsection{The Ising model}\label{sec:ising}

\begin{figure}
  \centering
  \includegraphics[scale=0.3]{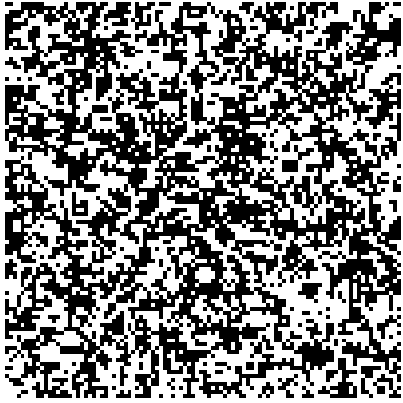} 
  \caption{\label{fig:isingData}Ising example: Data simulated from the
    Ising model in (\ref{eq:Ising}).}
\label{eq:isingModel}
\end{figure}

In this section we used a perfect sampler \citep{propp1996} to generate one realisation
from an Ising model on a $100\times 100$ lattice with torus boundary conditions.
The Ising model used is given by
\begin{equation}\label{eq:Ising}
p(x)=\exp\left (0.4 \sum_{(i,j)\sim (k,l)}I(x_{i,j}=x_{k,l})\right), 
\end{equation}
where the sum is over all pairs of first order neighbour
nodes, taking the torus boundary assumption into account. The
realisation used is shown in Figure \ref{fig:isingData}.
This Ising
model corresponds to the set of clique types $M=\{\{\emptyset\},\ABBB,\AABB,\ABAB\}$ with
the partition either being
$\mathscr{S}=\{\{\{\emptyset\}\},\{\ABBB\},\{\AABB,\ABAB\}\}$ or
$\mathscr{S}=\{\{\{\emptyset\}\},\{\ABBB\},\{\AABB\},\{\ABAB\}\}$. The
true parameter values for these two model formulations are 
$\varphi_{\{\{\emptyset\}\}}=1.3333$, $\varphi_{\{\ABBB\}}=-0.2666$
$\varphi_{\{\AABB,\ABAB\}}=-1.0666$, and 
$\varphi_{\{\{\emptyset\}\}}=1.6$, $\varphi_{\{\ABBB\}}=0$,
$\varphi_{\{\AABB\}}=-0.8$, $\varphi_{\{\ABAB\}}=-0.8$ respectively. 
To simulate from the posterior in (\ref{eq:posterior}) we
ran our sampler for $10^6$
iterations, and based on convergence plots we could see that the
MCMC sampler had converged after about $10^4$ iterations. The
acceptance rates for each of the proposals changing $z$ and $M$
were for this example $0.102, 10^{-5}, 0.002, 10^{-4}, 10^{-4}$
and $0.001$, given in the same order as in $p_o$. The reason for 
the very low figures for some of these acceptance rates is in 
fact not slow mixing, but just that the posterior distribution
has essentially all its mass in one state for $M$ and $\mathscr{S}$.
The acceptance rates for the three updates of $\eta$, $p_\star$ and $\sigma_{\varphi}^2$ were
$0.278$, $0.499$ and $0.462$ respectively. 

Discarding the burn-in period of $10^4$ iteration, we now
investigate the posterior distribution of $M$,
$\mathscr{S}$, the
parameter values $\varphi$, and the hyper-parameters
$\eta$, $p_\star$ and $\sigma_{\varphi}^2$. For
$M$ the estimated posterior probability of the correct state
becomes $0.999$, so the sampler just very briefly visit other states. If we
consider only the realisations that consist of the correct state of $M$,
there are only two partitions that are simulated, namely the two partitions
given above. The three and four cell partitions have
estimated posterior probabilities equal to $0.991$ and $0.009$
respectively, so the model with less parameters is preferred.   

\begin{figure}
  \centering
\subfigure[][$\phi^{\ABBB}-\phi^{\emptyset}$]{\label{fig:c2}\includegraphics[scale=0.3]{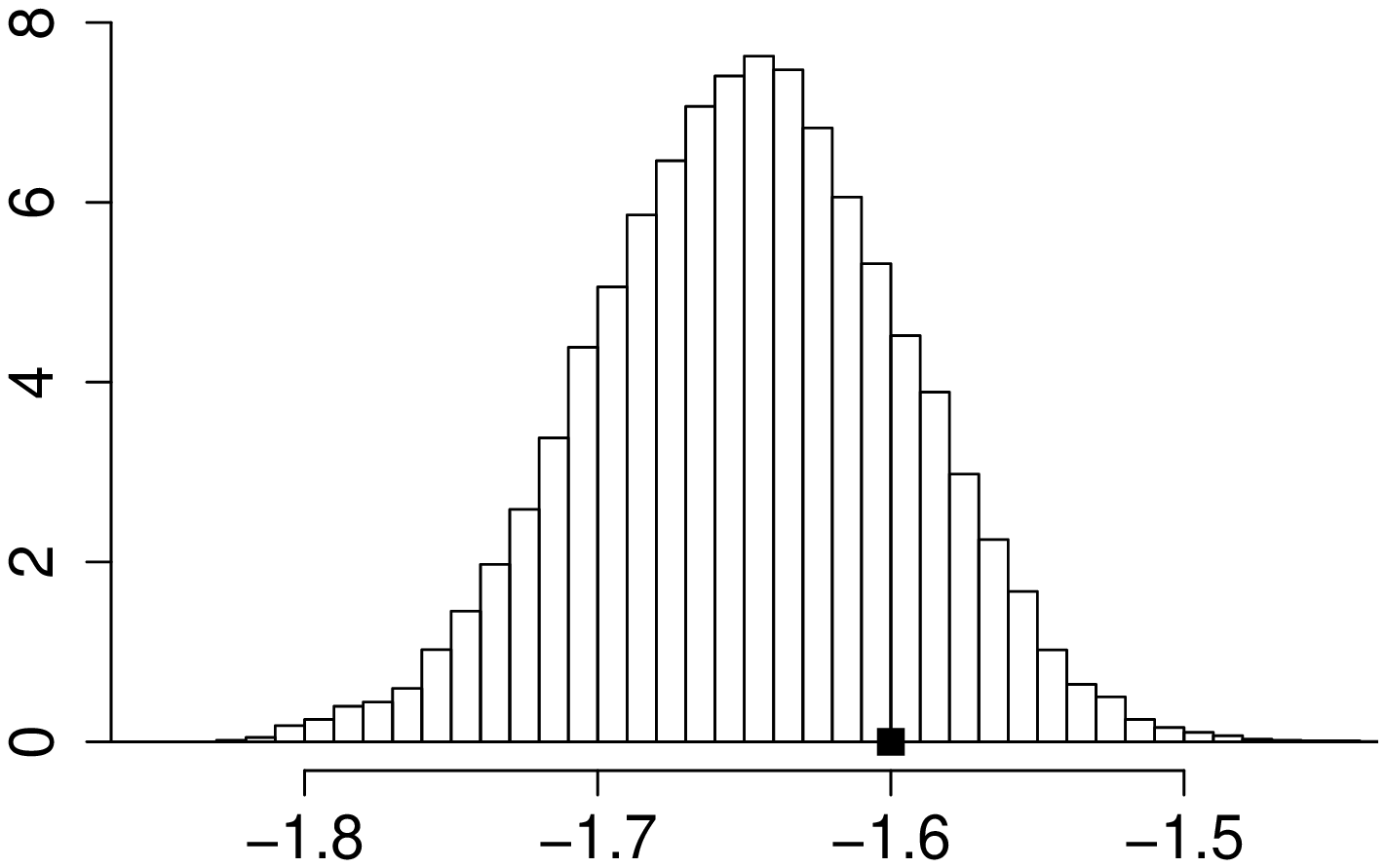}}
\subfigure[][$\phi^{\AABB}-\phi^{\emptyset}$]{\label{fig:c2}\includegraphics[scale=0.3]{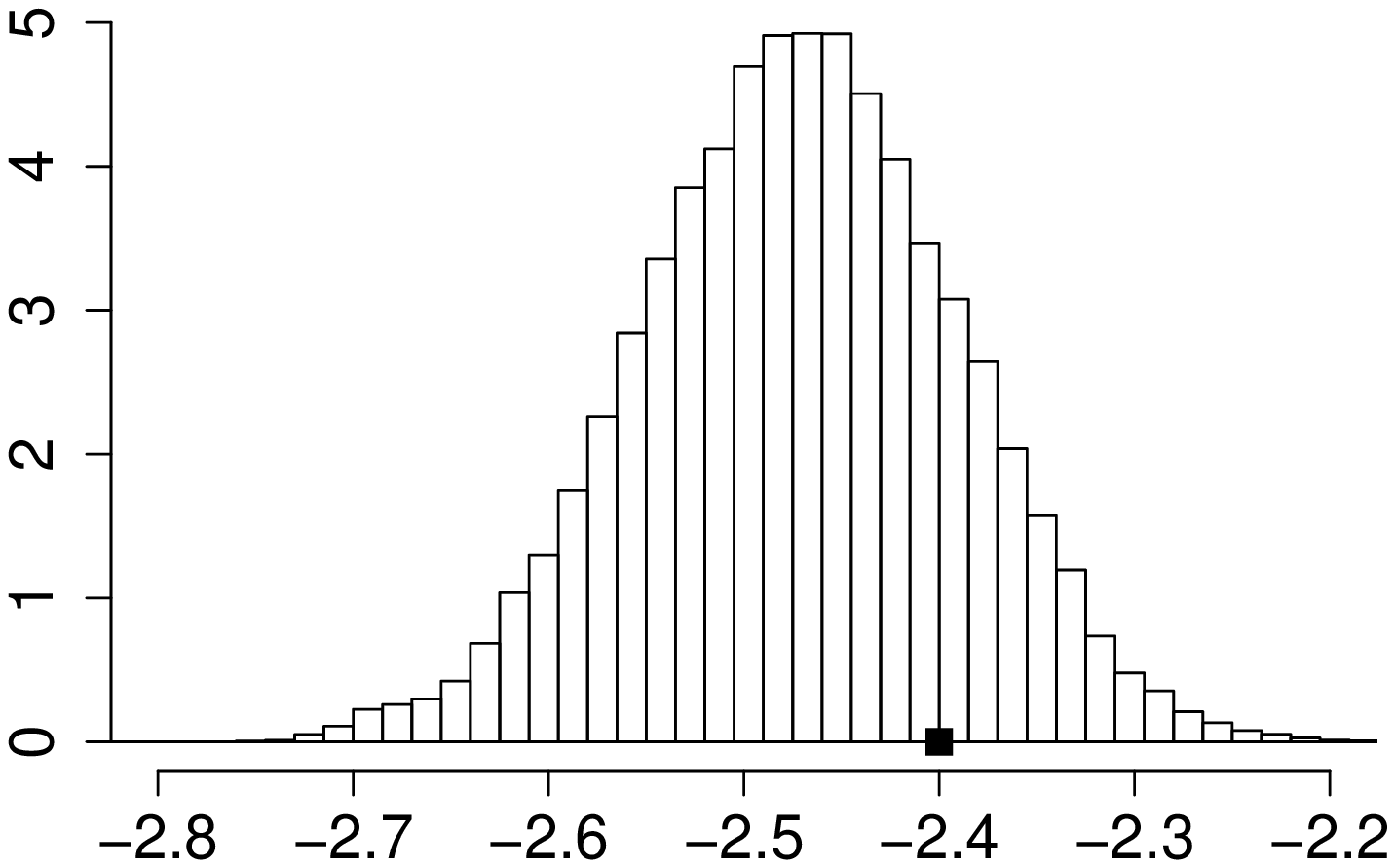}}
\subfigure[][$\phi^{\ABAB}-\phi^{\emptyset}$]{\label{fig:c3}\includegraphics[scale=0.3]{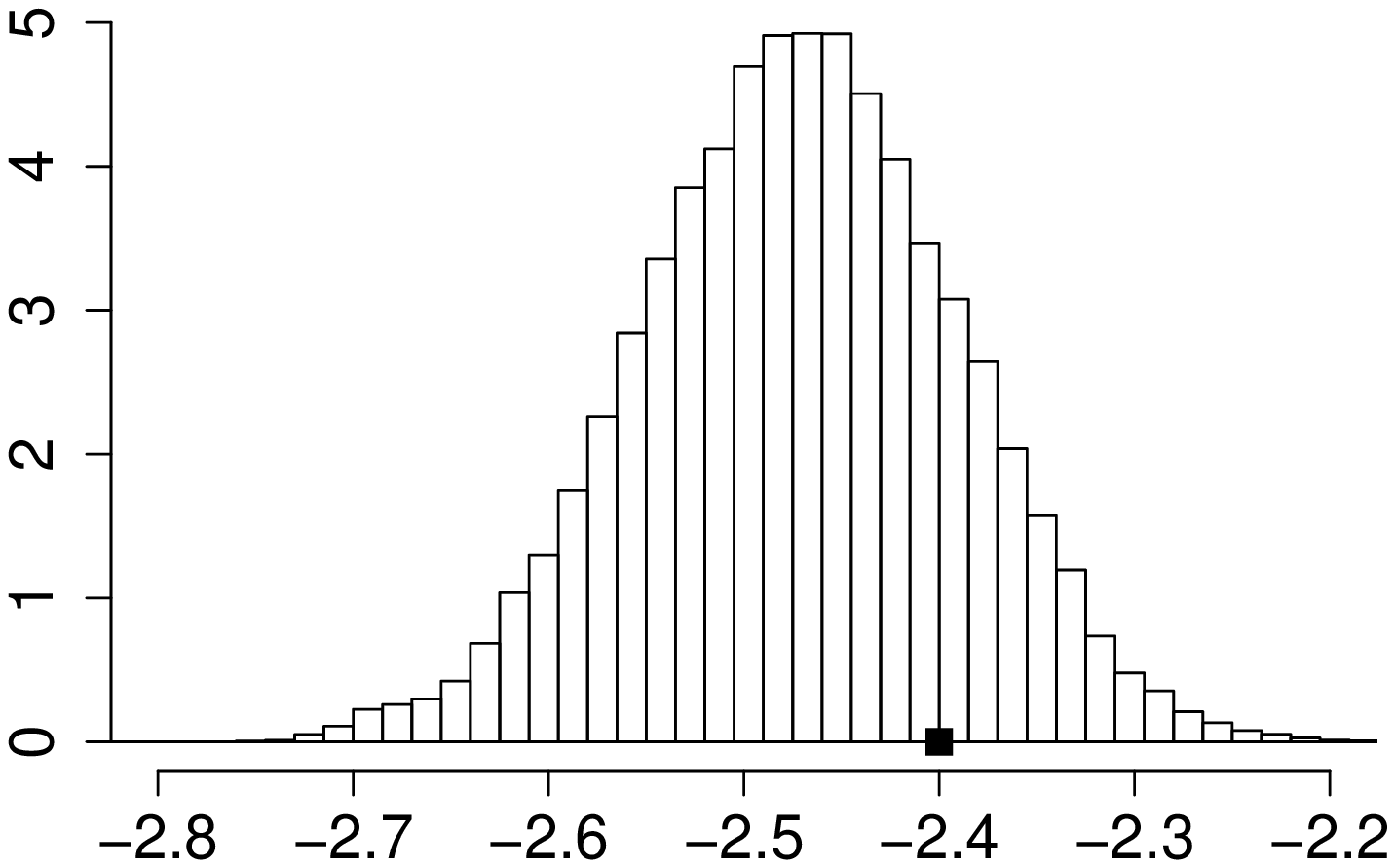}}
\\
\subfigure[][$\phi^{\ABBA}-\phi^{\emptyset}$]{\label{fig:c2}\includegraphics[scale=0.3]{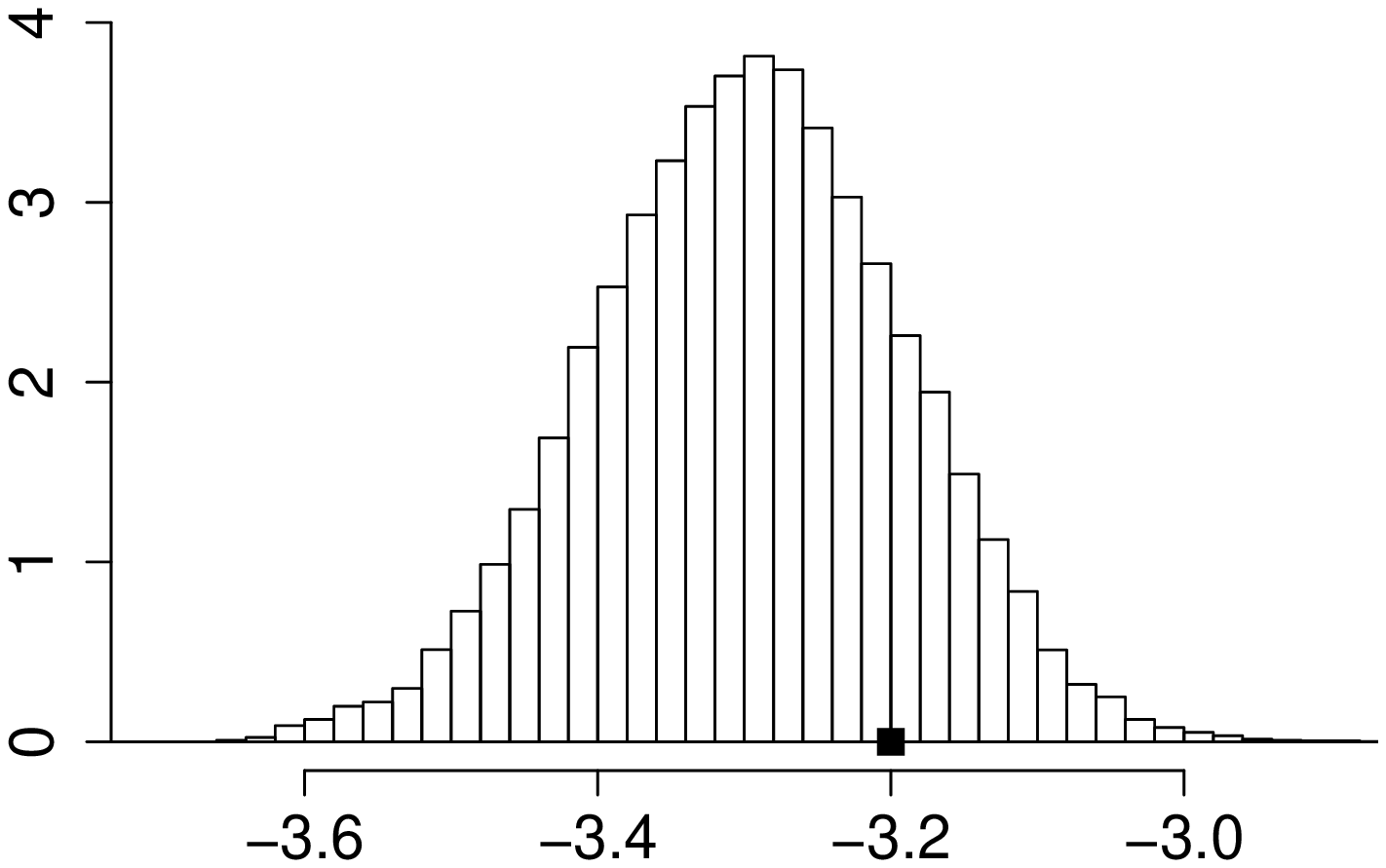}}
\subfigure[][$\phi^{\BAAB}-\phi^{\emptyset}$]{\label{fig:c4}\includegraphics[scale=0.3]{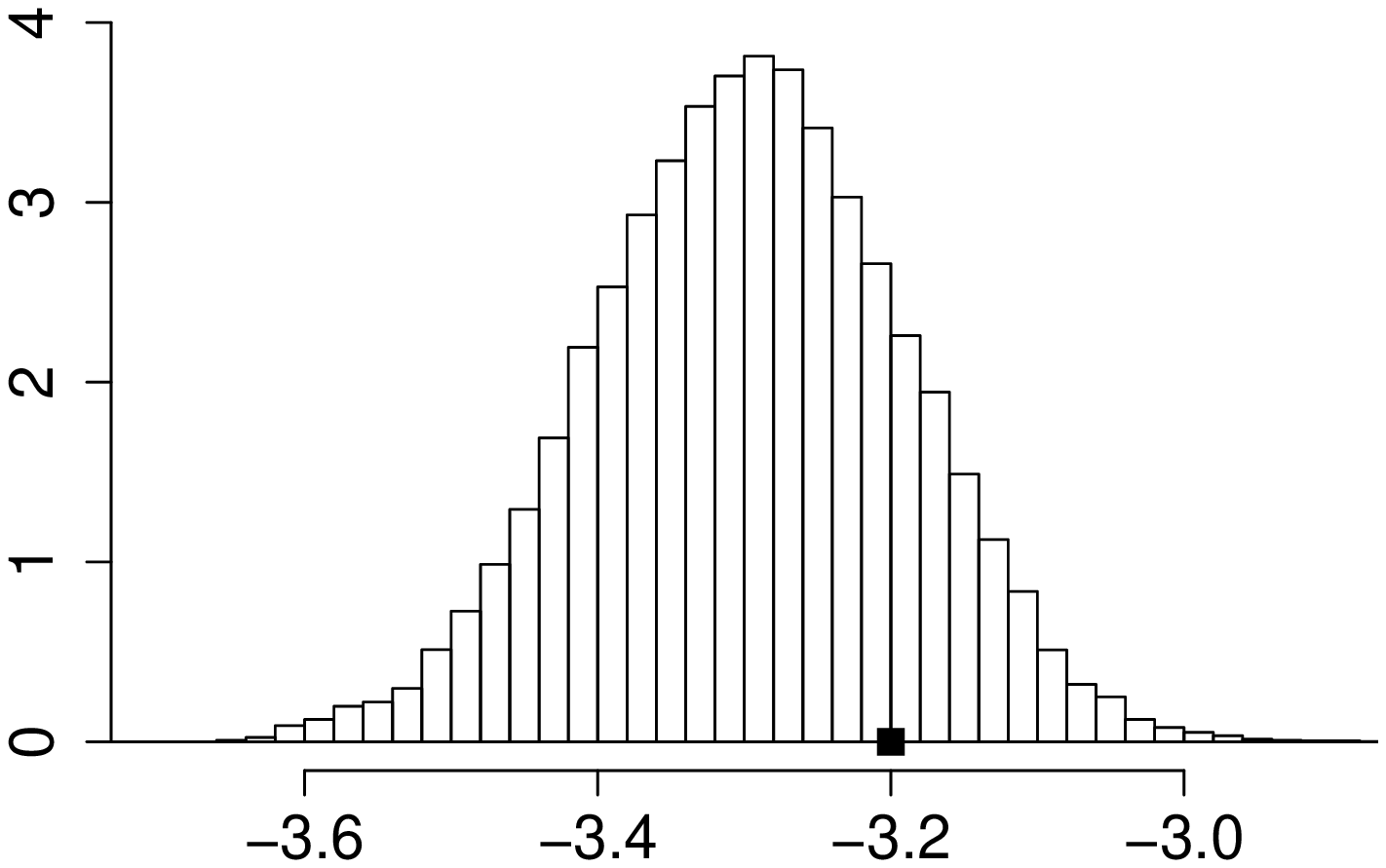}}
\subfigure[][$\phi^{\AAAB}-\phi^{\emptyset}$]{\label{fig:c3}\includegraphics[scale=0.3]{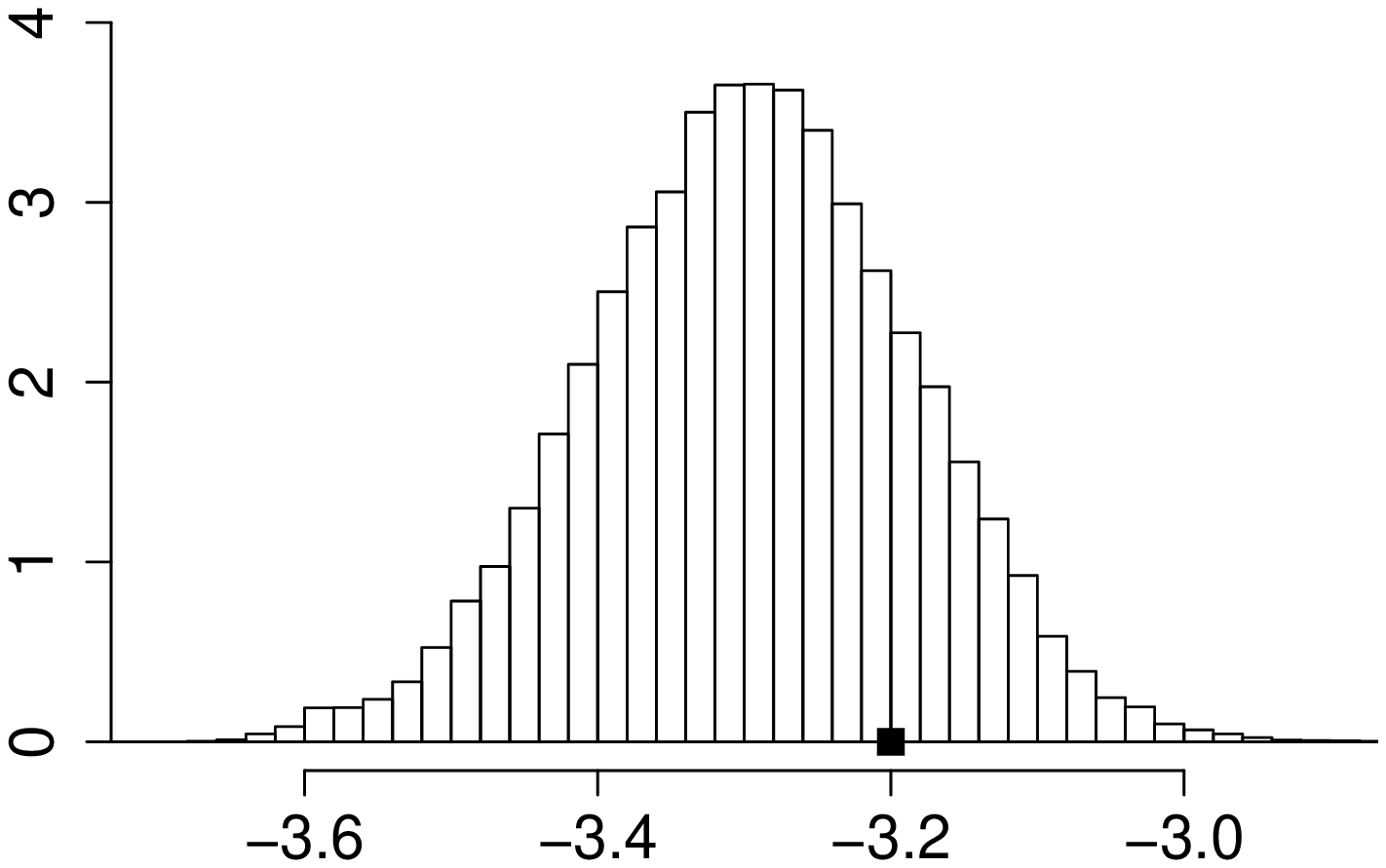}}
\\
\subfigure[][$\phi^{\AABA}-\phi^{\emptyset}$]{\label{fig:c2}\includegraphics[scale=0.3]{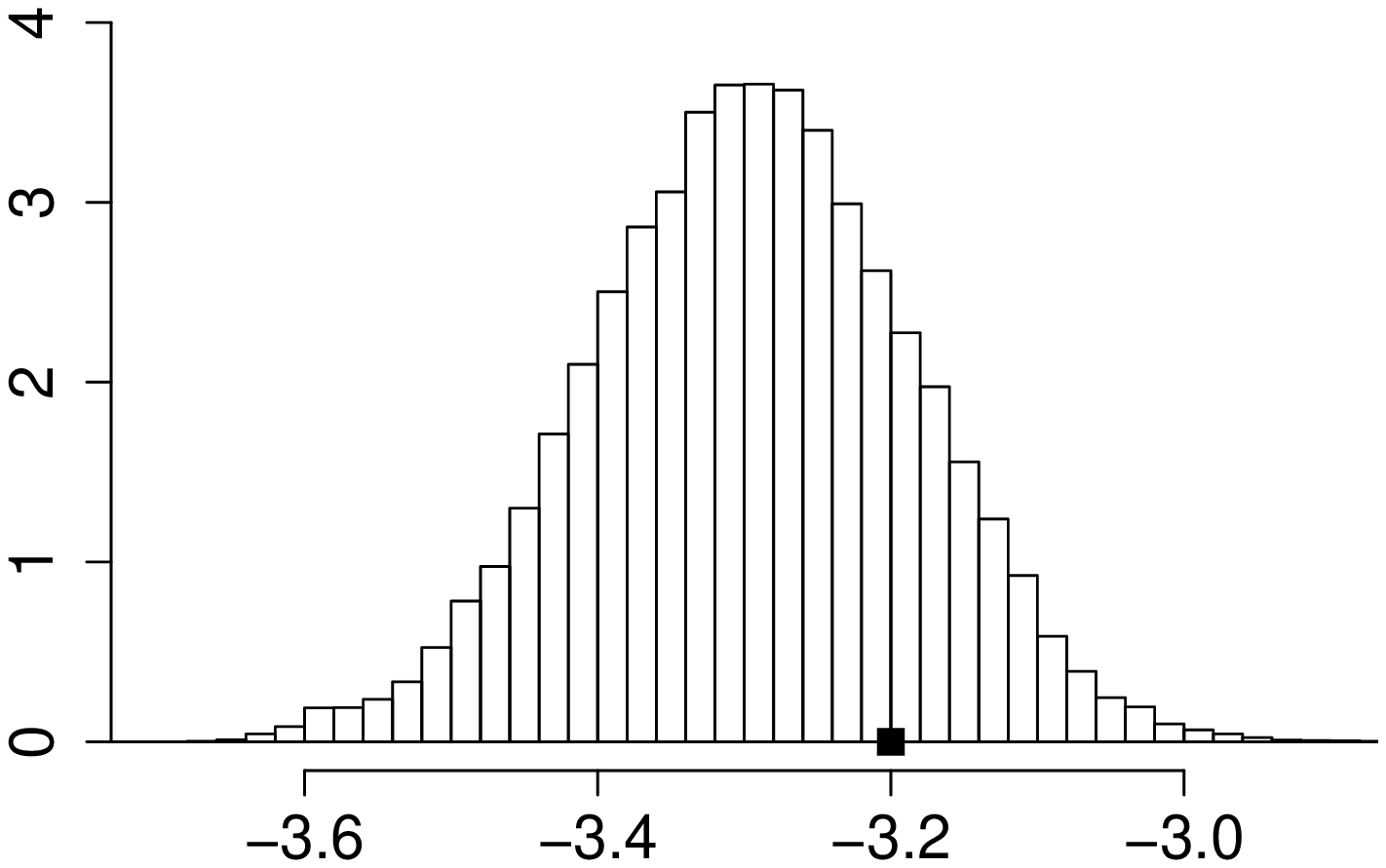}}
\subfigure[][$\phi^{\ABAA}-\phi^{\emptyset}$]{\label{fig:c4}\includegraphics[scale=0.3]{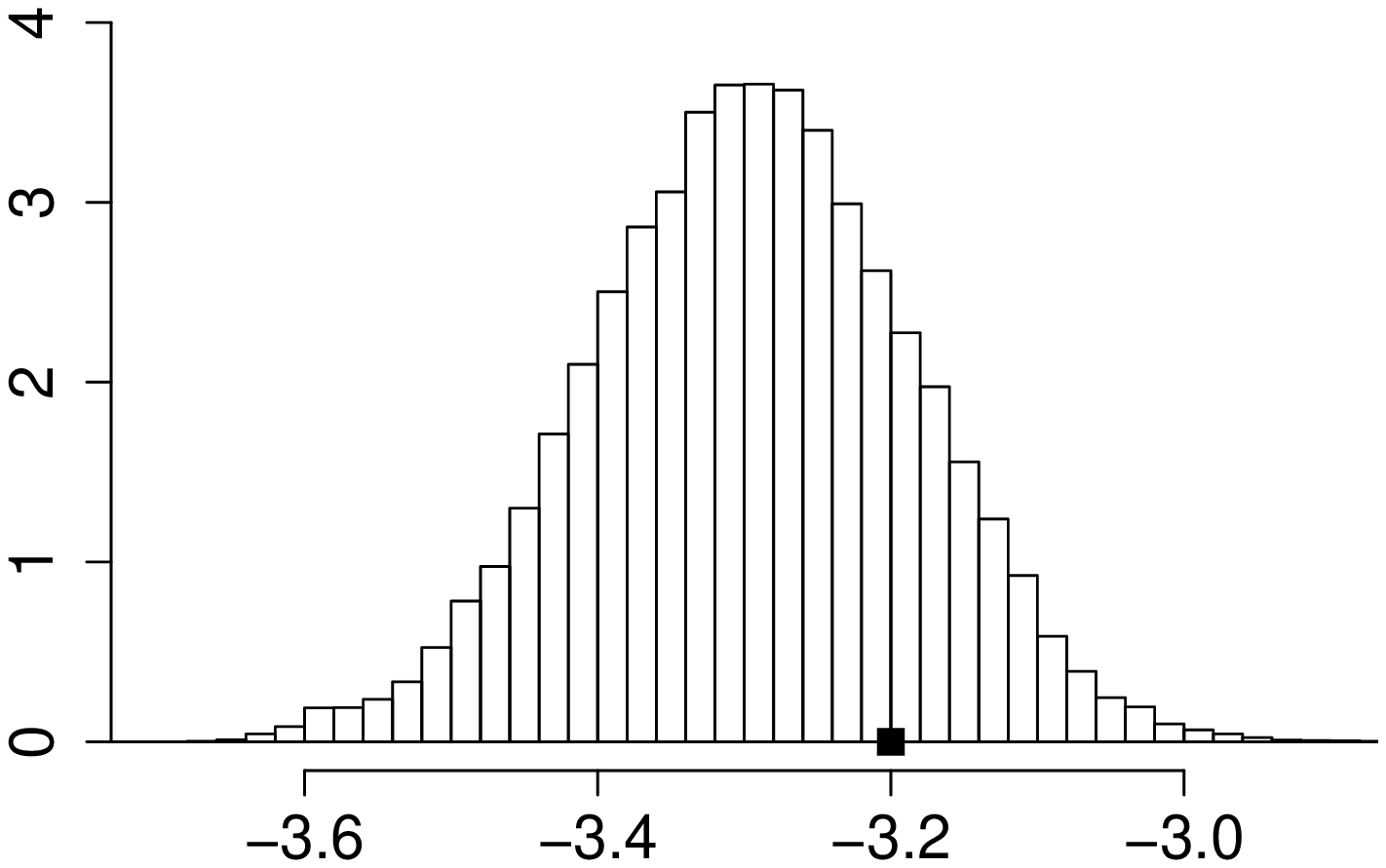}}
\subfigure[][$\phi^{\BAAA}-\phi^{\emptyset}$]{\label{fig:c3}\includegraphics[scale=0.3]{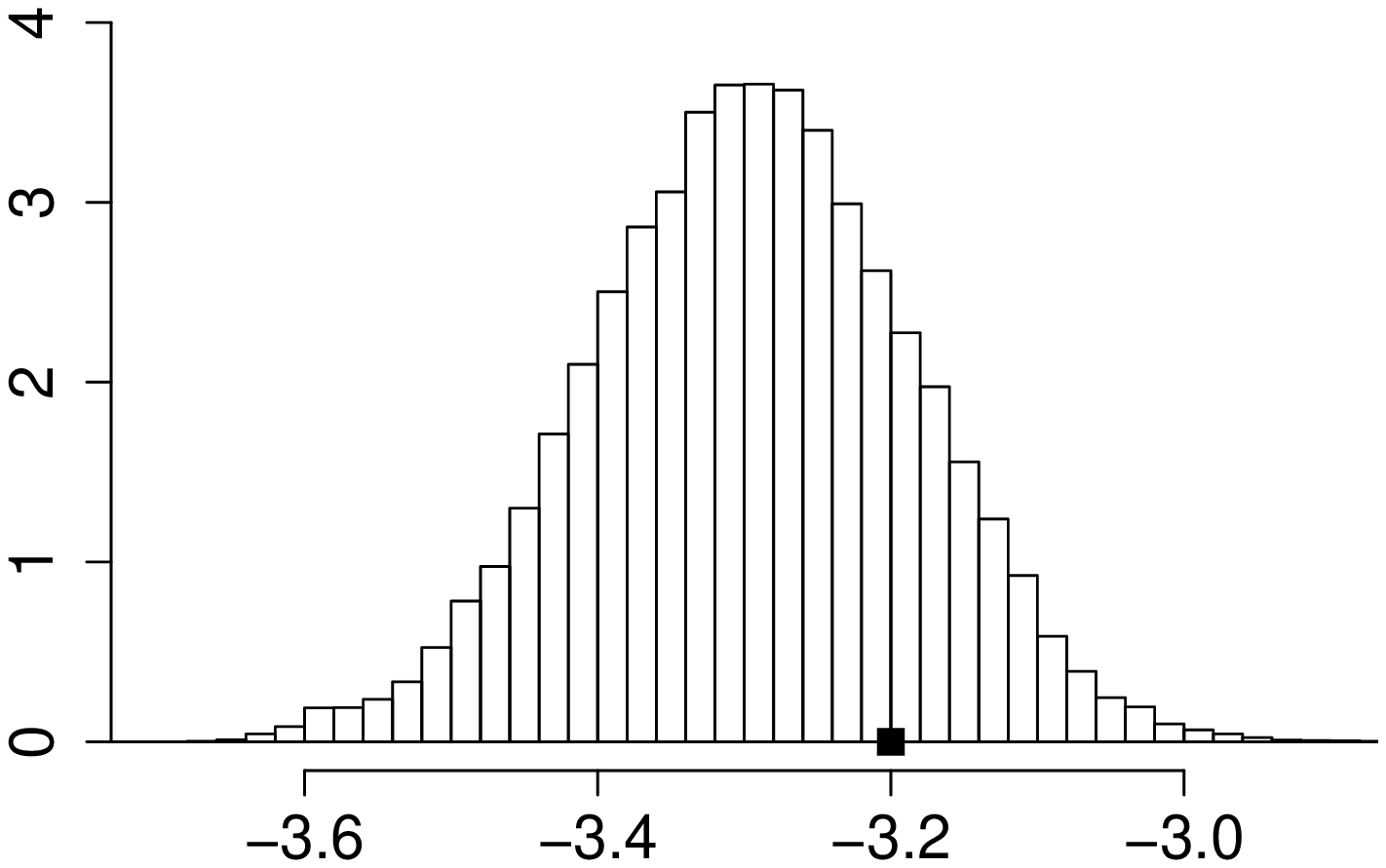}}
\\
\subfigure[][$\phi^{\AAAA}-\phi^{\emptyset} $]{\label{fig:c2}\includegraphics[scale=0.3]{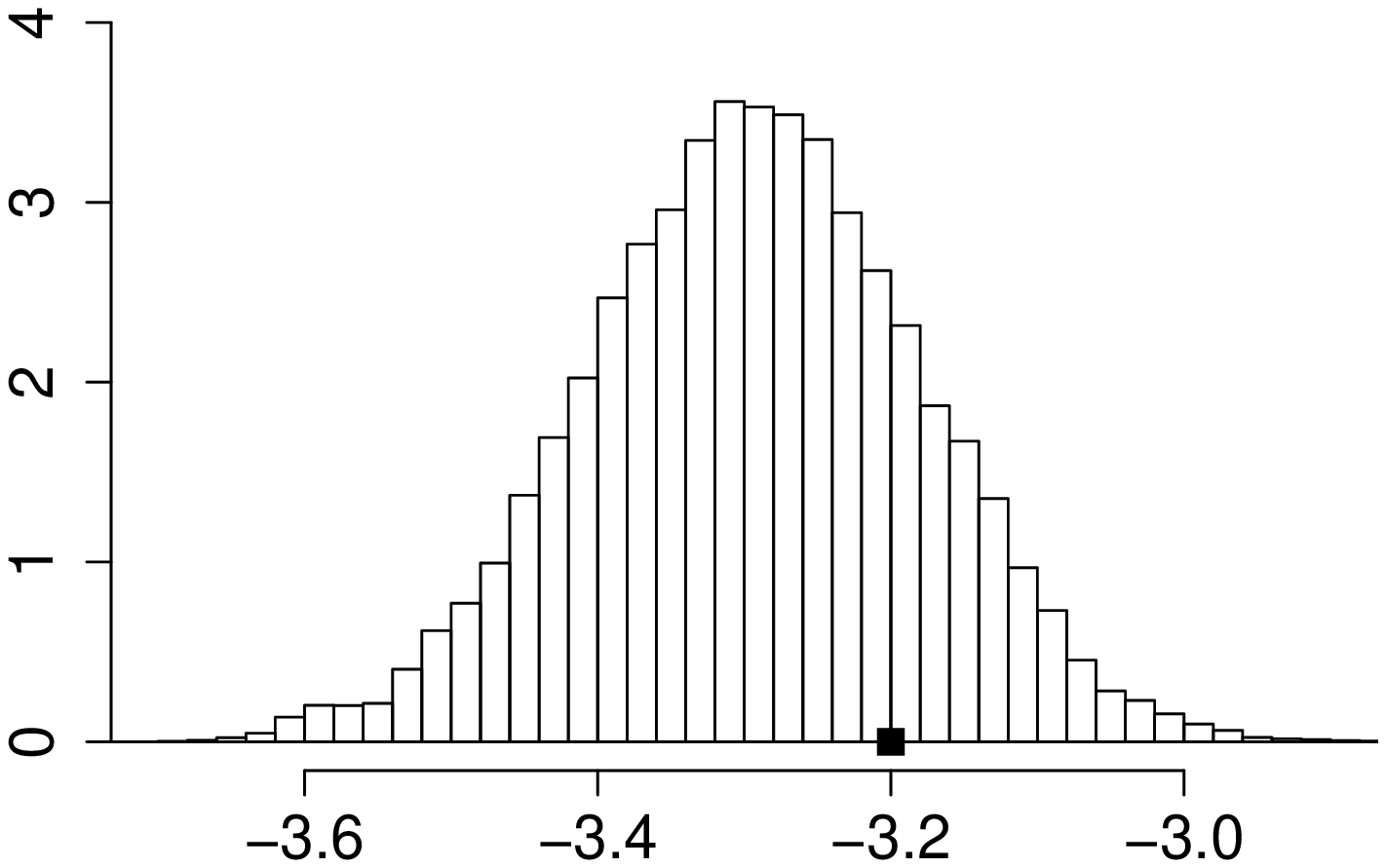}}
\caption{\label{fig:isingPhi}Ising example: Probability histograms for the posterior
simulated values for $\phi^\Lambda-\phi^{\{\emptyset\}}$ for ten values of $\Lambda$. True
  values are shown with a black box.}
\end{figure}

To evaluate the posterior distribution for the parameters is
complicated when simulating from a distribution where not only the 
parameter values, but also the dimension and interpretation of the 
parameters vary, see for example the discussion in 
\citet{Celeux2000}. In our model the interpretation of both 
$\varphi$ and $\{\phi^\Lambda,\Lambda\in M\}$ vary with $M$ and 
$\mathscr{S}$, so to study the posterior distribution of the
elements of these vectors across different states for $M$ and 
$\mathscr{S}$ is of no value. Recalling that $\phi^\Lambda$ is well 
defined also when $\Lambda$ is off, $\Lambda\not\in M$, we instead 
focus on $\phi^\Lambda-\phi^{\{\emptyset\}},\Lambda\in L\setminus\{\{\emptyset\}\}$ 
which has 
the same interpretation for all $M$ and $\mathscr{S}$. From (\ref{eq:phiParam}) we see 
that $\phi^\Lambda - \phi^{\{\emptyset\}} = U(\bold 1^\lambda) - U(\bold 1^\emptyset)$
for any $\lambda\in\Lambda$. We in particular choose to focus on 
$\phi^\Lambda-\phi^{\{\emptyset\}}$ for $\Lambda\in
\{\ABBB,\AABB,\ABAB,\ABBA,\BAAB,\AAAB,\AABA,\ABAA,\BAAA,\AAAA\}$. Probability
histograms of the simulated values for these ten statistics are shown 
in Figure \ref{fig:isingPhi}. 
The corresponding values for the
Ising model in (\ref{eq:Ising}) that we used to generate the data set 
we are analysing now are
shown as black boxes on the $x$-axis, and we see in fact that the correct $\phi^\Lambda-\phi^{\{\emptyset\}}$
values are well recovered. 
%Note that the true value for $\phi^\Lambda$ corresponding to
%$\Lambda\in \{\ABBA,\BAAB,\AAAB,\AABA,\ABAA,\BAAA,\AAAA\}$ is exactly the same
%because the exponent in \eqref{eq:isingModel}, $U(\bold 1^\lambda)$,
%is for all members, $\lambda\in \Lambda$, of all
%these cells a sum over the same number of equal and unequal pair of nodes.
 %$U(\bold 1^\lambda)=U(\bold 1^{\lambda^\star})$ for every pair
 %$\lambda,\lambda^\star$ in any of these cells, see the exponent of \eqref{eq:isingModel}.
 
\begin{figure}
  \centering
\subfigure[][$\eta \ (0.931,3.331)$]{\label{fig:a1}\includegraphics[scale=0.3]{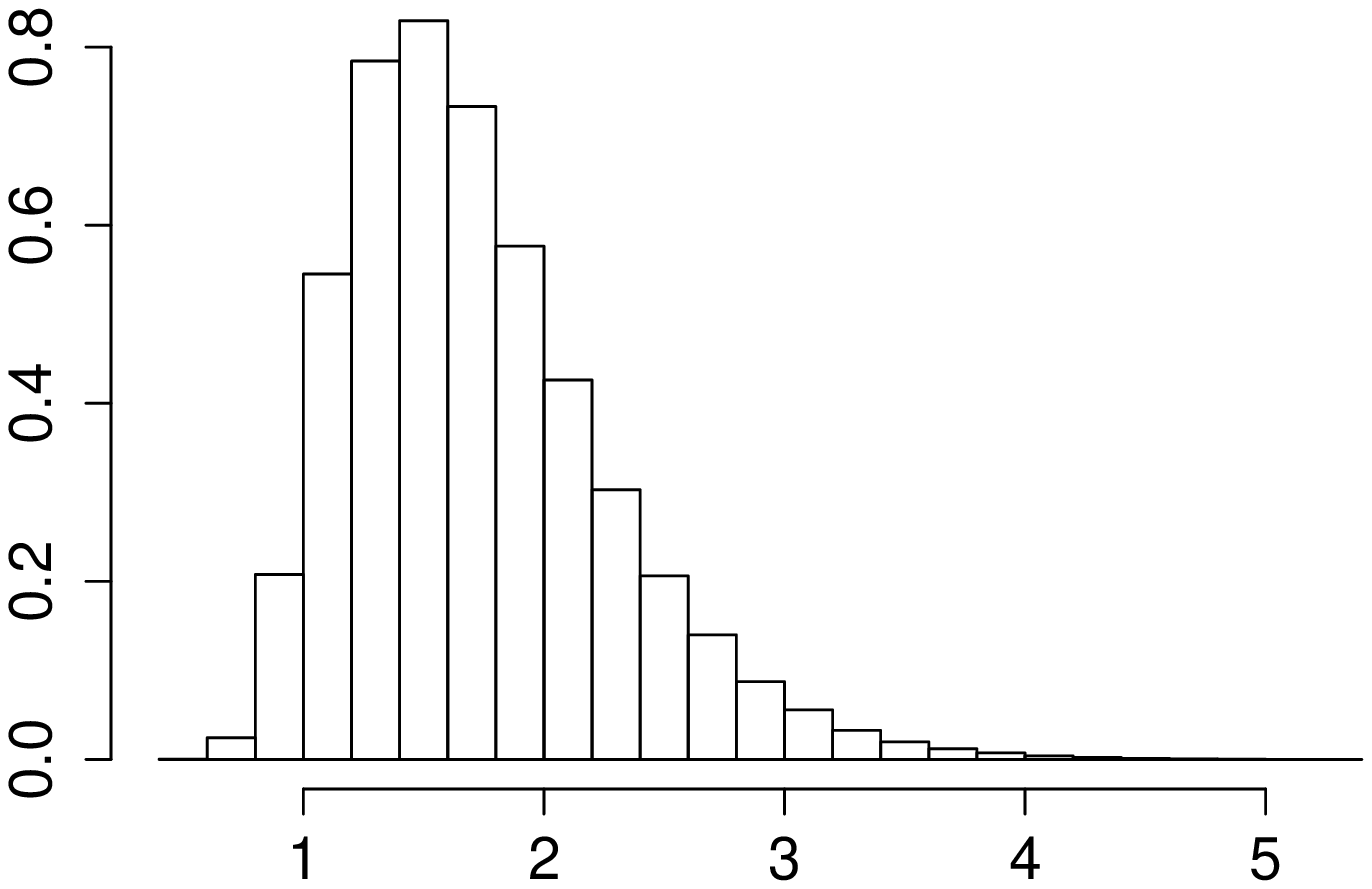}}
\subfigure[][$p_\star \ (0.025,0.975)$]{\label{fig:a2}\includegraphics[scale=0.3]{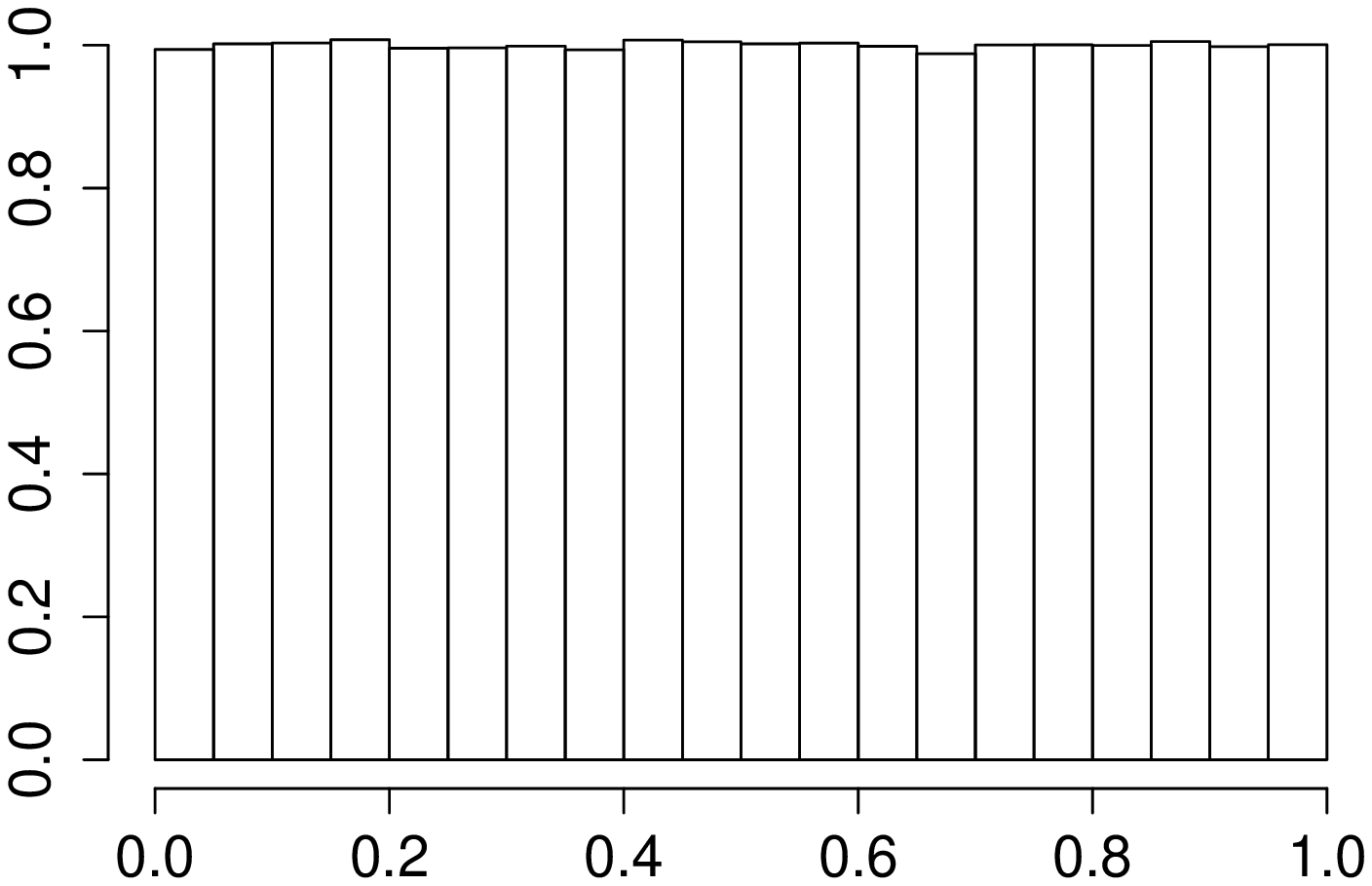}}
\subfigure[][$\sigma_\varphi^2 \ (0.665,21.694)$]{\label{fig:a3}\includegraphics[scale=0.3]{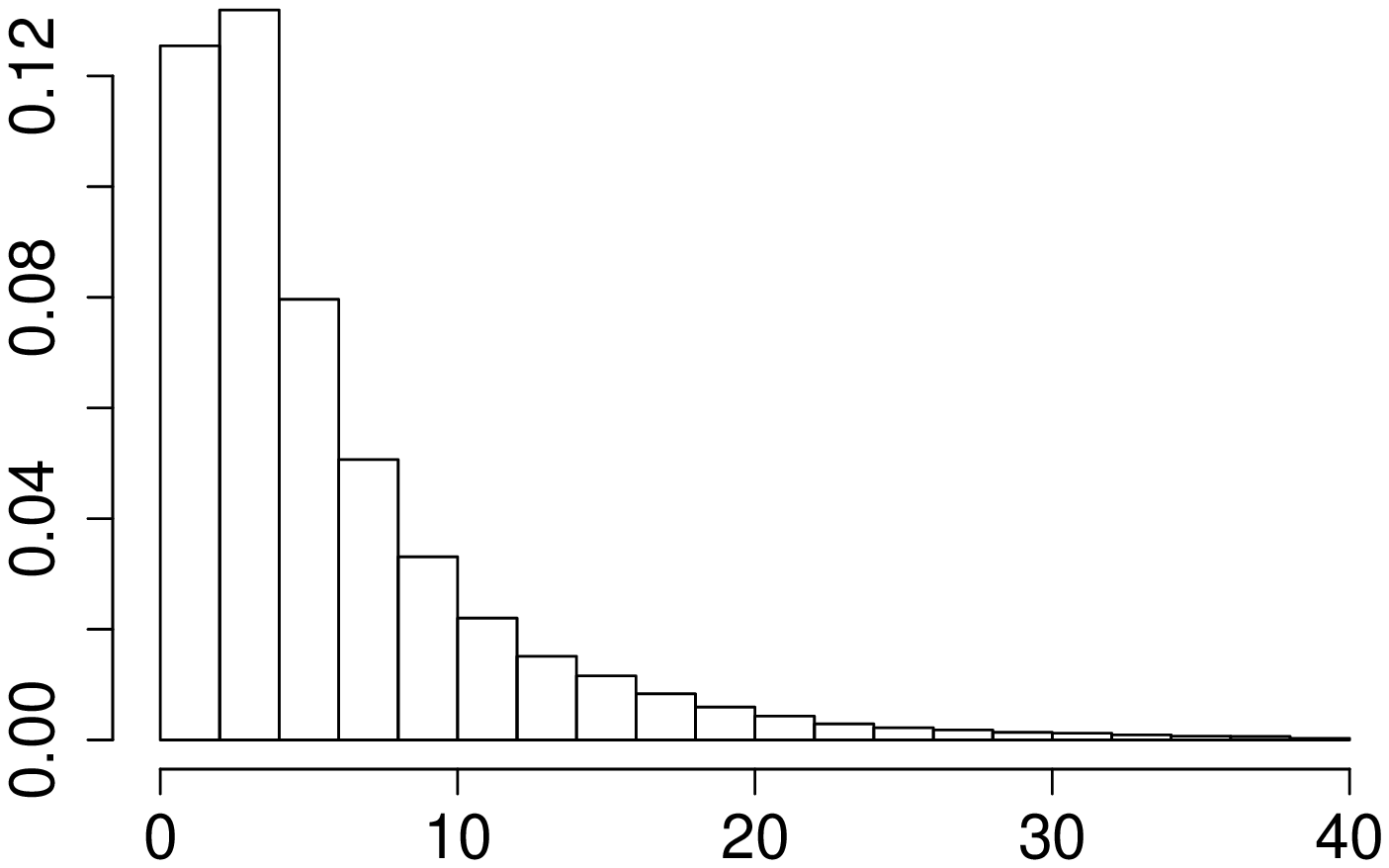}}
\caption{\label{fig:isingPriorParam}Ising example: Probability histograms for the
hyper-parameters $\eta$, $p_\star$ and $\sigma_\varphi^2$, with $95\%$ credible intervals.}
\end{figure}

For the prior parameters $\eta$, $p_\star$, and $\sigma_\varphi^2$ we get the results shown in
Figure \ref{fig:isingPriorParam}.
The median of the posterior samples of $\eta$ is 1.62
with $95\%$ credible interval
$(0.931,3.331)$. The posterior distribution for $p_\star$ is essentially
uniform, which is reasonable since there is no
possibility for higher order interactions in most of the posterior
realisations and thereby this parameter is in most of the iterations
just simulated from the prior. The
median of $\sigma_\varphi^2$ is 3.857 in this example.

\subsection{A $2\times 2$ maximal clique model}\label{sec:2t2}
\begin{figure}
  \centering
  \includegraphics[scale=0.3]{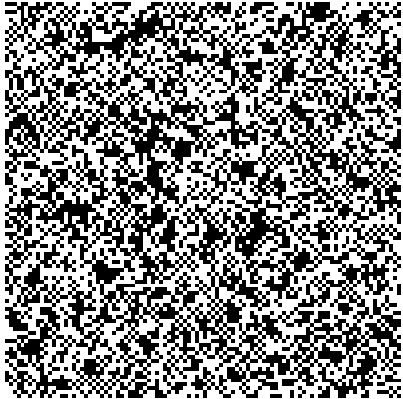} 
  \caption{\label{fig:2t2Data}MRF example with $2 \times 2$ cliques: Data simulated from the
    $2\times 2$ model with parameters given in Table
    \ref{tab:2t2paramSettings}.}
\end{figure}

In the next simulation experiment we generated a data set from 
an MRF with $2\times 2$ maximal cliques and $M$ as shown in 
Figure \ref{fig:exampleDAGmax} and $\mathscr{S}$ and parameter 
values as shown in Table \ref{tab:2t2paramSettings}. 
\def\arraystretch{0.8}
\begin{table}
\centering
\begin{tabular}{c|ccc|c}
$\Lambda\in L\setminus\{\{\emptyset\}\}$ & $\phi^\Lambda-\phi^{\{\emptyset\}}$ & 
\mbox{~~~~~~~~~~} &
$S\in\mathscr{S}$ &$\varphi_S$ \\ \cline{1-2}\cline{4-5}

$\ABBB$ & $-0.10$ & & $\{\{\emptyset\}\}$ & 0.4667 \\
$\AABB$, $\ABAB$ & $-0.75$ & & $\{\ABBB \}$ & 0.3667\\
$\ABBA$, $\BAAB$ & $0.20$ & & $\{\AABB\ ,\ \ABAB \}$ & -0.2833\\
$\BAAA$, $\ABAA$, $\AABA$, $\BAAA$ & $-1.05$ & & $\{\ABBA\ , \ \BAAB \}$ & 0.6667\\
$\AAAA$ & $-1.10$ & & $\{\AAAB \ , \ \AABA \ , \ \ABAA \ , \ \BAAA \}$ & -0.5833\\
\multicolumn{2}{c}{~} & & $\{\AAAA \}$ & -0.6333
\end{tabular}

\vspace*{0.2cm}

 \caption{\label{tab:2t2paramSettings} MRF example with $2\times 2$ cliques: True parameter
   setting with associated (minimal) partition.}
\end{table} 
Just as for the 
Ising model discussed above there are of course several 
possibilities for $\mathscr{S}$ specifying the model we are using, so in
Table \ref{tab:2t2paramSettings} we give $\mathscr{S}$ with the 
lowest number of elements. We also give the corresponding values
of $\phi^\Lambda-\phi^{\{\emptyset\}}$ for $\Lambda\in L\setminus\{\{\emptyset\}\}$
as the interpretation of these perhaps is easier, and because we want 
to compare these values with corresponding values for the posterior samples.
We were not able to obtain a perfect sample by coupling from the past for this
model, so we generated instead a realisation by a Gibbs sampling algorithm,
and the generated data is shown in Figure \ref{fig:2t2Data}.
This data set contains both 
small areas with clustering of black or white nodes and small areas with a checkerboard pattern.

\begin{figure}
  \centering
\subfigure[][$\phi^{\ABBB}-\phi^{\emptyset}$]{\label{fig:ea}\includegraphics[scale=0.3]{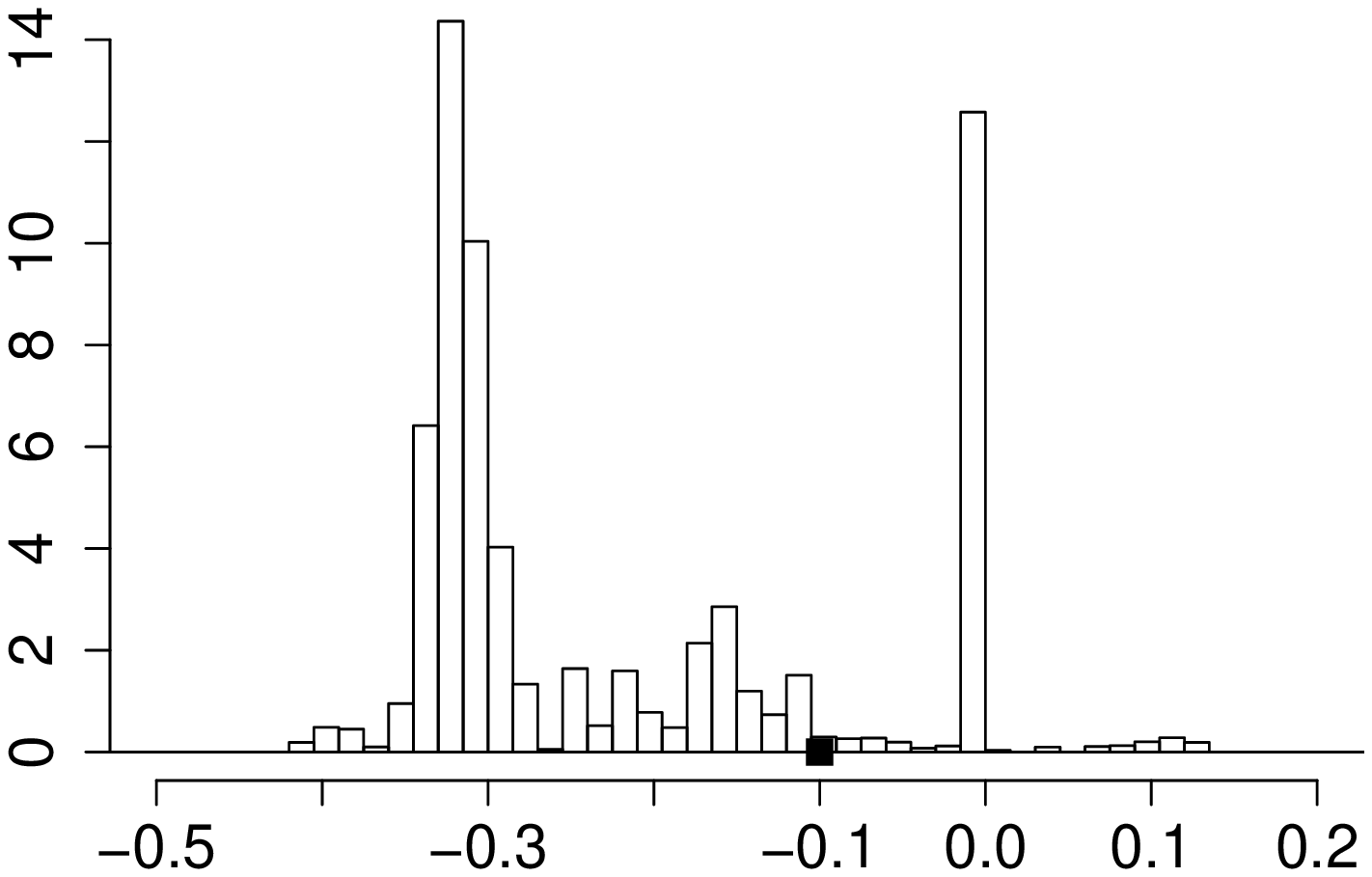}}
\subfigure[][$\phi^{\AABB}-\phi^{\emptyset}$]{\label{fig:eb}\includegraphics[scale=0.3]{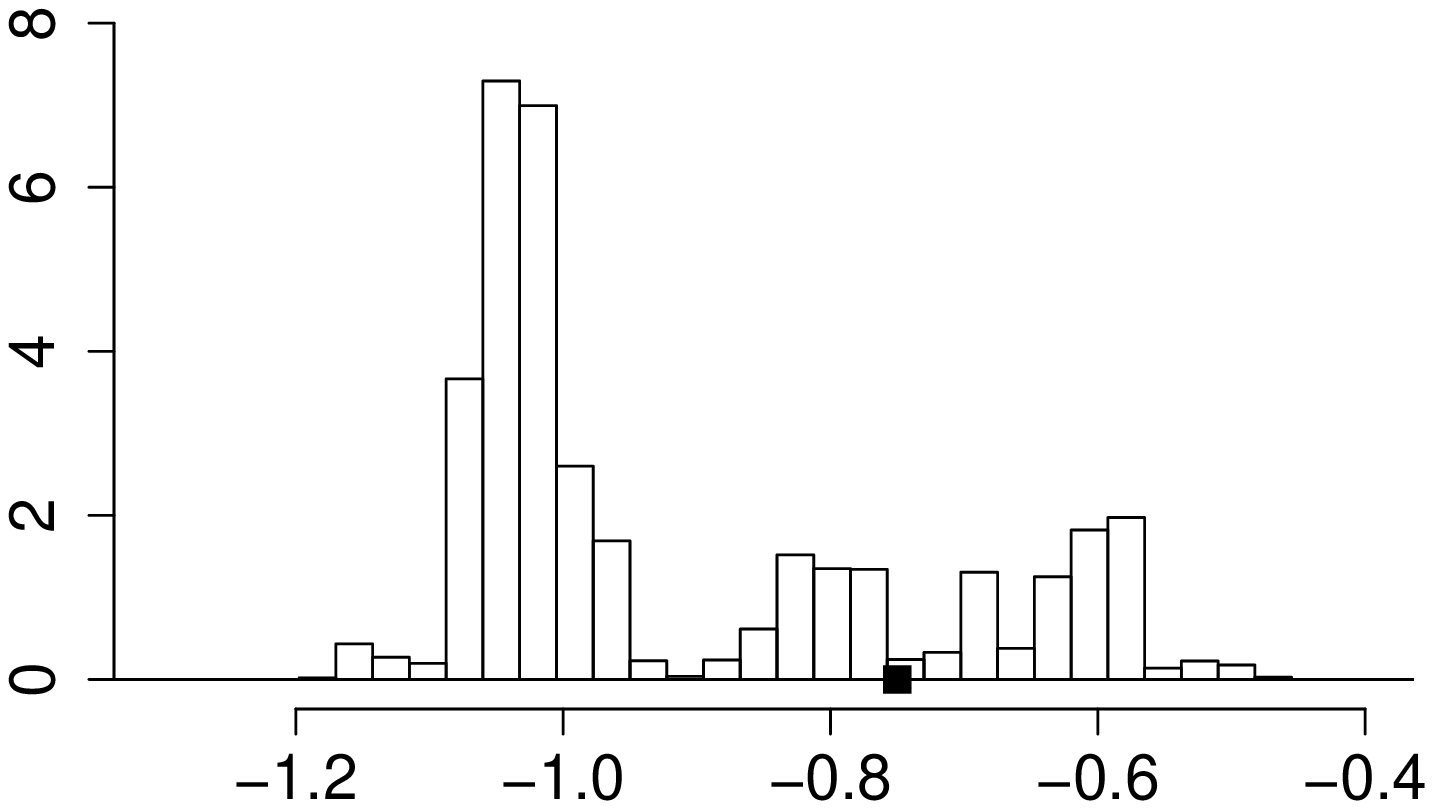}}
\subfigure[][$\phi^{\ABAB}-\phi^{\emptyset}$]{\label{fig:ec}\includegraphics[scale=0.3]{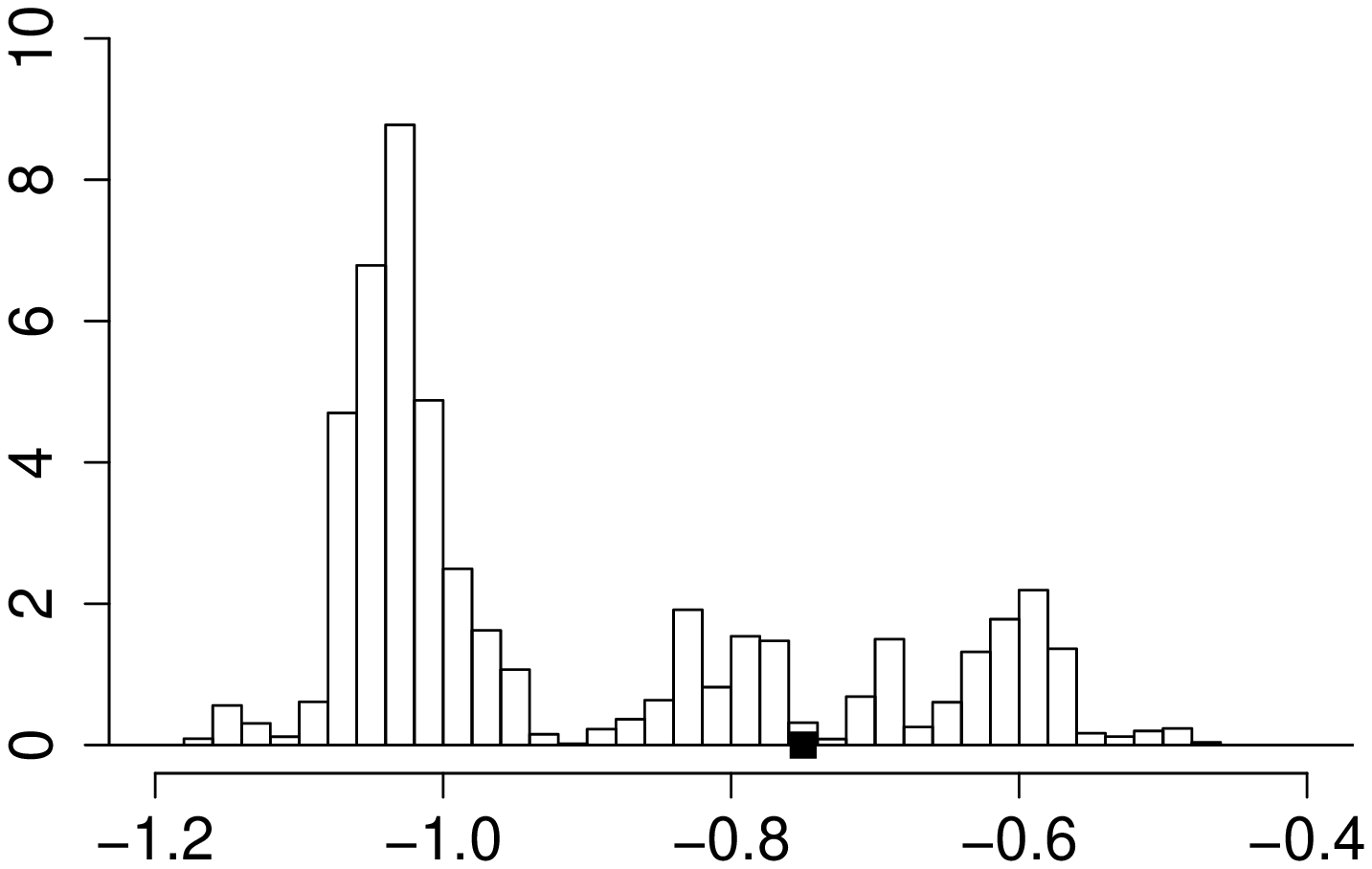}}
\\
\subfigure[][$\phi^{\ABBA}-\phi^{\emptyset}$]{\label{fig:ed}\includegraphics[scale=0.3]{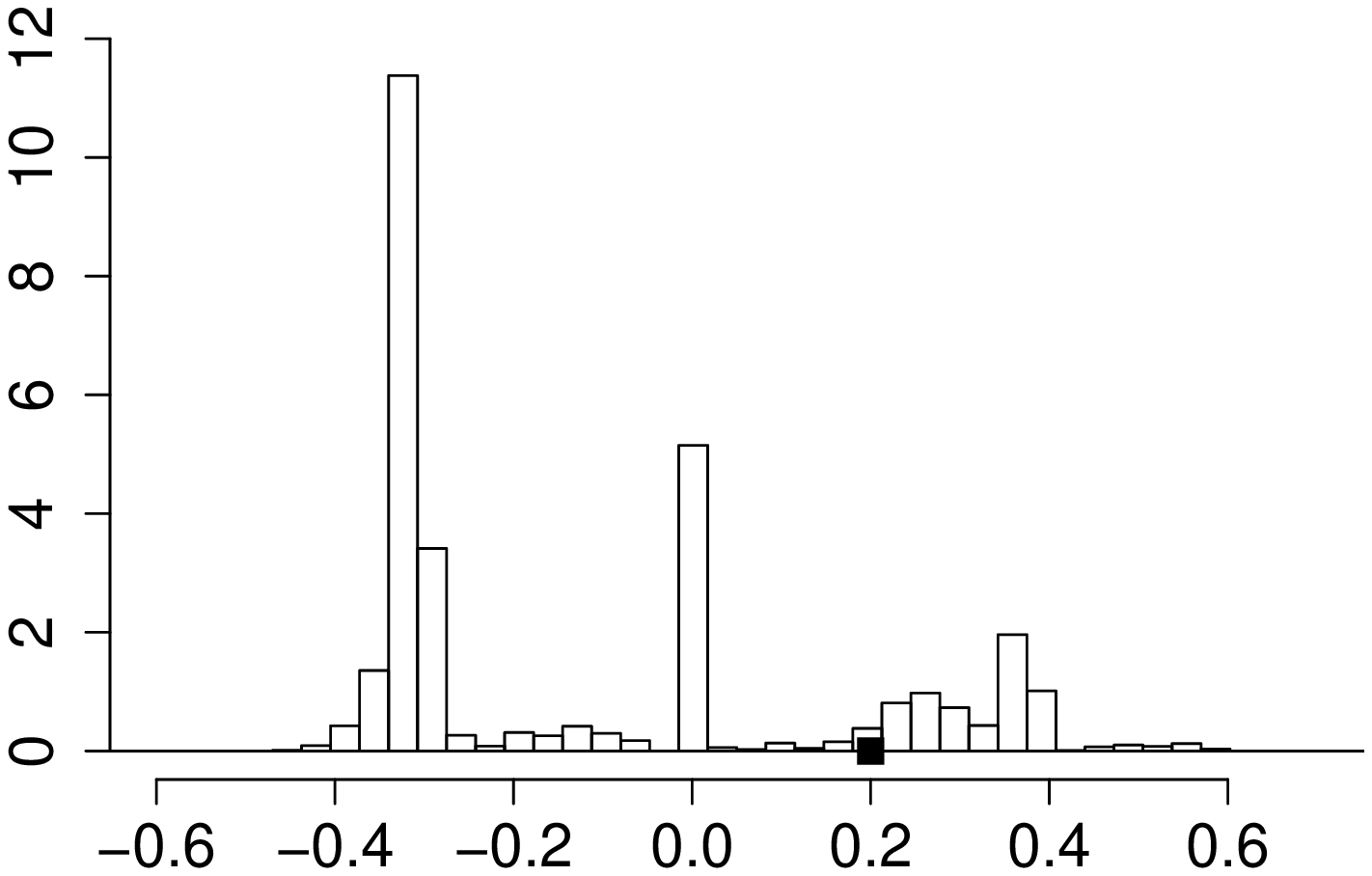}}
\subfigure[][$\phi^{\BAAB}-\phi^{\emptyset}$]{\label{fig:ee}\includegraphics[scale=0.3]{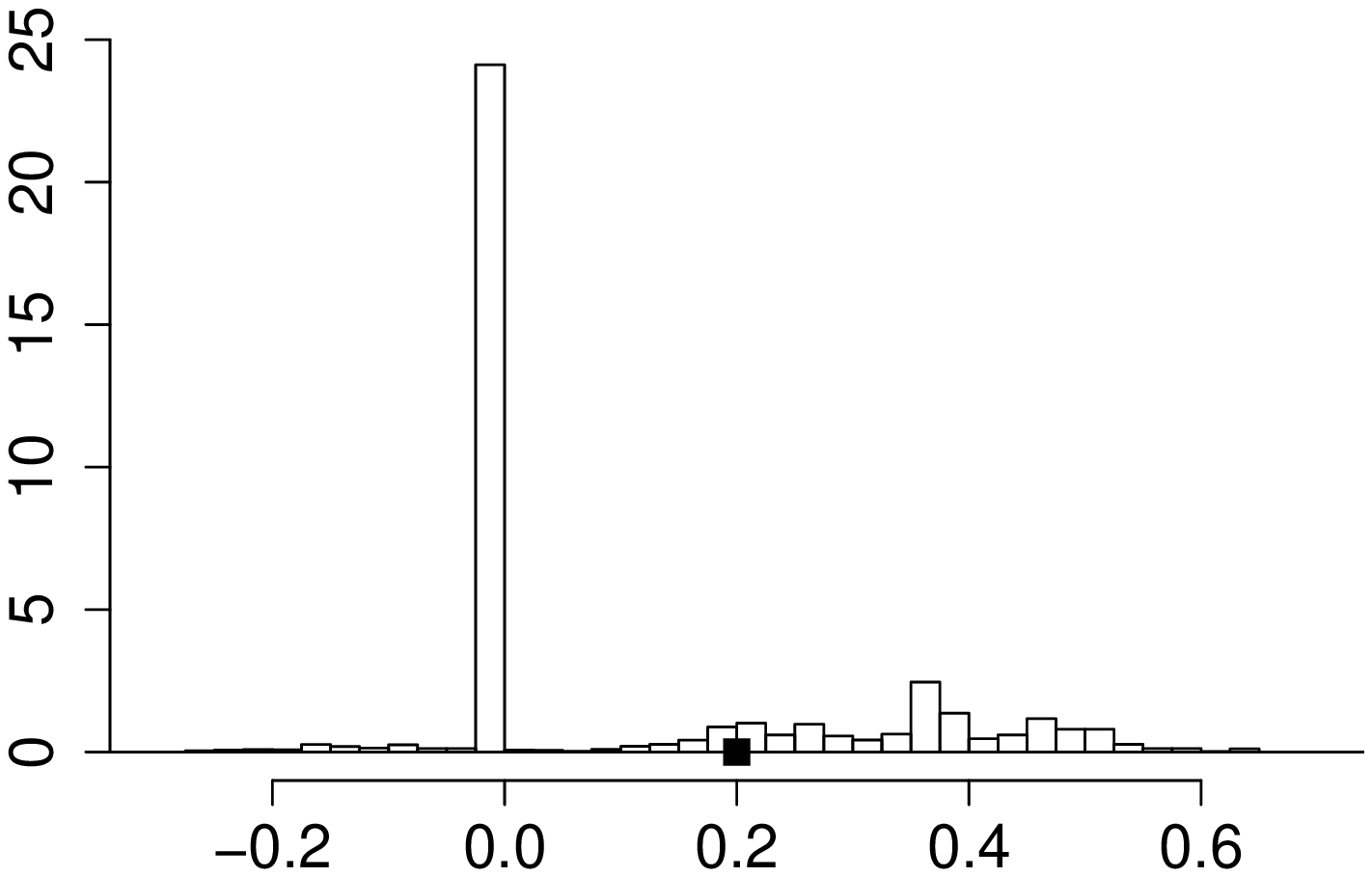}}
\subfigure[][$\phi^{\AAAB}-\phi^{\emptyset}$]{\label{fig:ef}\includegraphics[scale=0.3]{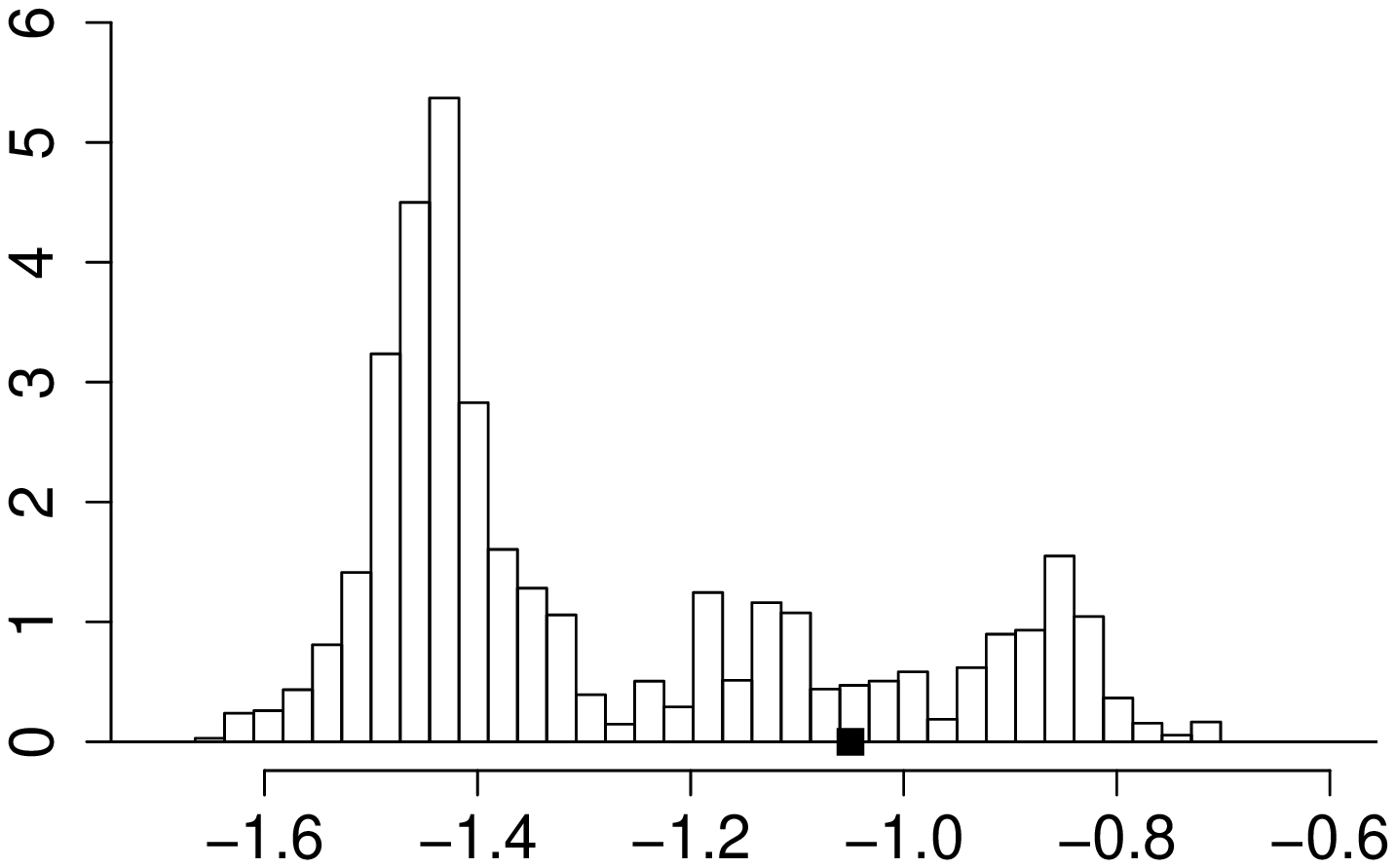}}
\\
\subfigure[][$\phi^{\AABA}-\phi^{\emptyset}$]{\label{fig:eg}\includegraphics[scale=0.3]{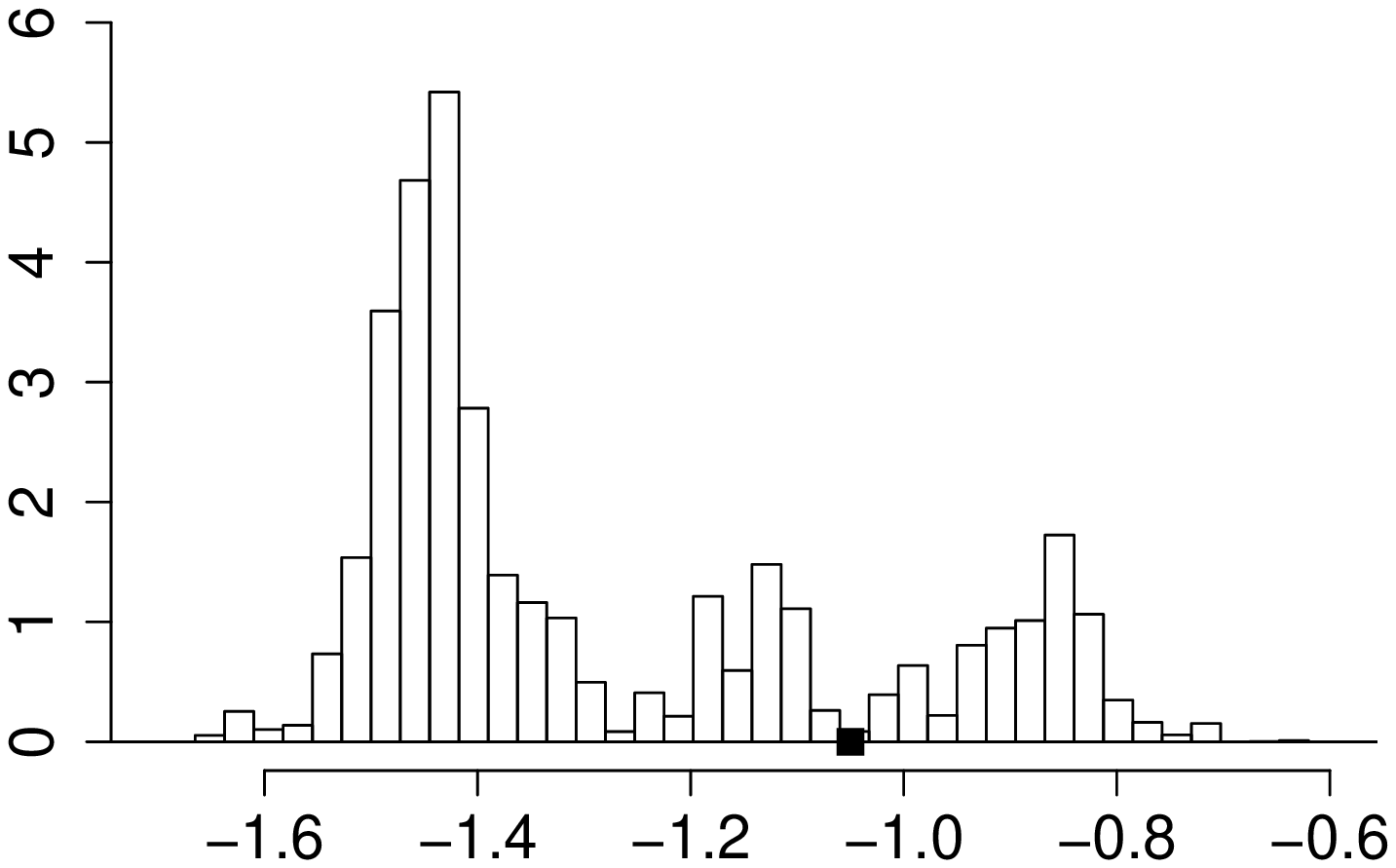}}
\subfigure[][$\phi^{\ABAA}-\phi^{\emptyset}$]{\label{fig:eh}\includegraphics[scale=0.3]{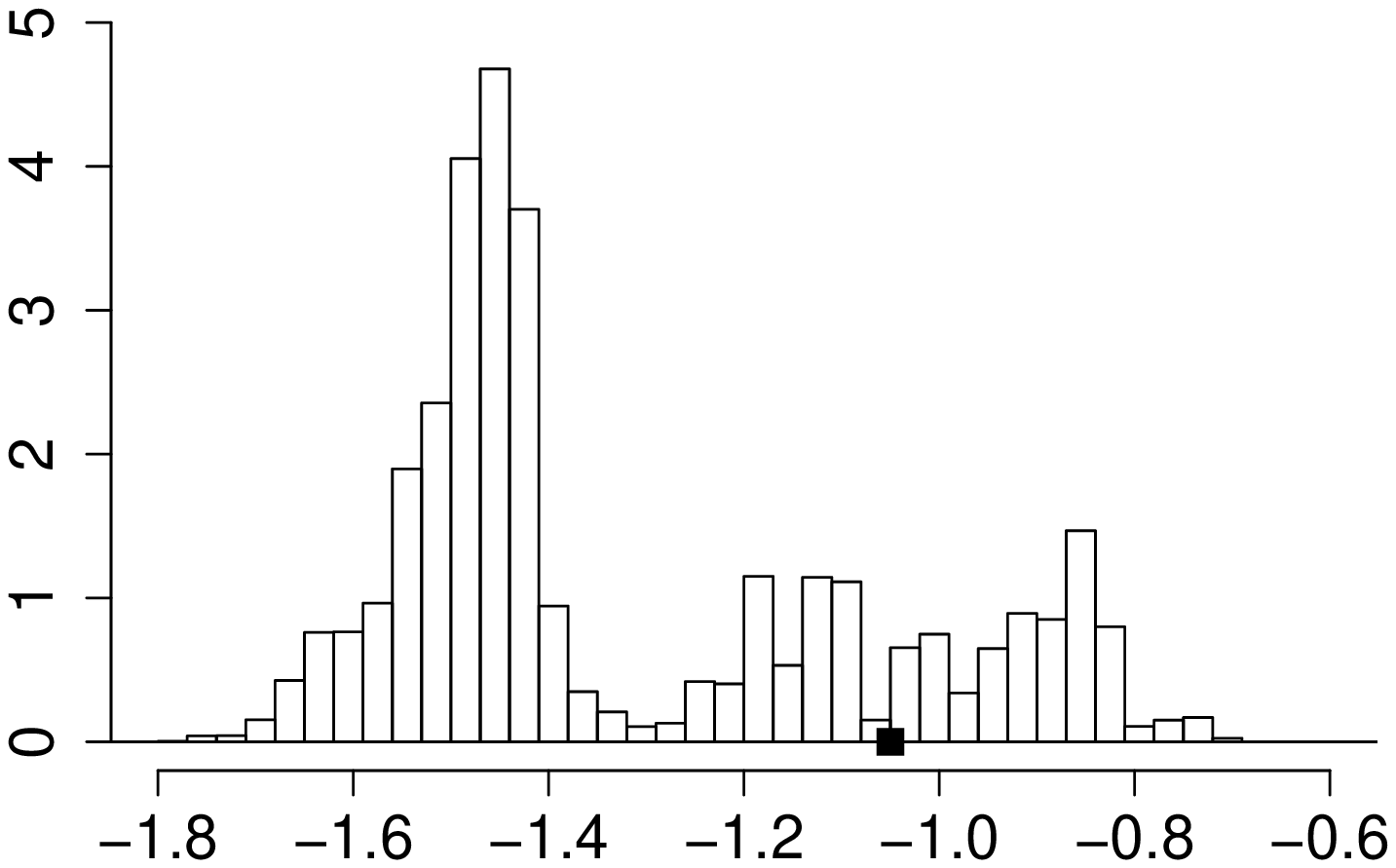}}
\subfigure[][$\phi^{\BAAA}-\phi^{\emptyset}$]{\label{fig:ei}\includegraphics[scale=0.3]{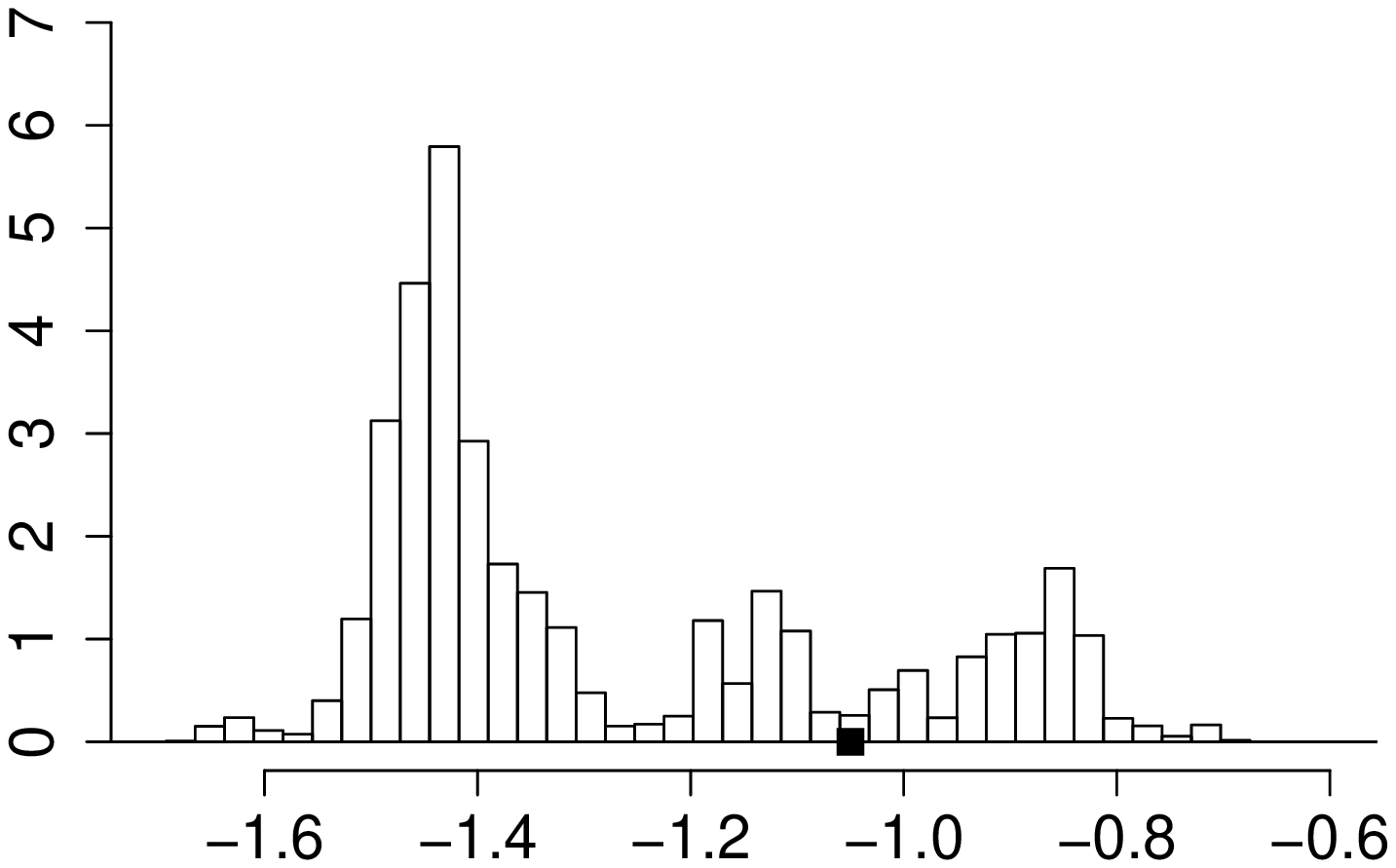}}
\\
\subfigure[][$\phi^{\AAAA}-\phi^{\emptyset} $]{\label{fig:ej}\includegraphics[scale=0.3]{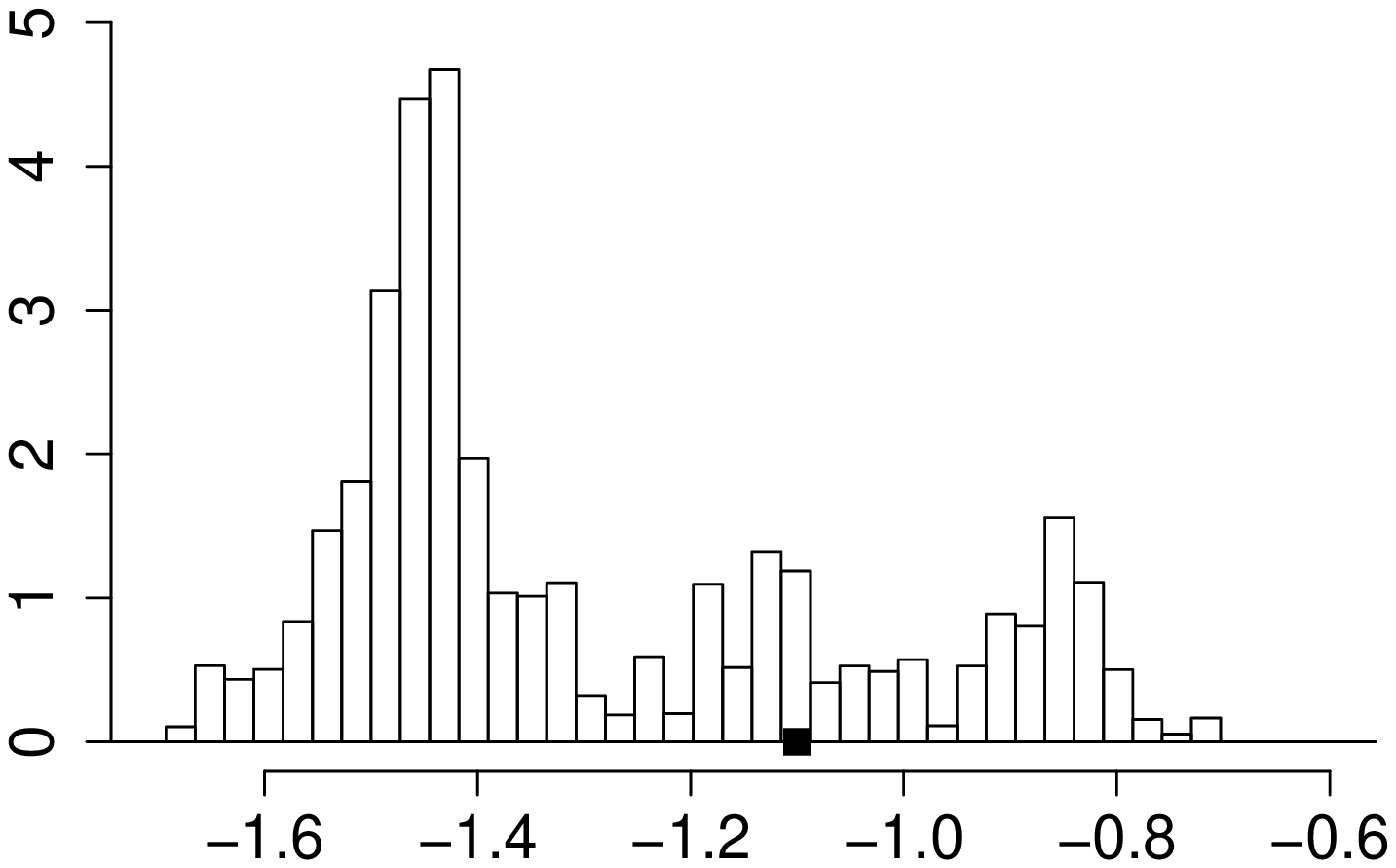}}
\caption{\label{fig:2t2Phi}MRF example with $2\times 2$ cliques: Probability histograms
for the posterior simulated values for $\phi^\Lambda-\phi^{\{\emptyset\}}$ for 
ten values of $\Lambda$. True
values are shown with a black box.}
\end{figure}

Also for this example we simulate from the posterior distribution by 
running our sampler for $10^6$ iterations. Based on convergence plots
we concluded that the sampler had converged
after about $10^5$ iterations. The acceptance rates
for the six proposals for $z$ and $M$ were $0.062, 0.022,
0.023, 10^{-4}, 10^{-4}$ and $10^{-4}$, again given in the same order
as in $p_o$,
while the acceptance rates for $\eta$, $p_\star$ and
$\sigma_\varphi^2$ was $0.203$, $0.333$ and $0.347$, respectively. 
As in the previous example we investigate the posterior distribution
of $M$, $\mathscr{S}$, $\phi^\Lambda-\phi^{\{\emptyset\}}$ for 
$\Lambda\in \{\ABBB,\AABB,\ABAB,\ABBA,\BAAB,\AAAB,\AABA,\ABAA,\BAAA,\AAAA\}$, 
and $\theta=(\eta,p_\star,\sigma_{\varphi}^2)$.
For $M$ we find that the correct model are
simulated with probability 0.99. The estimated most probable partition,
however, differ slightly from the true model. The aposteriori most 
probable partition is 
$\mathscr{S}=\{\{\{\emptyset\},\BAAB\},\{\ABBB,\ABBA\},
\{\AABB,\ABAB\},\{\AAAB,\AABA,\ABAA,\BAAA,\AAAA\}\}$, which has
probability $0.223$.
This partition differs from the true partition in that the 
cell $\{\ABBA,\BAAB\}$ is removed and 
$\ABBA$ and $\BAAB$ are instead 
inserted into the cells $\{\{\emptyset\}\}$ and $\{\ABBB\}$,
respectively, and in that the two cells $\{\AAAA\}$ and 
$\{\AAAB ,\AABA  , \ABAA  , \BAAA \}$ in the true partition 
are merged into one cell. In total eight partitions have probabilities larger
than $0.02$, and these are shown in Table \ref{tab:2times2Partitions}. 
Additional ten partitions have posterior probabilities larger that $0.01$ (not
shown). After the burn-in, the MCMC run visited a total $370$ partitions for 
the most probable state of $M$, and the partition used to generate 
the data set is the aposteriori
75th most probable partition, with an estimated posterior probability of
approximately $10^{-4}$. 
\def\arraystretch{1.0}
\begin{table}
\centering
\begin{tabular}{ c | c | c}
$\hat{P}(\mathscr{S}|M)$ &  $\mathscr{S}$ & $\card(\mathscr{S}$) \\
\hline
0.223 & $\{\{\{\emptyset\},\BAAB\},\{\ABBB,\ABBA\},\{\AABB,\ABAB\},\{\AAAB,\AABA,\ABAA,\BAAA,\AAAA\}\}$ &4\\
0.091 & $\{\{\{\emptyset\},\BAAB\},\{\ABBB,\ABBA\},\{\AABB,\ABAB\},\{\AAAB,\AABA,\BAAA,\AAAA\},\{\ABAA\}\}$&5\\
0.089 &  $\{\{\{\emptyset\},\ABBA\},\{\ABBB\},\{\BAAB\},\{\AABB,\ABAB\},\{\AAAB,\AABA,\ABAA,\BAAA,\AAAA\}\}$ &5\\
0.049 &  $\{\{\{\emptyset\},\ABBB\},\{\ABBA\},\{\BAAB\},\{\AABB,\ABAB\},\{\AAAB,\AABA,\ABAA,\BAAA,\AAAA\}\}$ &5\\
0.046 & $\{\{\{\emptyset\},\ABBB\},\{\ABBA,\BAAB\},\{\AABB,\ABAB\},\{\AAAB,\AABA,\ABAA,\BAAA,\AAAA\}\}$ &4\\
0.023 &  $\{\{\{\emptyset\}\},\{\ABBB\},\{\ABBA\},\{\BAAB\},\{\AABB,\ABAB\},\{\AAAB,\AABA,\ABAA,\BAAA,\AAAA\}\}$& 6\\
0.022 & $\{\{\{\emptyset\},\BAAB\},\{\ABBB,\ABBA\},\{\AABB,\ABAB\},\{\AAAB,\AABA,\ABAA,\AAAA\},\{\BAAA\}\}$ &5\\
0.021 &  $\{\{\{\emptyset\},\BAAB\},\{\ABBB,\ABBA\},\{\AABB,\ABAB\},\{\AAAB,\AABA,\BAAA\},\{\ABAA,\AAAA\}\}$&5\\
\end{tabular}

\vspace*{0.2cm}

 \caption{\label{tab:2times2Partitions} MRF example with $2\times 2$ cliques: 
The aposteriori ten most 
probable partitions.} 
\end{table} 
Looking at the most probable partitions given in Table \ref{tab:2times2Partitions} we notice that except one, all of these partitions have one or two
cells less than that of the true partition. Because of the relatively
week interactions in our data set the posterior prefers partitions
corresponding to a lower number of parameters. However, as shown in
Figure \ref{fig:2t2Phi},
 the true $\phi^\Lambda-\phi^{\{\emptyset\}}$ 
values are still not completely missed by the
posterior.

\begin{figure}
  \centering
  \includegraphics[scale=0.65]{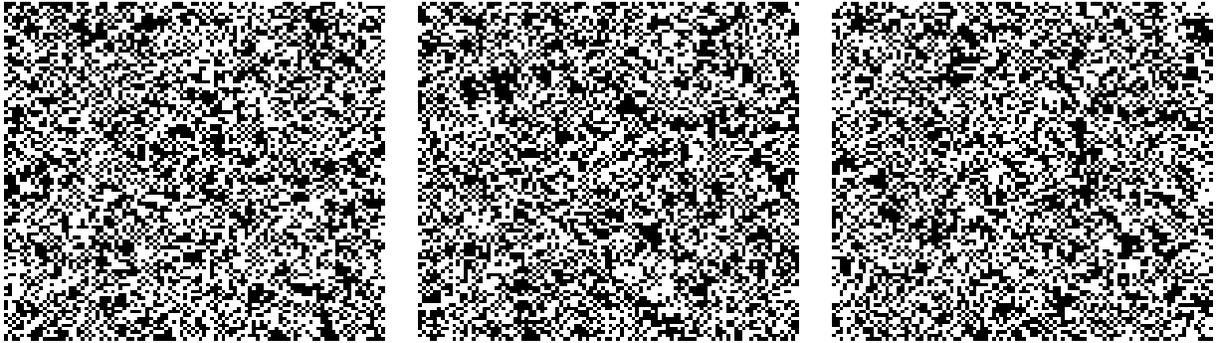} 
  \caption{\label{fig:2times2Real}MRF example with $2\times 2$ cliques: Simulations for three
    parameter realisation.}
\end{figure}

The multi-modality of the probability
histograms for $\phi^\Lambda-\phi^{\{\emptyset\}}$ can be understood. 
For example the spike at zero in Figure \ref{fig:ea} is due to
$\ABBB$ and $\{\emptyset\}$ being assigned to the same
cell. Similarly the spikes at zero in the distributions for $\ABBA$ and $\BAAB$ are the result
of these clique types being assigned in a cell with
$\{\emptyset\}$. Because of the weak interactions in the data, the tendency to
have too few cells in the partition, compared to the true
solution, makes the posterior samples often miss the true value of 
$\phi^\Lambda-\phi^{\{\emptyset\}}$. However one can argue
that this is done without a dramatic decrease in the
likelihood. This last claim can be
investigated for instance by simulating from the likelihood using some
of the model and associated parameters produced by the posterior in
the MCMC run. Three such simulations from the likelihood for three
different parameter settings are shown in
Figure \ref{fig:2times2Real}, where the last of these
realisations is from a parameter setting from the most probable
partition given in Table \ref{tab:2times2Partitions}. 
As we can see the same three phase patterns as in the original data
is present in all of these
simulations. In addition the strength of the interaction between the nodes seems similar as
well. 
 
\begin{figure}
  \centering
\subfigure[][$\eta \ (0.767,2.029)$]{\label{fig:a1}\includegraphics[scale=0.3]{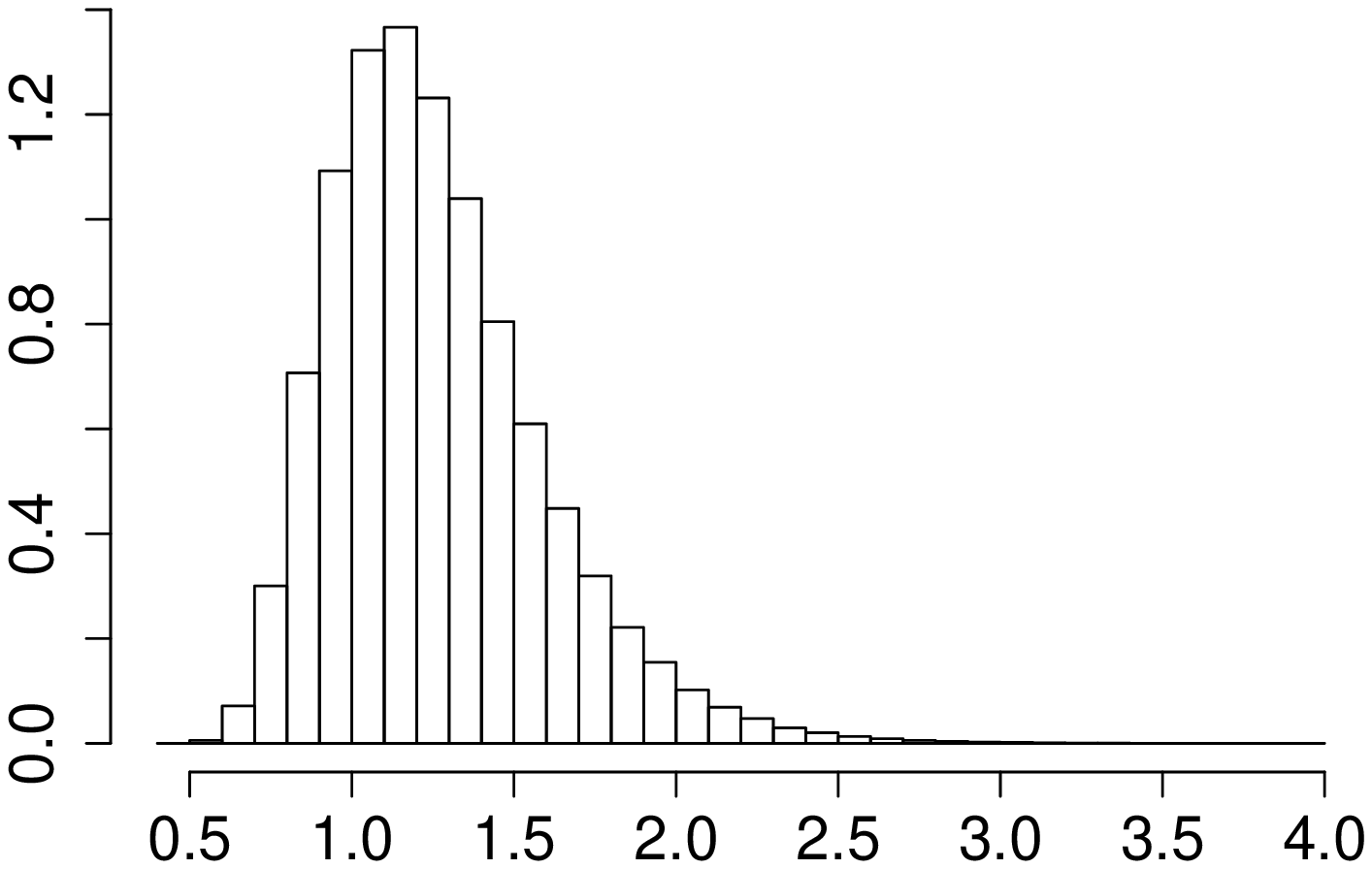}}
\subfigure[][$p_\star \ (0.216,0.994)$]{\label{fig:a2}\includegraphics[scale=0.3]{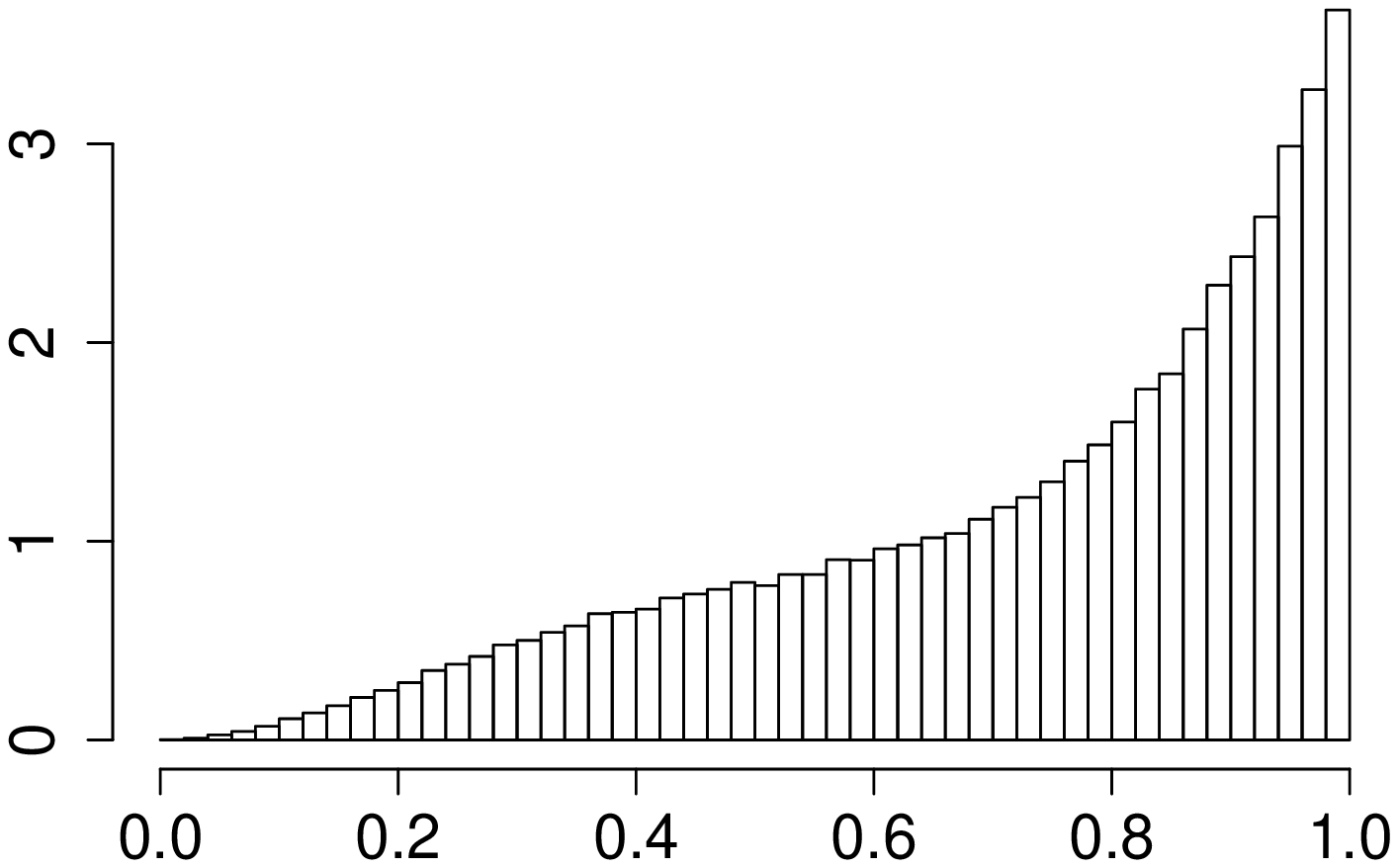}}
\subfigure[][$\sigma_\varphi^2 \ (0.210,10.437)$]{\label{fig:a3}\includegraphics[scale=0.3]{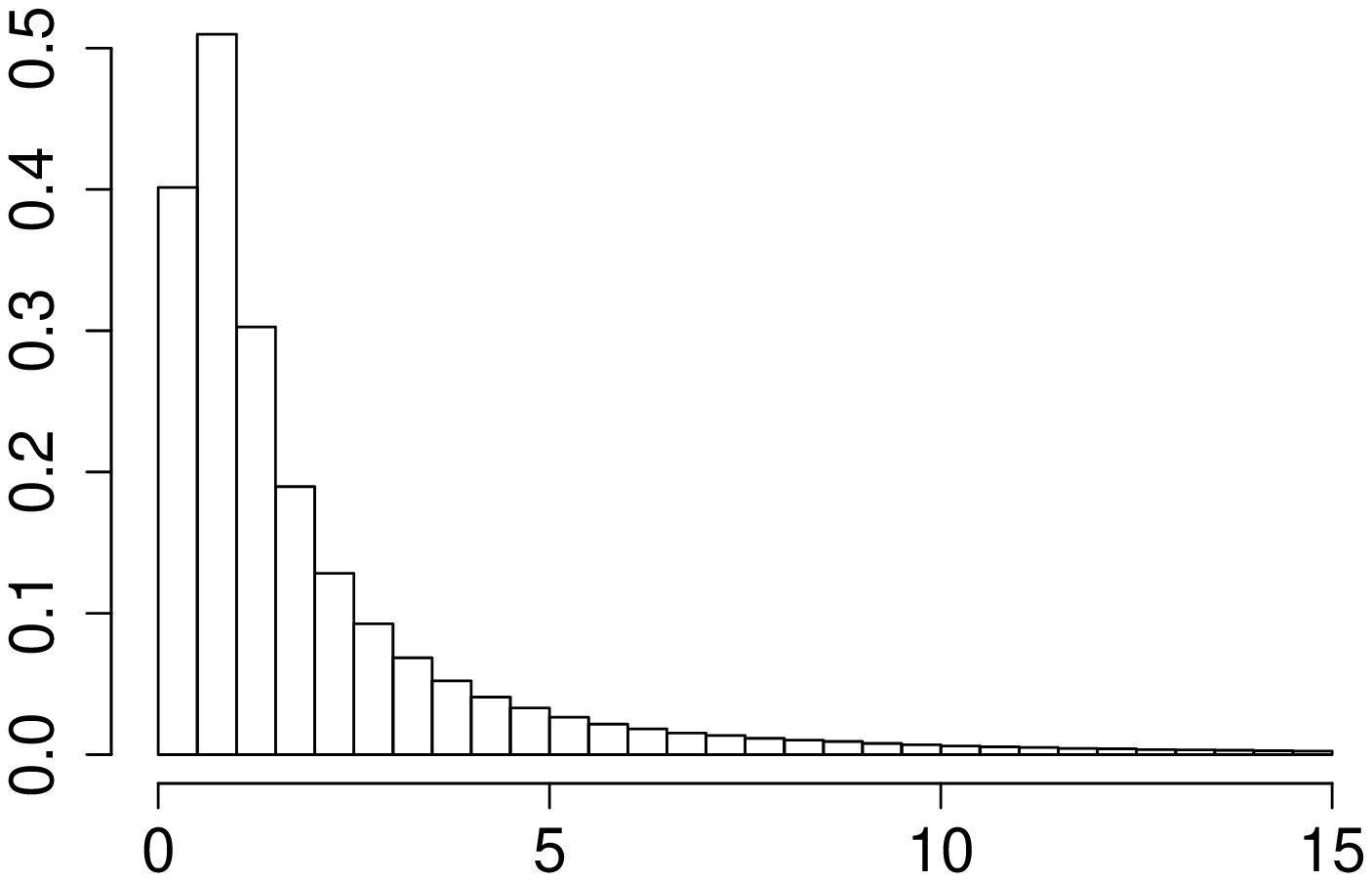}}
\caption{\label{fig:2t2PriorParam}MRF example with $2\times 2$ cliques:
  Prior-parameters with
  95\% credible interval.}
\end{figure}

For the prior parameters $\eta$, $p_\star$, and $\sigma_\varphi^2$ we get
the posterior probability histograms shown in
Figure \ref{fig:2t2PriorParam}.
The posterior median for $\eta$ is $1.207$ in this example.
%, which correponds to an expected number of second order cliques
%equal to about 1.790 with $95\%$ credible interval interval
%(0.477,4.919). 
The estimated posterior median for $p_\star$ is
$0.794$, which is reasonable given that this represents the
probability for turning on clique types $\Lambda$ where $\tau(\Lambda)>2$. 
Thus, the shape of the posterior distribution
have higher probability for values for $p_\star$ that favours
higher order clique types to be on. The estimated median for $\sigma_\varphi^2$ is equal
to $1.133$ in this example.

\subsection{Red deer example}\label{sec:deer}

\begin{figure}
  \centering
  \includegraphics[scale=0.5]{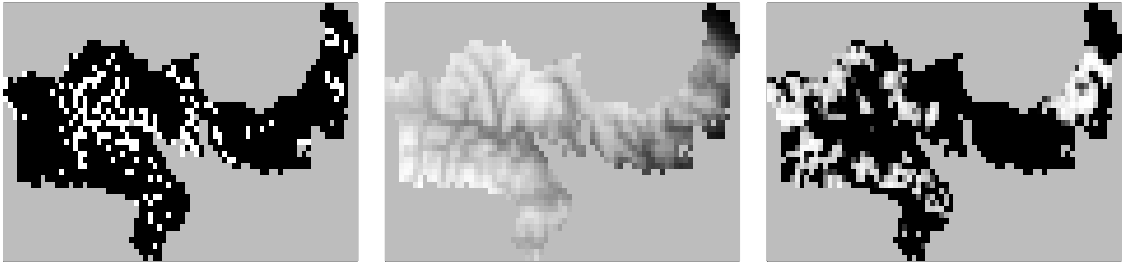} 
  \caption{\label{fig:deerData}Red deer example: The presence/absence of red deer (left),
    altitude (middle), and mires (right) in the Grampians Region of north-east Scotland.}
\end{figure}

In this section we consider a data set consisting of the classes
absence(0) and presence(1) of reed deer in the Grampians Region of
north-east Scotland, see Figure \ref{fig:deerData}.
A full description of the data set is given in \citet{augustin1996} and
\citet{buckland1993}, and the data set has previously been analysed in 
\citet{tjelmeland2012} and \citet{arnesen2013}. Our focus is to analyse
the spatial interaction in the observed values by adopting the prior
specified in Section \ref{sec:prior}.

The observed values do not lie 
on a rectangular lattice, and to cope with this 
we include the observed $x$ into
a larger lattice $x_l=(x,x_b)$ such that $x_l$ is a rectangular lattice
where every element of $x$ is separated from the boundary by at least
one node in $x_b$. For this extended lattice we then assume torus boundary
conditions. In simulation experiments not discussed here we have also tried to 
increase the size of the extended lattice,
but we then obtained essentially the same results as discussed below.

In addition to the binary values shown in Figure 
\ref{fig:deerData}, we also have four covariates available in each of the nodes, namely
altitude, mires, and north and east coordinates. 
We denote the covariate $k$ at a location $(i,j)$ by
$y_{i,j,k}, \ j=1,2,3,4$, and model them into the likelihood by 
\begin{equation*}
p(x_l|z,M,\kappa,y)=Z\exp{\left(
    U(x_l|z,M)+\sum_{(i,j)\in \chi}x_{i,j}\sum_{k=1}^4\kappa_ky_{i,j,k}\right )},
\end{equation*}  
where $U(x_l|z,M)$ is an energy function modelling spatial interaction as discussed in 
Sections \ref{sec:binaryMRF}, and $\kappa=(\kappa_1,...,\kappa_4)$ are parameters 
associated to each of the four covariates. 
On $z$ and $M$, and the associated 
hyper-parameter $\theta$, we assign a prior as discussed in Section 
\ref{sec:prior} and, independent of $z$, $M$ and $\theta$, 
we adopt independent zero mean Gaussian priors with
standard deviation equal to 10 on $\kappa_1,...,\kappa_4$.

The resulting posterior distribution of interest is
\begin{equation*}
p(z,M,\theta,\kappa,x_b|x)\propto p(x_l|z,M,\kappa,y) p(z|M,\theta)p(M|\theta)p(\theta)p(\kappa).
\end{equation*}
When simulating from this distribution we let each iteration consist of one of the 
six updates for $z$ and $M$ discussed in \ref{sec:proposal} or one update of a
randomly chosen element in $\kappa$, a Gibbs scan
for the elements in $x_b$, and updates of the elements in $\theta$ as discussed in 
Section \ref{sec:proposal}. Therefore we increase the
dimension of $p_o$, in this example, to include updates of the elements in $\kappa$ as
well. We set $p_o=(0.1,0.1,0.15,0.15,0.20,0.25,0.05)$ where the first
six entries are the probabilities as defined before and the last entry is
the probability to propose a change for one of the elements in $\kappa$.
To generate a potential new value for a chosen element in $\kappa$
we add to the current value a zero mean normally distributed random 
variable with standard deviation $0.2$. 
For this example we ran the sampler for $10^8$ iterations, and
investigation of convergence plots suggested convergence after about 
$10^5$ iterations. The acceptance rates
for the proposals of $z$, $M$ and $\kappa$ were $0.207, 0.026,
0.043, 0.005, 10^{-3}$, $0.002$ and $0.268$, where these rates are given in the same order
as in $p_o$, and in particular the last acceptance rate is for the
proposal of the covariates $\kappa$. The acceptance rates for $\eta$, $p_\star$ and
$\sigma_\varphi^2$ was $0.212$, $0.349$ and $0.469$ respectively in this example. 

Again we investigate the posterior distribution of $z$ and $M$ after
convergence. In total the chain visited more than $10\ 000$ different states of
$M$ after convergence, and there are nine models with estimated posterior
probability larger that $0.02$, all given in Table \ref{tab:deerDataTab}.
\def\arraystretch{1.6}
\begin{table}
\centering
\begin{tabular}{ c | c | c | c}
$\hat{P}(M)$ &  $M$ with partition
$\mathscr{S}_{\max}=\max_{\mathscr{S}}\hat{P}(\mathscr{S}|M)$ &
$\hat{P}(\mathscr{S}_{\max}|M)$ & Neighbourhood \\
\hline
0.284 & $\{\{\{\emptyset\}\}, \{\ABBB\}, \{ \ABAB ,  \AABB , \ABBA\} , \{ \AABA  ,  \ABAA \}\}$  &
0.870 &  $\neigs$ \\
0.061 & $\{\{\{\emptyset\}\}, \{\ABBB\}, \{\AABB  ,  \ABAB  ,  \cc  ,  \cd\}\}$  &
0.855 &  $\neigfo$ \\
0.061 & $\{\{\{\emptyset\}\}, \{\ABBB\}, \{ \ABAB  ,  \cc  ,  \cd\}\}$  &
0.953 &  $\neigf$ \\
0.041 & $\{\{ \{\emptyset\}\}, \{\ABBB\}, \{\ABBA  ,  \cc  ,  \cd\}\}$  &
0.923 &  $\neigt$ \\
0.030 & $\{\{\{\emptyset\}\}, \{\ABBB\}, \{ \AABB  ,  \ABBA ,  \cd \}\}$ &0.799 &  $\neigni$ \\
0.026 & $\{\{\{\emptyset\}\}, \{\ABBB\}, \{\AABB , \ABAB , \cc  \}\}$  &0.829 & $\neigte$ \\
0.026 & $\{\{\{\emptyset\}\}, \{\ABBB\}, \{\AABB  ,  \ABAB  ,  \cd\}\}$  & 0.797 &  $\neigEleven$ \\
0.025 & $\{\{\{\emptyset\}\}, \{\ABBB\}, \{ \AABB , \cc , \cd\}\}$  &0.901 &  $\neigfi$ \\
0.024 & $\{\{\{\emptyset\}\}, \{\ABBB\}, \{ \ABAB  ,  \ABBA ,  \cc \}\}$  &0.758 &  $\neigse$ \\
\end{tabular}

\vspace*{0.2cm}

 \caption{\label{tab:deerDataTab} Red deer example: The realisations of $M$ with 
posterior probability higher than $0.02$. First column:
   Estimated posterior probability of a specific $M$. Second column:
   Partition $\mathscr{S}_{\max}$ with the highest posterior probability
   for the given $M$. Third column: Estimated posterior
   probability for $\mathscr{S}_{\max}$ given $M$. Last column:
   Neighbours (shown with $\times$) for a node (shown with $\cdot$) 
   resulting from $M$.}
\end{table} 
Together with posterior probabilities for $M$ and the most probable
partition under $M$, the
associated neighbourhoods are also shown in this table. Note that all
the neighbourhoods in this table suggest some anisotropy from
up-left to down-right.  In addition to the nine states in Table
\ref{tab:deerDataTab}, $12$ additional states for $M$ have probabilities 
larger that $0.01$.

\begin{figure}
  \centering
\subfigure[][$\phi^{\ABBB}-\phi^{\{\emptyset\}}$]{\label{fig:c2}\includegraphics[scale=0.3]{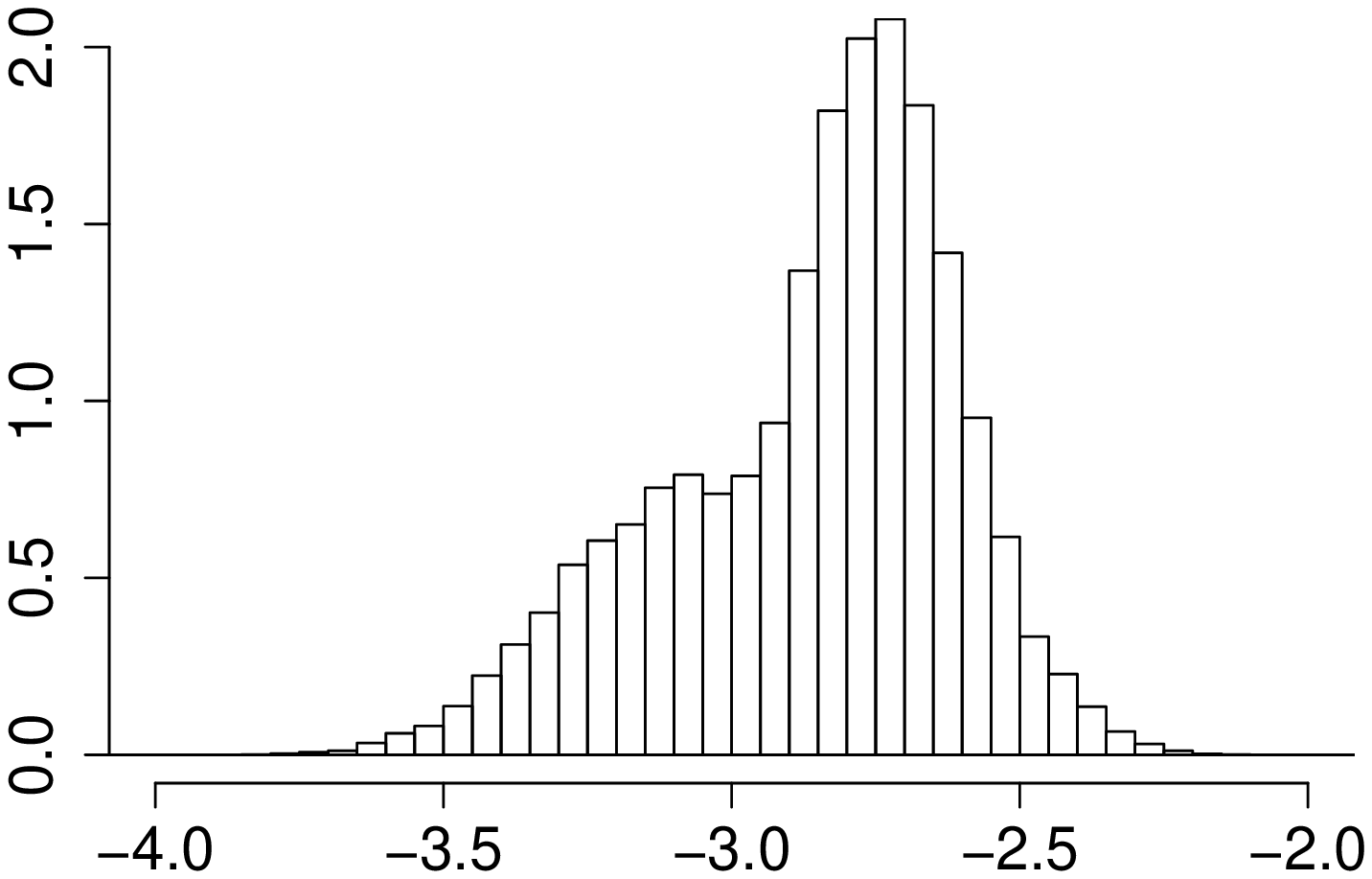}}
\subfigure[][$\phi^{\ABAB}-\phi^{\{\emptyset\}}$]{\label{fig:c2}\includegraphics[scale=0.3]{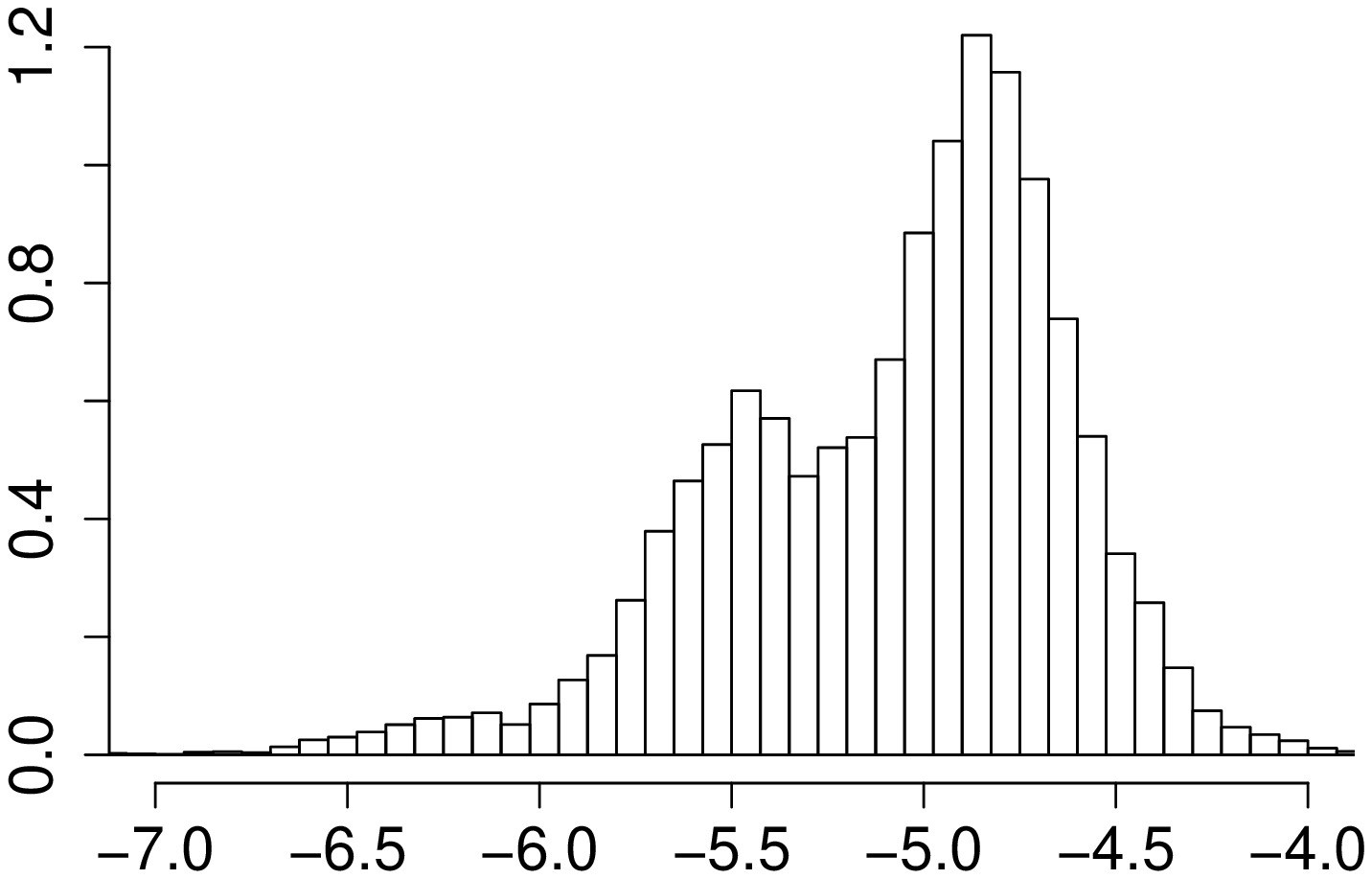}}
\subfigure[][$\phi^{\AABB}-\phi^{\{\emptyset\}}$]{\label{fig:c3}\includegraphics[scale=0.3]{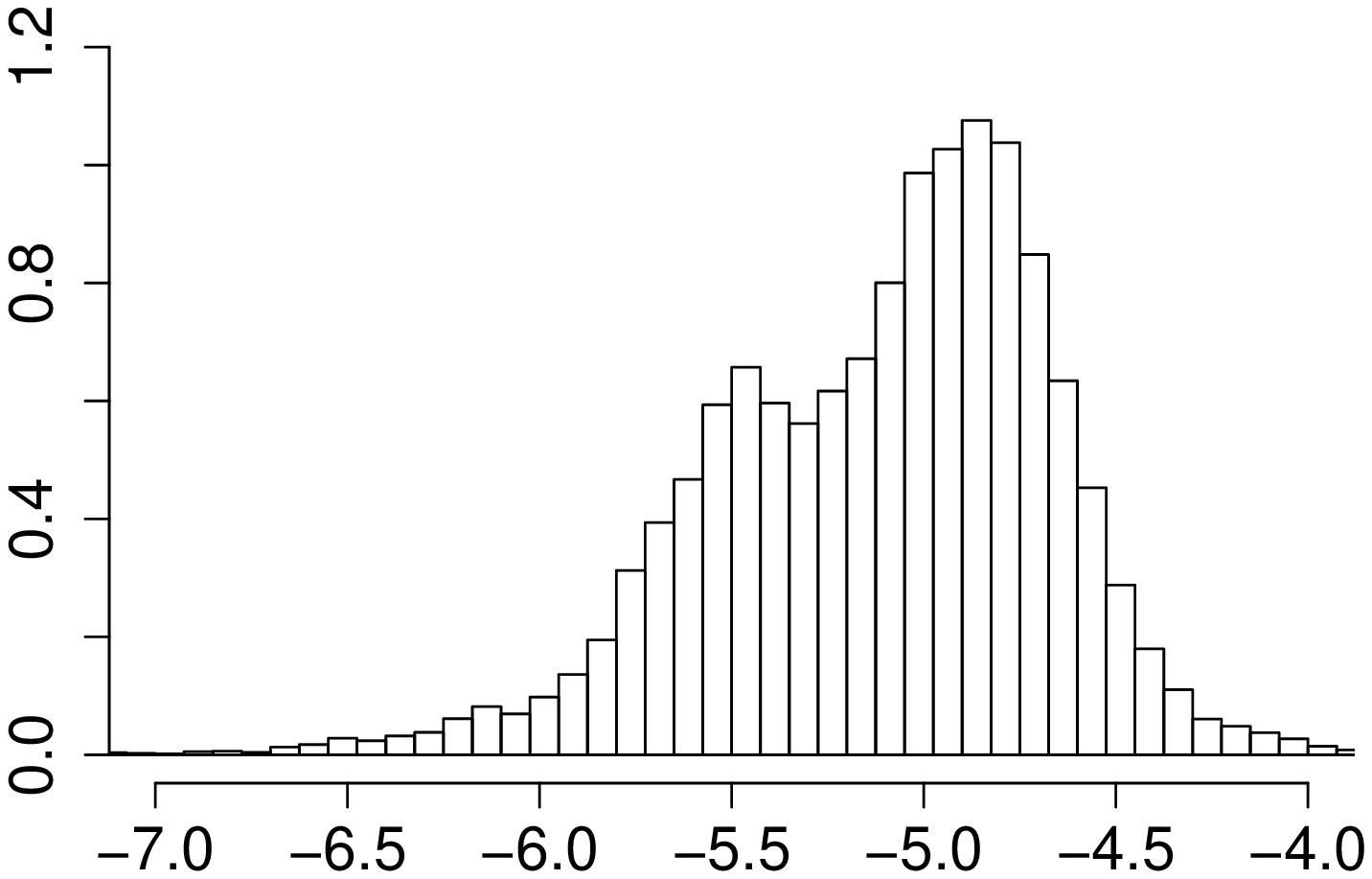}}\\
\subfigure[][$\phi^{\ABBA}-\phi^{\{\emptyset\}}$]{\label{fig:c3}\includegraphics[scale=0.3]{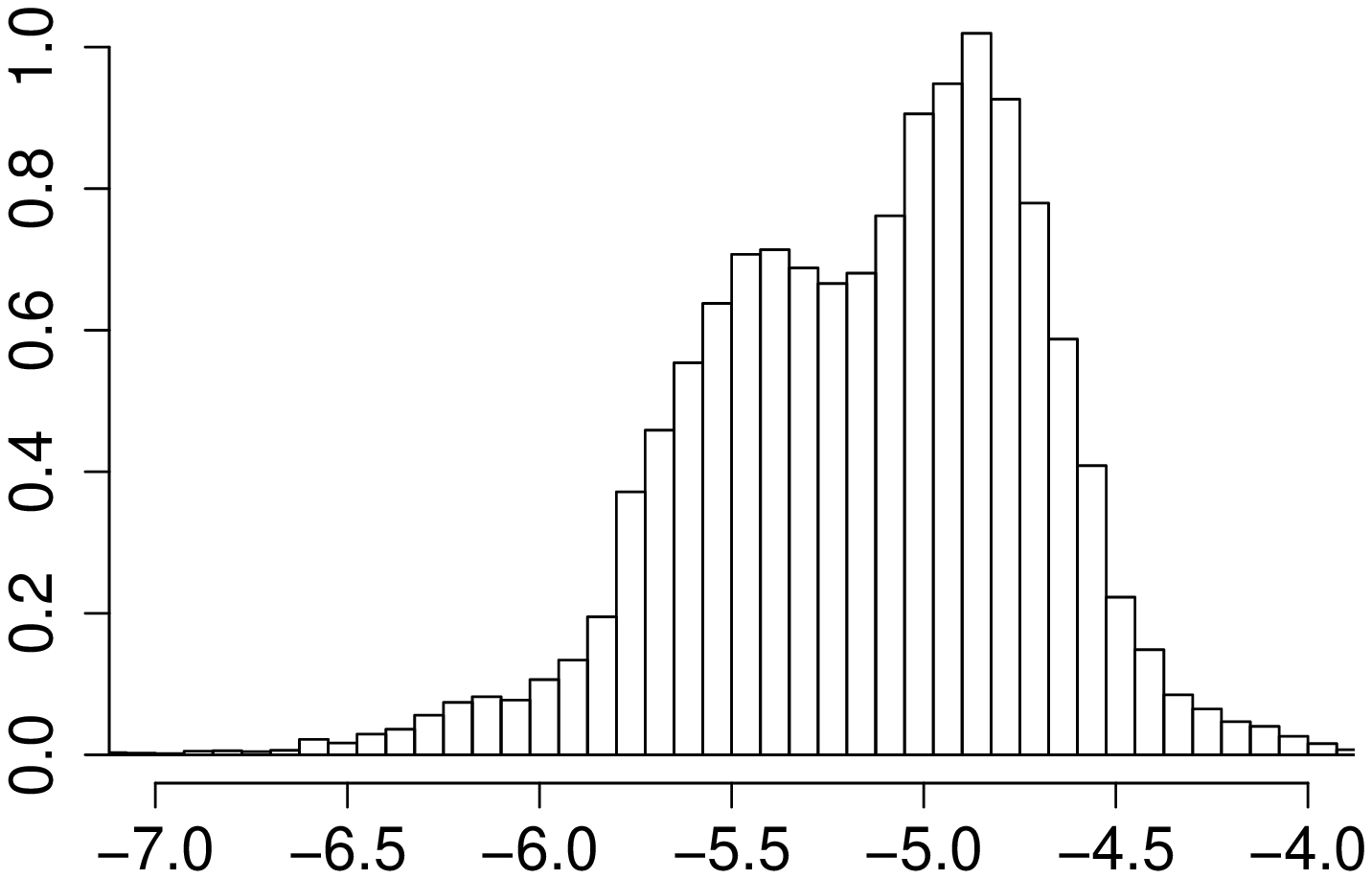}}
\subfigure[][$\phi^{\BAAB}-\phi^{\{\emptyset\}}$]{\label{fig:c2}\includegraphics[scale=0.3]{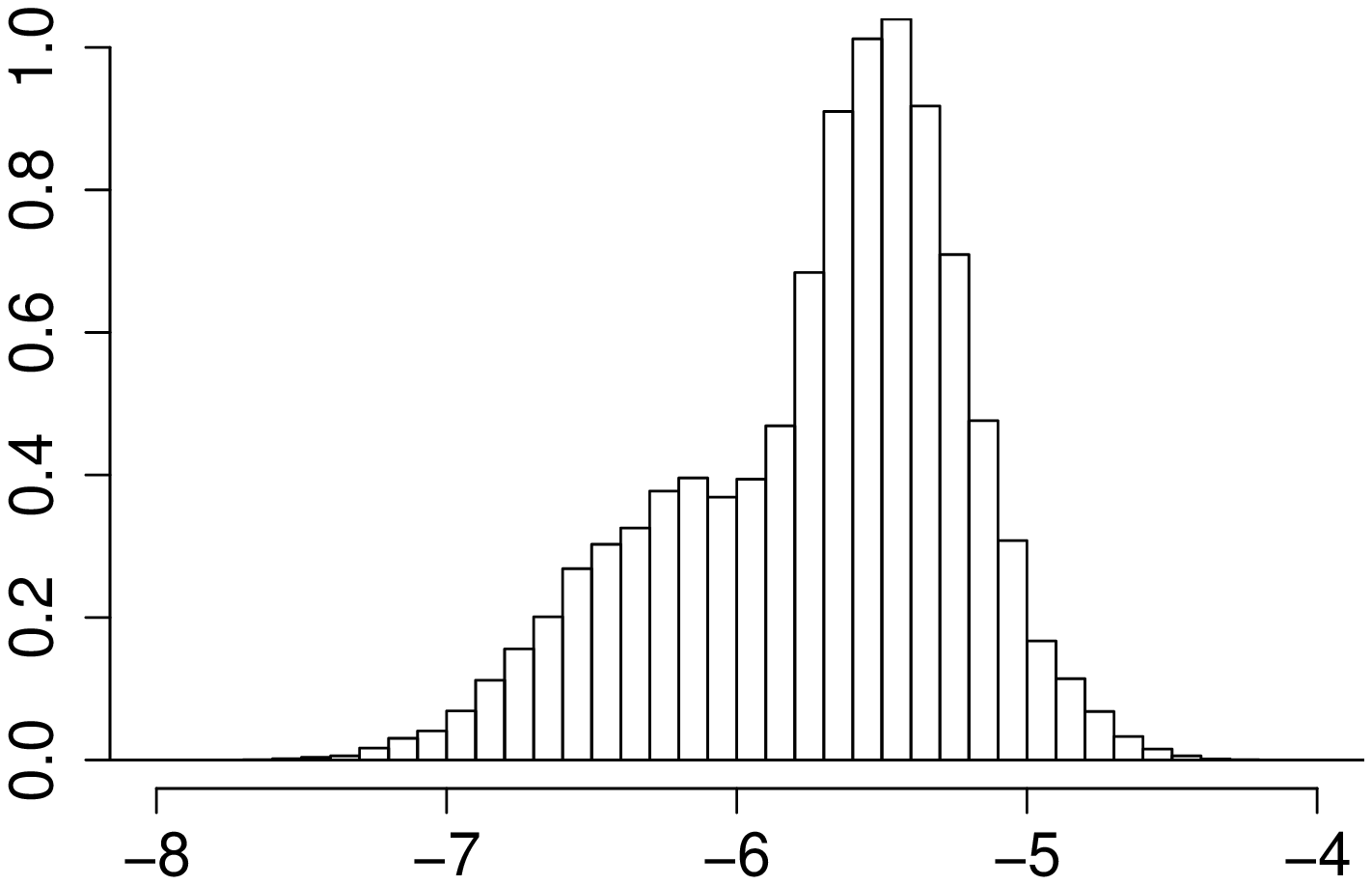}}
\subfigure[][$\phi^{\AAAB}-\phi^{\{\emptyset\}}$]{\label{fig:c2}\includegraphics[scale=0.3]{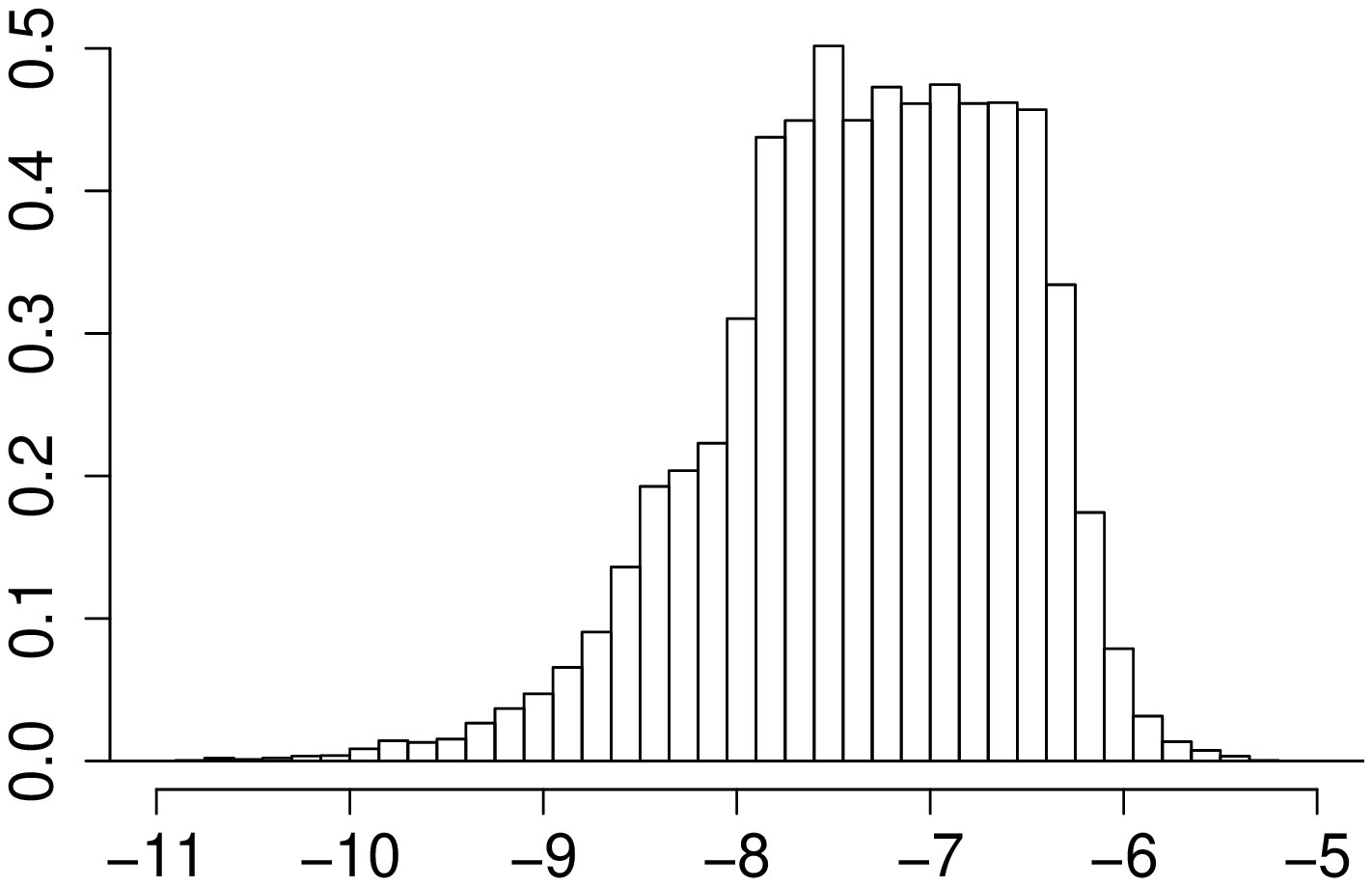}}\\
\subfigure[][$\phi^{\AABA}-\phi^{\{\emptyset\}}$]{\label{fig:c3}\includegraphics[scale=0.3]{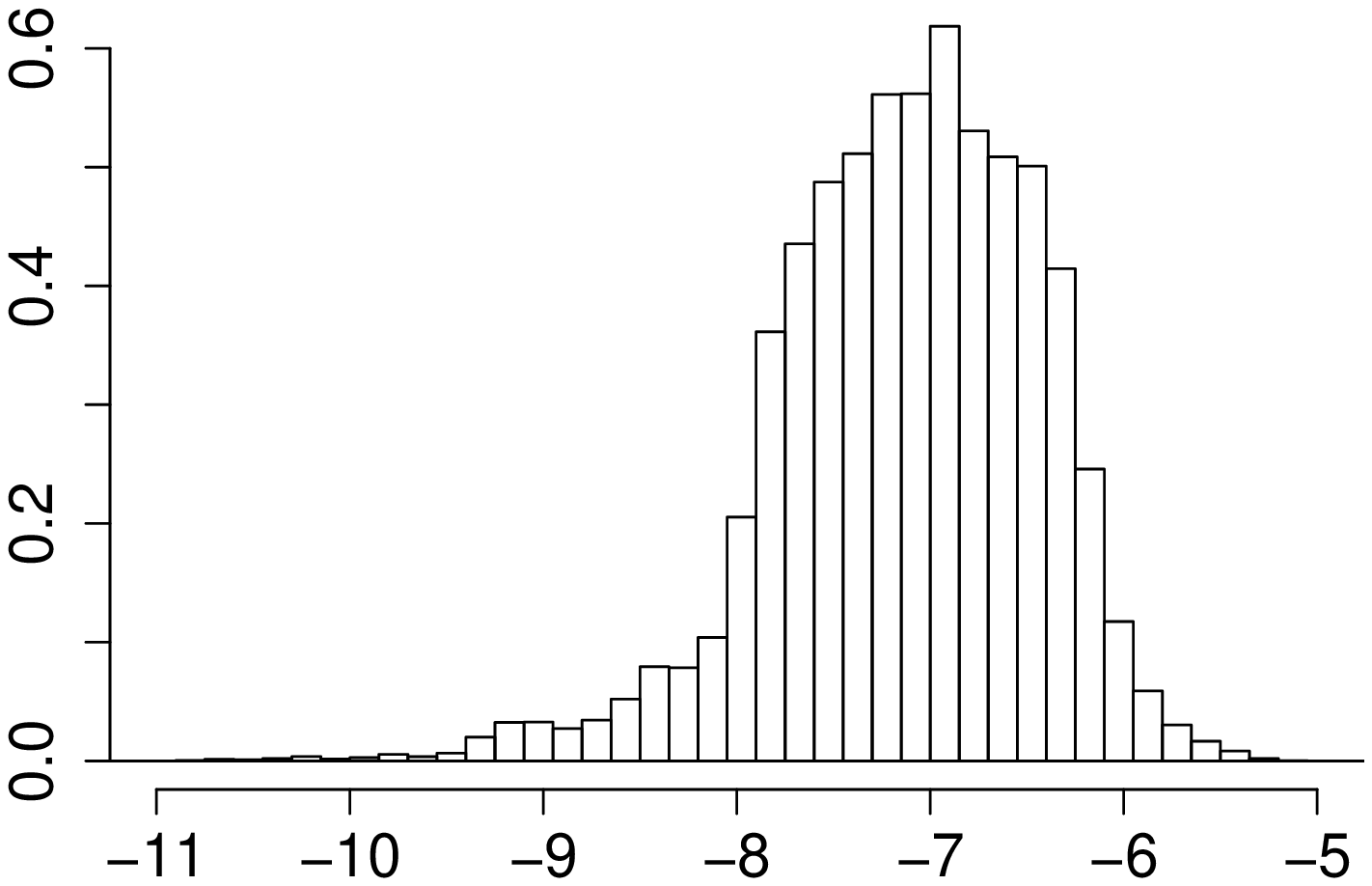}}
\subfigure[][$\phi^{\ABAA}-\phi^{\{\emptyset\}}$]{\label{fig:c3}\includegraphics[scale=0.3]{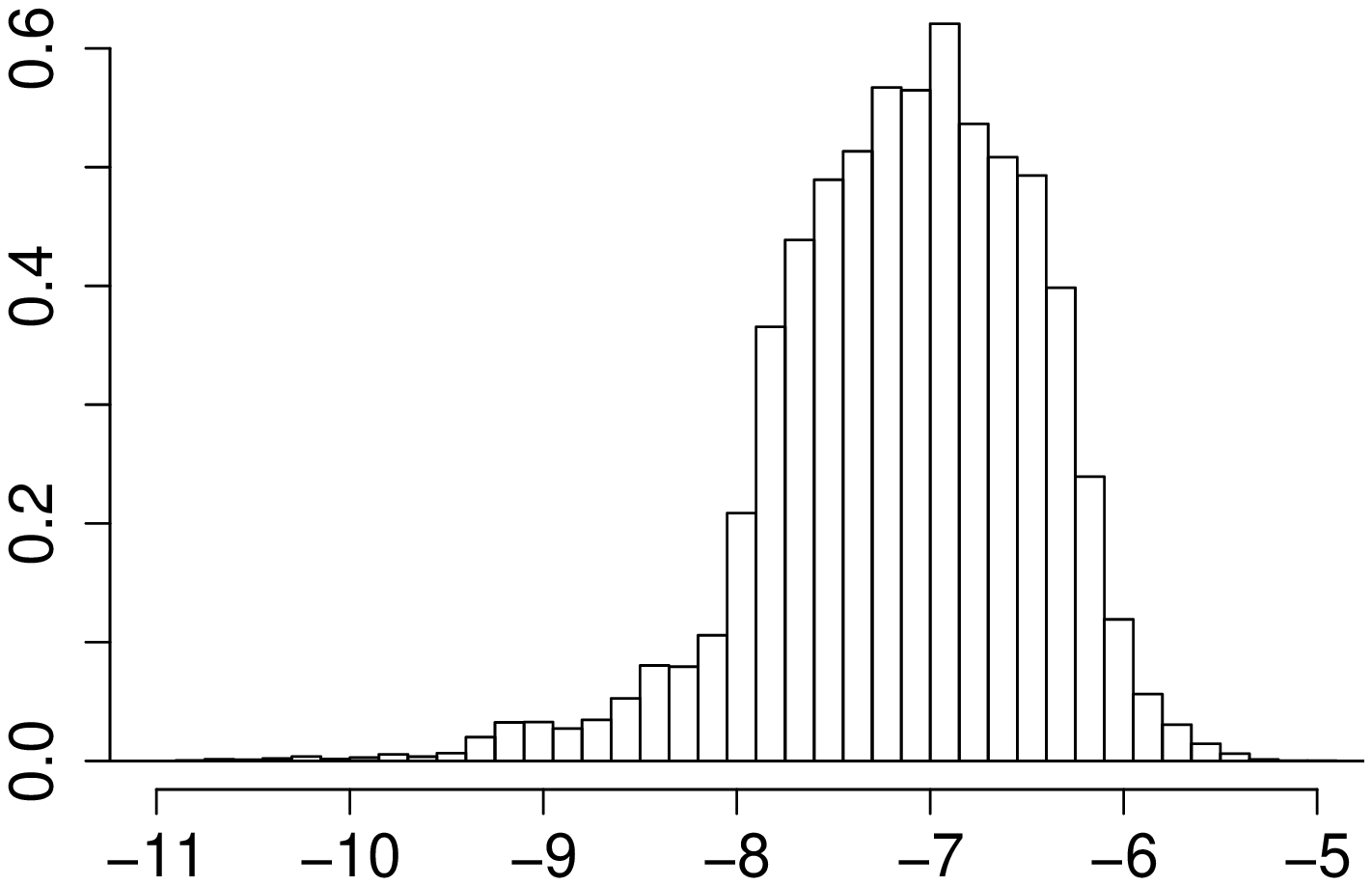}}
\subfigure[][$\phi^{\BAAA}-\phi^{\{\emptyset\}}$]{\label{fig:c3}\includegraphics[scale=0.3]{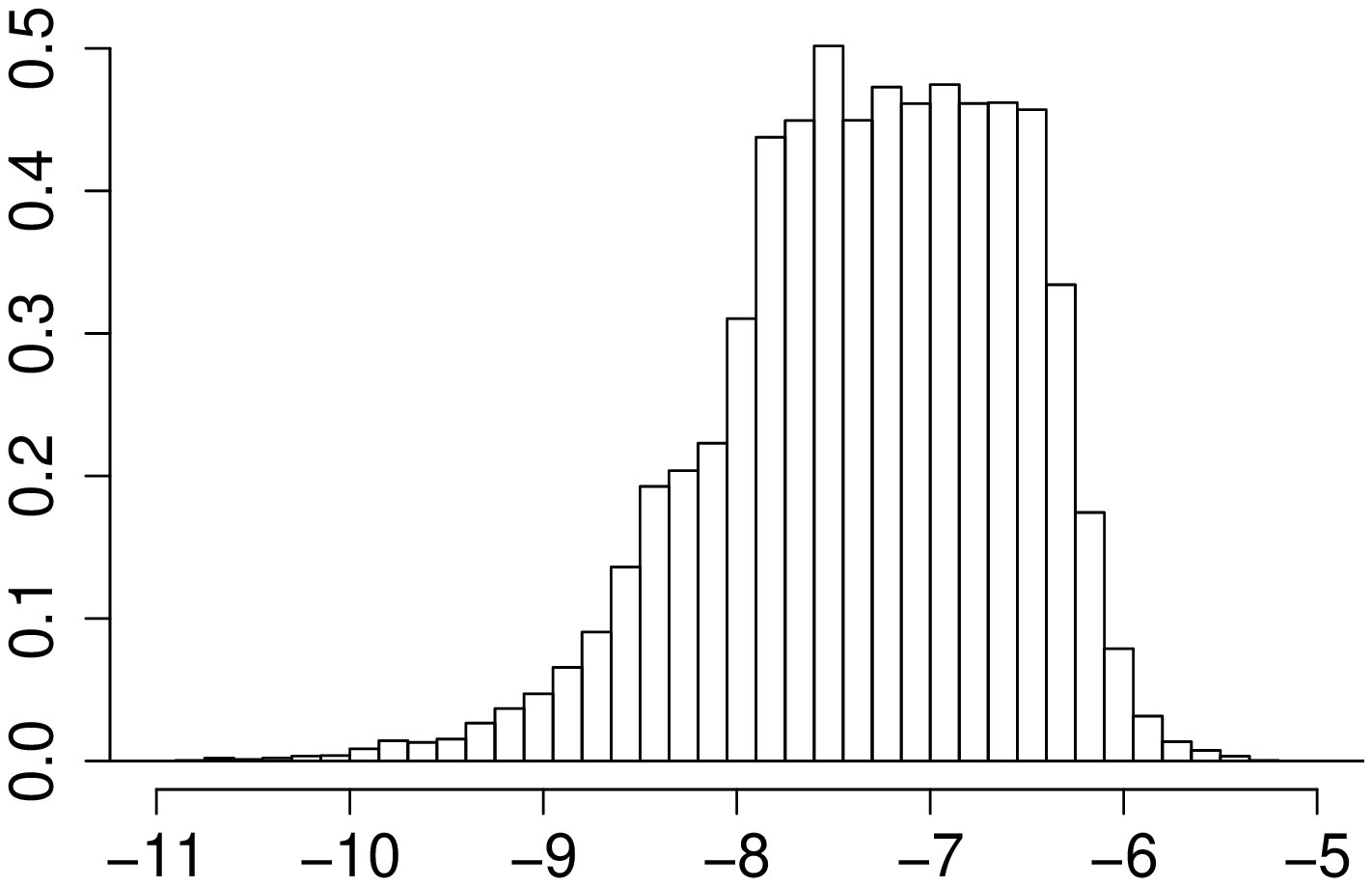}}\\
\subfigure[][$\phi^{\AAAA}-\phi^{\{\emptyset\}}$]{\label{fig:c3}\includegraphics[scale=0.3]{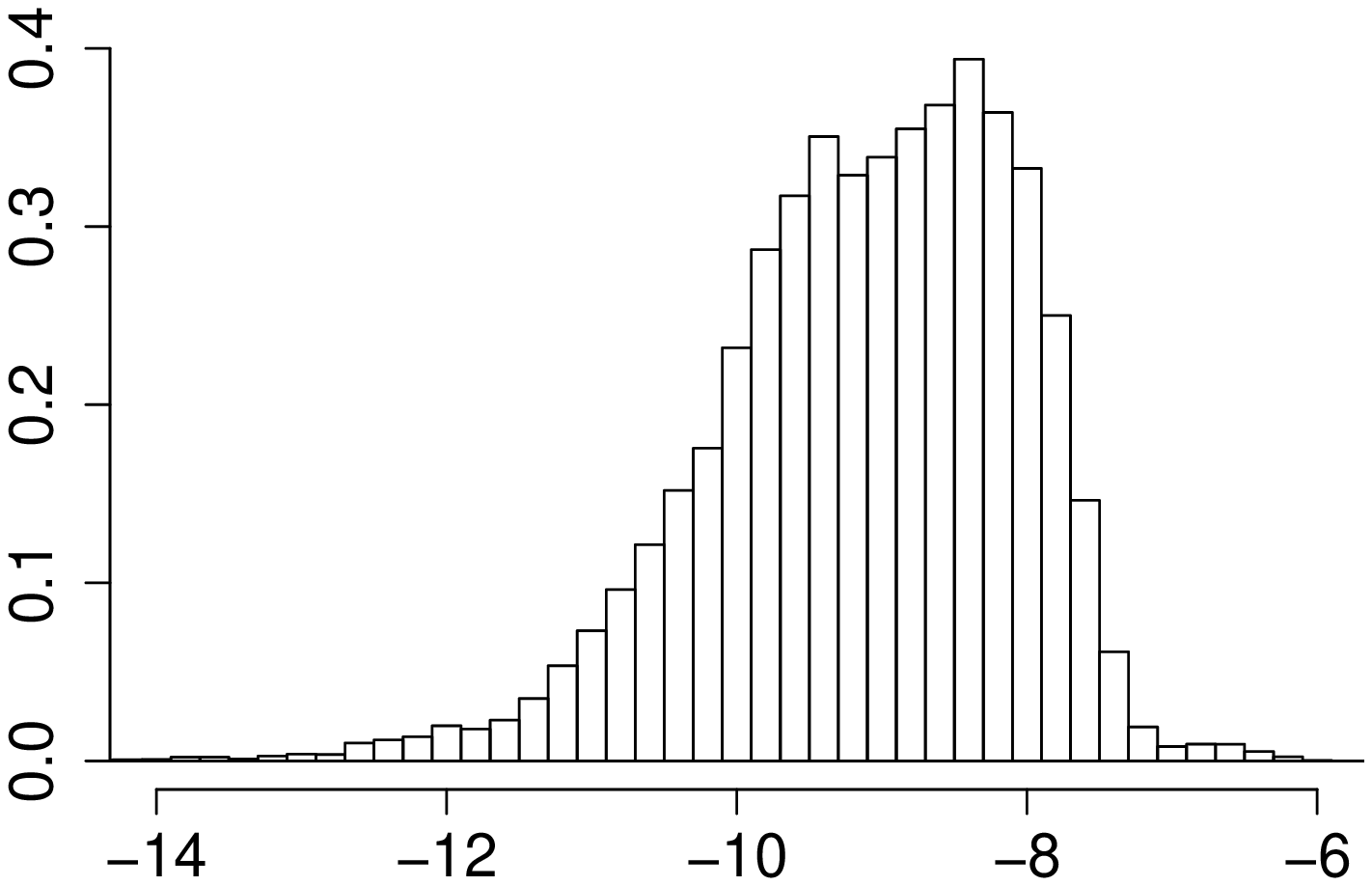}}
\subfigure[][$\phi^{\cc}-\phi^{\{\emptyset\}}$]{\label{fig:c4}\includegraphics[scale=0.3]{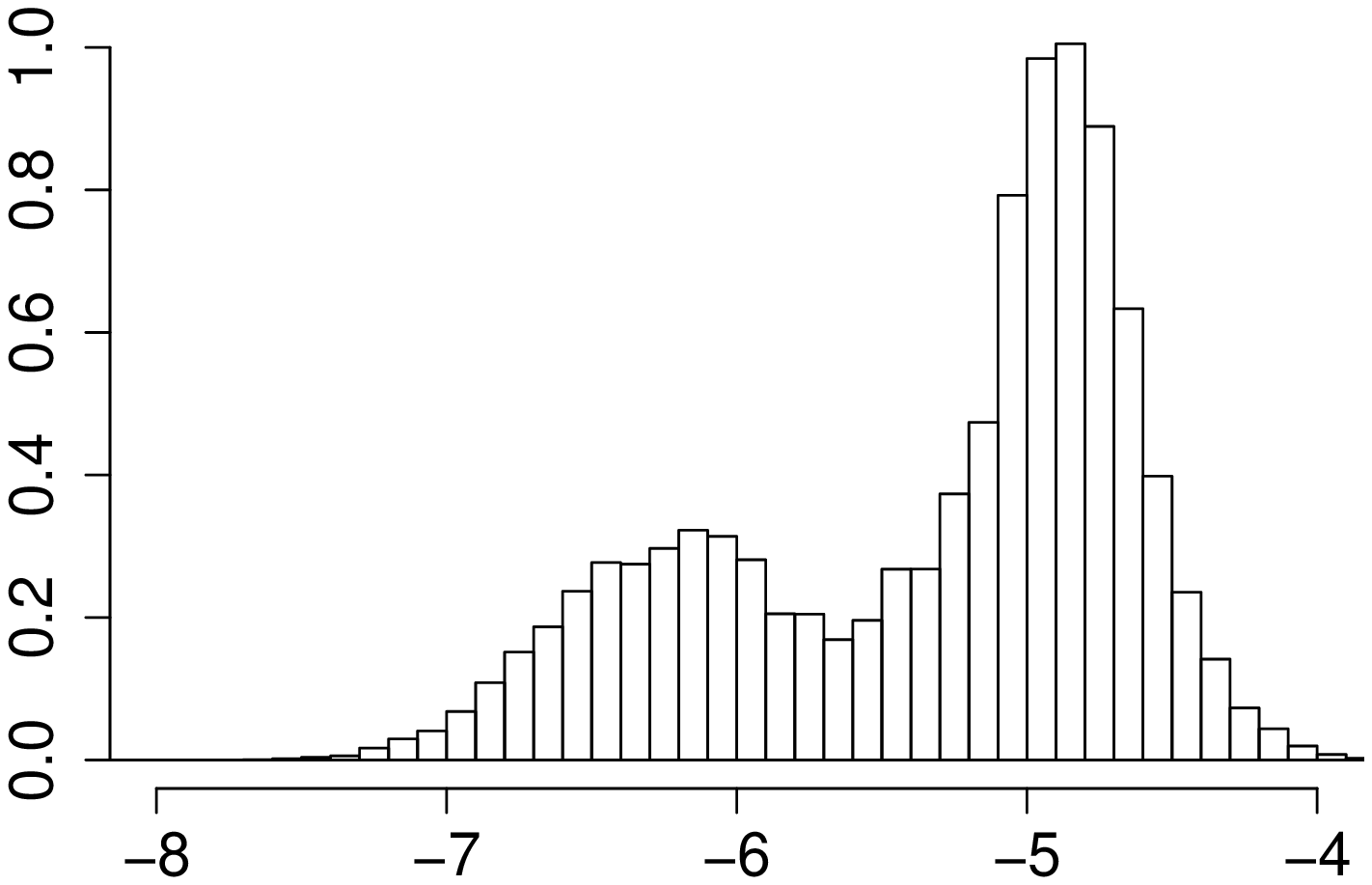}}
\subfigure[][$\phi^{\cd}-\phi^{\{\emptyset\}}$]{\label{fig:c3}\includegraphics[scale=0.3]{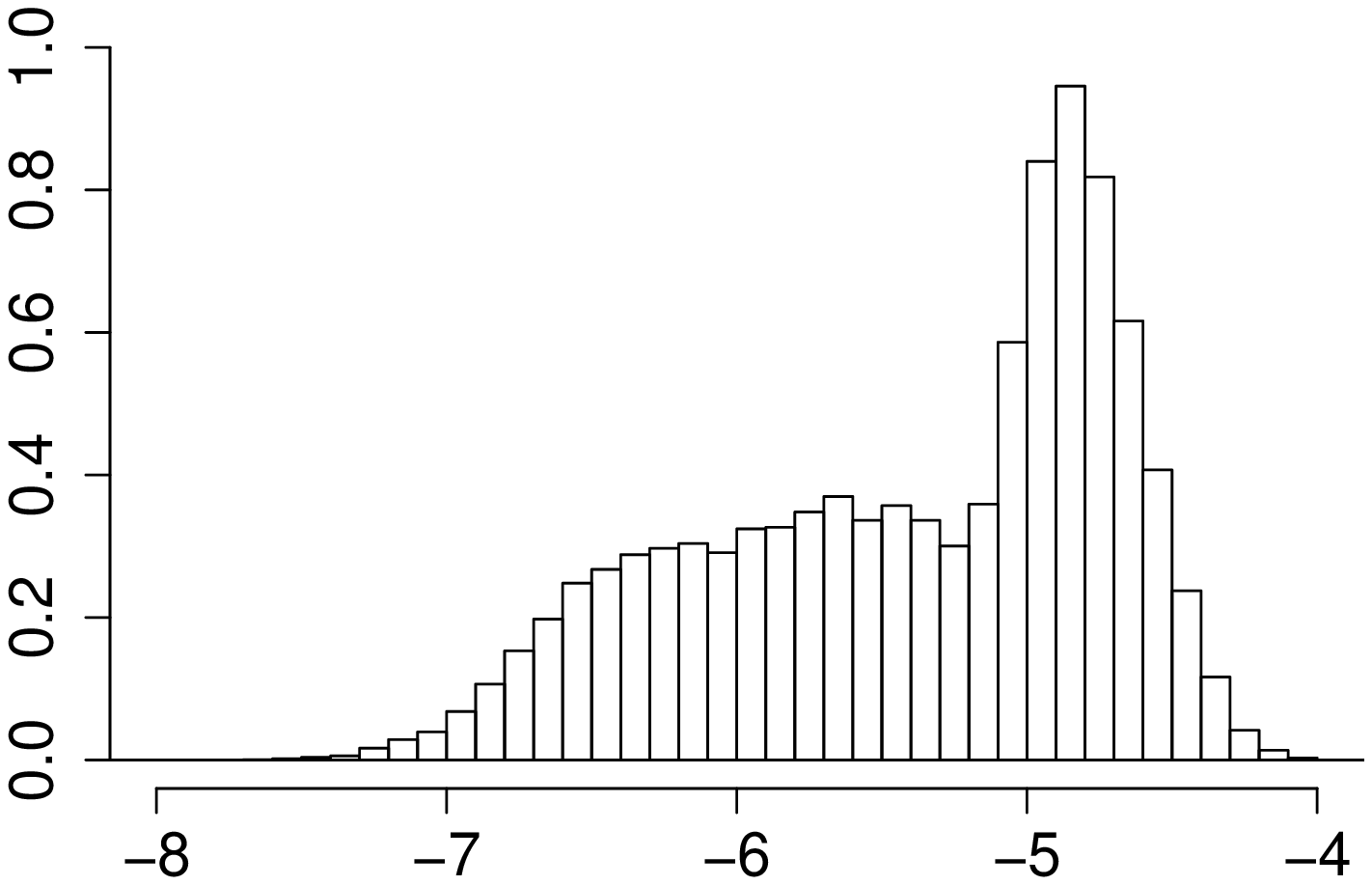}}
\caption{\label{fig:deerPhi}Red deer example: Probability histograms for the posterior
simulated values for $\phi^\Lambda-\phi^{\{\emptyset\}}$ for twelve values of $\Lambda$.}
\end{figure}

The estimated distribution for $\phi^\Lambda-\phi^{\{\emptyset\}}$ for the same $\Lambda$s as considered in the 
simulation examples, and in addition for $\Lambda \in \{\cc,\cd\}$, are shown in Figure \ref{fig:deerPhi}.  
Several of these histograms are bimodal. In Section S.2 in the supplementary material
we investigate this bimodality more in detail. In particular we show that the bimodality
can be explained by splitting the posterior
realisations into two groups. If we consider only realisations under the
most probable posterior set of clique types $M=\{\{\emptyset\},\ABBB,
\AABB,\ABAB,\ABBA,\AABA,\ABAA\}$, see Table
\ref{tab:deerDataTab}, we get unimodal distributions for the $\phi^\Lambda$
values where its mode belongs to the first mode in Figures
\ref{fig:deerPhi}(a), (e), (k) and  (l), and to the second mode in the
rest of these plots. The second group of realisations consists of the remaining 
realisations, and in fact the distribution of $\phi^\Lambda$ for this group
is again bimodal for some $\phi^\Lambda$. This bimodality can in turn
by explained by looking at the realisations separately for $\phi^\Lambda$ when
$\Lambda$ is either on or off, again see Section S.2 in the supplementary 
materials for details. It is also worth mentioning that most of the most 
probable set of clique types with posterior probability
smaller than $0.02$, i.e. the most probable realisations of $M$ not shown i Table
\ref{tab:deerDataTab}, are states containing different combinations of
the second order clique types $\AABB$, $\ABAB$, $\ABBA$, $\cc$ and
$\cd$, which are the ones also used for the most
probable set of clique types given in Table \ref{tab:deerDataTab}. In fact, 
the posterior probability that $M$ is a subset of
$\{\{\emptyset\},\ABBB, \AABB, \ABAB, \ABBA, \cc, \cd\}$ but not equal to 
$\{\{\emptyset\}\}$ or $\{\{\emptyset\},\ABBB\}$ is as high 
as $0.307$. Also, the posterior probability
that at least one clique type from the set $\{\AABB, \ABAB, \ABBA,
\cc, \cd\}$ is included in $M$ is very close to unity. In
other words, these second order clique types seem to be the building blocks of
the dependency structure in this data set.

\begin{figure}
  \centering
  \includegraphics[scale=0.5]{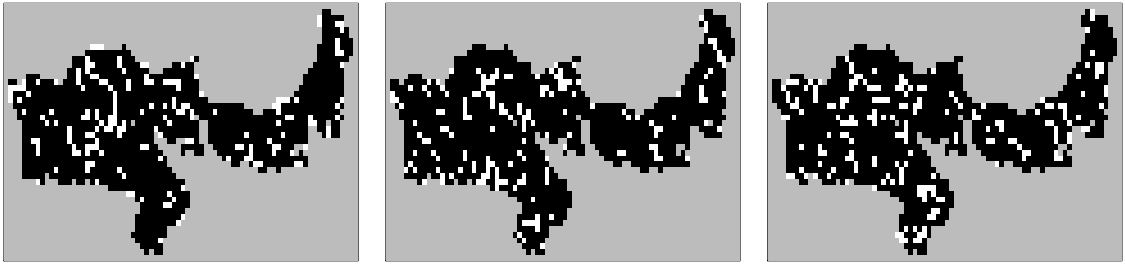} 
  \caption{\label{fig:deerReal}Red deer example: Simulations for three
    parameter realisation.}
\end{figure}

To investigate the fit of the likelihood to the data we did the following small
experiment, similar to what we did for the $2\times 2$ clique MRF example. For 
three randomly chosen realisations of $M$,
$z$, $x_b$ and $\kappa$ we simulated from the likelihood three
realisations of $x$. The results are shown in Figure \ref{fig:deerReal},
and we see that these realisation, at least visually, seem to have the same
spatial dependence structure as the data in Figure \ref{fig:deerData}.  

\begin{figure}
  \centering
\subfigure[][$\eta \ (0.739,2.357)$]{\label{fig:a1}\includegraphics[scale=0.3]{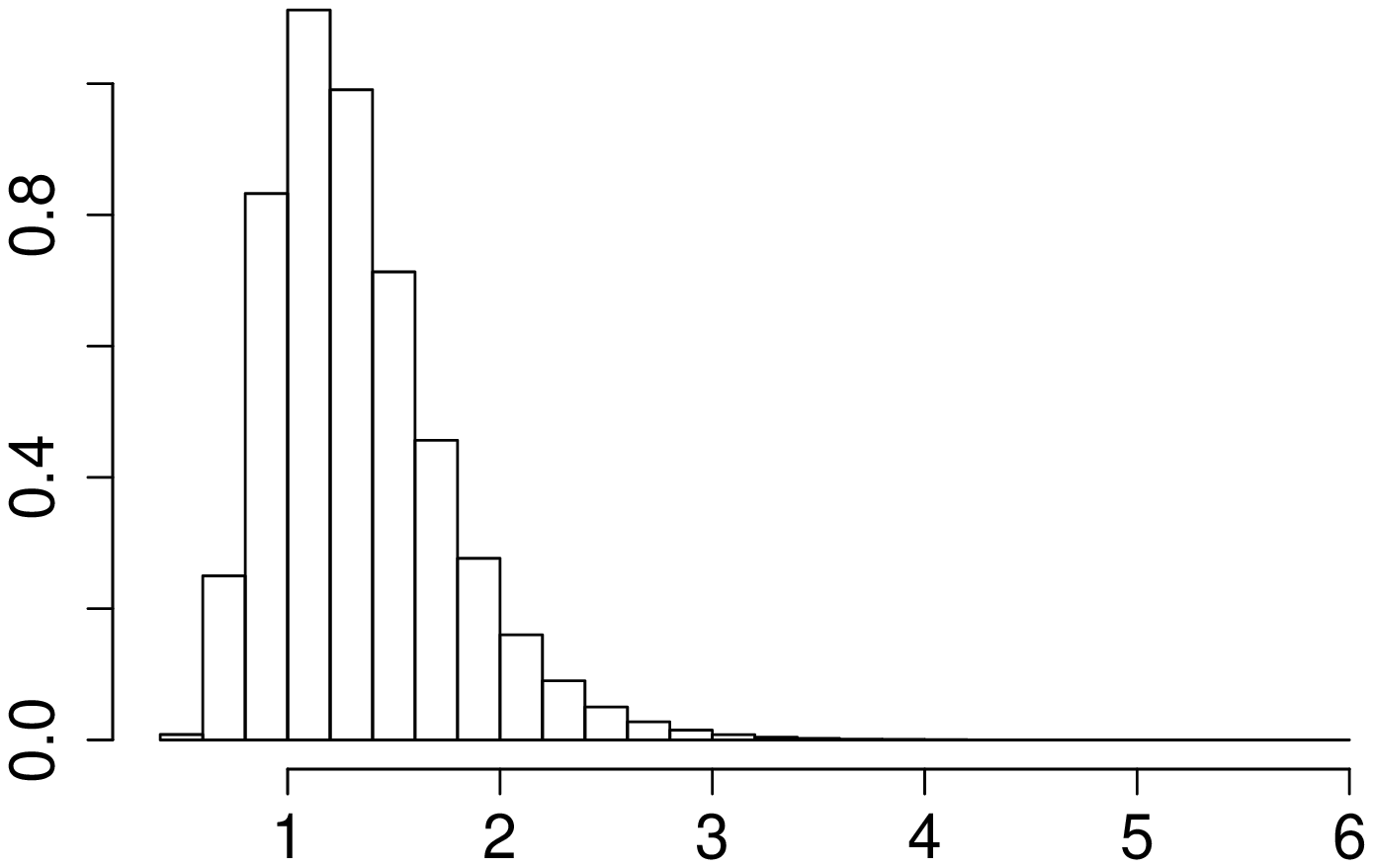}}
\subfigure[][$p_\star \ (0.034,0.978)$]{\label{fig:a2}\includegraphics[scale=0.3]{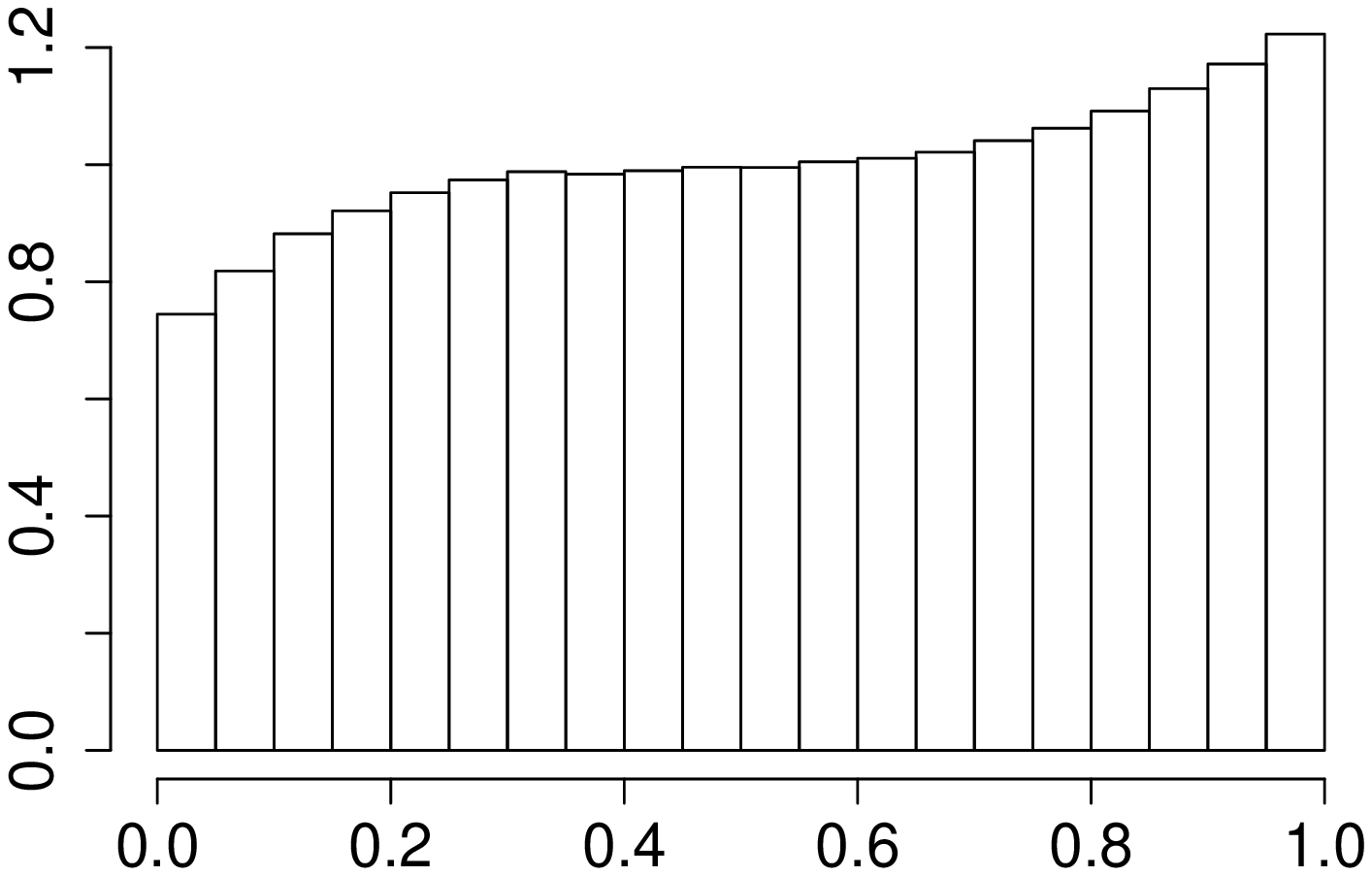}}
\subfigure[][$\sigma_{\varphi^2} \ (2.143,29.080)$]{\label{fig:a3}\includegraphics[scale=0.3]{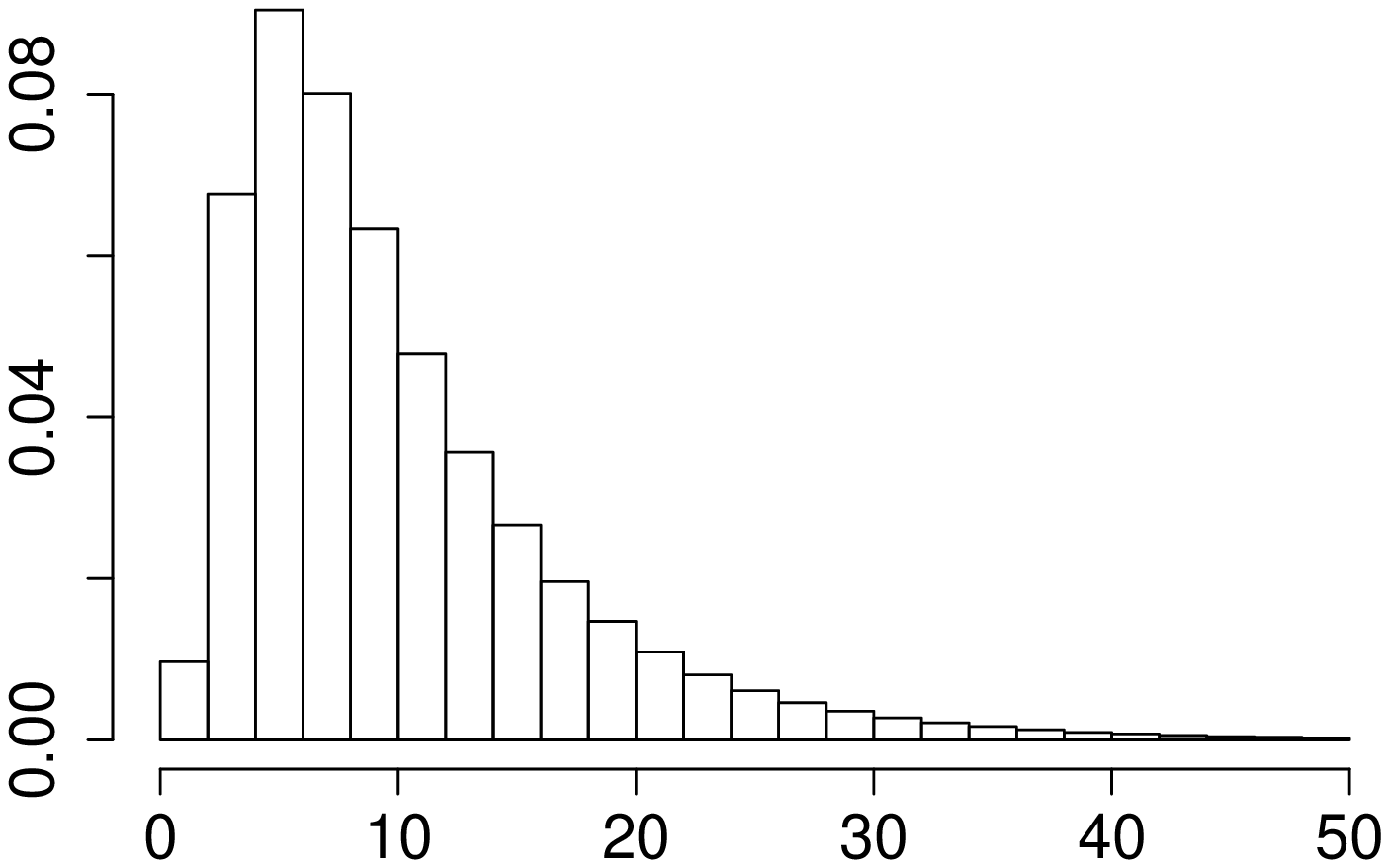}}
\caption{\label{fig:deerPriorParam}Red deer example: Prior-parameters with
  95\% credible interval.}
\end{figure}

\begin{figure}
  \centering
\subfigure[][$\kappa_1 \ (-0.549,-0.146)$]{\label{fig:c1}\includegraphics[scale=0.5]{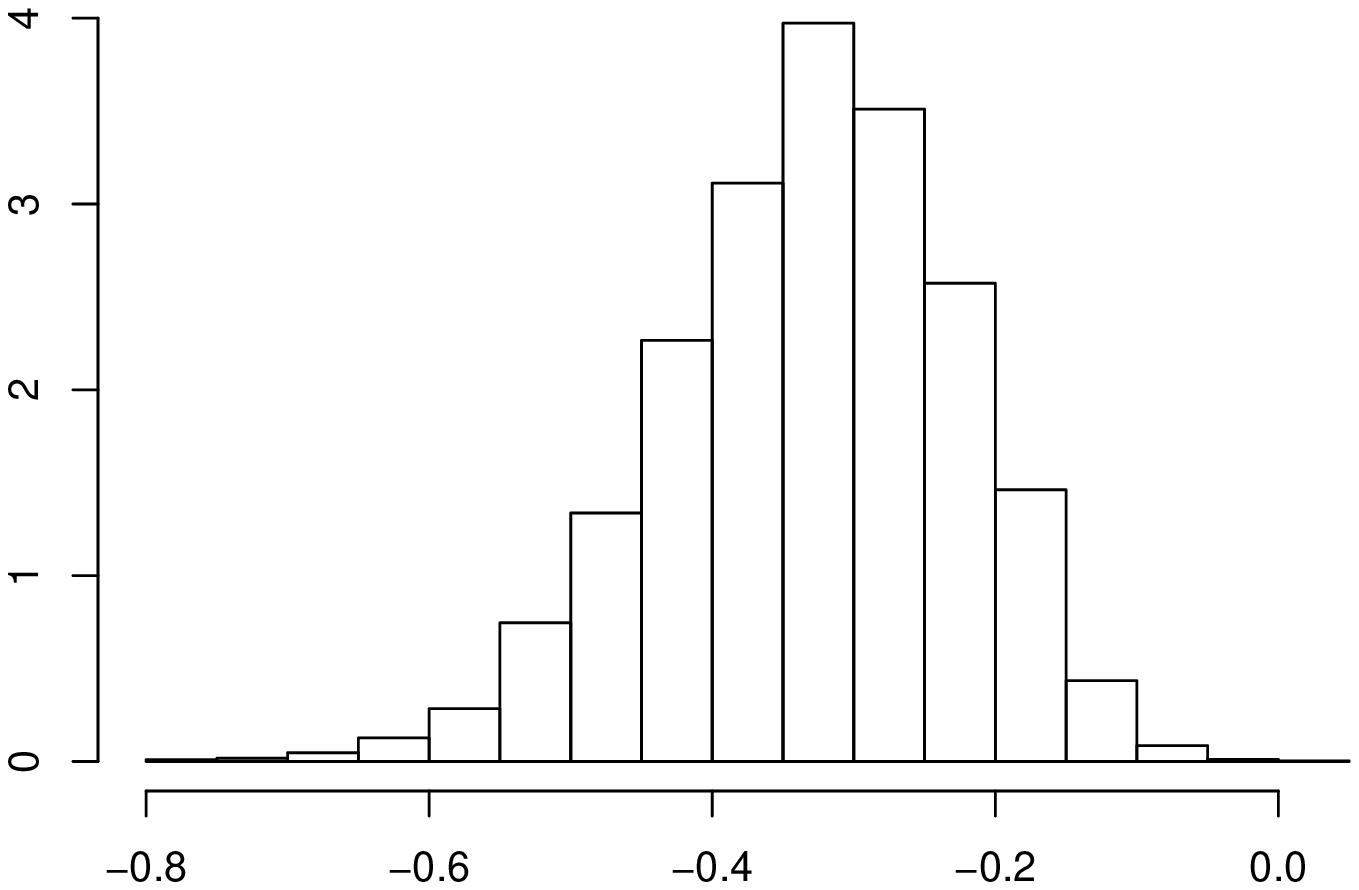}}
\subfigure[][$\kappa_2 \ (-0.441,-0.058)$]{\label{fig:c2}\includegraphics[scale=0.5]{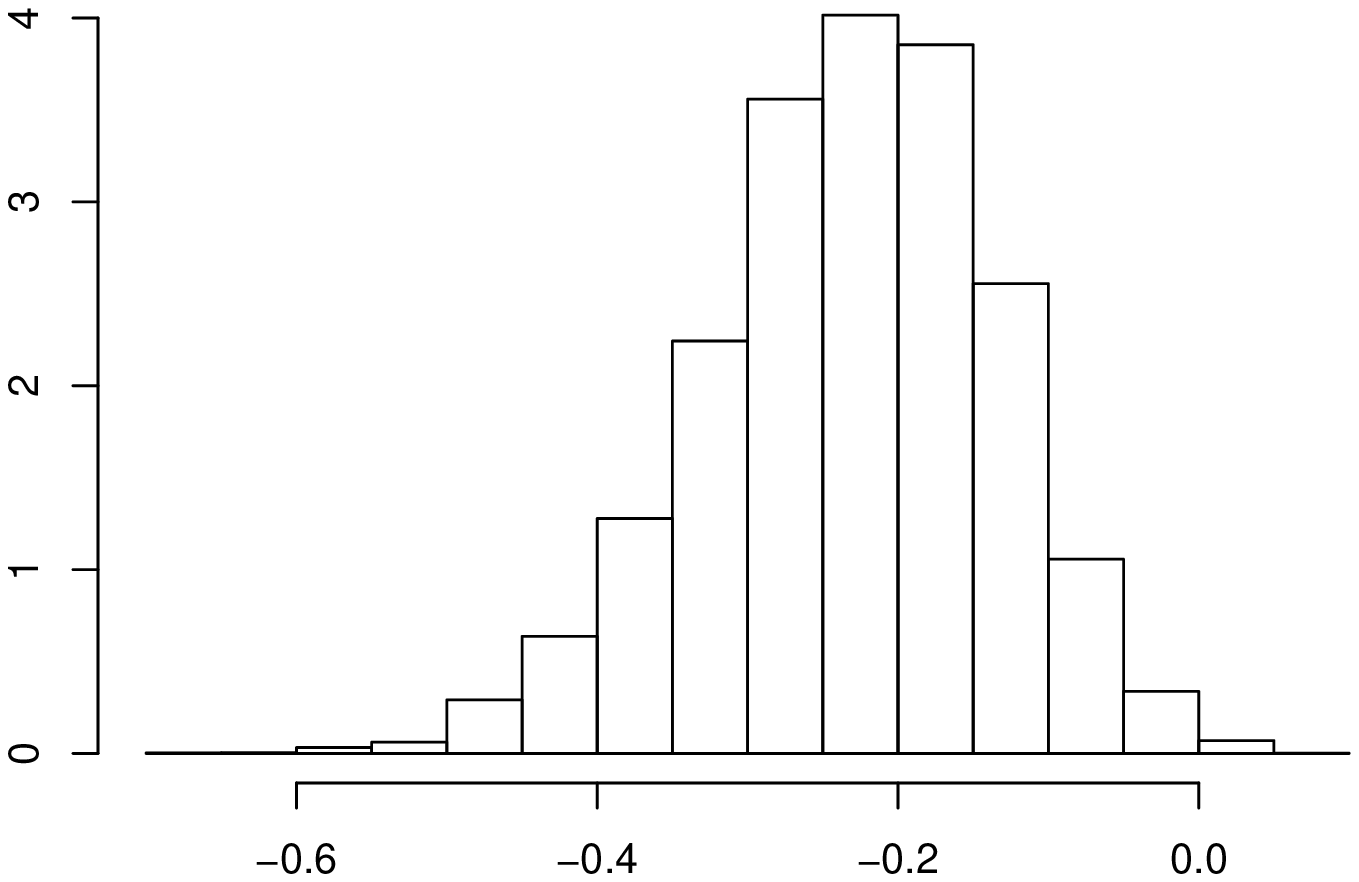}}\\
\subfigure[][$\kappa_3 \ (-0.266,-0.002)$]{\label{fig:c3}\includegraphics[scale=0.5]{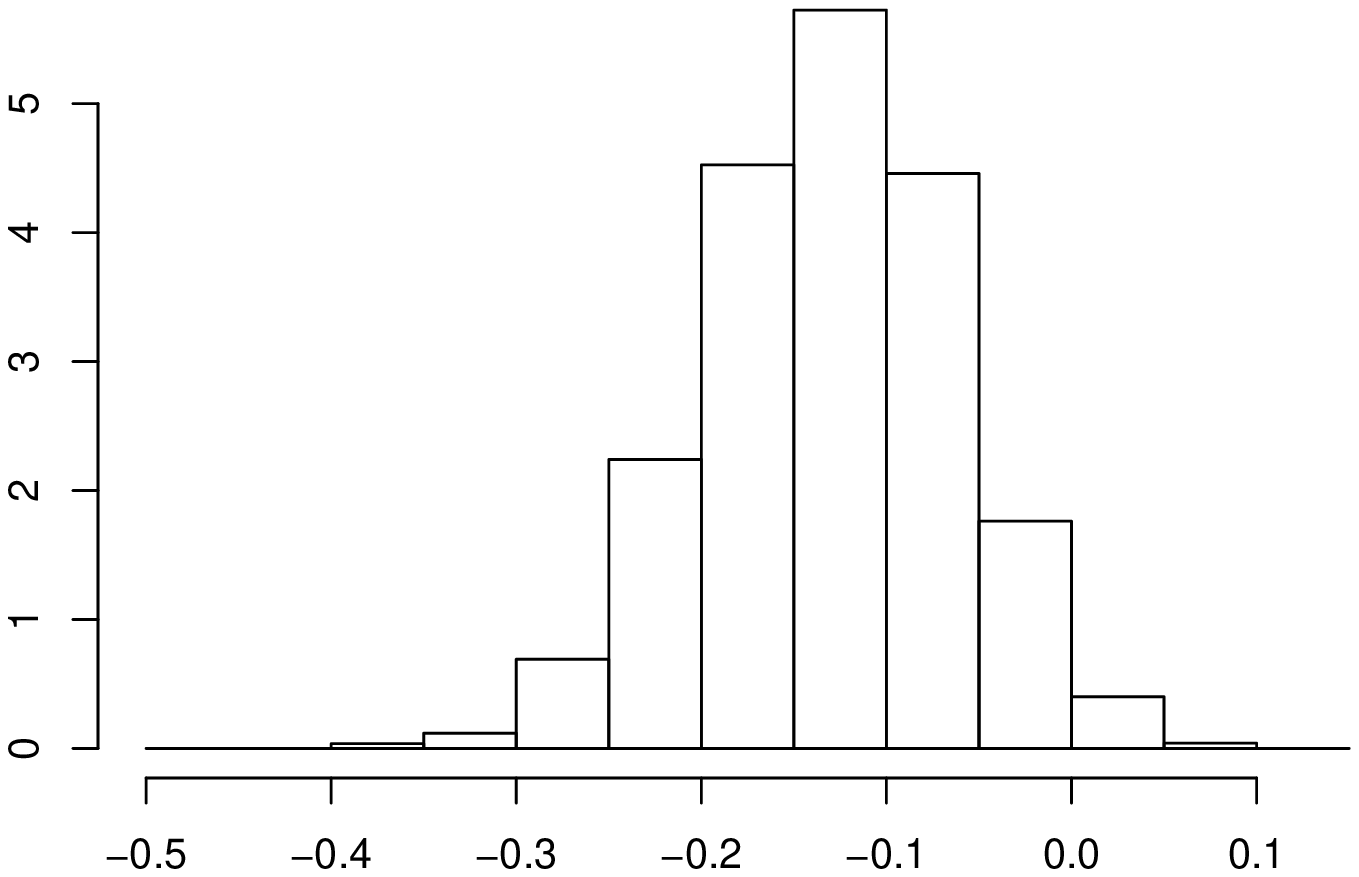}}
\subfigure[][$\kappa_4 \ (-0.546,-0.164)$]{\label{fig:c4}\includegraphics[scale=0.5]{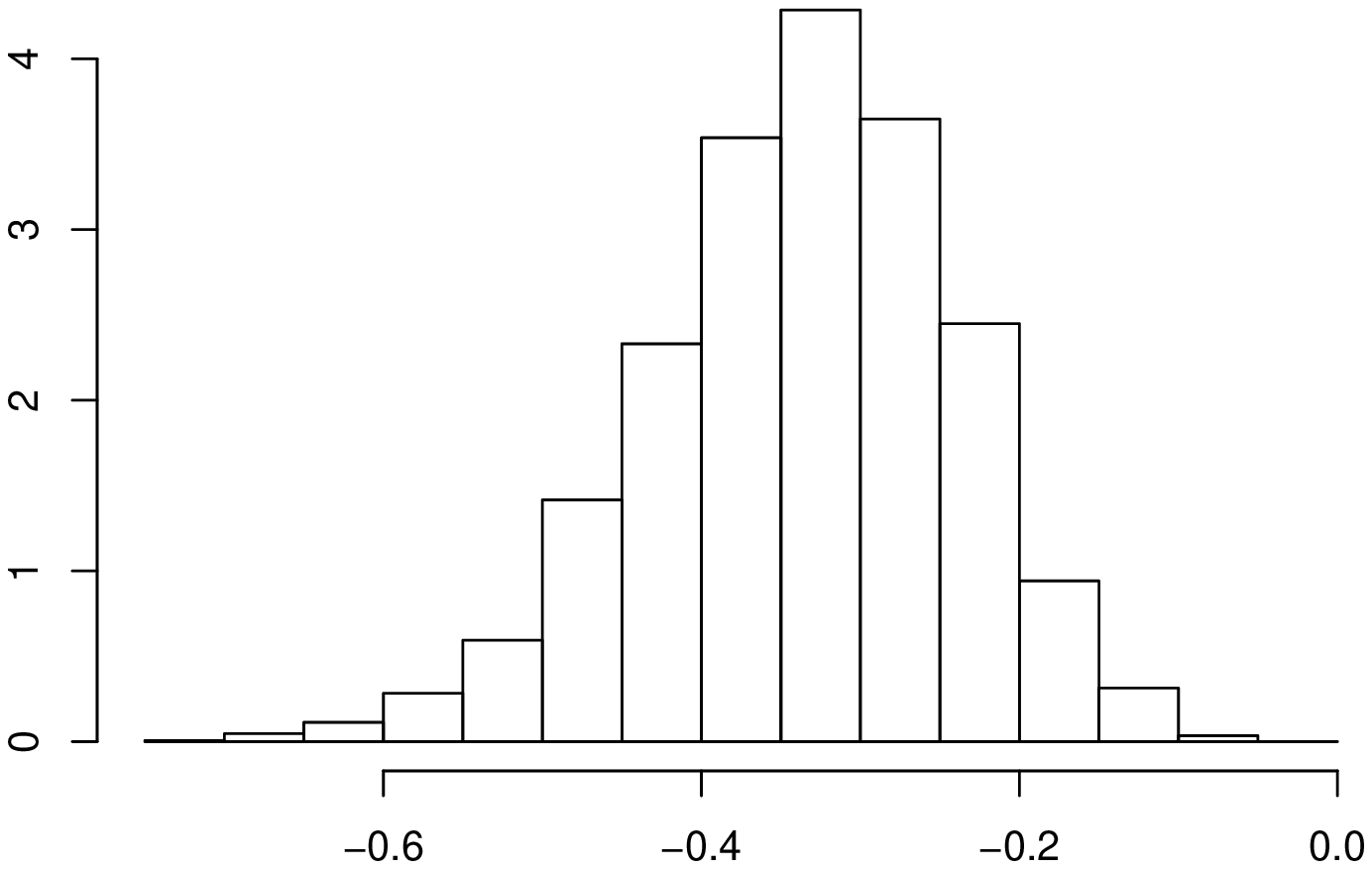}}
\caption{\label{fig:covariatesDeer}Red deer example: 
  Estimated marginal posterior distributions 
  for the parameters of the
  covariates. Estimated 95\%
  credible interval is given for each parameter.}
\end{figure}

Figure \ref{fig:deerPriorParam} shows posterior probability histograms for 
the prior-parameters $\eta$, $p_\star$ and $\sigma_{\varphi^2}$.
The median of $\eta$ is in this example $1.255$. The median
of $p_\star$ is $0.538$, and as we can see from the estimated posterior
distribution of $p_\star$ we slightly prefer models where interactions
of order higher than two are turned on. The median of
$\sigma_{\varphi^2}$ is $8.058$ for this example.
Lastly the estimated posterior distributions of the covariate parameters with 
95\% credible interval are given i Figure \ref{fig:covariatesDeer}. As
we can see all of these have credible intervals that does not
include zero, supporting the need to include them in the model. 

\section{Closing remarks}\label{sec:remarks}

In this paper we propose a prior distribution for the dependence
structure of a stationary binary MRF. Our prior is formulated 
in three levels. First we define a prior for the neighbourhood
system. Next, given the neighbourhood system, and thereby the 
set of cliques, we formulate a prior for the parametric form of the 
potential functions. In particular this includes a prior for 
whether higher-order interactions should be included in the MRF,
and non-zero prior probabilities are assigned for the event that groups of 
parameters have exactly the same value. The latter serves both 
as a way to reduce the effective number of parameters and 
to allow properties like symmetry and isotropy.
Lastly, given the neighbourhood system and 
parametric potential functions we define a prior for the parameter 
values. To our knowledge this is the first time a prior is 
formulated for all three levels in the dependency structure 
specification of an MRF. In particular it is thereby the first attempt
to learn the neighbourhood system of an MRF from data via 
a fully Bayesian approach. 

%Assigning prior distribution to only the parameter values of
%an MRF is done already in \cite{art109} and
%\cite{ryden1998}, and \cite{arnesen2013} assign a prior distribution
%for both the parametric form of the potential functions and the 
%parameter values. This is done by constructing a prior for the
%parameters such that groups of parameters have a non-zero
%probability to be assigned the same value. This serves as a way to both
%reduce the number of parameters and to detect properties like
%symmetry, isotropy etc. in the data. We adopt this prior, but in
%addition we place a prior also on the dependency structure of the
%MRF. In particular we assing separate priors to the neighbourhood system and
%the appearance of higher order interactions in the MRF. 
%%This is
%%done by combining two parametric models of MRF, one being a natural
%%choise to use for the prior distribution of the dependency structure,
%%and the other being a natural chiose for the prior of the parameter
%%values. Following \cite{arnesen2013} we construct the prior for the
%%parameters such that groups of parameters have a non-zero
%%probability to be assigned the same value. This serves as a way to both
%%reduce the number of parameters and to detect properties like
%%symmetry, isotropy etc. in the data. 
To sample from the resulting posterior distribution we adopt
an RJMCMC approach, and in particular we formulate
proposal distributions that can be used in such an algorithm. 
To cope with the computationally intractable normalising constant we adopt an
auxiliary variable algorithm  when computing acceptance probabilities, similar
to what is done in \cite{Everitt2012}.

We present three examples of data in this paper. Two examples with
simulated data, one from the Ising model and one from an MRF with
$2\times 2$ maximal cliques. Both the dependency structure and parameter
values in these two examples are reasonably well recovered. The last example is a
data set of presence/absence of red deer in a region in
Scotland. The results show a clear spatial dependency within
the data. In particular the spatial dependency involves higher order
interaction, which demonstrates the importance of exploring higher 
order interaction MRFs
also for data sets where the spatial dependency is week. Clearly
more experience with our proposed prior is needed, both on synthetic
and real world data sets.  
% simulation from the fitted likelihood seems to agree
%with the initial data.

In the current article we restrict ourselves to consider only binary
data. However a generalisation to the discrete case is straightforward. 
The bottleneck of this approach is clearly computation
time, as we need to simulate an MCMC sample from the likelihood for each proposed
change in the parameter vector. A possible
strategy to reduce this problem would be to replace the MRF with
a partially ordered Markov model (POMM), see \citet{qian1991}, 
\cite{davidson1998} and \cite{phd7}. The POMM is known to be
computationally superior to MRF, and we are
currently exploring such an option. 

Another important direction for future research is to improve
the proposal distributions in order to obtain better convergence and mixing of
the MCMC algorithm, which in turn also would improve the overall computation time. We
believe that better proposal distributions would help the MCMC
algorithm to converge also for more complex data sets than what we
have presented here. With our current proposal distributions, we have 
in particular found it hard to get satisfactory convergence and mixing 
properties for data sets with strong spatial dependence.
One improvement that we believe would help the convergence is in how to
propose new parameter values when adding or removing a parameter. 
For example, when proposing to add a new parameter in our current 
algorithm we just propose a new value for this new parameter and keep all the other 
parameter values unchanged, except for the sum-to-zero adjustment.
A better alternative would be to propose a change also in 
the other parameter values so that the new model is closer, in some sense, 
to the current model. Fixing the value of the potential new parameter,
one possibility would be to define the potential new values of the remaining parameters by defining the 
potential new model as
a least squares approximation of the current model. This strategy has strong 
similarities with an approximation technique used in \citet{phd6}, but the 
results there can not directly be used in our setting because we have adopted
a partition of $M$.

%However the assumption of 
%stationarity will now longer be
%valid using POMMs.
%  \begin{figure}
%  \centering
%\subfigure[][$\theta^C_1 \ (-0.50,-0.21)$]{\label{fig:c1}\includegraphics[scale=0.5]{c1}}
%\subfigure[][$\theta^C_2 \ (-0.36,-0.07)$]{\label{fig:c2}\includegraphics[scale=0.5]{c2}}\\
%\subfigure[][$\theta^C_3 \ (-0.55,-0.27)$]{\label{fig:c3}\includegraphics[scale=0.5]{c3}}
%\subfigure[][$\theta^C_4 \ (-0.26,-0.05)$]{\label{fig:c4}\includegraphics[scale=0.5]{c4}}
%\caption{\label{fig:covariatesDeer}Red deer example: 
%  Estimated marginal posterior distributions 
%  for the parameters of the
%  covariates. Estimated 95\%
%  credible interval is given for each parameter.}
%\end{figure}

%\begin{figure}
%  \centering
%  \includegraphics[scale=0.5]{simulationsDeer_gamma05} 
%  \caption{\label{fig:simulationsDeer}Red deer example: 
%    Three realisations from the likelihood for three random
%    samples of $z$ from the posterior distribution.}
%\end{figure}

%\appendix
%\section{Appendix A}

\section*{Supporting Information}

Additional Supporting Information may be found in the online version
of this article:\\

{\noindent \bf Section S.1:} Details for the MCMC sampling algorithm.\\
{\noindent \bf Section S.2:} Investigating the modality in
the red deer example.

\bibliographystyle{jasa}
\bibliography{bibl}
%\clearpage

%\vspace{1cm} \noindent Petter Arnesen, Department of Mathematical Sciences,
%Norwegian University of Science and Technology, Trondheim 7491, Norway.
%\\
%E-mail: petterar@math.ntnu.no

\end{document}

\begin{center}
{\noindent{\Large \textbf{Supplementary materials to the paper: \\Prior specification of neighbourhood
 and interaction\\[-0.1cm] structure 
in binary Markov random fields} \vspace{0.6cm} \\} 
%{\noindent{\LARGE \textbf{Fully Bayesian prior specification of
%      dependency structure in stationary binary Markov random fields} \vspace{1cm} \\} 
{\large \textsc{Petter Arnesen}} and {\large \textsc{H\aa kon Tjelmeland}}\\{\it Department of
    Mathematical Sciences, Norwegian University of Science and
    Technology}
\vspace{0.4cm}
%\\{\large \textsc{H\aa kon Tjelmeland}}\\{\it Department of
%    Mathematical Sciences, Norwegian University of Science and
%    Technology}\vspace{1cm}
%\let\thefootnote\relax\footnote{Petter Arnesen is PhD student, Department of Mathematical Sciences, 
%Norwegian University of Science and Technology, Trondheim 7491, Norway 
%(E-mail: \textit{petterar@math.ntnu.no}).
%H\aa kon Tjelmeland is Professor, Department of Mathematical Sciences, 
%Norwegian University of Science and Technology, Trondheim 7491, Norway 
%(E-mail: \textit{haakont@math.ntnu.no}).}
}
\end{center}	
%\topmargin -22mm
	
%\numberwithin{equation}{section}
%\numberwithin{figure}{section}
%\numberwithin{table}{section}

%\hspace{1cm} \vspace{-0.cm}

\renewcommand{\baselinestretch}{1.5} \small\normalsize

\renewcommand{\thesection}{S.\arabic{section}}
\renewcommand{\thesubsection}{\thesection.\arabic{subsection}}

\section{Details for the MCMC sampling algorithm}
\label{sec:S1}

In this section we provide details of the proposal distributions that we use when sampling from the posterior distribution
\begin{equation*}
p(z,M,\theta|x)\propto p(x|z,M)p(z|M,\theta)p(M|\theta)p(\theta),
\end{equation*}
where $p(x|z,M)$ is the likelihood MRF and
$p(z|M,\theta)$, $p(M|\theta)$ and $p(\theta)$ are the priors defined in Section 5 in the paper. All of the proposal distributions are also briefly discussed in Section 6 in the paper. The proposals given in Sections \ref{sec:S1-3} and \ref{sec:S1-6} are 
transdimensional proposals, resulting in reversible 
jump Markov chain Monte Carlo (RJMCMC) updates. Note that all
proposals cannot be applied to all states, or that some of these
proposals could result in non-reversible states. For these cases we
simply regard the proposal as a reject. In this section we use $S, S^\star$ and $S^{\star\star}$ to denote specific elements of
the partition $\mathscr{S}$.

\subsection{Random walk for parameter updates}
Assume the current state to be $z=\{(S,\varphi_S);S\in \mathscr{S}\}$. The first proposal is simply a random walk proposal
for the $\varphi$ parameters, but correcting the update so that the parameter values
still sum to zero. We do this by first uniformly at random selecting one of the 
$\varphi$ parameters, $\varphi_{S^\star}$ for $S^\star\in \mathscr{S}$ say, and add to 
it a random change $\varepsilon \sim
N(0,\sigma_w^2)$, where $\sigma_w$ is an algorithmic tuning
parameter. Next, for the proposed new parameter values to comply with the sum-to-zero 
constraint, we subtract the same value from all the parameters. The potential 
new state $\tilde z$ becomes
\begin{equation*}
\tilde z= \left \{ \left(S,\varphi_S-\frac{\varepsilon}{\card(\mathscr{S})} \right
  );S\in \mathscr{S}\setminus \{S^\star\} \right \}\cup \left \{ \left (
    S^\star,\varphi_{S^\star}+\varepsilon-\frac{\varepsilon}{\card(\mathscr{S})} \right )\right \}.
\end{equation*}

\subsection{Switching cell for one clique type}
\label{sec:swappingProposal}
Assume again the current state to be
$z=\{(S,\varphi_S);S\in \mathscr{S}\}$. Note that this proposal is not
possible if the number of cells equals one or all the cells contains only one clique
type each. 
To generate a potential new state we
first sample a pair of distinct cells $S^\star,S^{\star\star}\in\mathscr{S}$ 
from the distribution
\begin{equation*}
\label{eq:cellProposal}
q(S^\star,S^{\star\star})\propto\left\{
\begin{array}{ll}
\exp\left\{-(\varphi_{S^\star}-\varphi_{S^{\star\star}})^2\right\} & \mbox{if $S^\star\neq
  S^{\star\star}$ and $\card(S^\star)>1$,}\\
0  & \mbox{otherwise.}
\end{array}\right.
\end{equation*}
Thus, we have that the number of clique types in $S^\star$ is strictly larger than one,
and pairs $S^\star,S^{\star\star}$ where the corresponding parameter values are similar
have a higher probability to be sampled. Given $S^\star$ and
$S^{\star\star}$ the potential new state is formed by drawing at random a clique type $\Lambda\in S^\star$
and moving $\Lambda$ over to $S^{\star\star}$. The potential new state thereby becomes
\begin{equation*}
\tilde{z}=\{(S,\varphi_S);S\in \mathscr{S}\setminus \{S^\star,S^{\star \star}\}\}\cup\{(S^\star\setminus
\{\Lambda\},\varphi_{S^\star}),(S^{\star\star}\cup \{\Lambda\},\varphi_{S^{\star\star}})\}.
\end{equation*}

\subsection{To propose a change in the partition}
\label{sec:S1-3}
Again let the current state be $z=\{(S,\varphi_S);S\in \mathscr{S}\}$. In this proposal 
we propose to increase or decrease the number of cells by one, with
probability a half for each. We refer to these proposals as split and merge proposals,
respectively. We start by explaining how to propose to increase the
number of cells by one (split), before we explain the opposite move
(merge). Note that the spilt proposal is not possible if the number of
cells equals the number of clique types, and that the merge proposal is
not possible if there is no
cell with only one clique type or if $M=\{\{\emptyset\}\}$.

To propose to increase the number of cells by one we start by drawing
uniformly at random one of the cells where the number of clique types in
the cell are larger than one. Assuming cell $S^\star$ is sampled, we next
uniformly at random draw one of the clique types in this cell, $\Lambda\in S^\star$ say, and use
this to form a new cell including only this clique type, i.e. $\{\Lambda\}$. This
new cell must be assigned an associated 
parameter value, $\varphi_{\{\Lambda\}}$, and to do this we add an 
$\varepsilon\sim \mbox{N}(0,\sigma_g^2)$, where $\sigma_g^2$ is an algorithmic tuning
parameter, to the old parameter value for $\Lambda$, $\varphi_{S^\star}$.
Finally, for the potential new parameter values to comply with the sum-to-zero 
constraint, we subtract the same value from all the parameters, so that the 
potential new state becomes
\begin{eqnarray*}
\tilde z &=& \left \{ \left (S,\varphi_S - \frac{\varphi_S +
    \varepsilon}{\card(\mathscr{S})+1}\right );
S\in \mathscr{S}\setminus \{S^\star\}\right \} \cup \\
&&\left \{
\left (S^\star\setminus \{\Lambda\},\varphi_{S^\star} - \frac{\varphi_{S^\star} + \varepsilon}{\card(\mathscr{S})+1}\right),
\left(\{ \Lambda\},\varphi_{S^\star} + \varepsilon - \frac{\varphi_{S^\star} + \varepsilon}{\card(\mathscr{S})+1}\right)\right \}.
\end{eqnarray*}

To propose to decrease the number of cells, we can only consider to
merge a cell, $S^\star$ say, consisting of only one clique type,
i.e. $\card(S^\star)=1$,  into another
cell, $S^{\star\star}$ say, in order for this proposal to
conform with the split move specified above. We therefore first sample 
two distinct cells $S^\star,S^{\star\star}\in\mathscr{S}$ from the distribution
\begin{equation*}
\label{eq:cellProposal2}
q(S^\star,S^{\star\star})\propto\left\{
\begin{array}{ll}
\exp\left\{-(\varphi_{S^\star}-\varphi_{S^{\star\star}})^2\right\} & \mbox{if $S^\star\neq
  S^{\star\star}$ and $\card(S^\star)=1$,}\\
0  & \mbox{otherwise.}
\end{array}\right.
\end{equation*}
Given $S^\star$ and $S^{\star\star}$ the potential new state is formed by first
merging $S^\star$ into $S^{\star\star}$ and thereafter correcting all parameter values
with the same amount to obtain a proposal that comply with the sum-to-zero 
constraint. The potential new state then becomes
\begin{equation*}
\tilde z = \left \{ \left (S,\varphi_S + \frac{\varphi_{S^\star}}{\card(\mathscr{S})-1}\right
 ); S\in \mathscr{S}\setminus \{S^\star,S^{\star\star}\}\right \}
\cup \left \{ \left (S^{\star}\cup S^{\star\star},\varphi_{S^{\star\star}} +
    \frac{\varphi_{S^\star}}{\card(\mathscr{S})-1}\right )\right \}.
\end{equation*}

The split and merge proposals give a change in the dimension of 
the parameter space, and in the acceptance probability for these
RJMCMC steps we then need to include a Jacobi
determinant. For the split and merge proposals specified above it is straightforward
to show that the Jacobi determinants 
become $\frac{\card(\mathscr{S})}{\card(\mathscr{S})+1}$ and 
$\frac{\card(\mathscr{S})}{\card(\mathscr{S})-1}$, respectively.

%[[In order to get a one-to-one transition
%between these two models we need to define
%$\epsilon=\epsilon(\varphi^*_1,...,\varphi^*_{r+1})$, From the
%equation giving the $\varphi^*_{r+1}$ and the
%equation giving $\varphi_{i}$ we find that
%
%\begin{equation*}
%\epsilon=\varphi^*_{r+1}-\varphi^*_{i}.
%\end{equation*} 
%To calculate the acceptance probability of such proposals,
%it is required to calculate the Jacobi determinant of the transition
%between the models. Calculating the Jacobi determinant for the merge
%step gives  
%
%\[ |J|=\left | \begin{array}{ccccccc}
%1 & 0 & 0 & \dots &  0 & 0 & 1/r \\
%0 & 1 & 0 & \dots &  0 & 0 & 1/r \\
%&  &  & \vdots &  & & \\
%0 & 0 & 0 & \dots &  1 & 0 & 1/r \\
%0 & 0 & 0 & \dots &  0 & 1 & 1/r\\
%0 & 0 & 0 & \dots &  0 & -1 & 1  \end{array} \right |, \]
%where the last row represents the equation for $\epsilon$. This is a
%determinant of size $(r+1)\times(r+1)$, and by iteratively
%expanding the determinant along the first column we are left with the
%$2\times2$ determinant
%
%\[ |J|=\left| \begin{array}{cc}
% 1 & 1/r \\
% -1 & 1  \end{array} \right|=\frac{r+1}{r}.\]  
%Lastly, we note that the Jacobi
%determinant of the split step is the inverse of this determinant.]]

\subsection{To propose a change in $M$ by replacing a clique type}

In this step we propose to replace one of the clique types that is on
with another clique type. Note that this proposal is not possible if
there are no clique types with $\tau(\Lambda)>1$, or if the
proposed clique type either consists of two equal nodes or is equal to
a clique type already in $M$, in both cases giving a non-reversible state. Assuming the current state to be
$M$ and
$z=\{(S,\varphi_S);S\in\mathscr{S}\}$, we start this proposal by uniformly drawing one
of the clique types $\Lambda \in M$ where $\tau(\Lambda)> 1$. In the following we
let $S^\star$ denote the cell in $\mathscr{S}$ that contains $\Lambda$. 
Then consider an arbitrary element in $\Lambda$, $\lambda\in\Lambda$ say,
and draw at random one of the elements in $\lambda$, $(i,j)$ say.
We then generate a modified version of $\lambda$ by replacing $(i,j)$ with 
one of the first order neighbours of $(i,j)$.
More formally, we first draw uniformly at random a $(k,l)$ from the set 
$\{ (0,1),(1,0),(0,n-1),(m-1,0)\}$, 
%We then have two possibilities, either
%$(i,j)\oplus (k,l)\in\lambda$ or $(i,j)\oplus (k,l)\not\in \lambda$.
%If $(i,j)\oplus (k,l)\in\lambda$ we are not able to generate the modified 
%version of $\lambda$ by as we want here, so in that case we simply propose 
%unchanged values for $M$ and $z$. Otherwise, if $(i,j)\oplus
%(k,l)\not\in\lambda$,
and let the modified version of $\lambda$ be given by 
$\lambda^\star=(\lambda\setminus\{(i,j)\})\cup \{ (i,j)\oplus (k,l)\}$.
A corresponding modified version of $\Lambda$ is then
$\Lambda^\star = \left\{ \lambda^\star\oplus (t,u);(t,u)\in \chi\right \}$.
The potential new state is then generated by just replacing $\Lambda$
with $\Lambda^\star$ in the definition of $M$ and $z$. The 
potential new values of $M$ and $z$ becomes
$\tilde{M}=(M
\setminus \{\Lambda\})\cup \{\Lambda^\star\}$ and
$\tilde{z}=\{(S,\varphi_S);S\in \mathscr{S}\setminus \{S^\star\}\} \cup \{((S^\star\setminus \{\Lambda\})\cup \{\Lambda^\star\},\varphi_{S^\star})\}$.

\subsection{To propose a change in $M$ by adding/deleting a clique type}
\label{sec:S1-5}
Again let the current state be $M$ and
$z=\{(S,\varphi_S);S\in\mathscr{S}\}$. In this proposal we either add a clique type to $M$ or delete
a clique type from $M$, but keeping the same number of $\varphi$ 
parameters. For the adding proposal note that a clique type that is
already in $M$ could be proposed or that the
proposed clique type could consist of two equal nodes, in both cases
giving a non-reversible state. Also, the deleting proposal is not possible if there are no cells with
more than one clique type. First we draw at random whether to increase or decrease the number of
clique types, with probability a half for each. In the following we first
explain the increase proposal and thereafter the decrease
proposal.

We now explain how we propose to increase the number of clique
types, keeping the same number of $\varphi$ parameters. We start 
by drawing uniformly at random one of the existing clique types in $M$, 
$\Lambda$ say. A new clique type $\Lambda^\star$ is then generated
in three steps. 
First we take out an arbitrary element of $\Lambda$, $\lambda$ say,
and draw uniformly at random a node $(i,j)$ from the set of nodes in $\lambda$. 
Secondly we draw uniformly at random a $(k,l)\in\chi\setminus \{(0,0)\}$ from the
distribution
\begin{equation*}
\label{eq:distDist}
q(k,l)\propto \exp(-\rho(|k|+|l|)) \text{ for } (k,l)\in
\chi\setminus \{(0,0)\},
\end{equation*}
where $\rho$ is an algorithmic tuning parameter. Given $(k,l)$ a new modified version 
of $\lambda$, $\lambda^\star$ is constructed by
$\lambda^\star=\lambda\cup\{(i,j)\oplus(k,l)\}$. Thirdly, a modified version of
$\Lambda$, $\Lambda^\star$ is constructed by $\Lambda^\star=\Lambda\cup \{\lambda^\star\oplus(t,u);(t,u) \in
\chi\}$,
% We then have two possibilities, either $\Lambda^\star\in M$ or $\Lambda^\star\not\in M$.
%If $\Lambda\in M$ we are not able to generate the potential new state as we want here,
%so we simply propose unchanged values of $M$ and $z$. Otherwise, if $\Lambda\not\in M$,
and we generate the potential new value of $M$ by $\tilde{M} = M \cup \{ \Lambda^\star\}$.
To generate a corresponding potential new value for $z$ we first draw
uniformly at random a cell 
$S^\star$ from all the cells in $\mathscr{S}$, and then let the potential new value of 
$z$ be given by
$\tilde{z}=\{(S,\varphi_S);S\in \mathscr{S}\setminus 
\{S^\star\}\}\cup\{(S^\star\cup \{\Lambda^\star\},\varphi_{S^\star})\}$.

Next we explain the proposal we make when the number of clique types is to be
decreased by one, but keeping the same number of $\varphi$ parameters. We start by uniformly
drawing one of the clique types that belongs to a cell with more than
one element. Let $\Lambda$ denote the clique type we draw and let $S^\star$ denote 
the cell that contains $\Lambda$. The potential new state is then generated simply
by removing $\Lambda$ from the current state, so that the potential new values
for $M$ and $z$ become $\tilde{M}=M\setminus
\{\Lambda\}$ and $z^\star=\{(S,\varphi_S);S\in \mathscr{S}\setminus \{S^\star\}\}\cup\{(S^\star\setminus
\{\Lambda\},\varphi_{S^\star})\}$, respectively.

\subsection{To propose a change in $M$ and in the partition} 
\label{sec:S1-6}
Let again the current state be $M$ and
$z=\{(S,\varphi_S);S\in\mathscr{S}\}$. In this proposal we either add a clique type to $M$ or delete
a clique type from $M$, and increase or decrease the number of $\varphi$ 
parameters correspondingly. For the adding proposal note again that a clique type that is
already in $M$ could be proposed or that the
proposed clique type could consist of two equal nodes, in both cases
giving a non-reversible state. Also, the deleting proposal is not
possible if there are no cells with only one clique type or $M=\{\{\emptyset\}\}$. We first draw at random which of these two possible proposals 
to perform, with probability a half for each.
In the following we start by explaining the increasing proposal, and thereafter
move on to the decreasing proposal. 

We start by generating a $\Lambda^\star$ in exactly the same way as we did for 
the increase proposal in Section \ref{sec:S1-5}. 
%If $\Lambda^\star\in M$ we are again 
%not able to generate the potential new state as we want, so in that case we again 
%simply propose unchanged values for $M$ and $z$. Otherwise, if $\Lambda\not\in M$,
Next, we form a new cell containing only $\Lambda^\star$. To generate the associated 
parameter value $\varphi_{\{\Lambda^\star\}}$ we follow a strategy similar to what we did
in Section \ref{sec:S1-3}. We first add an 
$\varepsilon \sim N(0,\sigma_c^2)$, where $\sigma_c$ is an
algorithmic tuning parameter, to the current value $\phi^{\Lambda^\star}$, and thereafter
subtract the same value from all the $\varphi$ parameters to get a proposal that 
comply with the sum-to-zero constraint. The proposed new values for $M$ and $z$ then 
becomes $\tilde{M}=M\cup 
\{\Lambda^\star\}$ and
\begin{equation*}
\tilde{z}=\left \{
    \left(S,\varphi_S-\frac{\phi^{\Lambda^\star}+\varepsilon}{\card(\mathscr{S})+1}
    \right );S\in \mathscr{S}\right \}\cup \left \{
    \left(\{\Lambda^\star\},\phi^{\Lambda^\star}+\varepsilon-\frac{\phi^{\Lambda^\star}+\varepsilon}{\card(\mathscr{S})+1}
    \right )\right \},
\end{equation*}
respectively.

Next we explain how we generate a proposal where the number of clique types
is decreased by one, and the number of $\varphi$ parameters is correspondingly reduced.
We start by drawing uniformly at random one of the clique types that is the only member of a
cell. 
%Clearly it may happen that no such clique types exist and in that case we 
%simply propose unchanged values for $M$ and $z$. If at least one such clique
%type exist,
Let $\Lambda$ denote the one we sample and let $S^\star=\{\Lambda\}$ denote 
the corresponding cell. This clique type and its corresponding cell is now
simply deleted from the current state resulting in the proposed
values $\tilde{M}=M\setminus \{\Lambda\}$ and 
\begin{equation*}
\tilde{z}=\left\{ \left(   S,\varphi_S+\frac{\varphi_{S^\star}}{\card(\mathscr{S})-1}\right);S\in
 \mathscr{S}\setminus \{S^\star\} \right\},
\end{equation*}
where we again has subtracted the same value from all the $\varphi$ parameters to 
comply with the sum-to-zero constraint. As in Section \ref{sec:S1-3} it is straightforward to show that the Jacobi determinants
for the increase and decrease proposals become 
$\frac{\card(\mathscr{S})}{\card(\mathscr{S})+1}$ and 
$\frac{\card(\mathscr{S})}{\card(\mathscr{S})-1}$  respectively.  
%%%%%%%%%

%[[In order to
%calculate the acceptance probability of such proposals we need to
%determine the $\epsilon$ value that is needed to be simulated in a
%proposal to go back to the old model, i.e.
%
%\begin{equation*}
%\epsilon=\varphi_{r+1}-\mathcal D(\phi,\Lambda^\star)=\varphi_{r+1}+\sum_{i=1}^%ra_i\varphi_i,
%\end{equation*}
%where we have from \eqref{eq:sumCoeff} that $\sum_{i=1}^ra_i=-1$.
%The jakobi for this will be 
%\begin{eqnarray*} 
%\left | \begin{array}{ccccc}
%1 & 0 & \cdots & 0 & 1/r \\
%0 & 1 & \cdots & 0 & 1/r \\
%  &   & \vdots &   &  \\
%0 & 0 & \cdots & 1 & 1/r\\
%a_1 & a_2 & \cdots & a_r  & 1\end{array} \right | &=&
%\left | \begin{array}{ccccc}
%1 & 0 & \cdots & 0 & 1/r \\
%0 & 1 & \cdots & 0 & 1/r \\
%  &   & \vdots &   &  \\
%0 & 0 & \cdots & 1 & 1/r\\
%a_2 & a_3 & \cdots & a_r  & 1\end{array} \right | + \underbrace{\frac{(-1)^r}{r%}
%\left | \begin{array}{ccccc}
%0 & 1 & 0 & \cdots & 0 \\
%0 & 0 & 1 & \cdots & 0 \\
%  &   & \vdots &   &  \\
%0 & 0 & \cdots & 0 & 1\\
%a_1 & a_2 & \cdots & a_{r-1}  & a_r\end{array} \right
%|}_{\frac{(-1)^r}{r}(-1)^{r-1} \left | \begin{array}{cc} 0 & 1\\a_1 &
%    a_r \end{array}\right |=-\frac{a_1}{r}},
%\end{eqnarray*}
%where similar calculations for the first term ultimately gives
%\begin{equation*}
%|J|=\left | \begin{array}{cc} 1 & 1/r\\a_r & 1 \end{array}\right |-\frac{1}{r}(%a_1+...+a_{r-1})=1-\frac{1}{r}\underbrace{(a_1+...+a_r)}_{=-1}=\frac{r+1}{r}.
%\end{equation*}
%The Jacobi determinant for the opposite proposal, adding one clique
%type, will be the inverse of this determinant. ]]

\subsection{Proposal for the model parameters}

In total we have three
model parameters that we assign prior
distributions to, namely $\eta$, $p_\star$ and $\sigma_\varphi^2$. We make one
proposal for each of these parameters for every iteration of the
sampling algorithm. Firstly we
update $\eta$ by proposing a zero mean Gaussian change with standard
deviation $\sigma_\eta$. Secondly we propose a new value for $p_\star$ simply
by sampling a new value from a uniform distribution on the interval
$(0,1)$. For the last parameter $\sigma_\varphi^2$ we propose a zero mean Gaussian
change with standard deviation $\sigma_{\sigma_\varphi}$. 
%Since we have a sum to zero
%constrain on the elements of $\varphi$, the usual conjugacy properties
%of the normal and inverse gamma distribution cannot be used here.         

\section{The multi-modality in the red deer example}\label{sec:S2}

\begin{figure}
\begin{tabular}{m{2.3cm}|m{4cm}m{4cm}m{4cm}}
& \hspace{1.8cm}$\Lambda=\ABBB$
&\hspace{1.8cm}$\Lambda=\AABB$
&\hspace{1.8cm}$\Lambda=\ABAB$\\
\hline
All
& \subfigure{\includegraphics[scale=0.28]{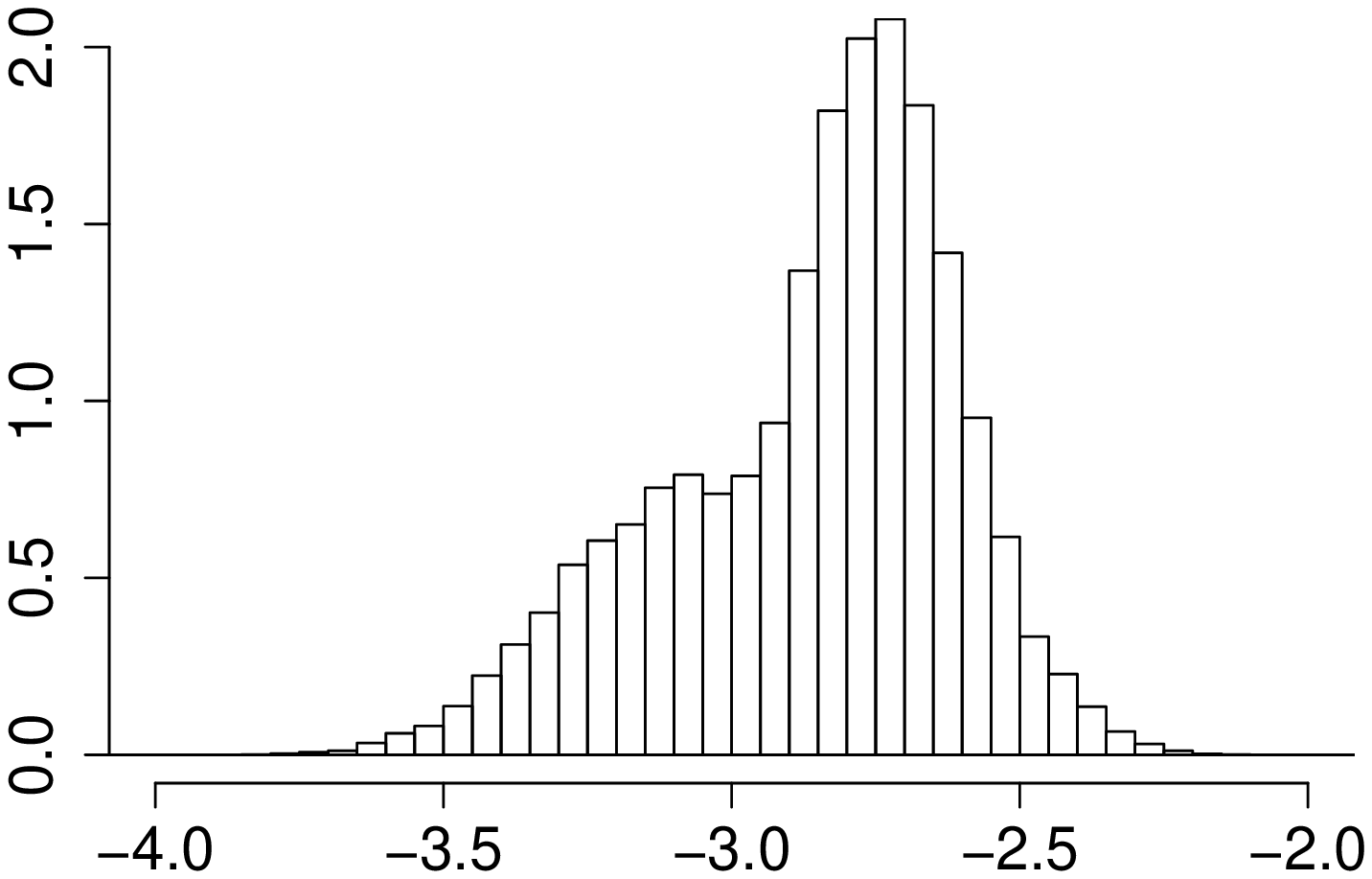}}
& \subfigure{\includegraphics[scale=0.28]{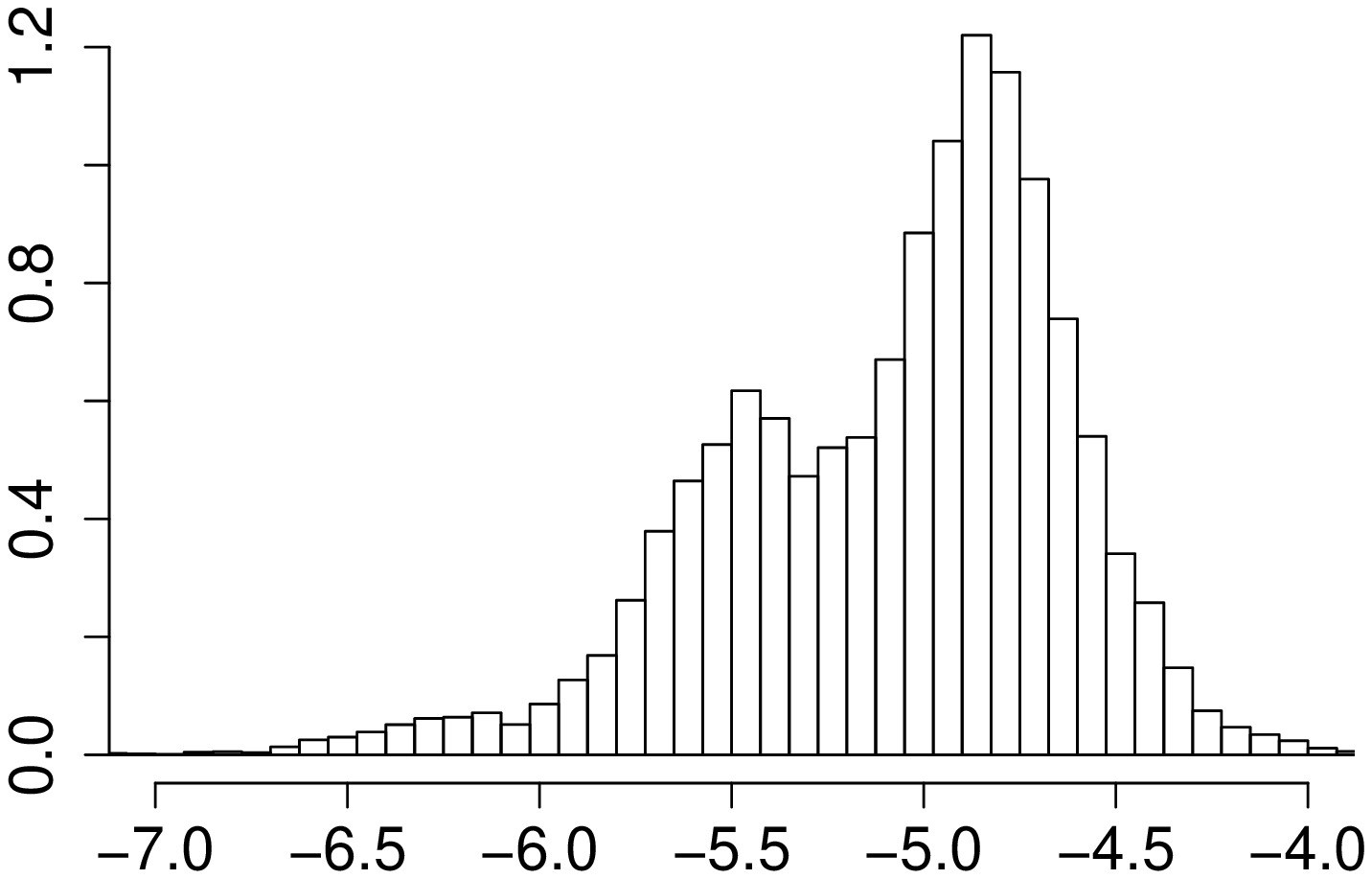}}
& \subfigure{\includegraphics[scale=0.28]{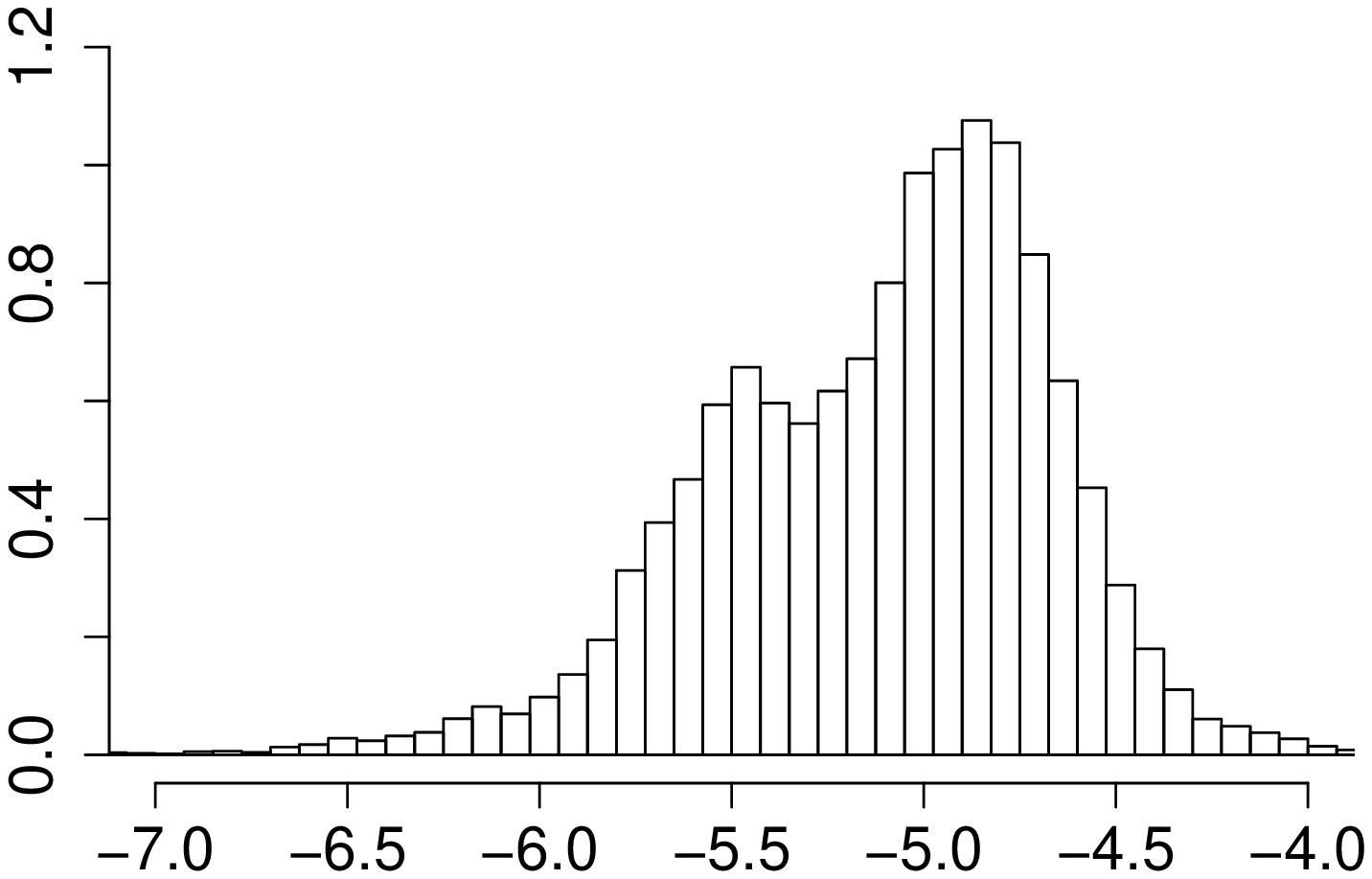}}\\
$M=M_{\max}$ 
& \subfigure{\includegraphics[scale=0.28]{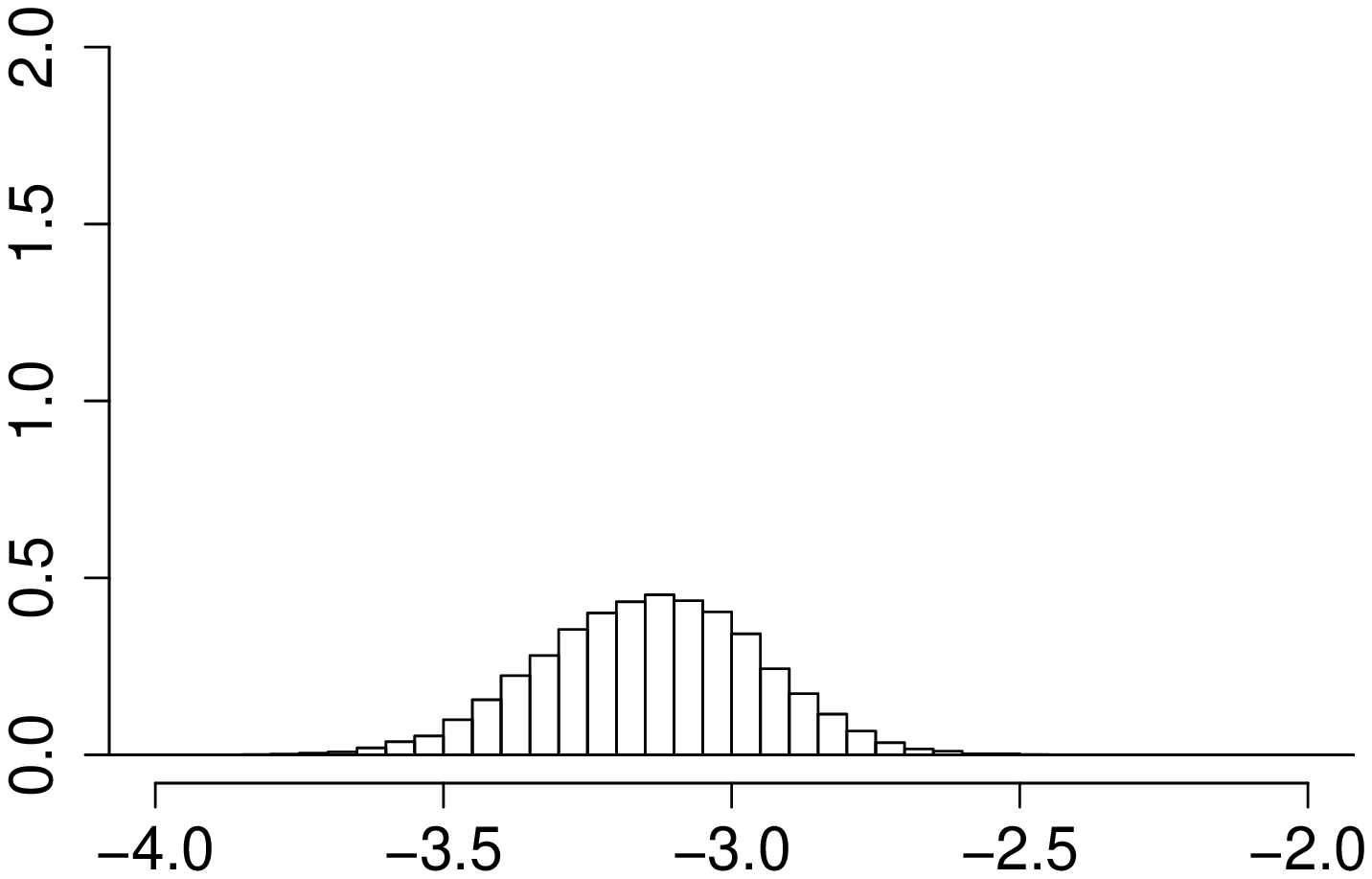}}
& \subfigure{\includegraphics[scale=0.28]{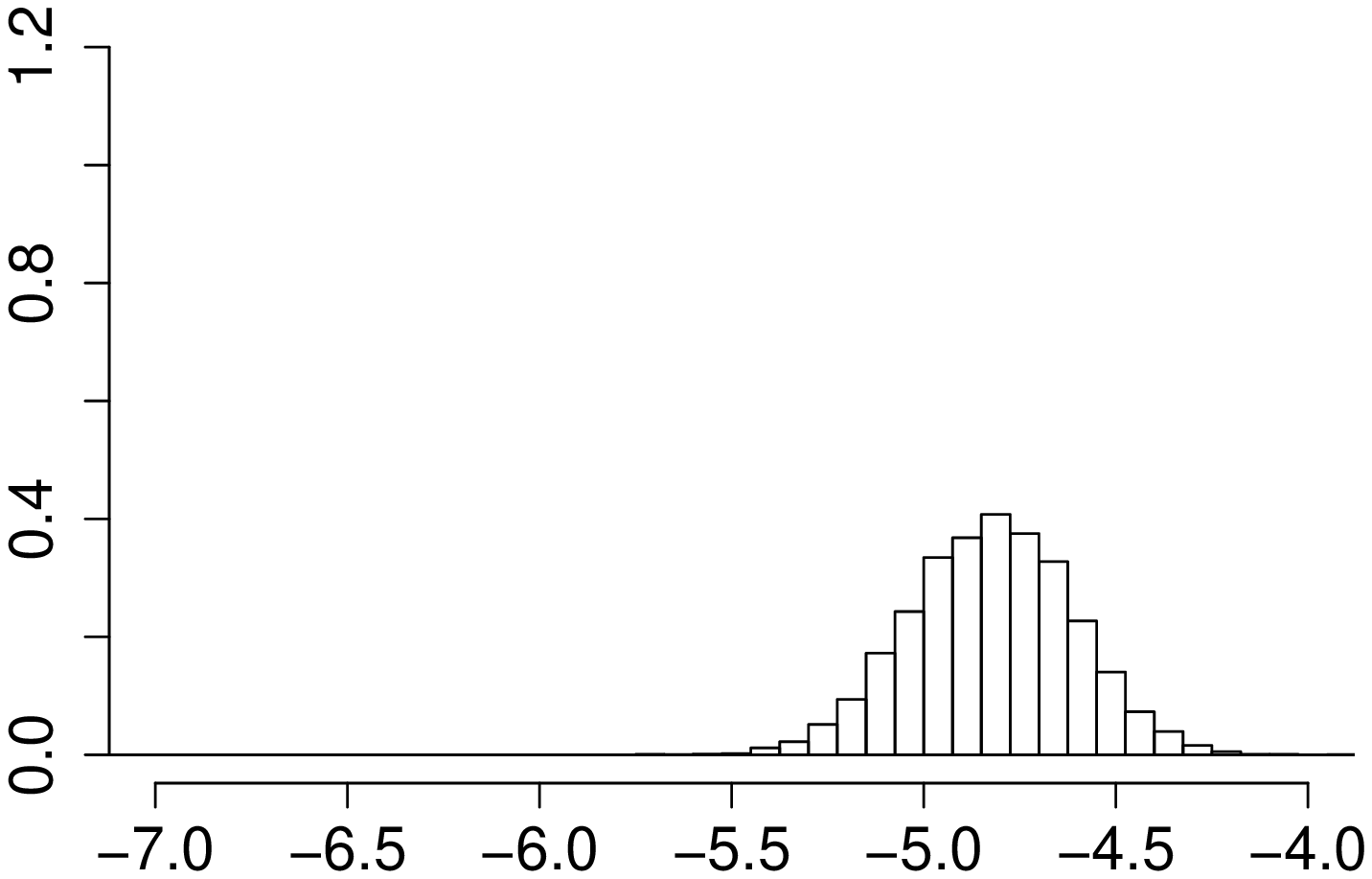}}
& \subfigure{\includegraphics[scale=0.28]{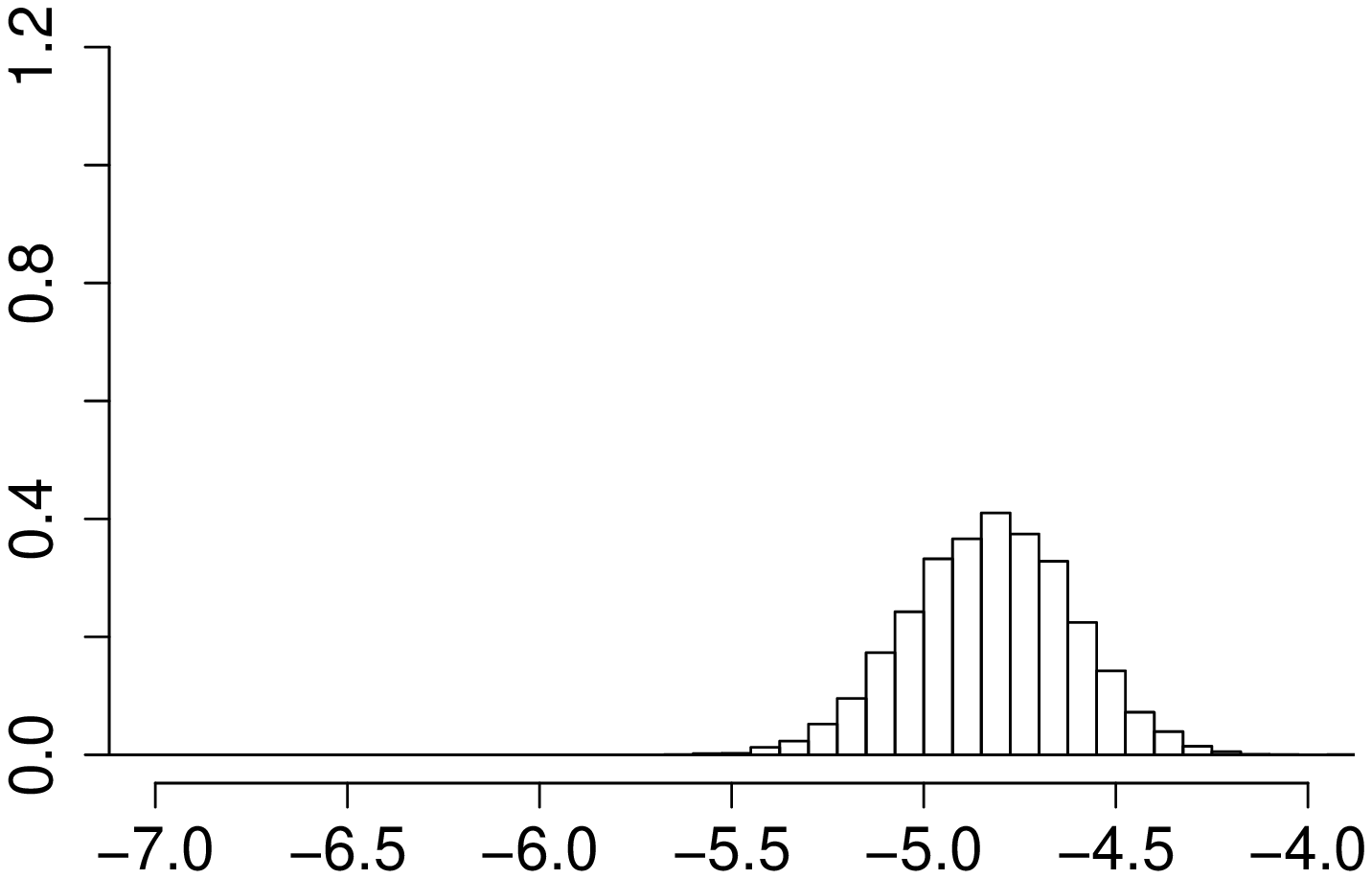}}\\
$M\neq M_{\max}$ 
& \subfigure{\includegraphics[scale=0.28]{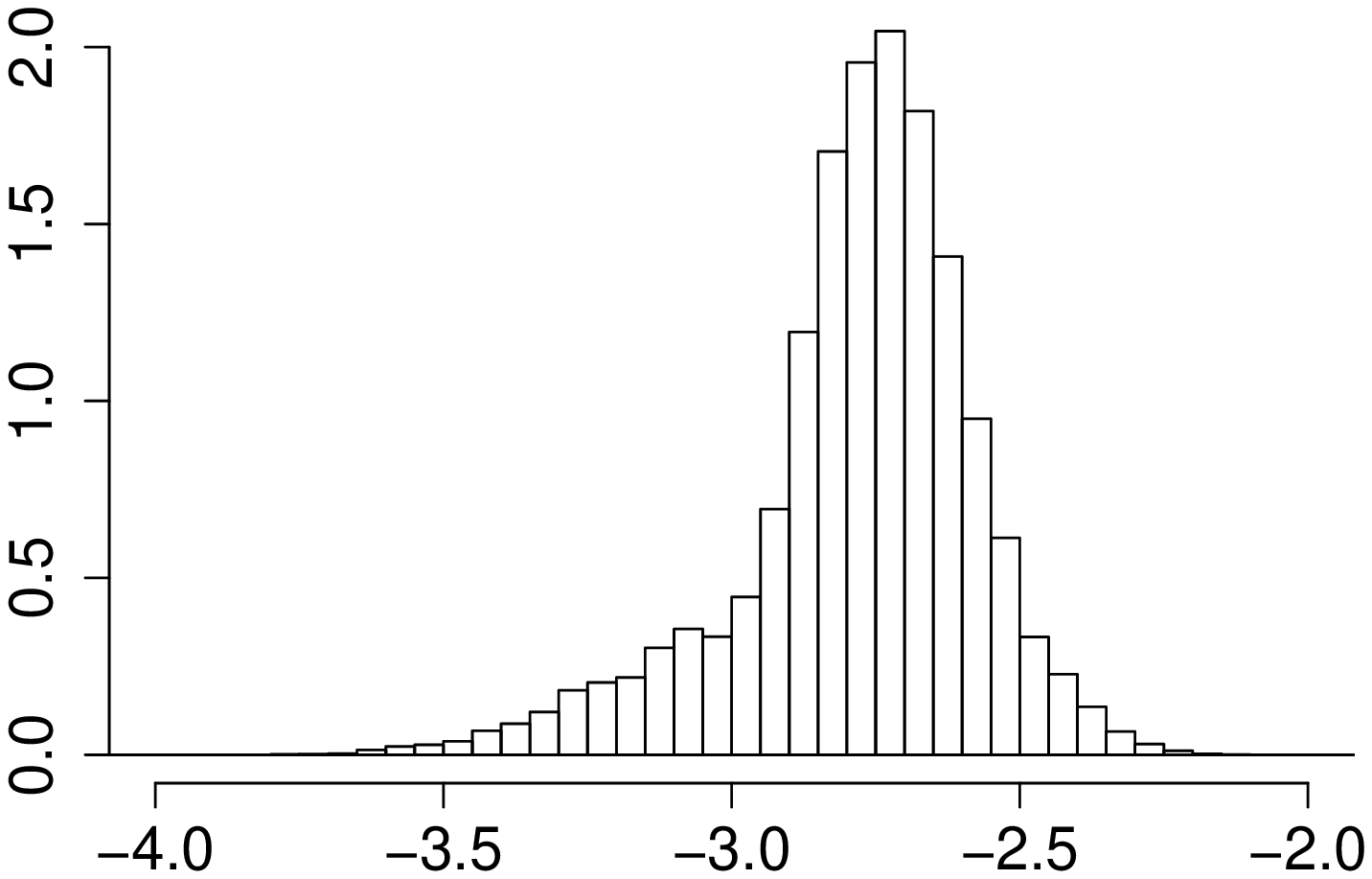}}
&\subfigure{\includegraphics[scale=0.28]{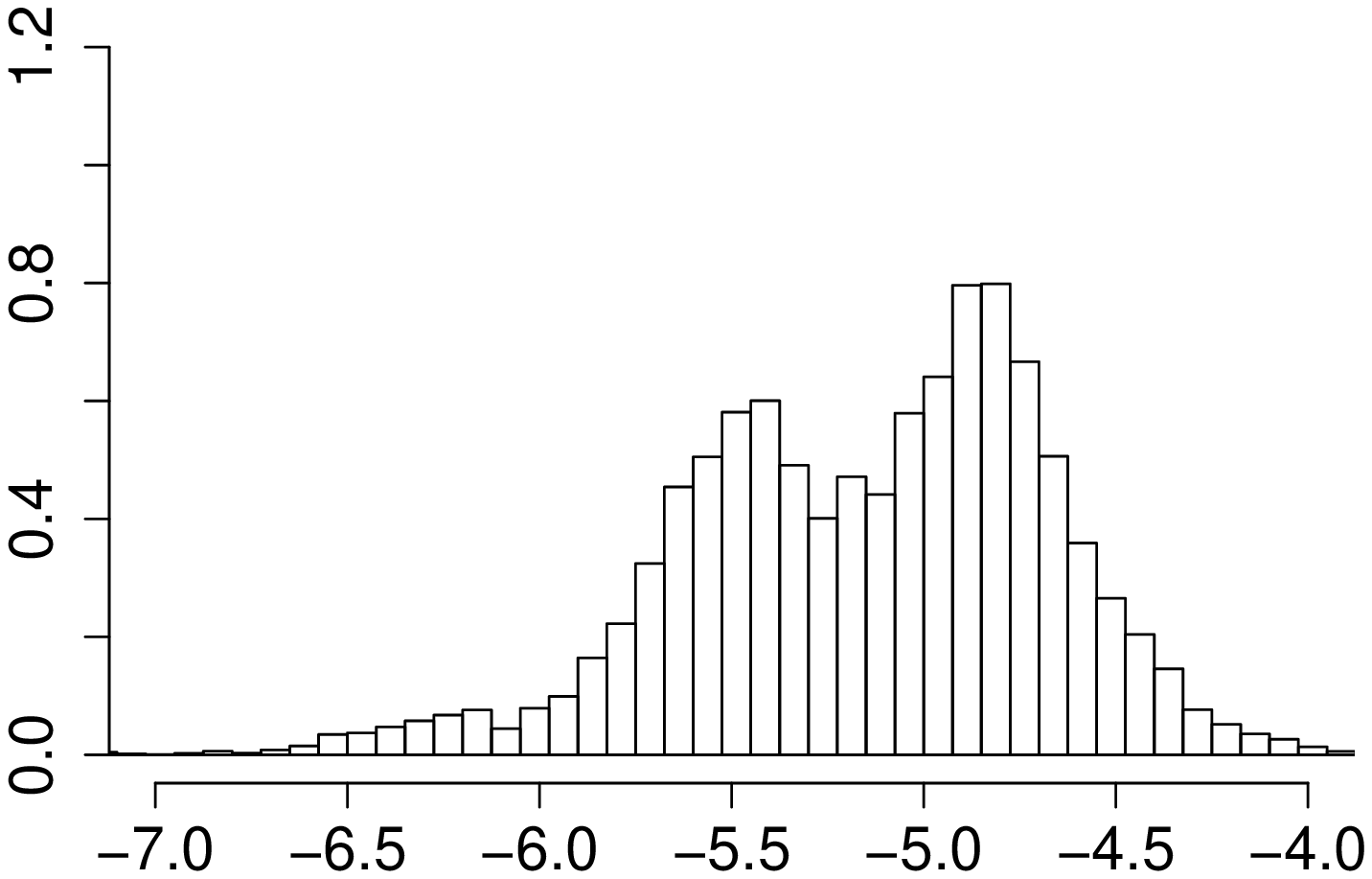}}
&\subfigure{\includegraphics[scale=0.28]{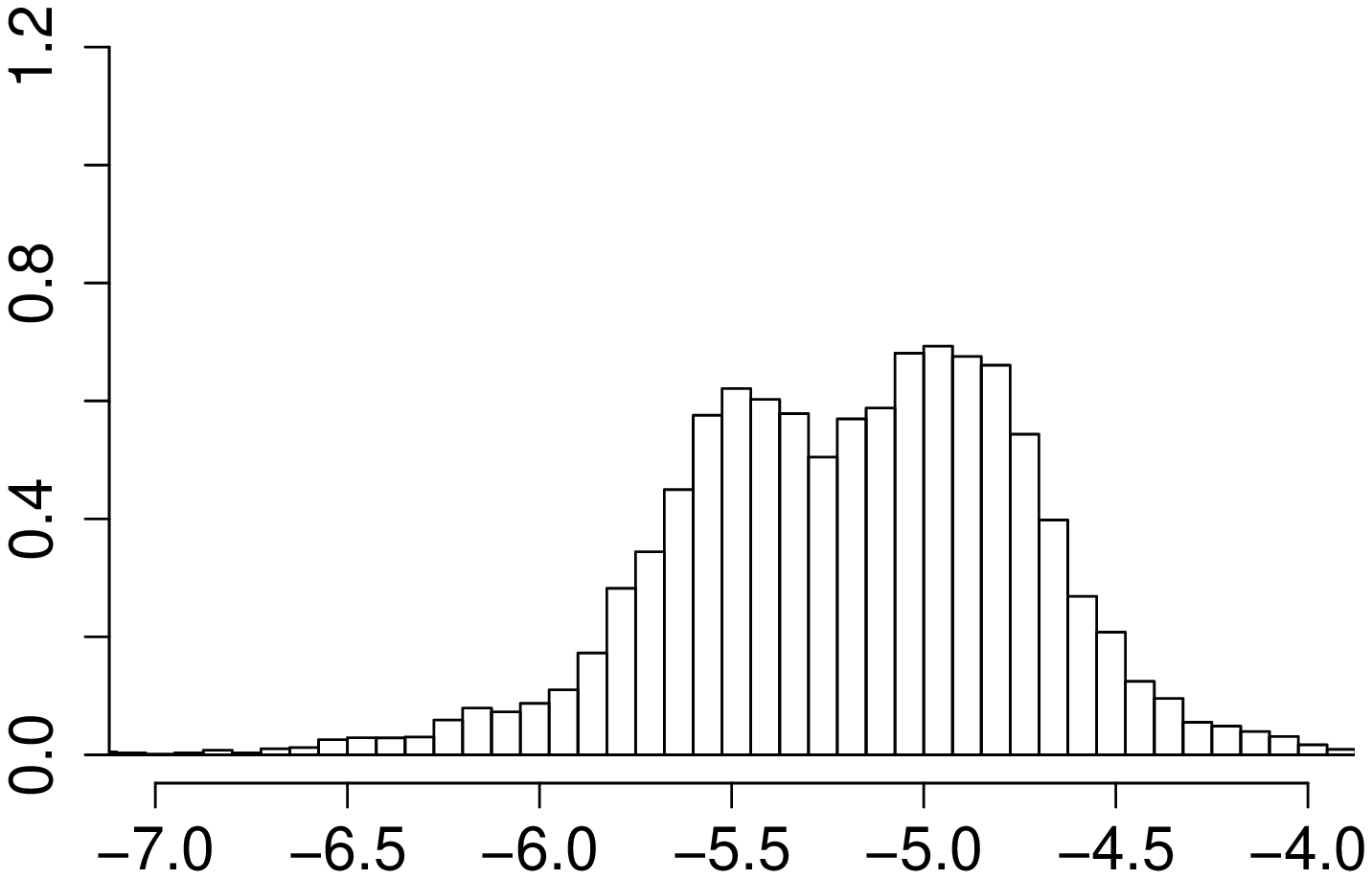}}\\
$M\neq M_{\max}$ and $\Lambda\in M$ 
&\subfigure{\includegraphics[scale=0.28]{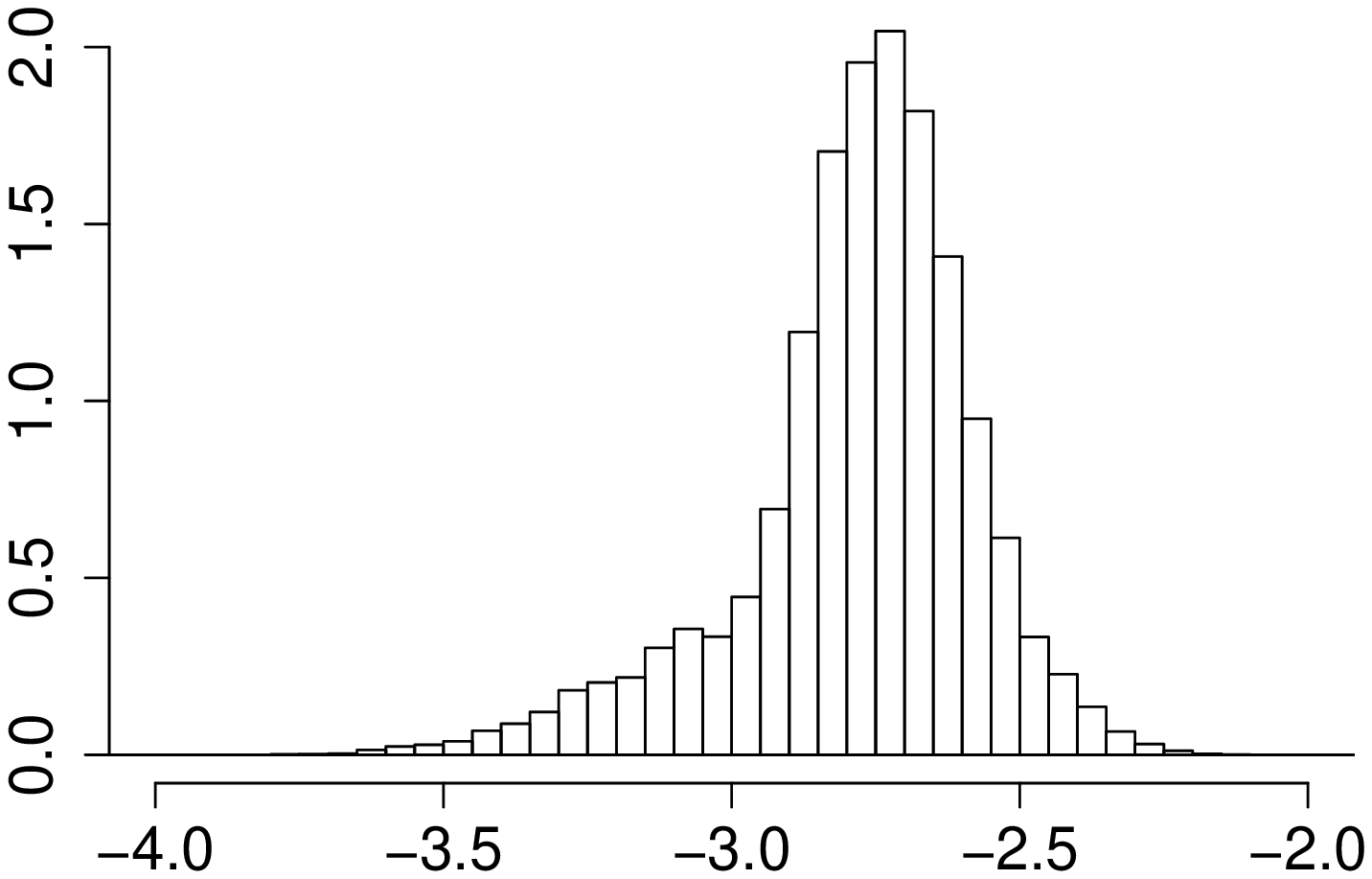}}
& \subfigure{\includegraphics[scale=0.28]{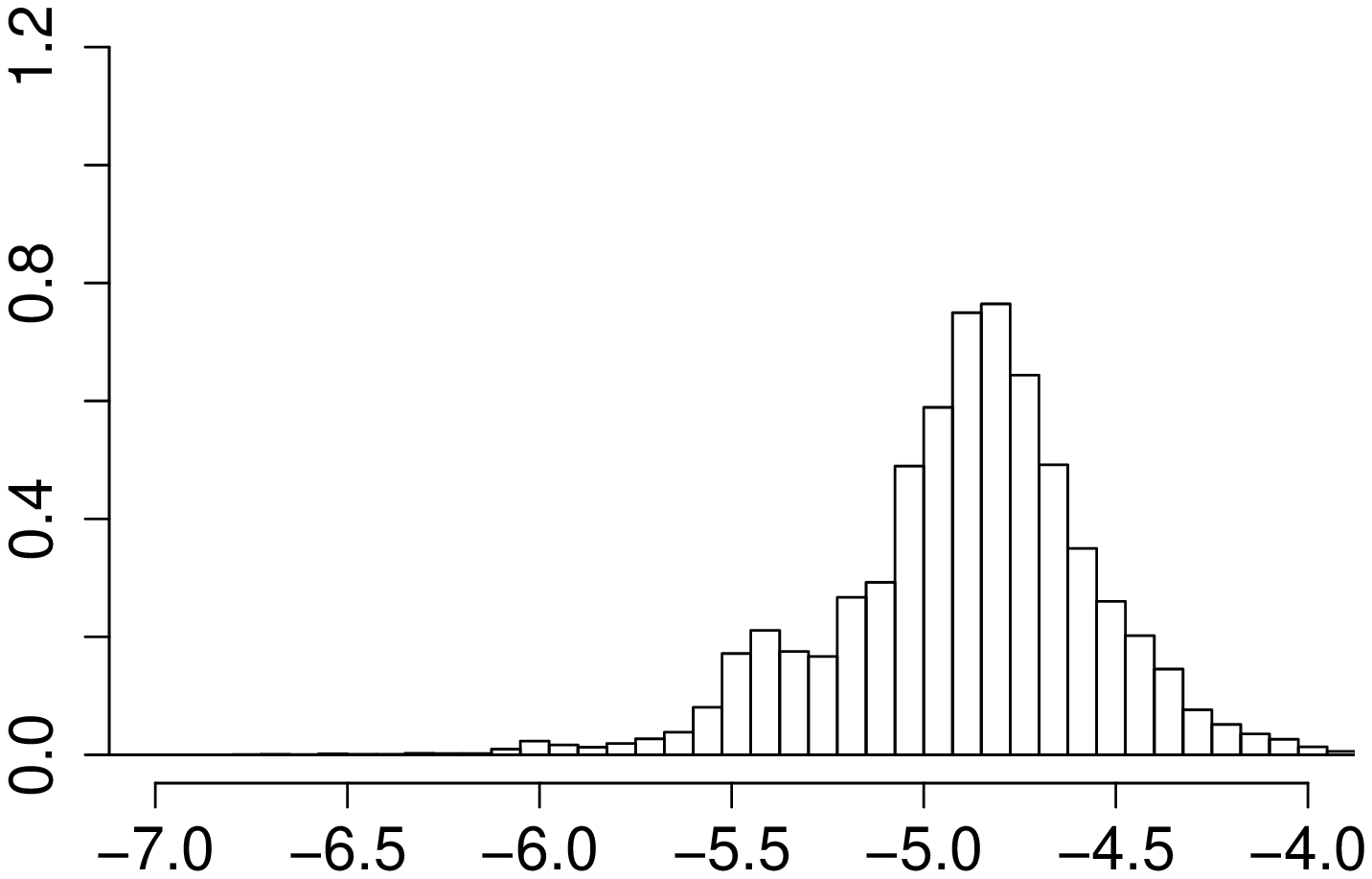}}
& \subfigure{\includegraphics[scale=0.28]{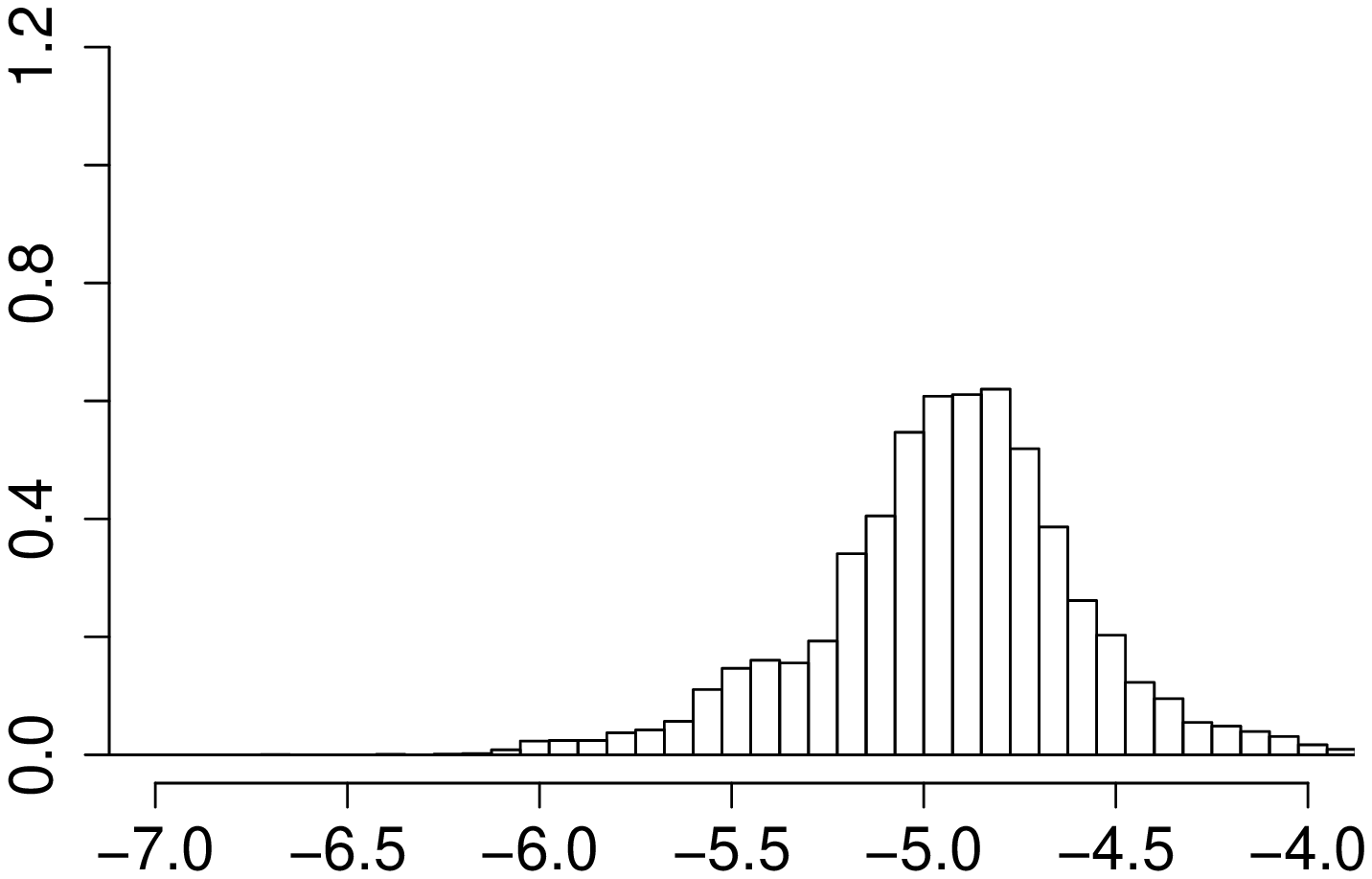}}\\
$M\neq M_{\max}$ and $\Lambda\not\in M$
&\hspace{1.8cm} NA 
& \subfigure{\includegraphics[scale=0.28]{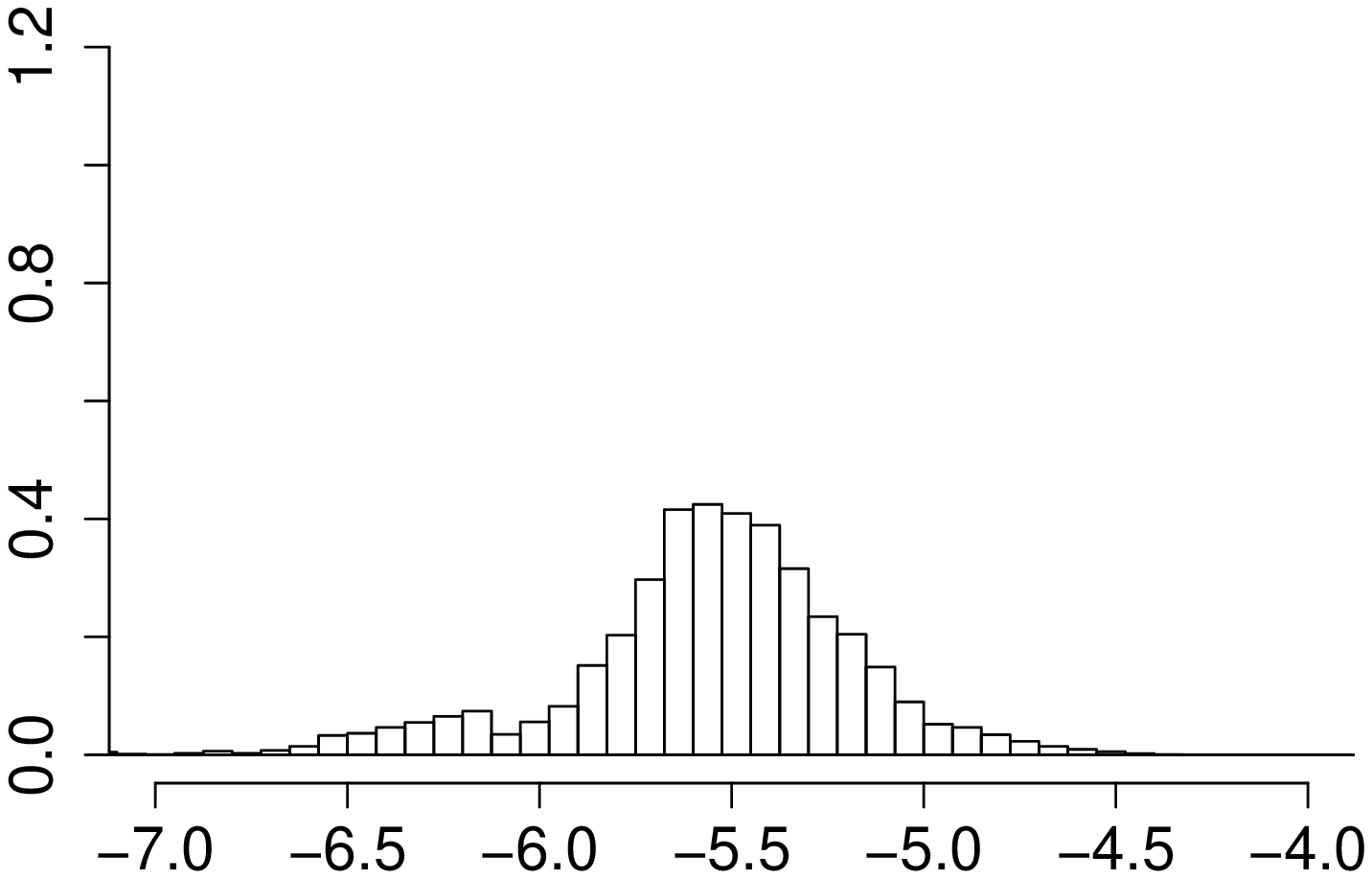}}
& \subfigure{\includegraphics[scale=0.28]{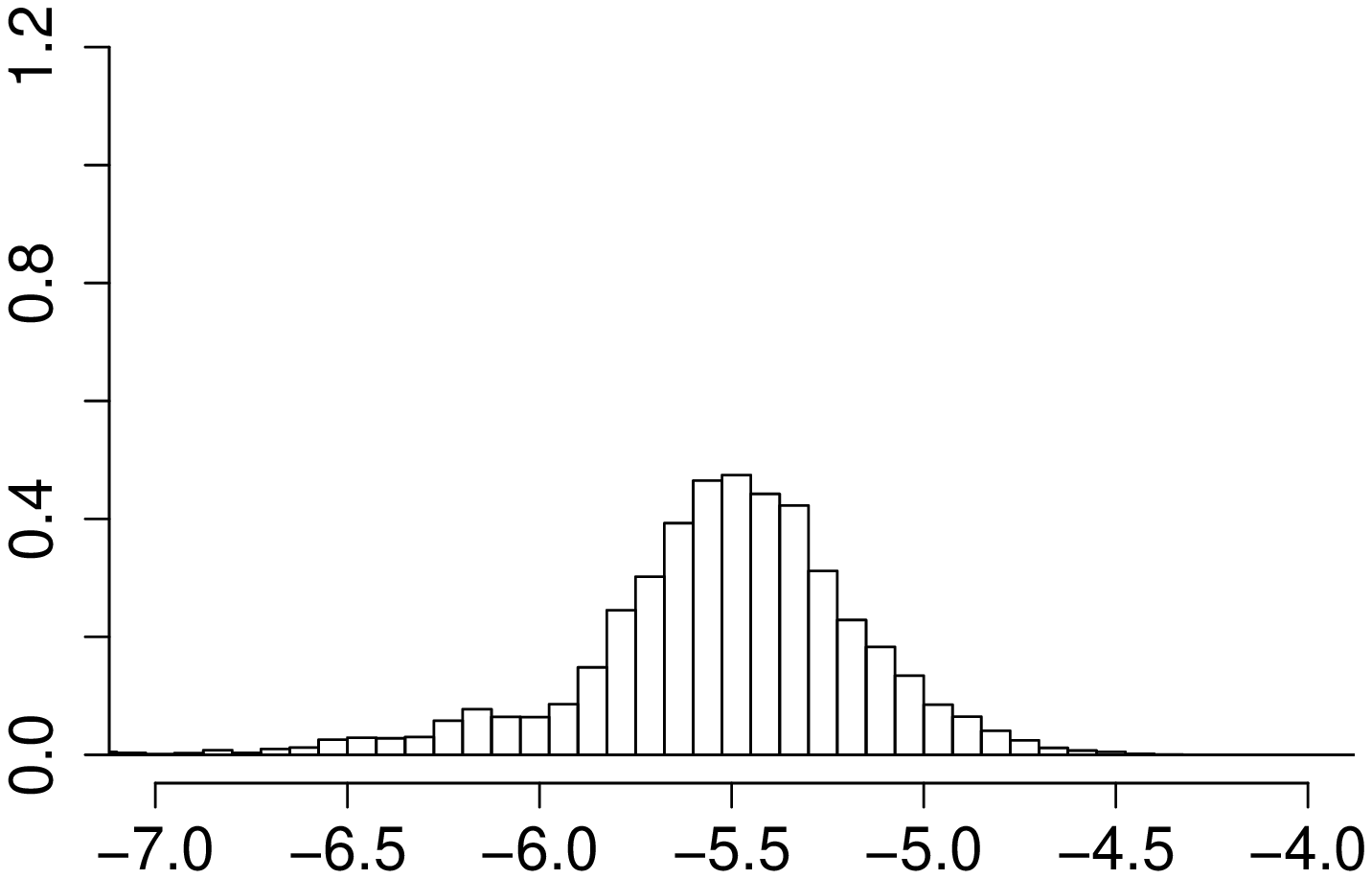}}\\
\end{tabular}
\caption{\label{fig:Sphi1}Red deer example: Histograms for 
$\phi^\Lambda$ for some $\Lambda\in L$ for different subgroups of the 
posterior samples. Results for different clique types $\Lambda$ in 
each column, and different subgroups in each row. The 
conditions given in the first column are defining the different subgroups.}
\end{figure}

\begin{figure}
\begin{tabular}{m{2.3cm}|m{4cm}m{4cm}m{4cm}}
& \hspace{1.8cm}$\Lambda=\ABBA$
&\hspace{1.8cm}$\Lambda=\BAAB$
&\hspace{1.8cm}$\Lambda=\BAAA$\\
\hline
All
& \subfigure{\includegraphics[scale=0.28]{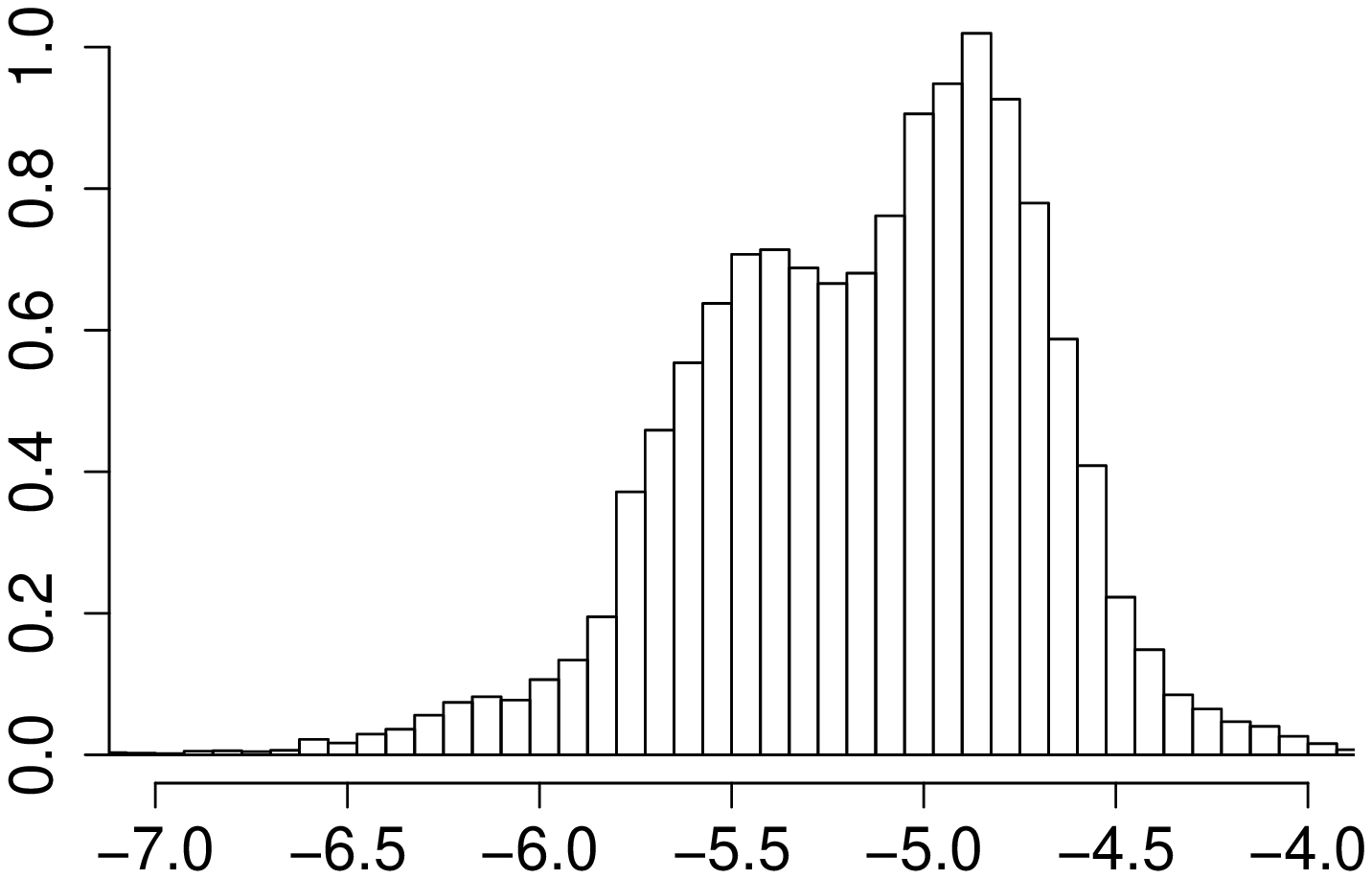}}
& \subfigure{\includegraphics[scale=0.28]{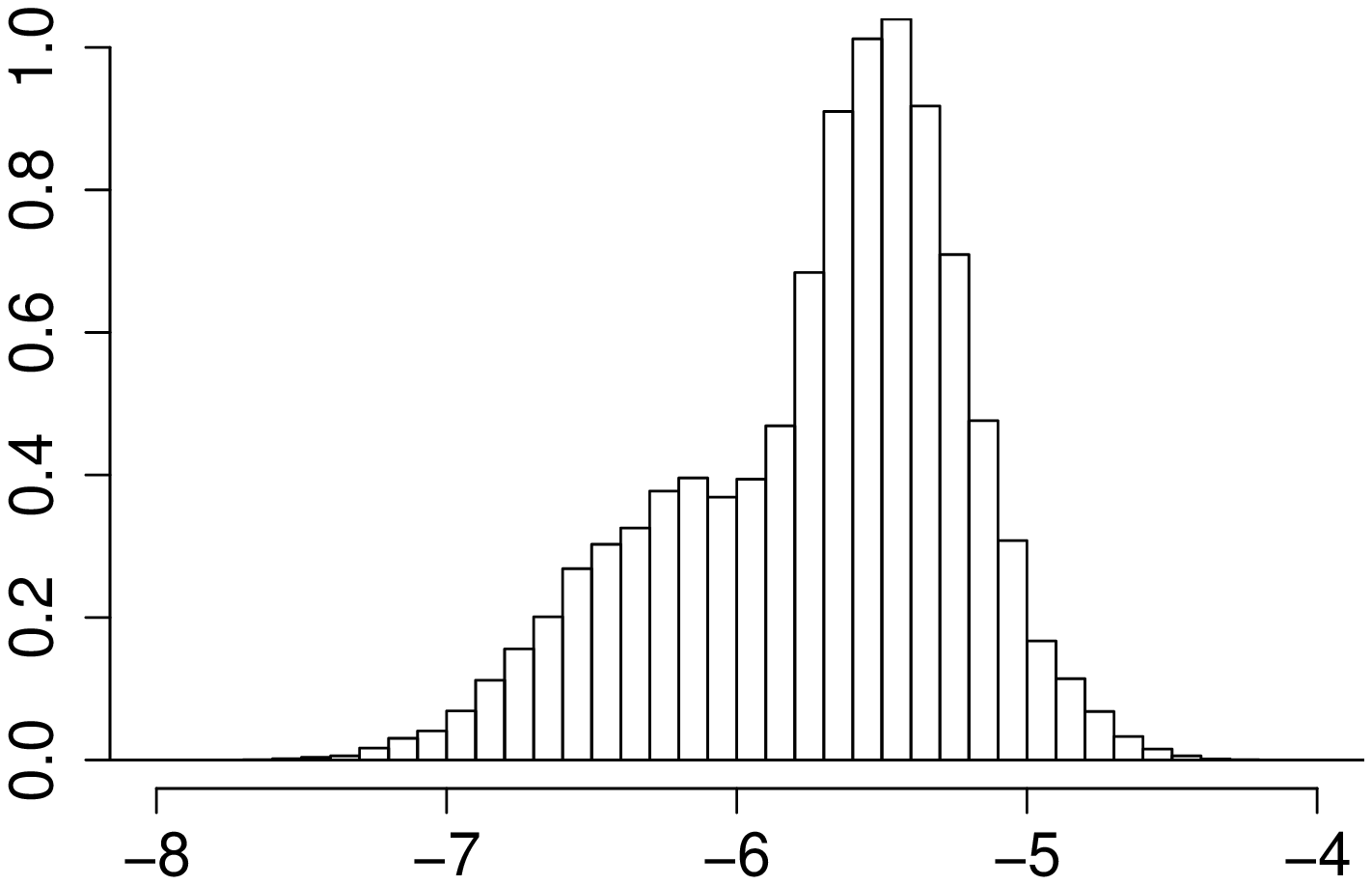}}
& \subfigure{\includegraphics[scale=0.28]{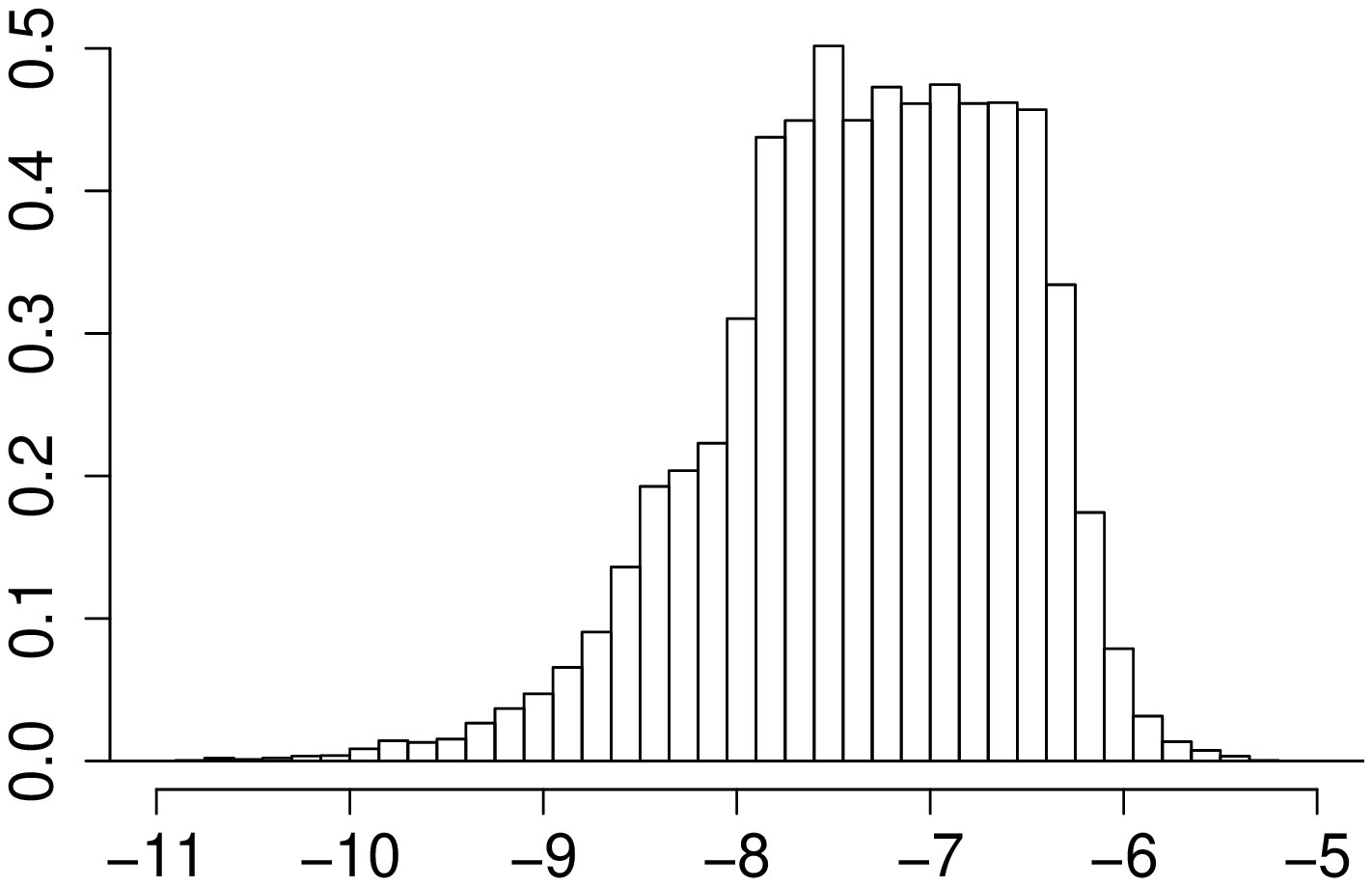}}\\
$M=M_{\max}$ 
& \subfigure{\includegraphics[scale=0.28]{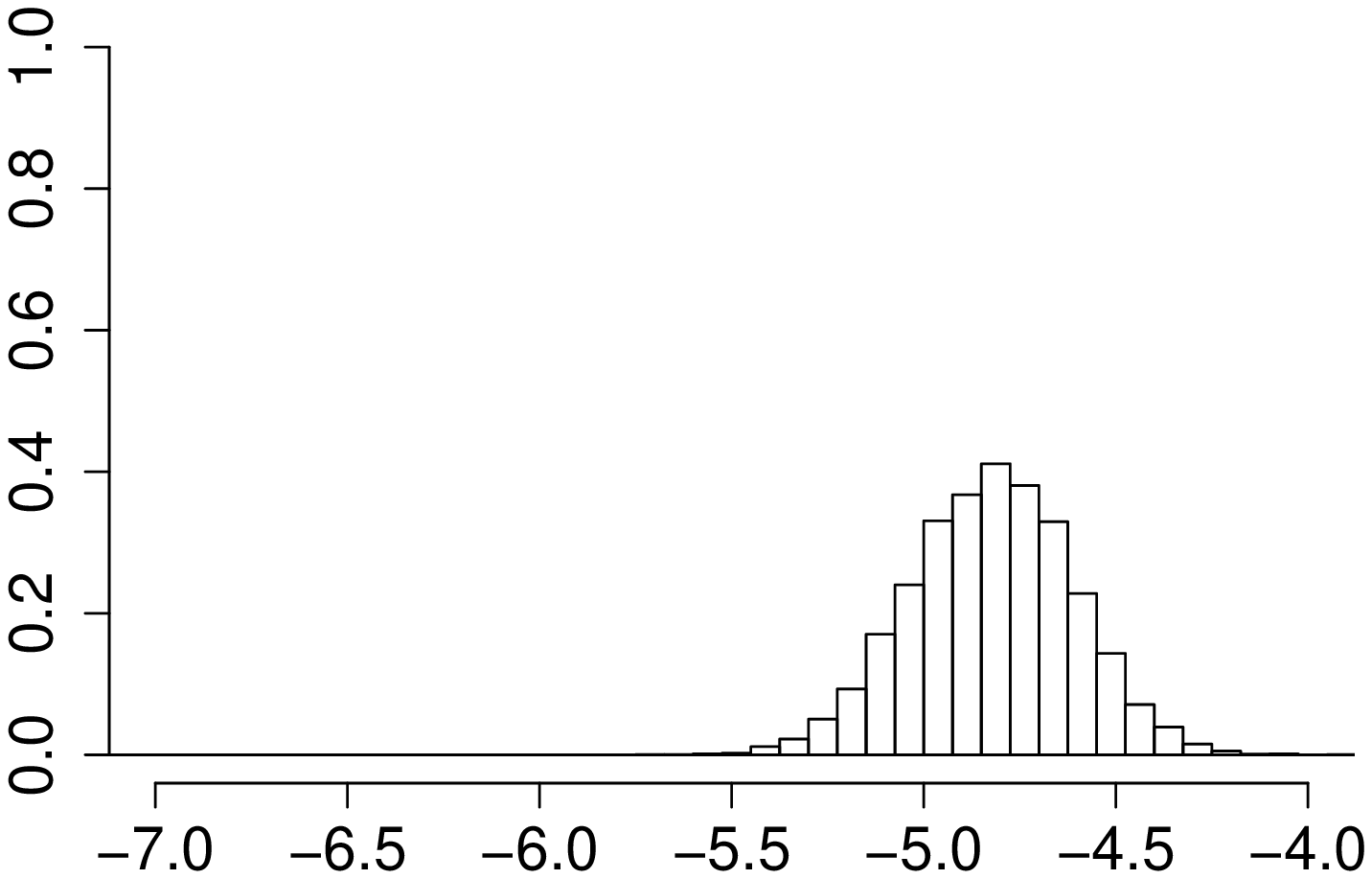}}
& \subfigure{\includegraphics[scale=0.28]{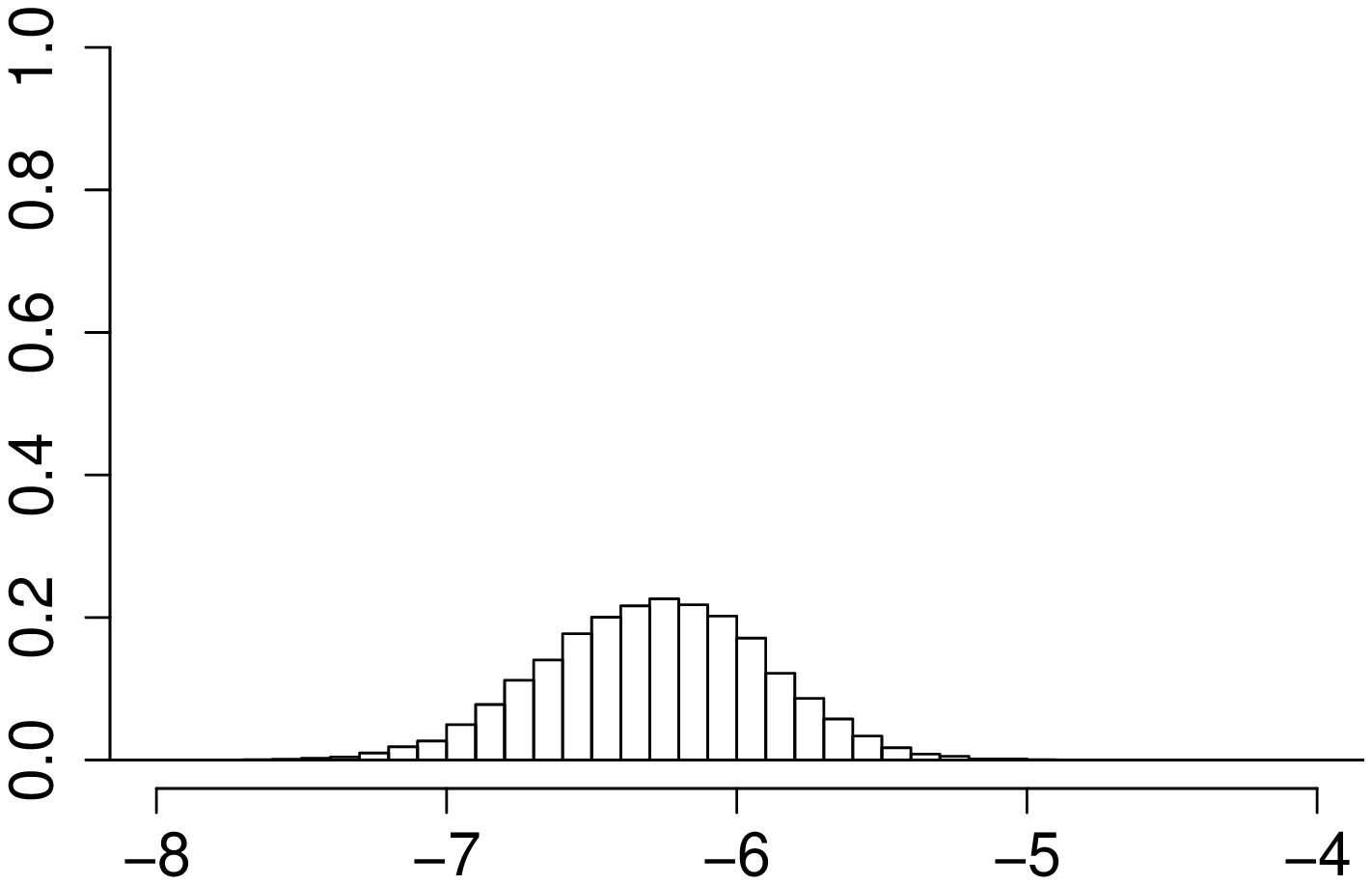}}
& \subfigure{\includegraphics[scale=0.28]{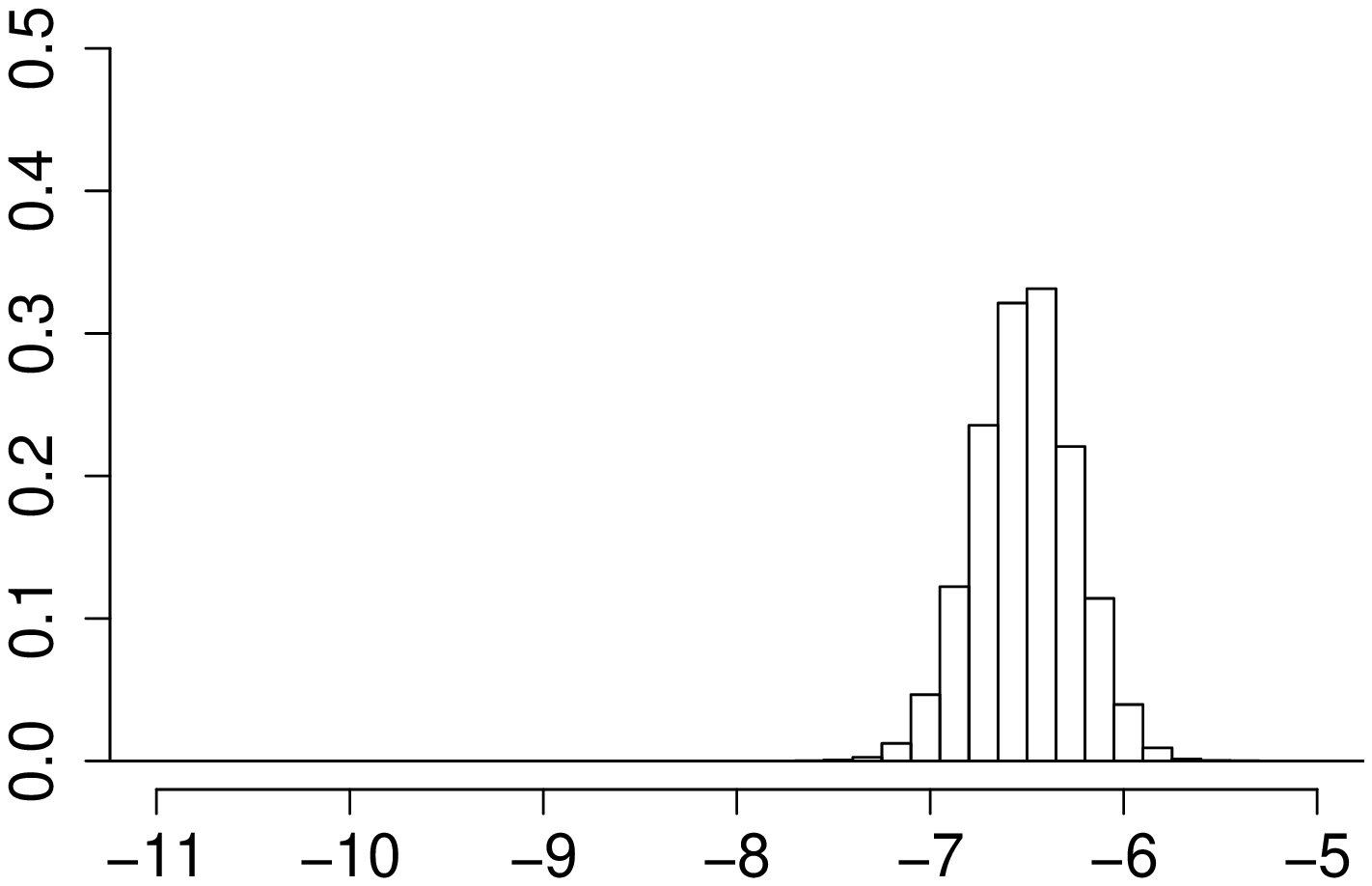}}\\
$M\neq M_{\max}$ 
& \subfigure{\includegraphics[scale=0.28]{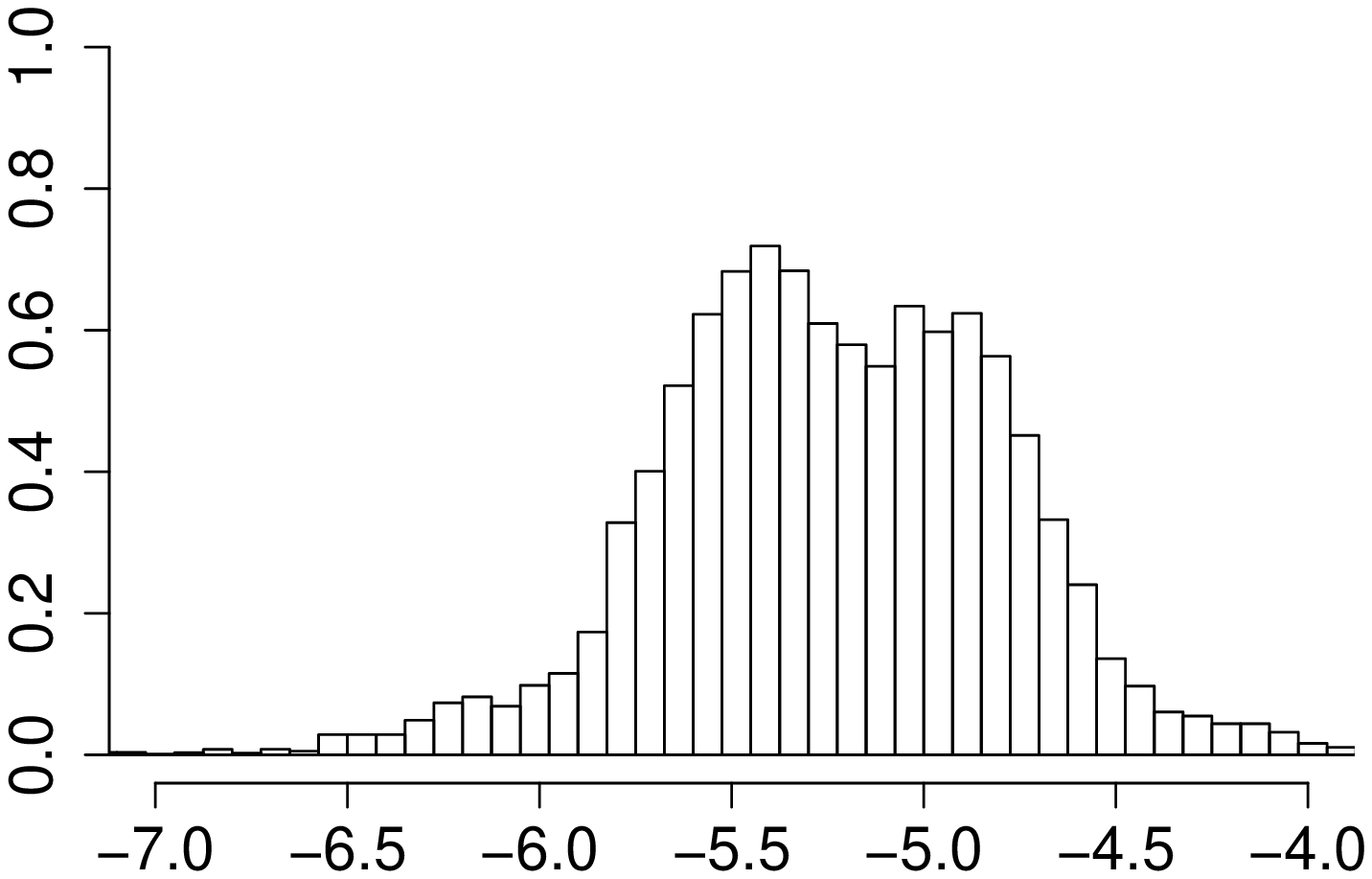}}
&\subfigure{\includegraphics[scale=0.28]{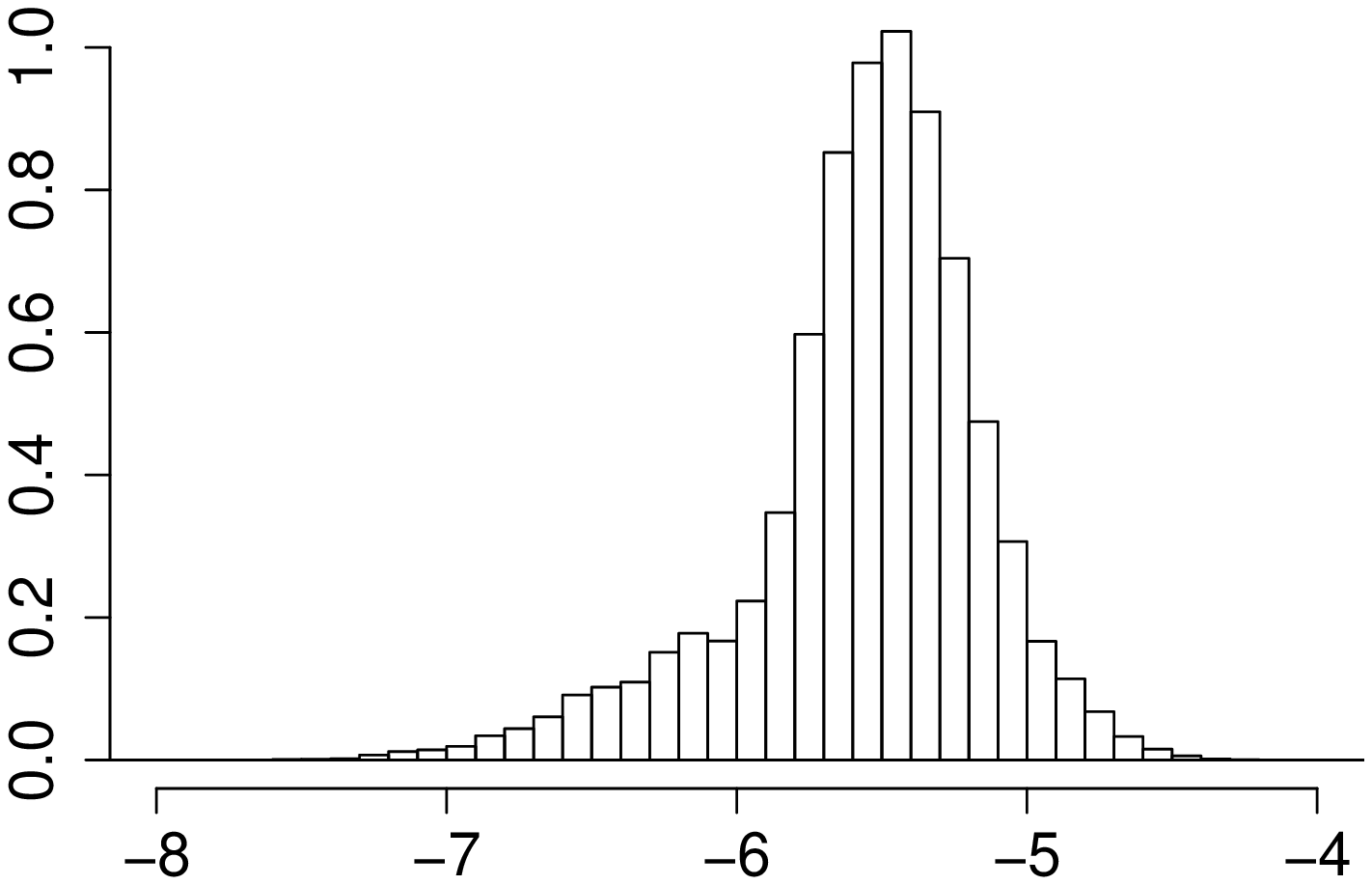}}
&\subfigure{\includegraphics[scale=0.28]{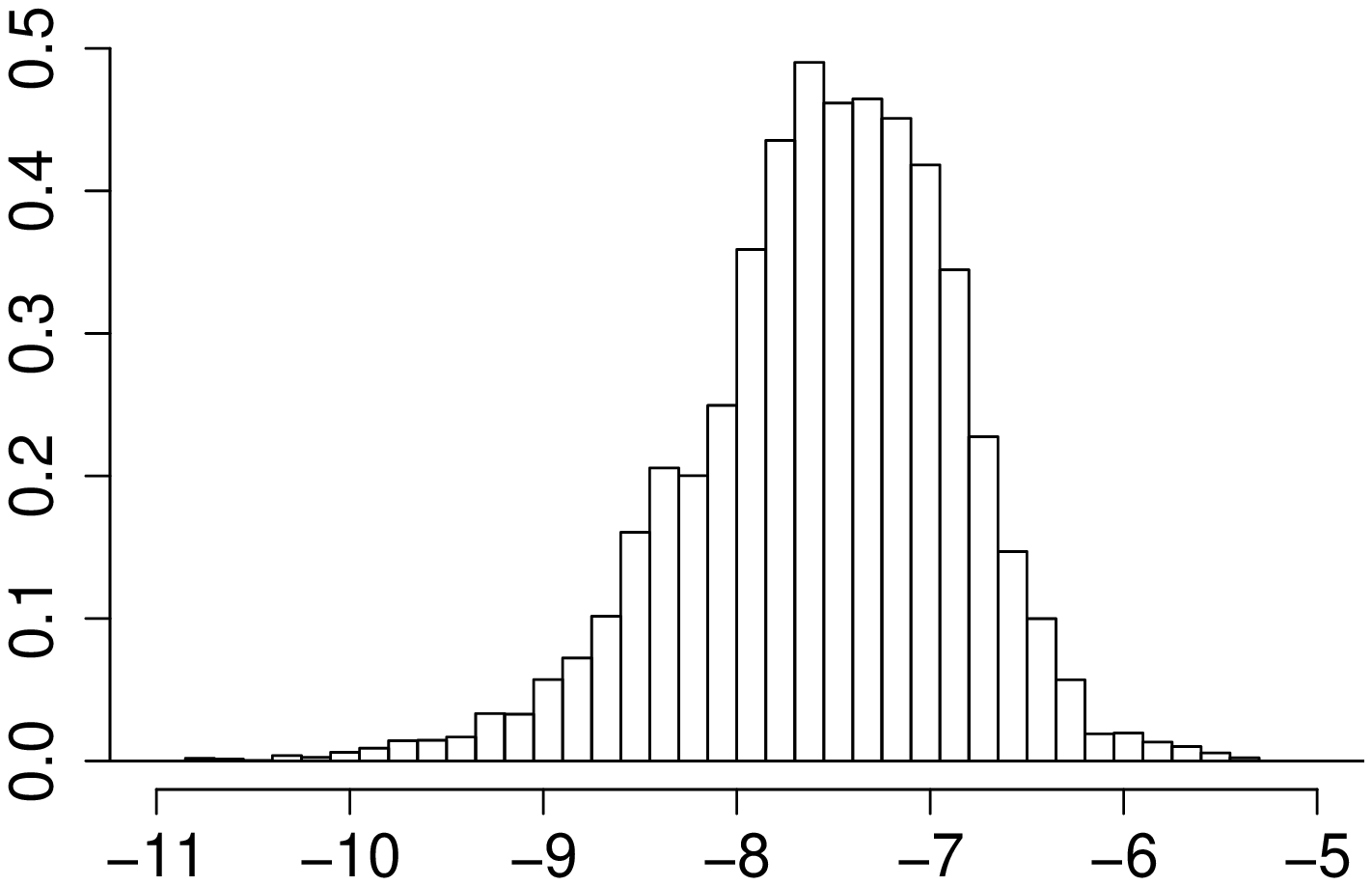}}\\
$M\neq M_{\max}$ and $\Lambda\in M$
&\subfigure{\includegraphics[scale=0.28]{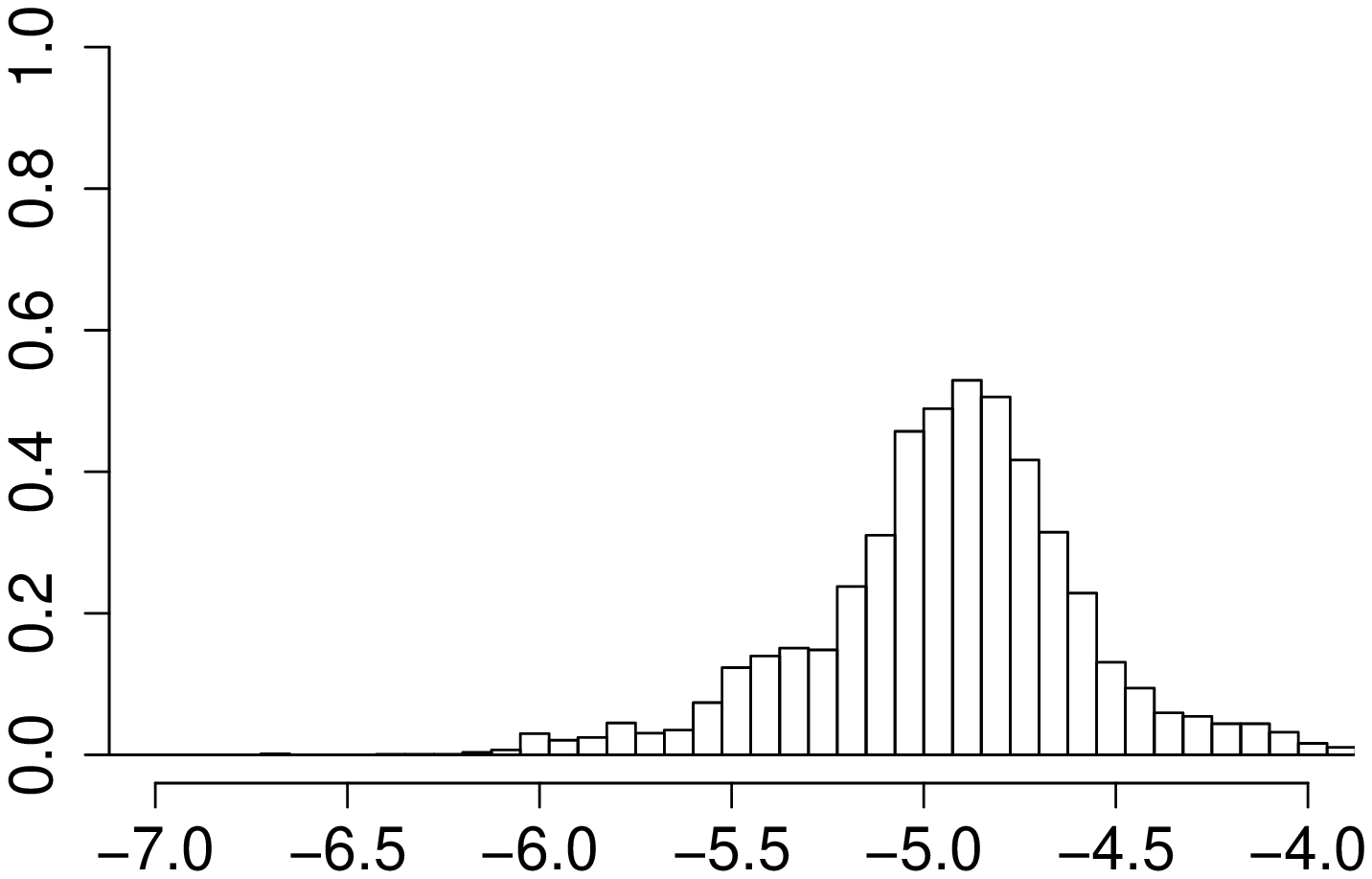}}
&\hspace{1.8cm} NA % \subfigure{\includegraphics[scale=0.28]{Newpno0phi22.eps}}
& \hspace{1.8cm} NA
% \subfigure{\includegraphics[scale=0.28]{Newpno0phi24.eps}}
\\
$M\neq M_{\max}$ and $\Lambda\not\in M$
& \subfigure{\includegraphics[scale=0.28]{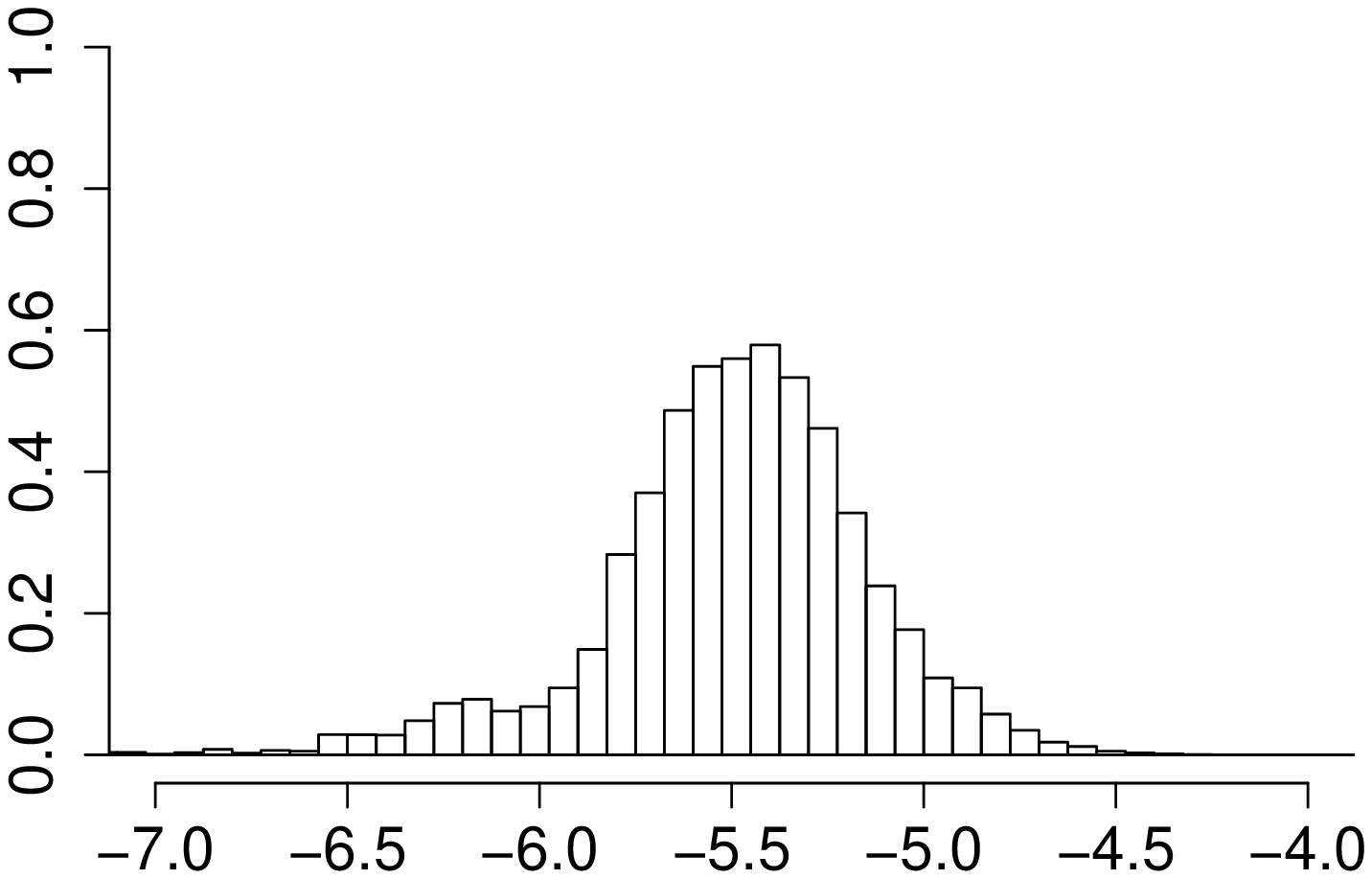}}
& \subfigure{\includegraphics[scale=0.28]{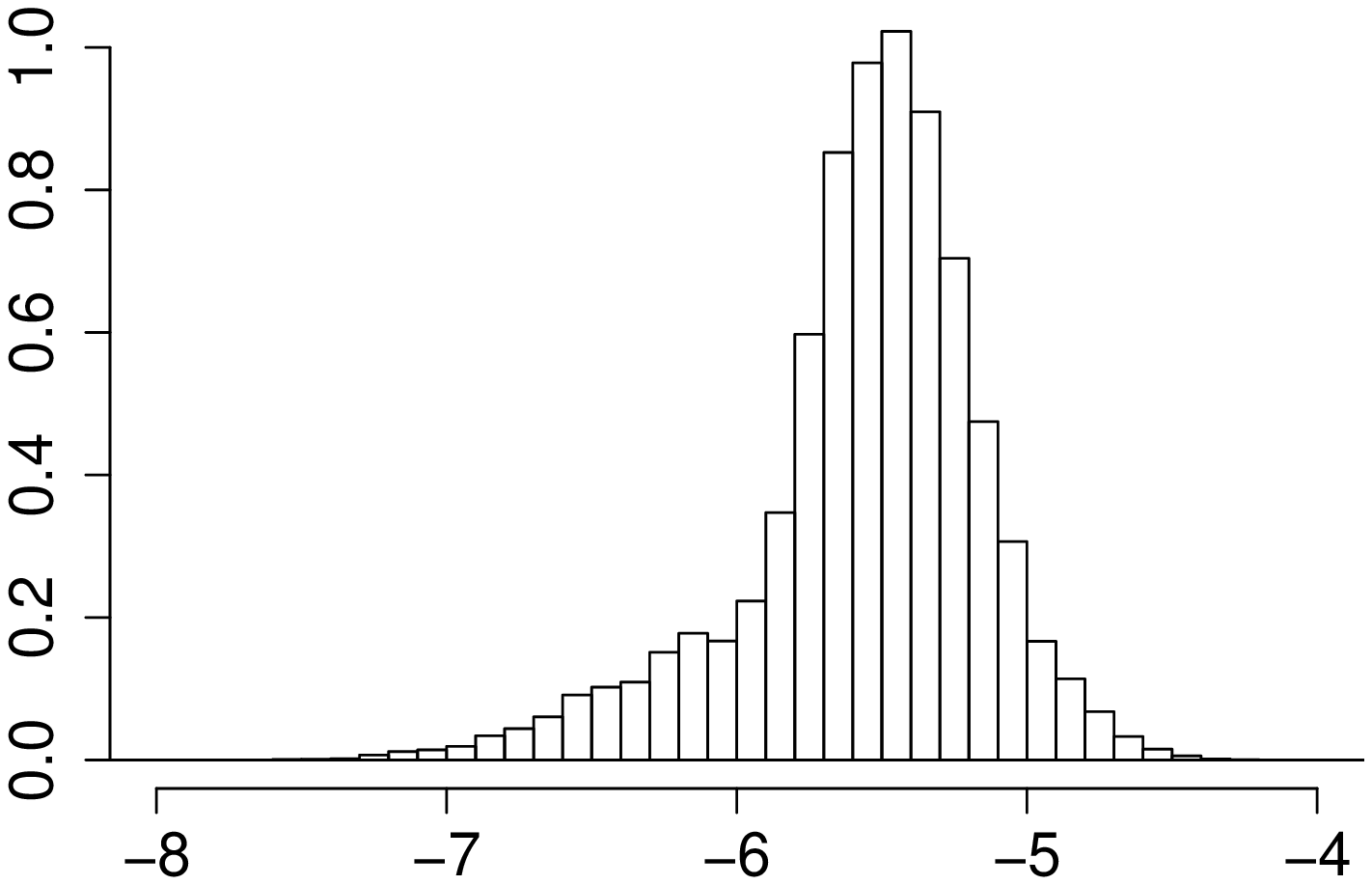}}
& \subfigure{\includegraphics[scale=0.28]{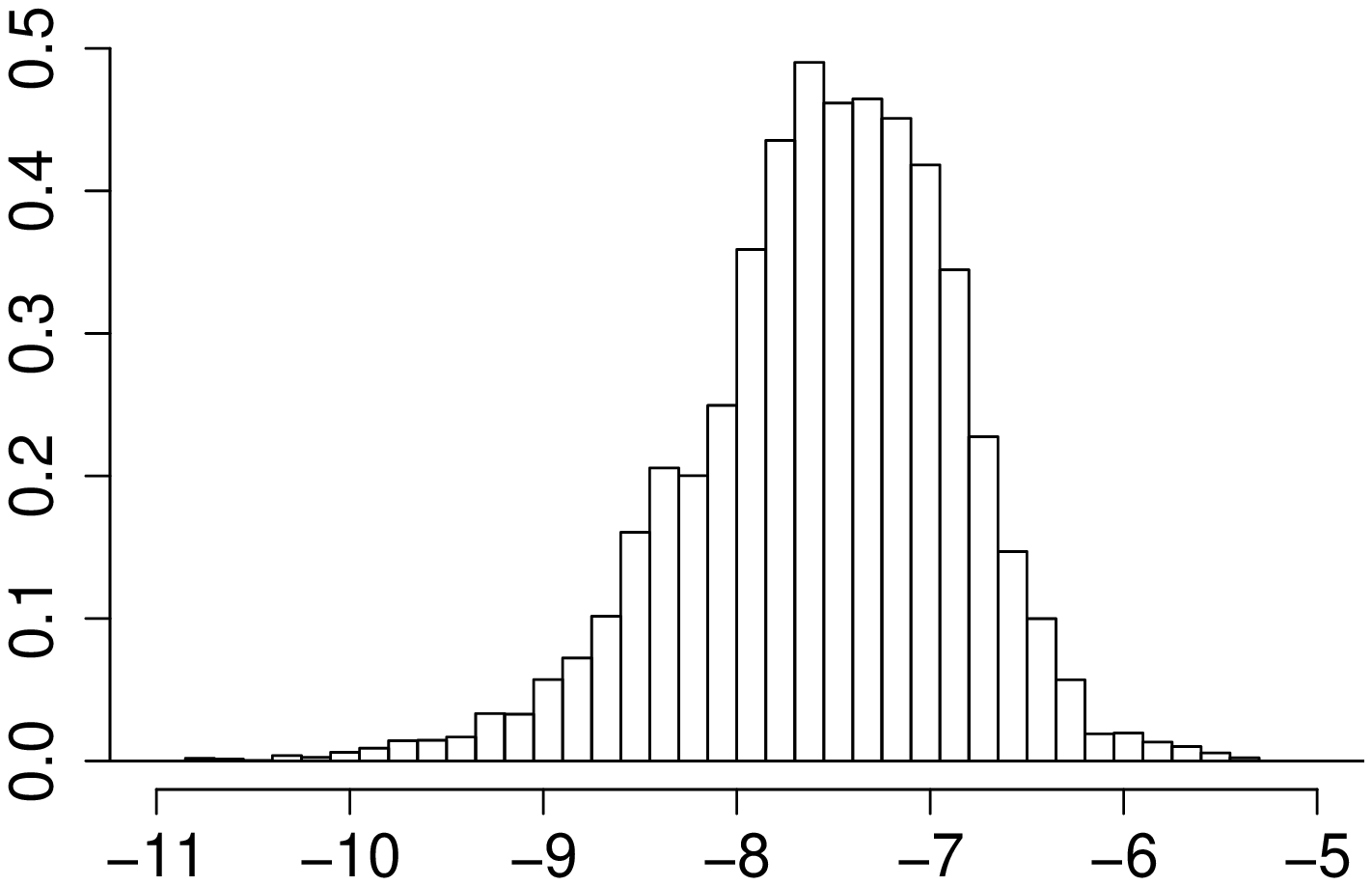}}\\
\end{tabular}
\caption{\label{fig:Sphi2}Red deer example: Histograms for 
$\phi^\Lambda$ for some $\Lambda\in L$ for different subgroups of the 
posterior samples. Results for different clique types $\Lambda$ in 
each column, and different subgroups in each row. The 
conditions given in the first column are defining the different subgroups.}
\end{figure}

\begin{figure}
\begin{tabular}{m{2.3cm}|m{4cm}m{4cm}m{4cm}}
& \hspace{1.8cm}$\Lambda=\ABAA$
&\hspace{1.8cm}$\Lambda=\AABA$
&\hspace{1.8cm}$\Lambda=\AAAB$\\
\hline
All
& \subfigure{\includegraphics[scale=0.28]{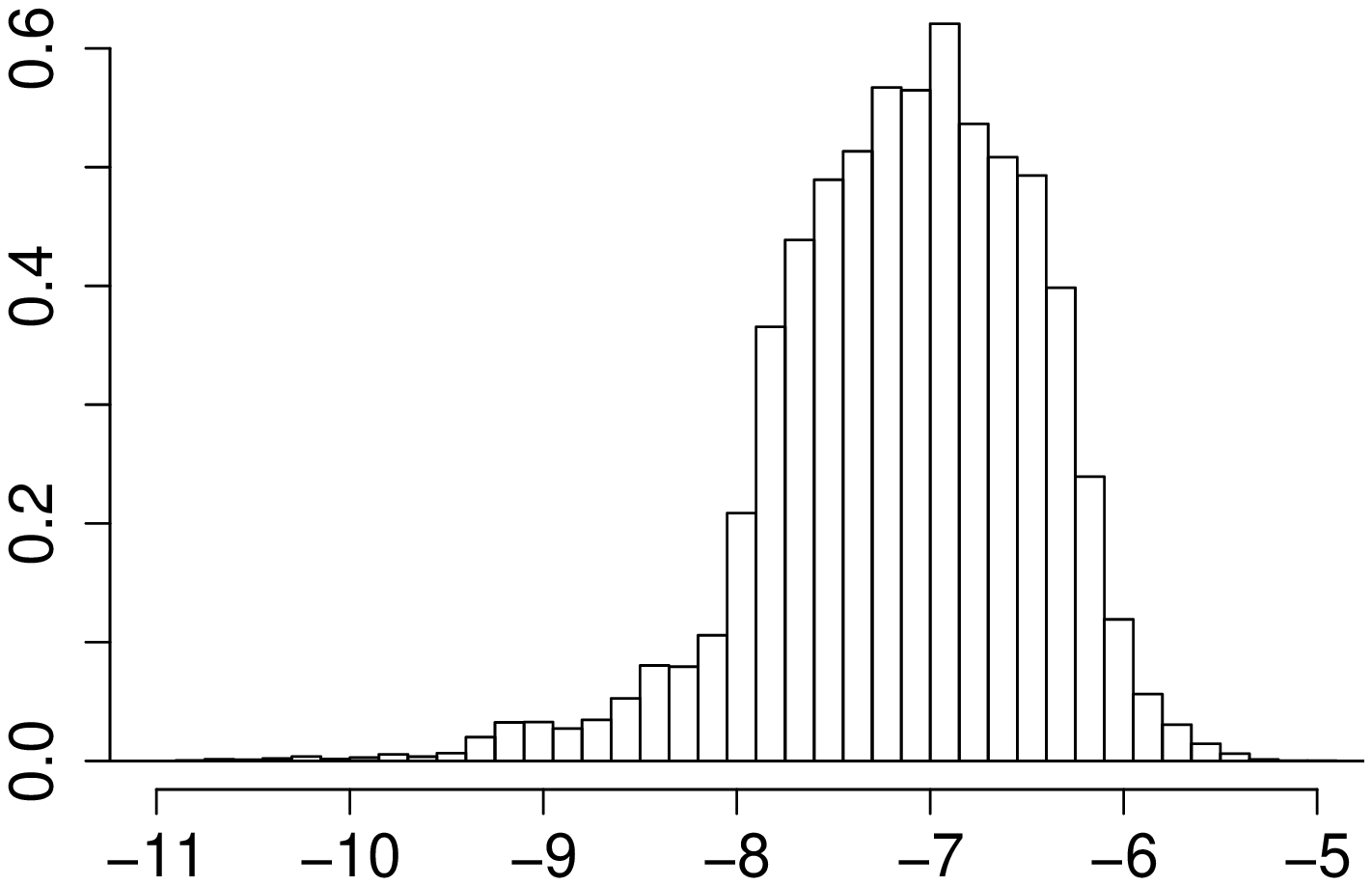}}
& \subfigure{\includegraphics[scale=0.28]{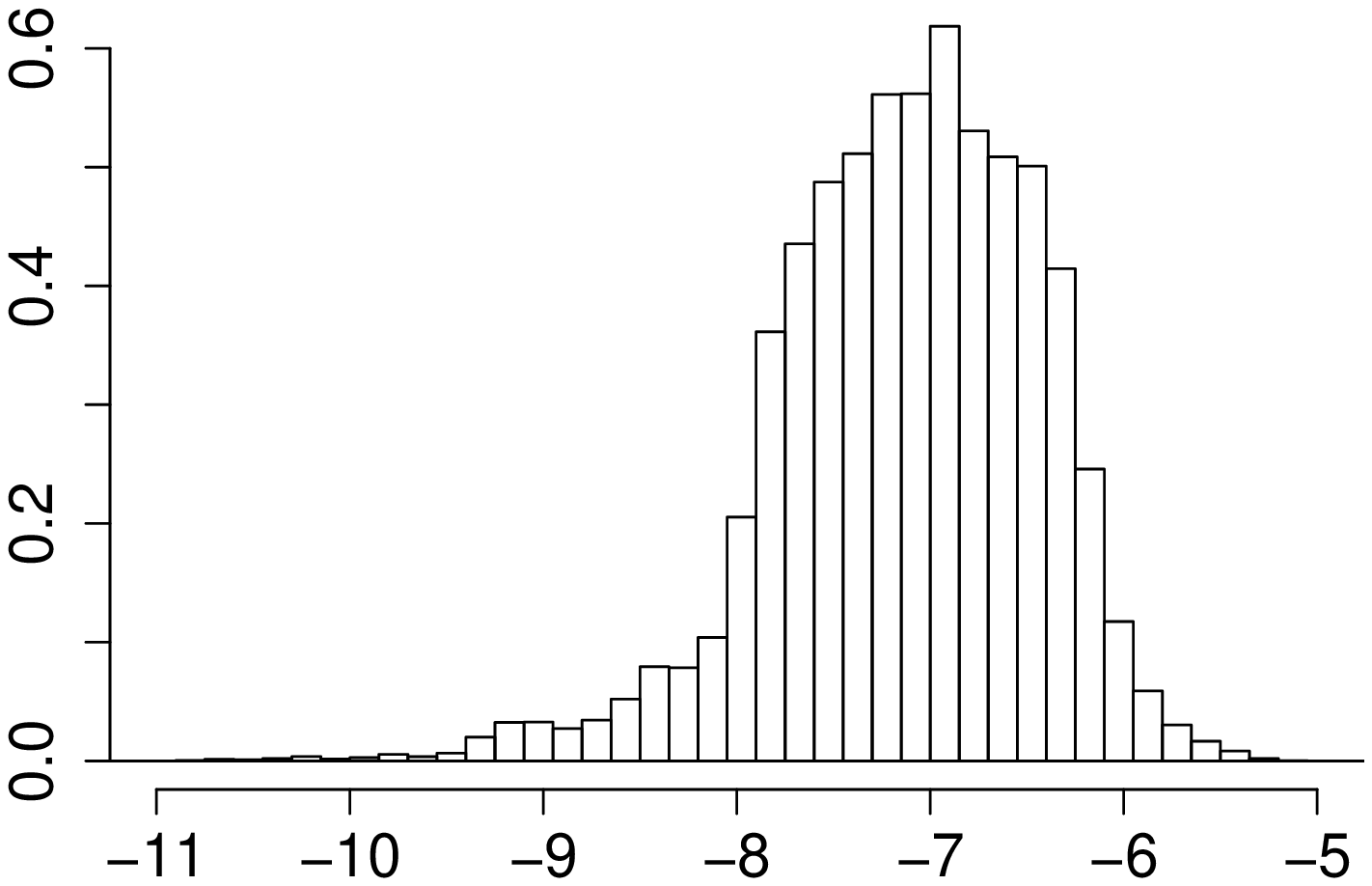}}
& \subfigure{\includegraphics[scale=0.28]{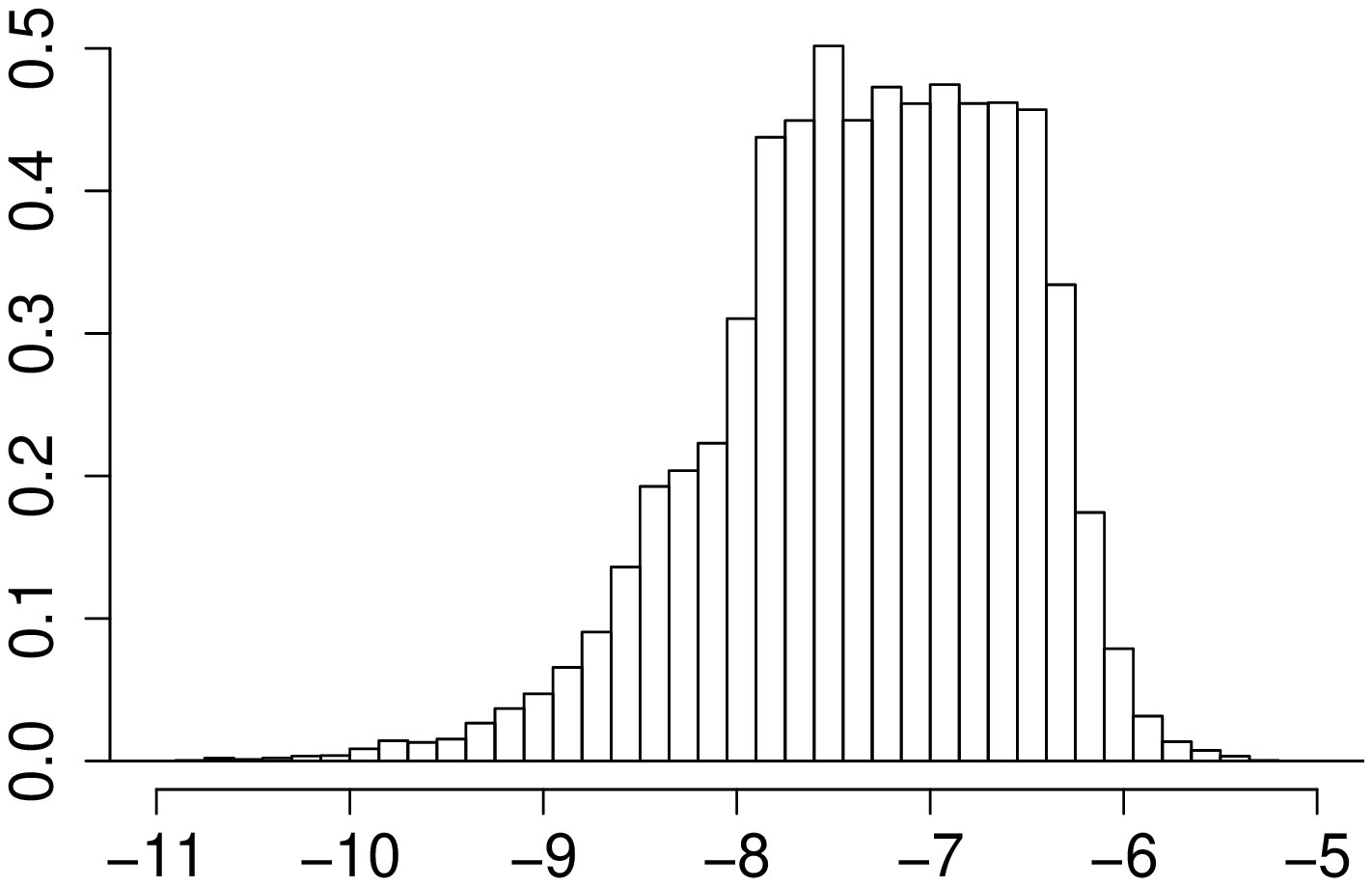}}\\
$M=M_{\max}$ 
& \subfigure{\includegraphics[scale=0.28]{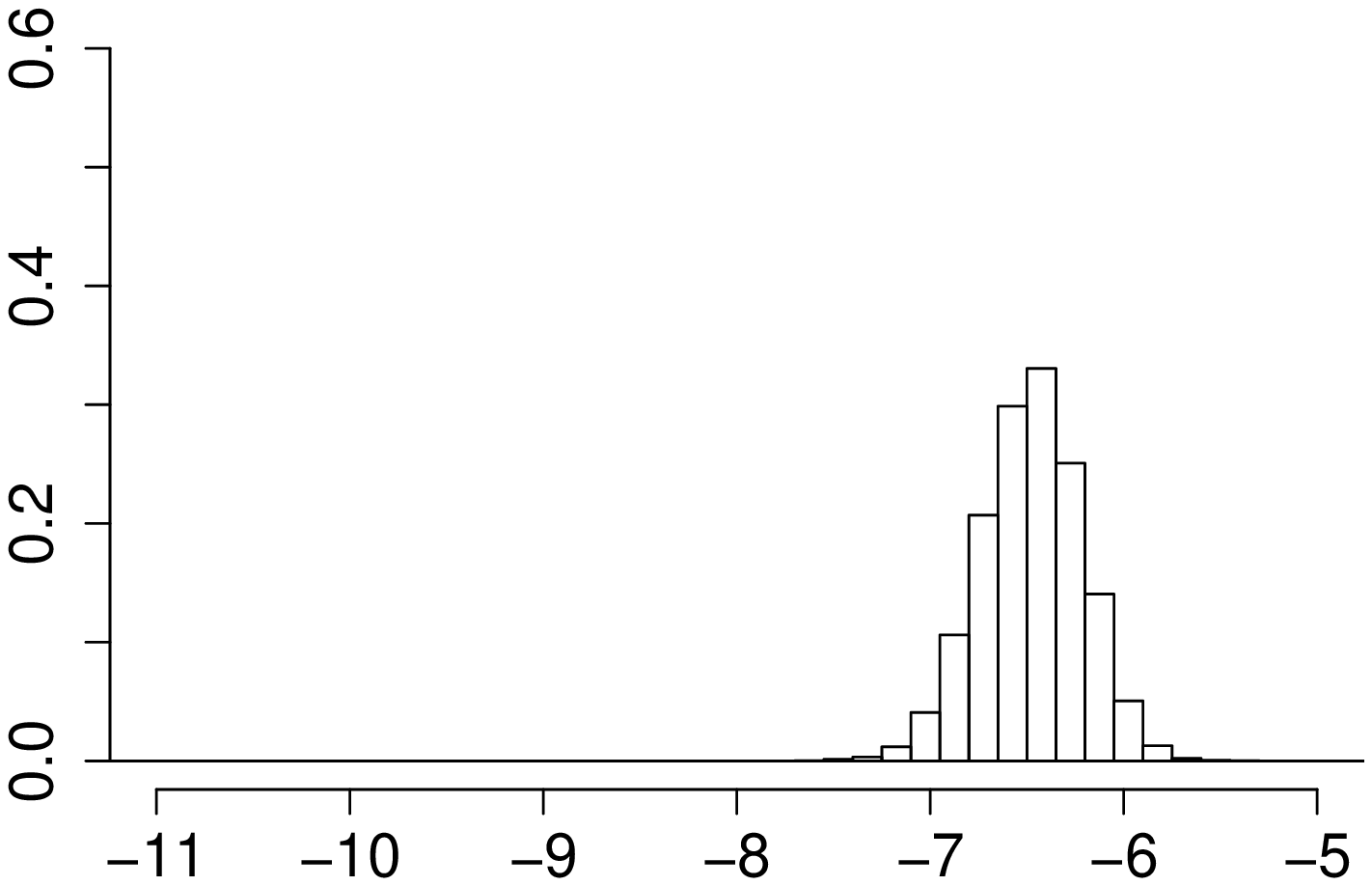}}
& \subfigure{\includegraphics[scale=0.28]{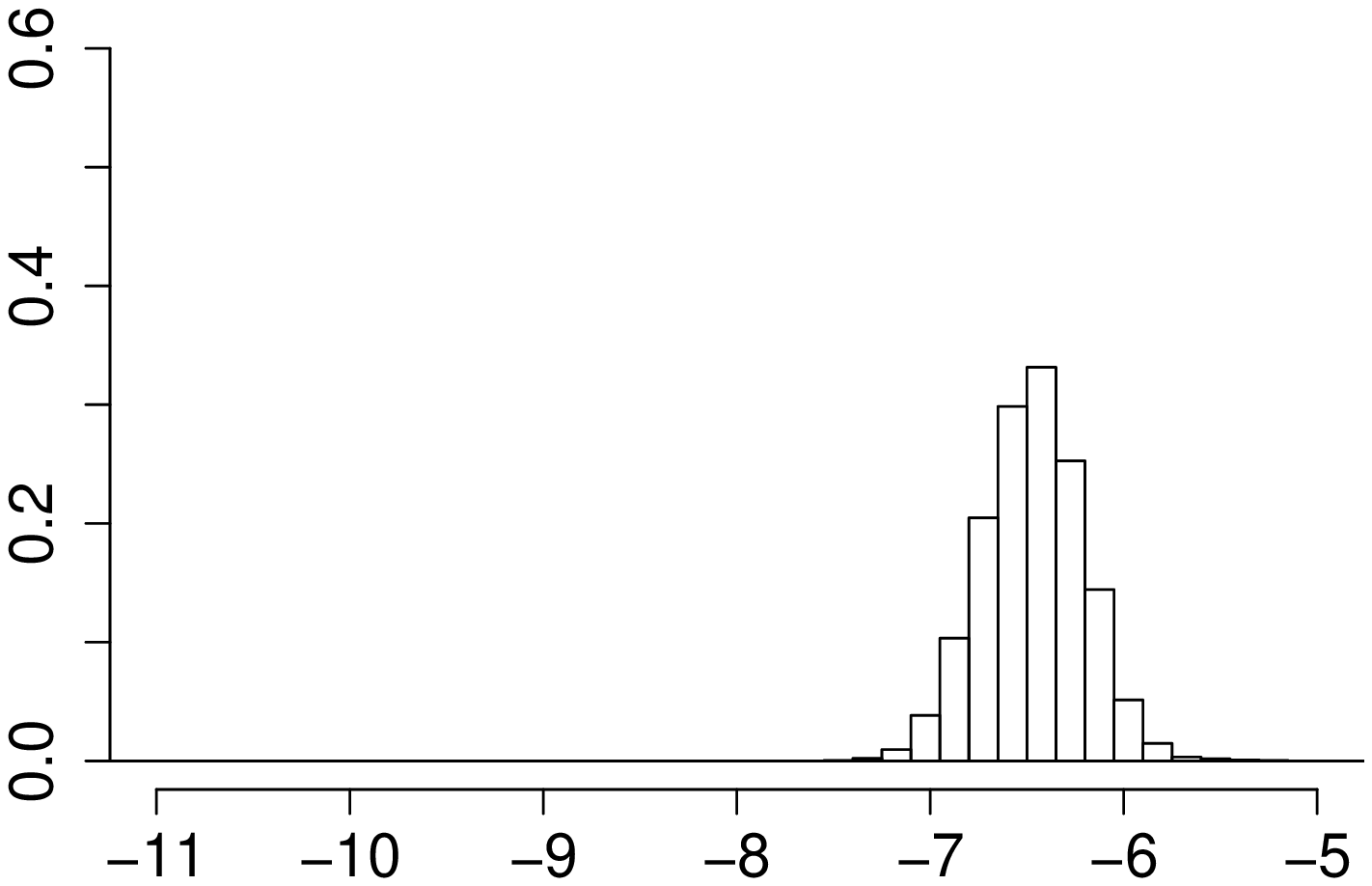}}
& \subfigure{\includegraphics[scale=0.28]{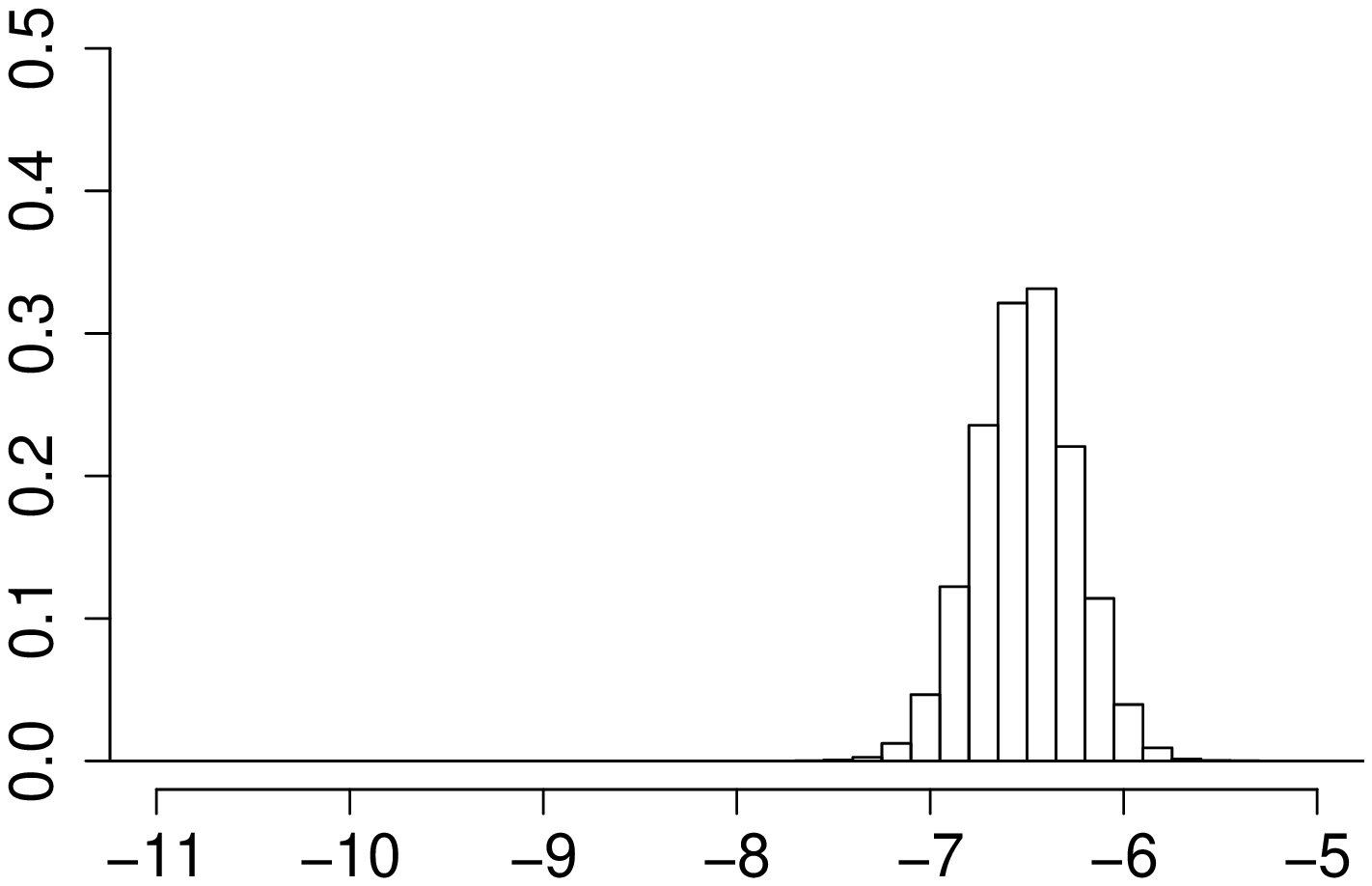}}\\
$M\neq M_{\max}$ 
& \subfigure{\includegraphics[scale=0.28]{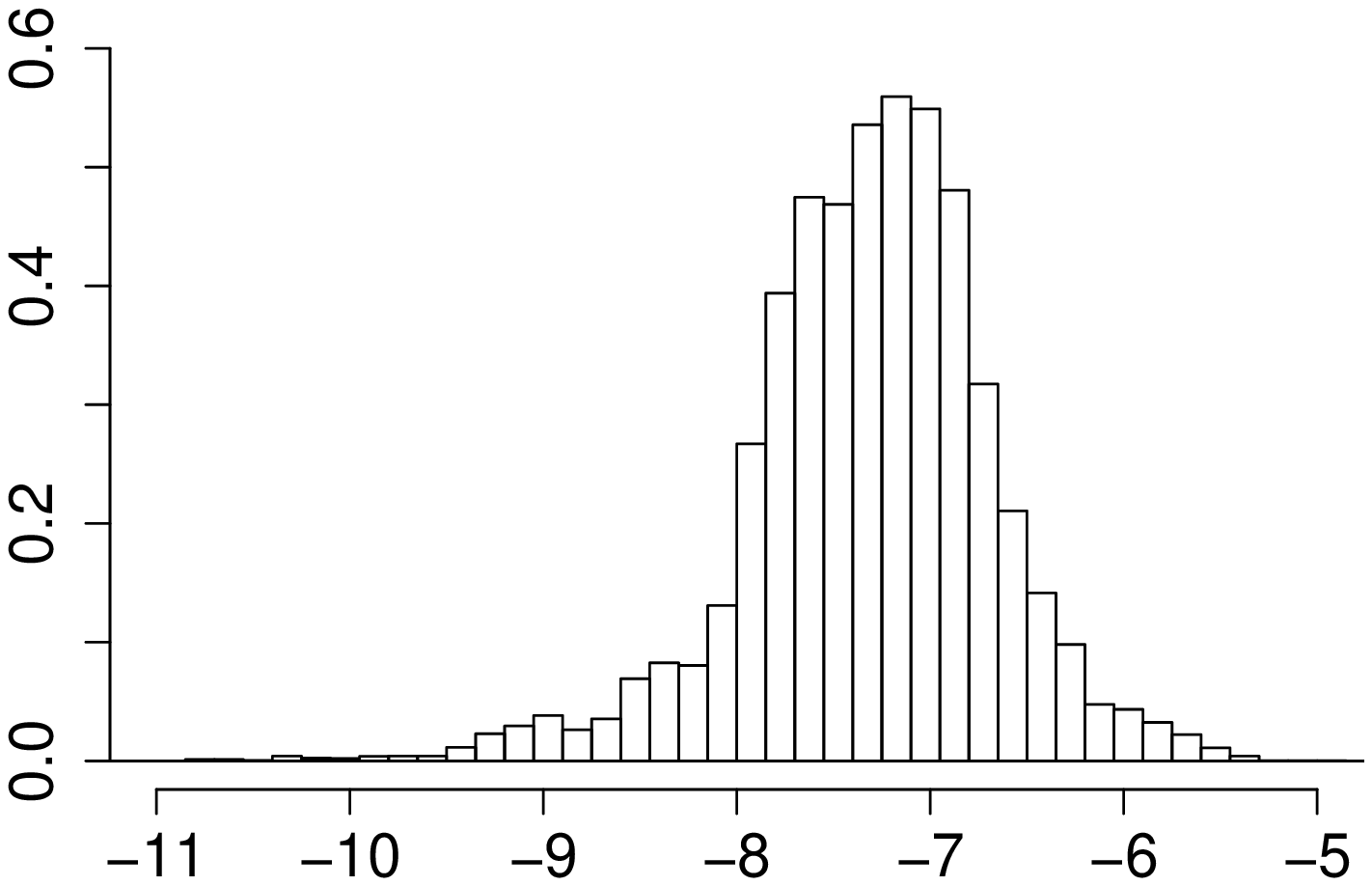}}
&\subfigure{\includegraphics[scale=0.28]{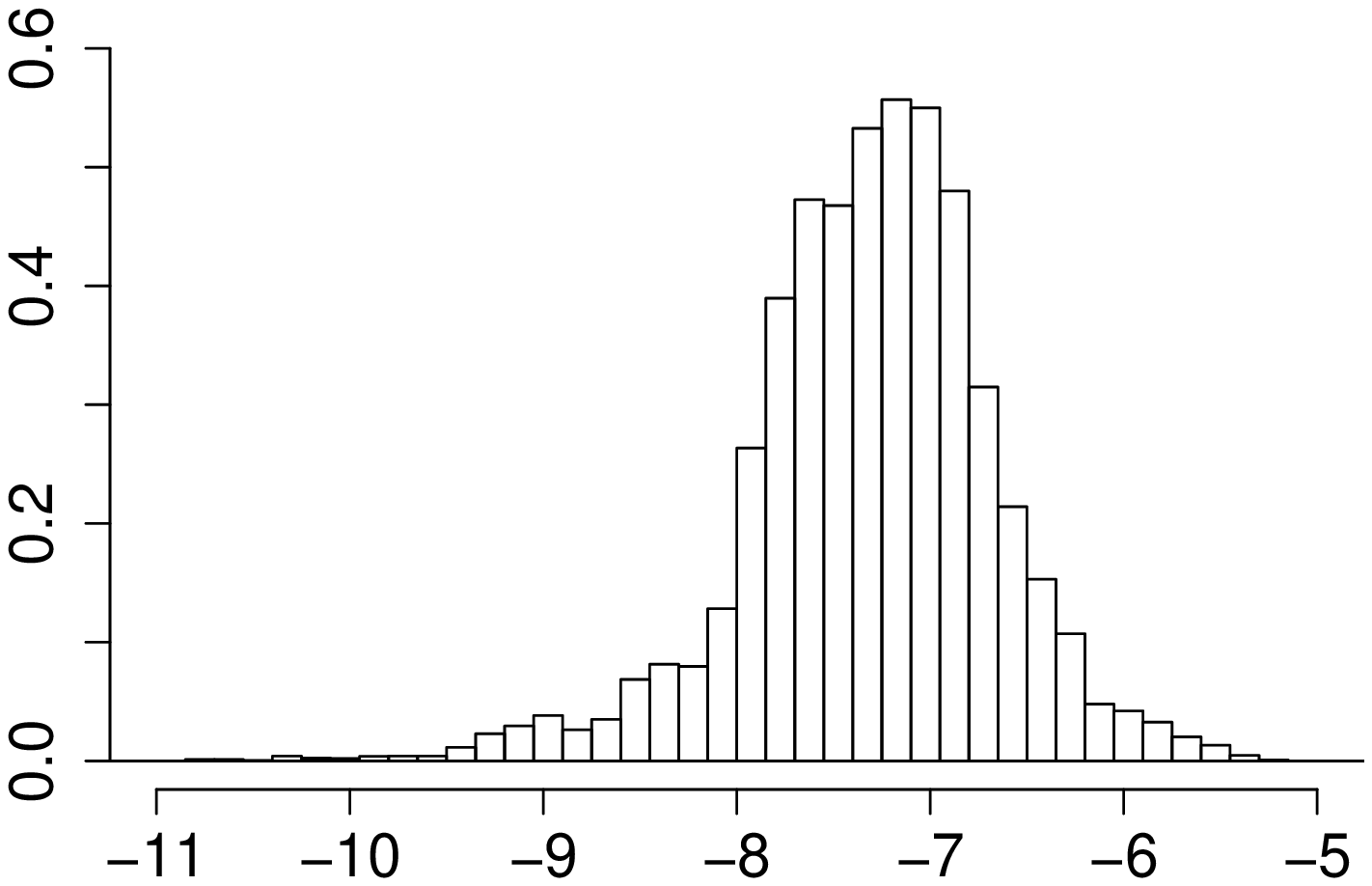}}
&\subfigure{\includegraphics[scale=0.28]{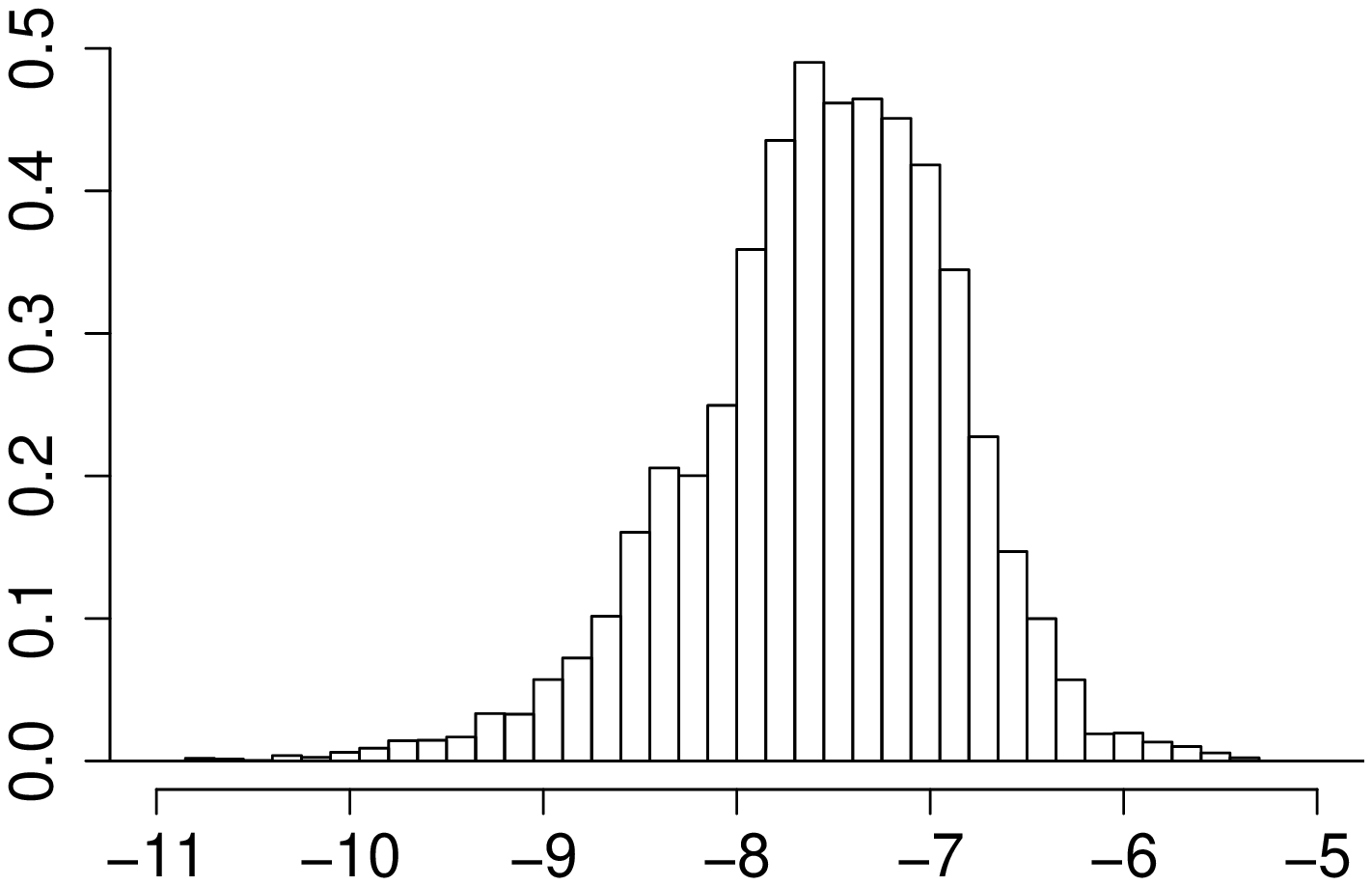}}\\
$M\neq M_{\max}$ and $\Lambda\in M$
&\subfigure{\includegraphics[scale=0.28]{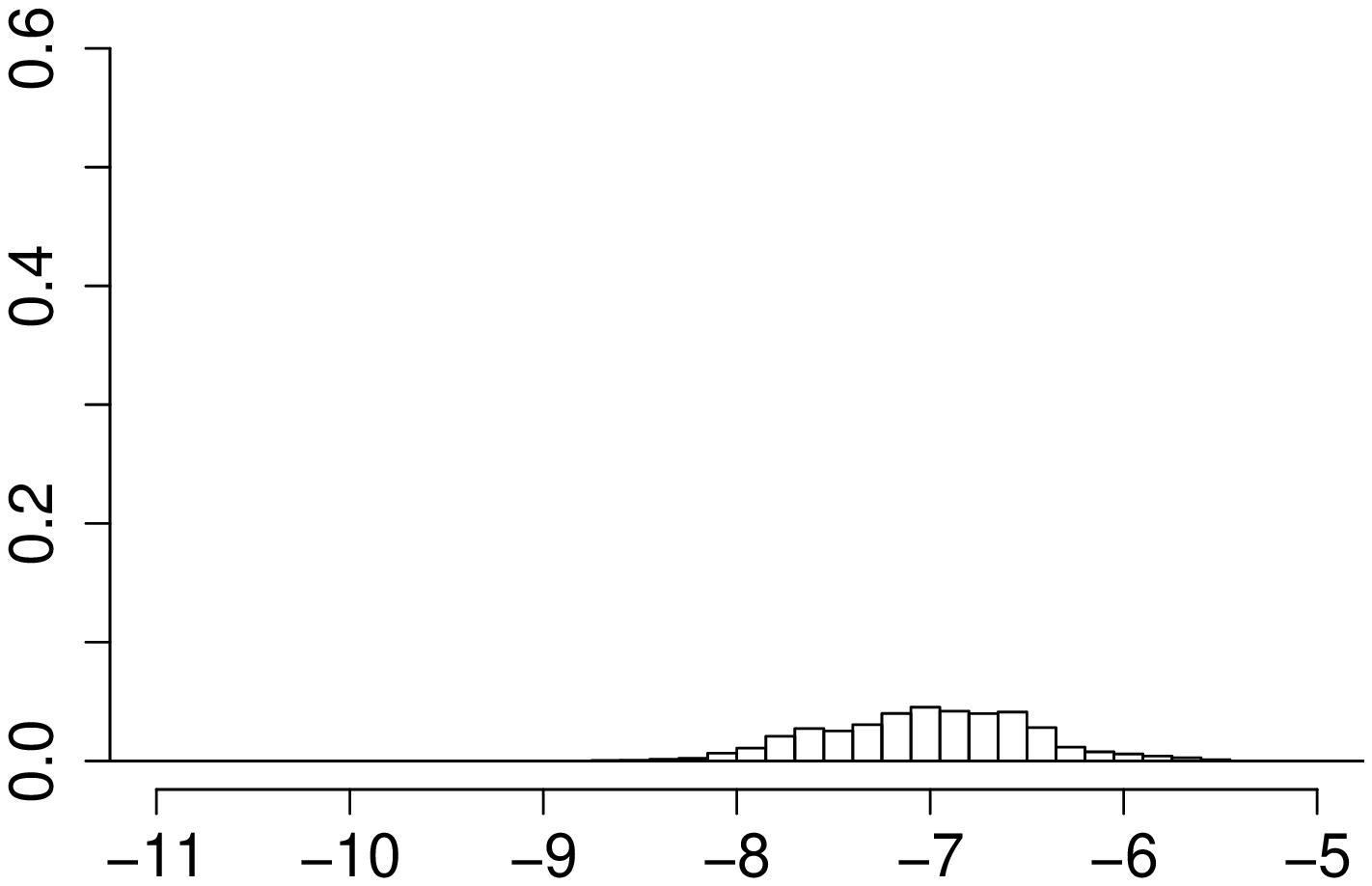}}
& \subfigure{\includegraphics[scale=0.28]{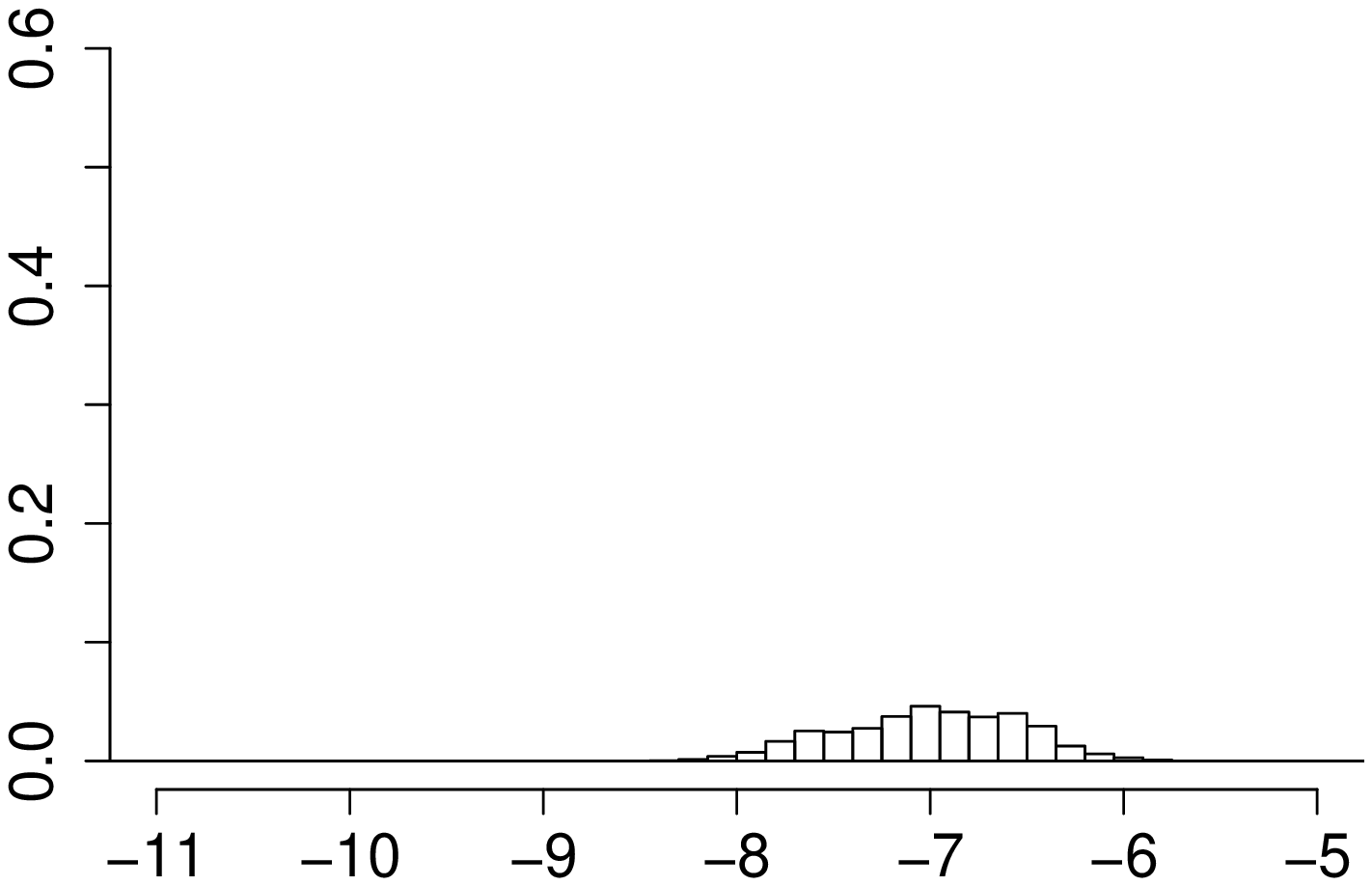}}
&\hspace{1.8cm} NA % \subfigure{\includegraphics[scale=0.28]{Newpno0phi23.eps}}
\\
$M\neq M_{\max}$ and $\Lambda\not\in M$
& \subfigure{\includegraphics[scale=0.28]{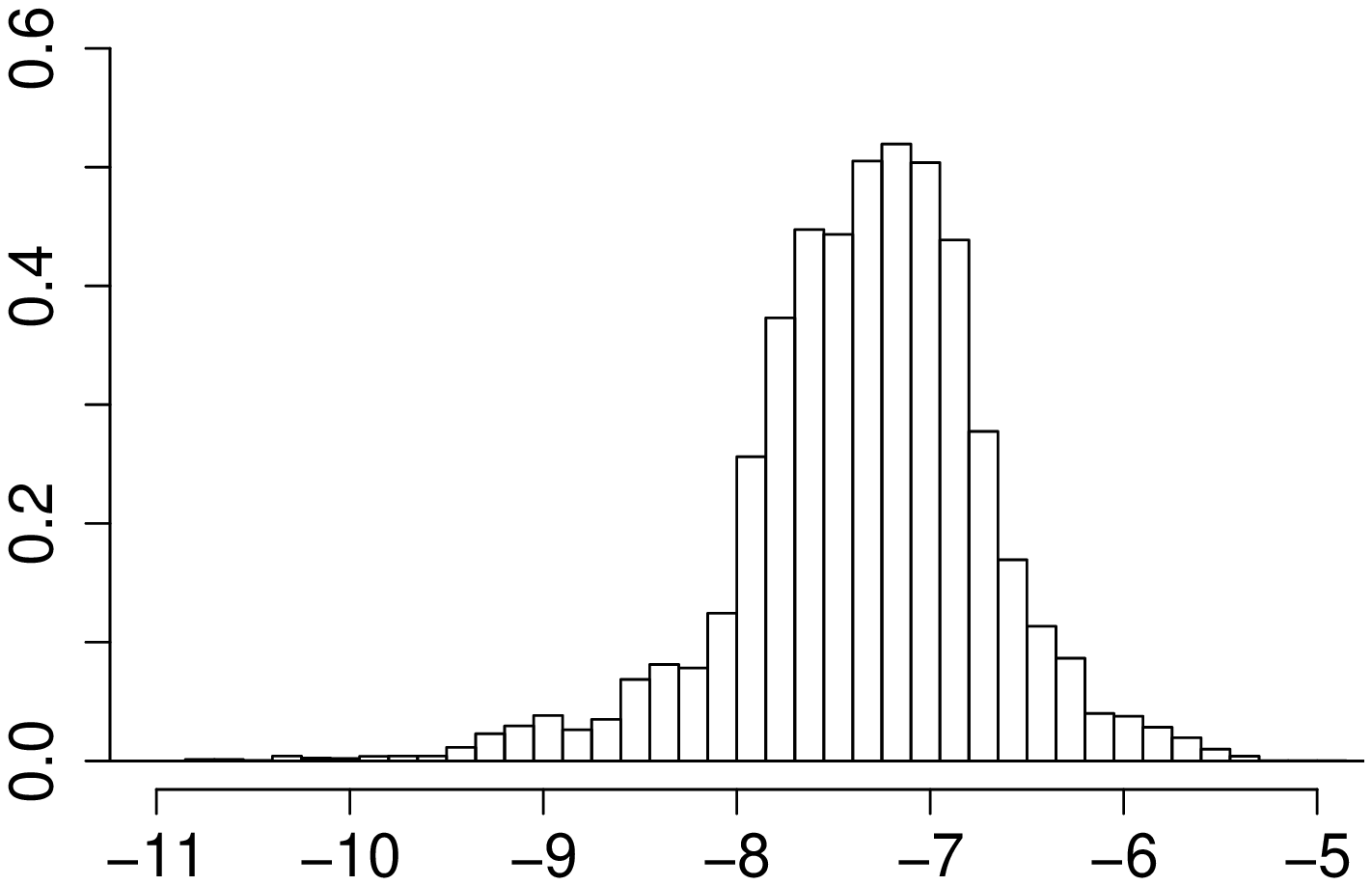}}
& \subfigure{\includegraphics[scale=0.28]{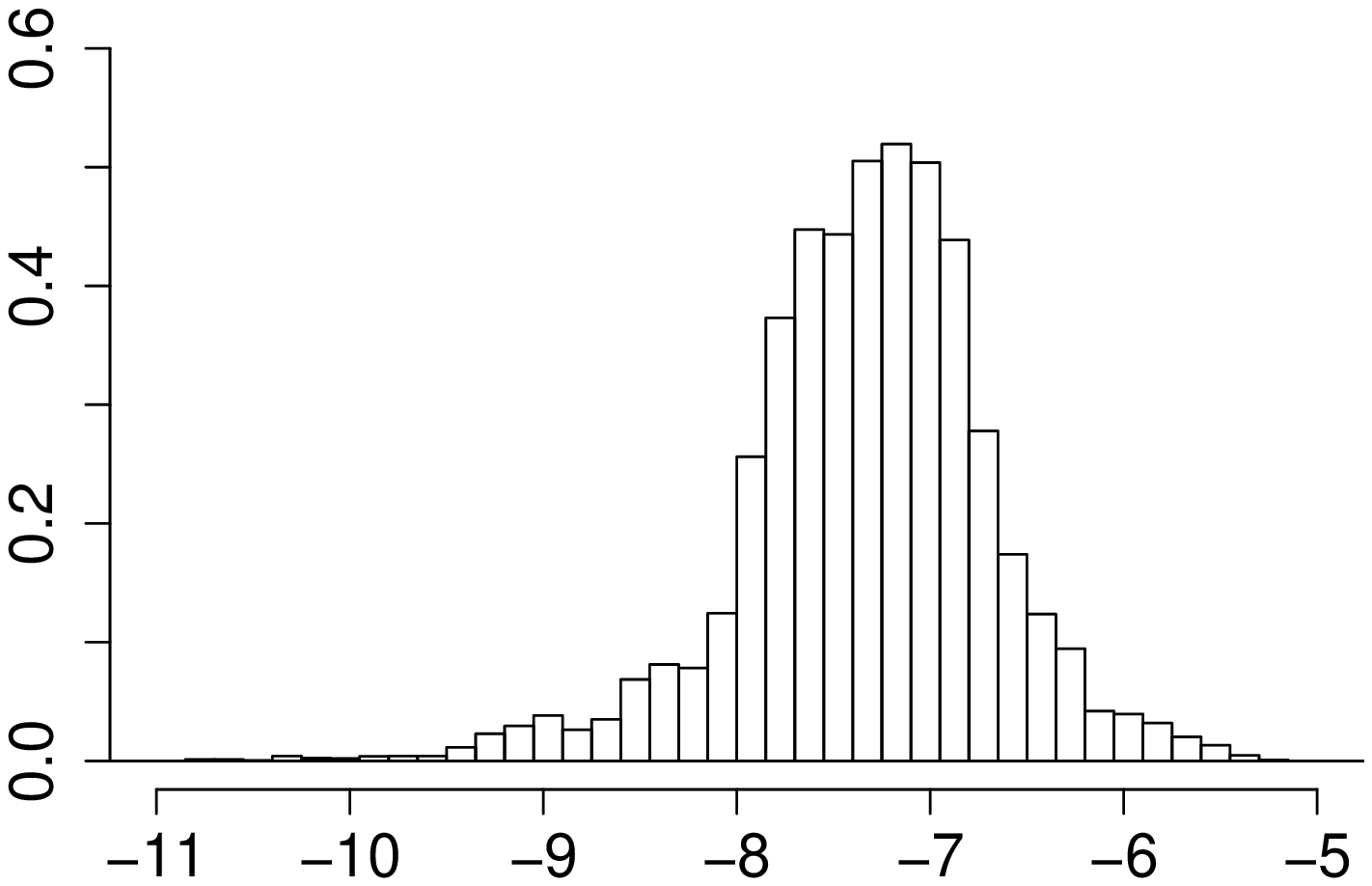}}
& \subfigure{\includegraphics[scale=0.28]{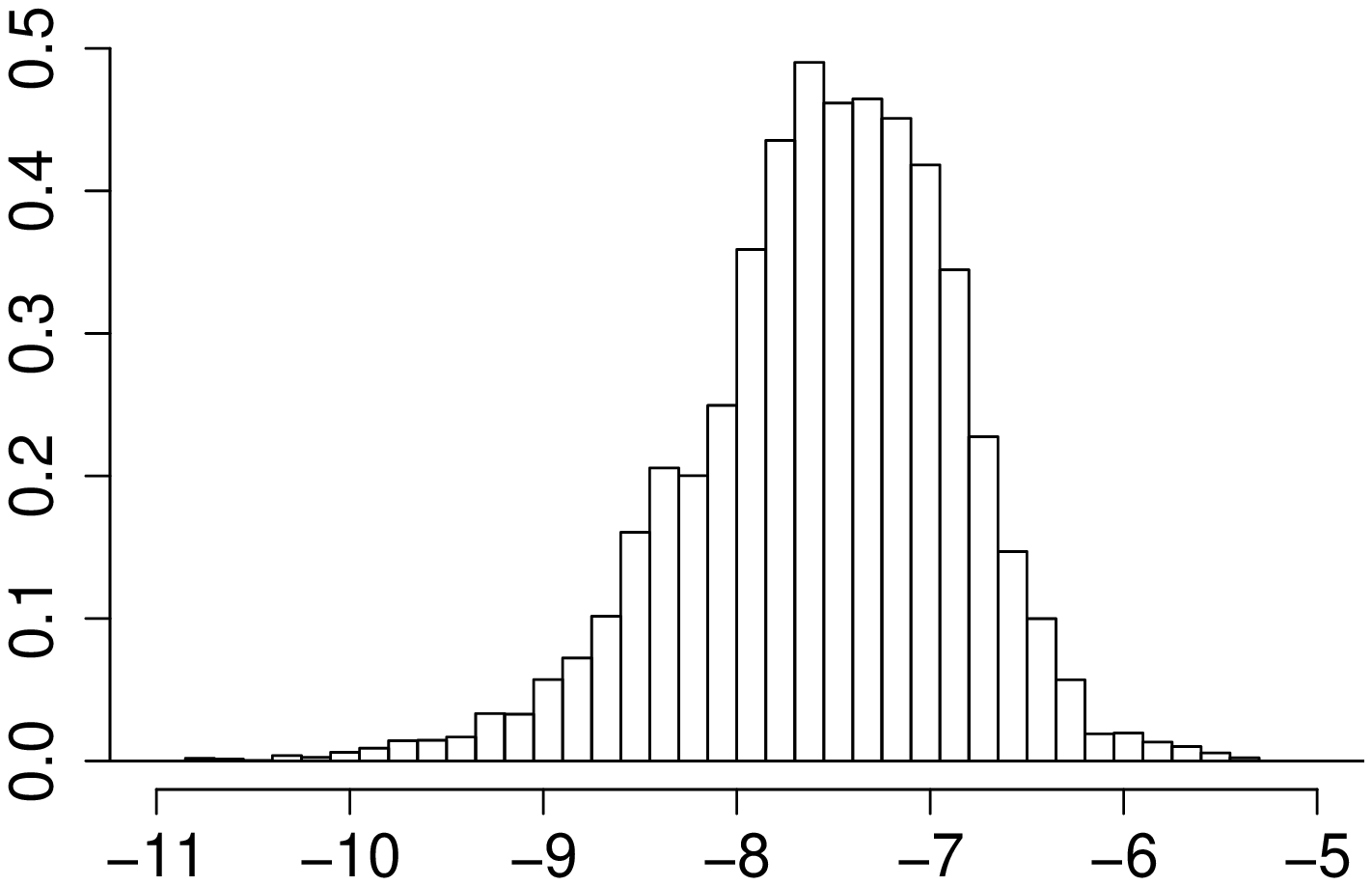}}\\
\end{tabular}
\caption{\label{fig:Sphi3}Red deer example: Histograms for 
$\phi^\Lambda$ for some $\Lambda\in L$ for different subgroups of the 
posterior samples. Results for different clique types $\Lambda$ in 
each column, and different subgroups in each row. The 
conditions given in the first column are defining the different subgroups.}
\end{figure}

\begin{figure}
\begin{tabular}{m{2.3cm}|m{4cm}m{4cm}m{4cm}}
& \hspace{1.8cm}$\Lambda=\AAAA$
&\hspace{1.8cm}$\Lambda=\cc$
&\hspace{1.8cm}$\Lambda=\cd$\\
\hline
All
& \subfigure{\includegraphics[scale=0.28]{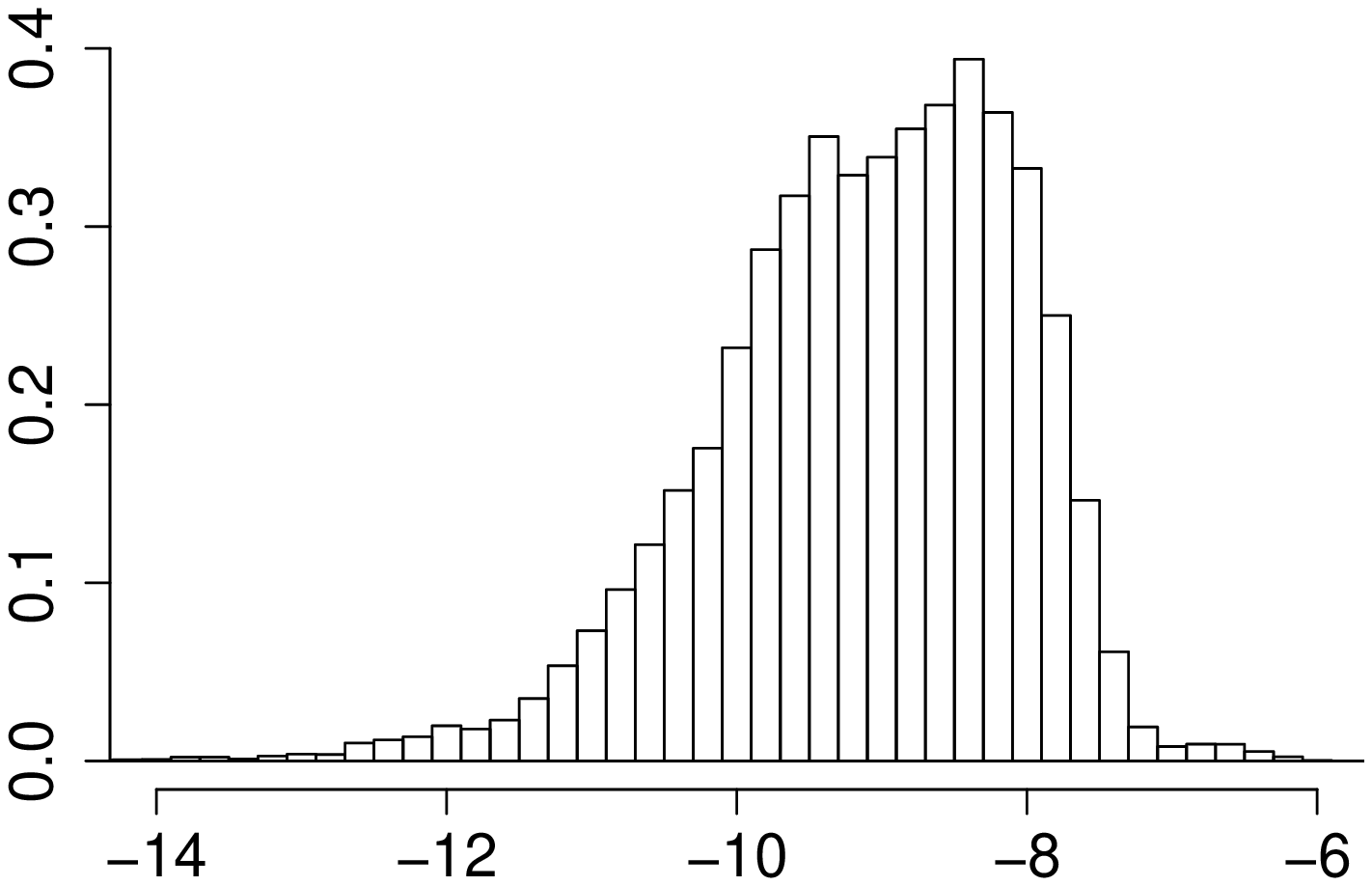}}
& \subfigure{\includegraphics[scale=0.28]{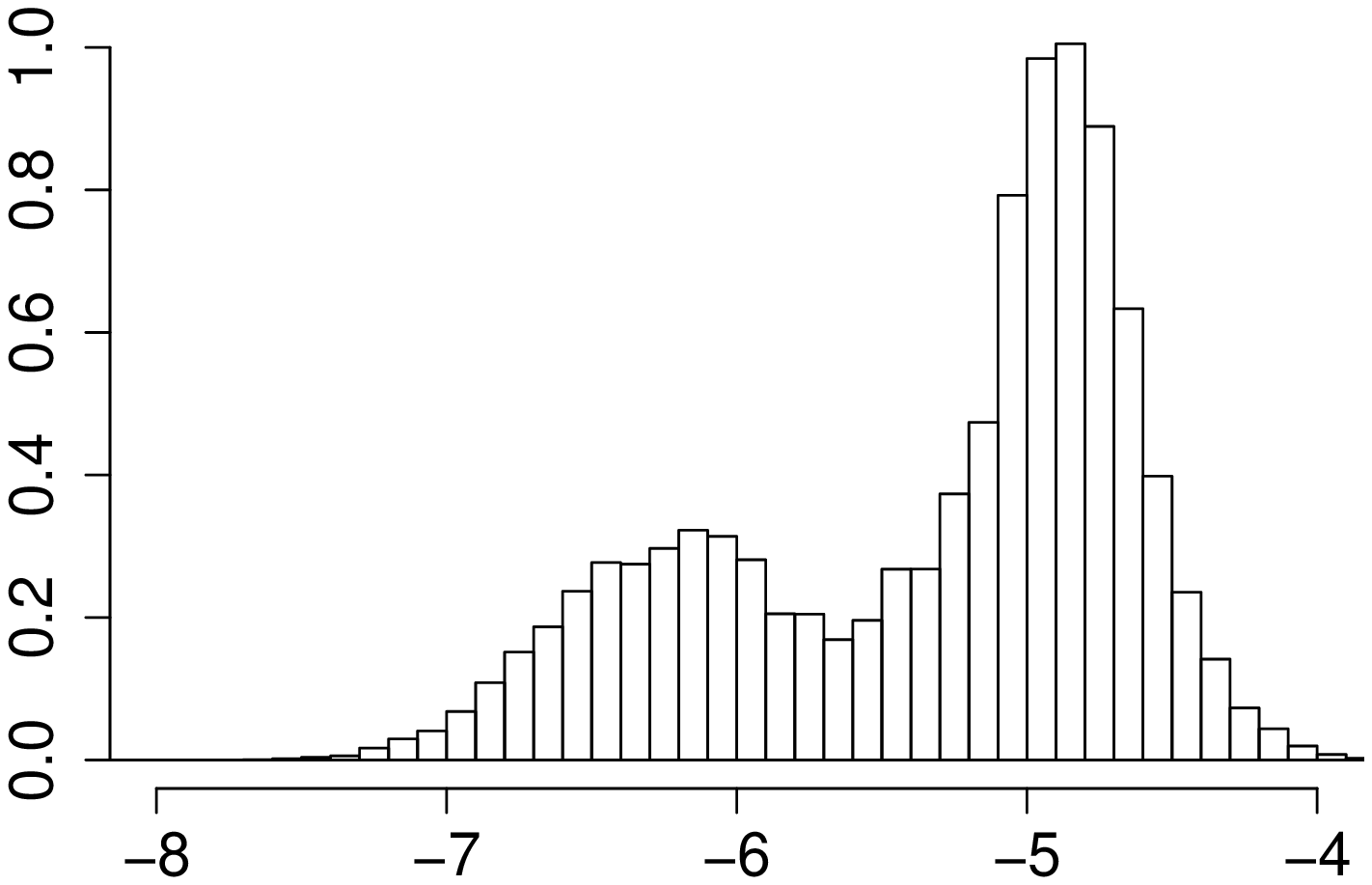}}
& \subfigure{\includegraphics[scale=0.28]{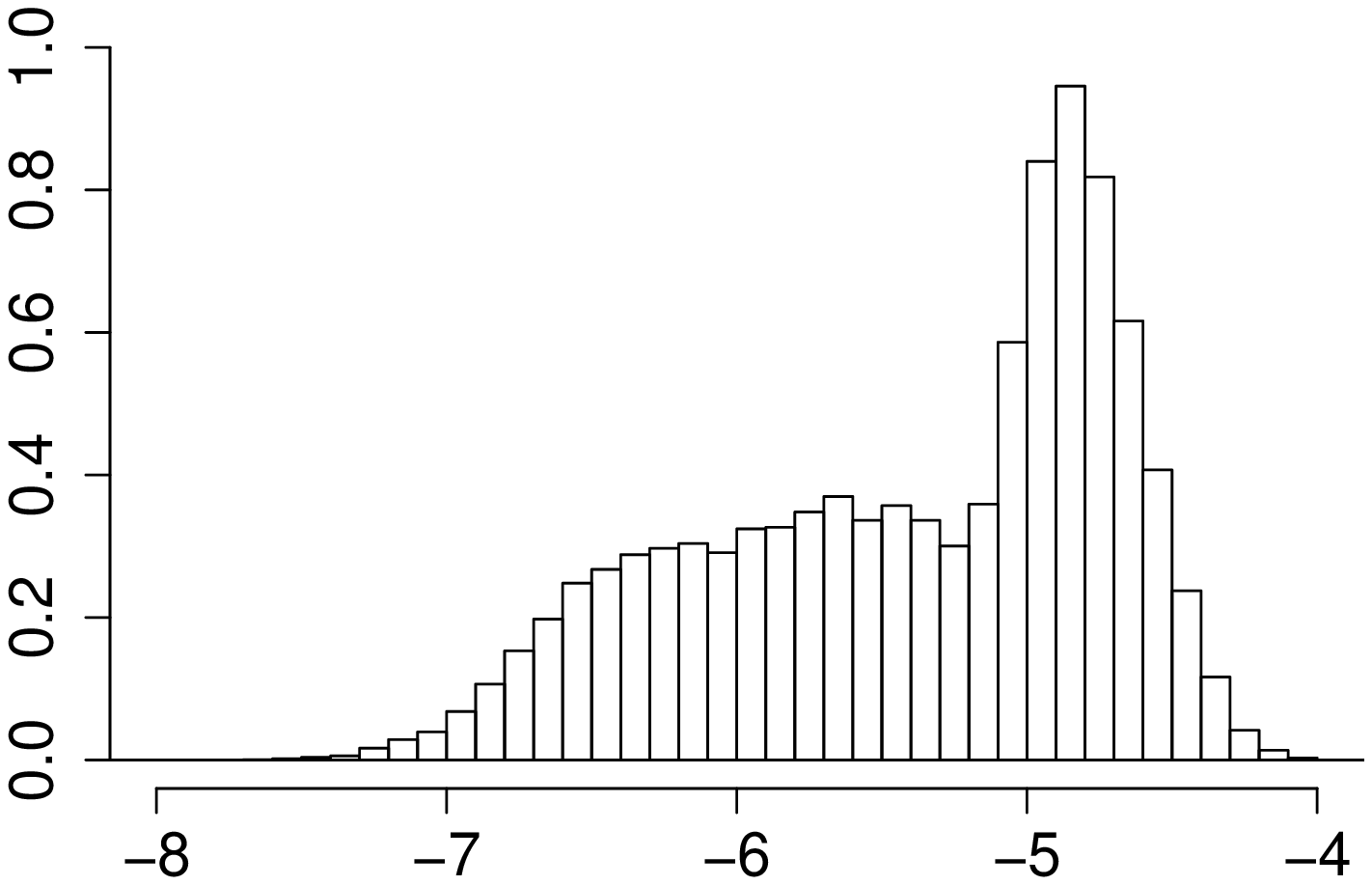}}\\
$M=M_{\max}$ 
& \subfigure{\includegraphics[scale=0.28]{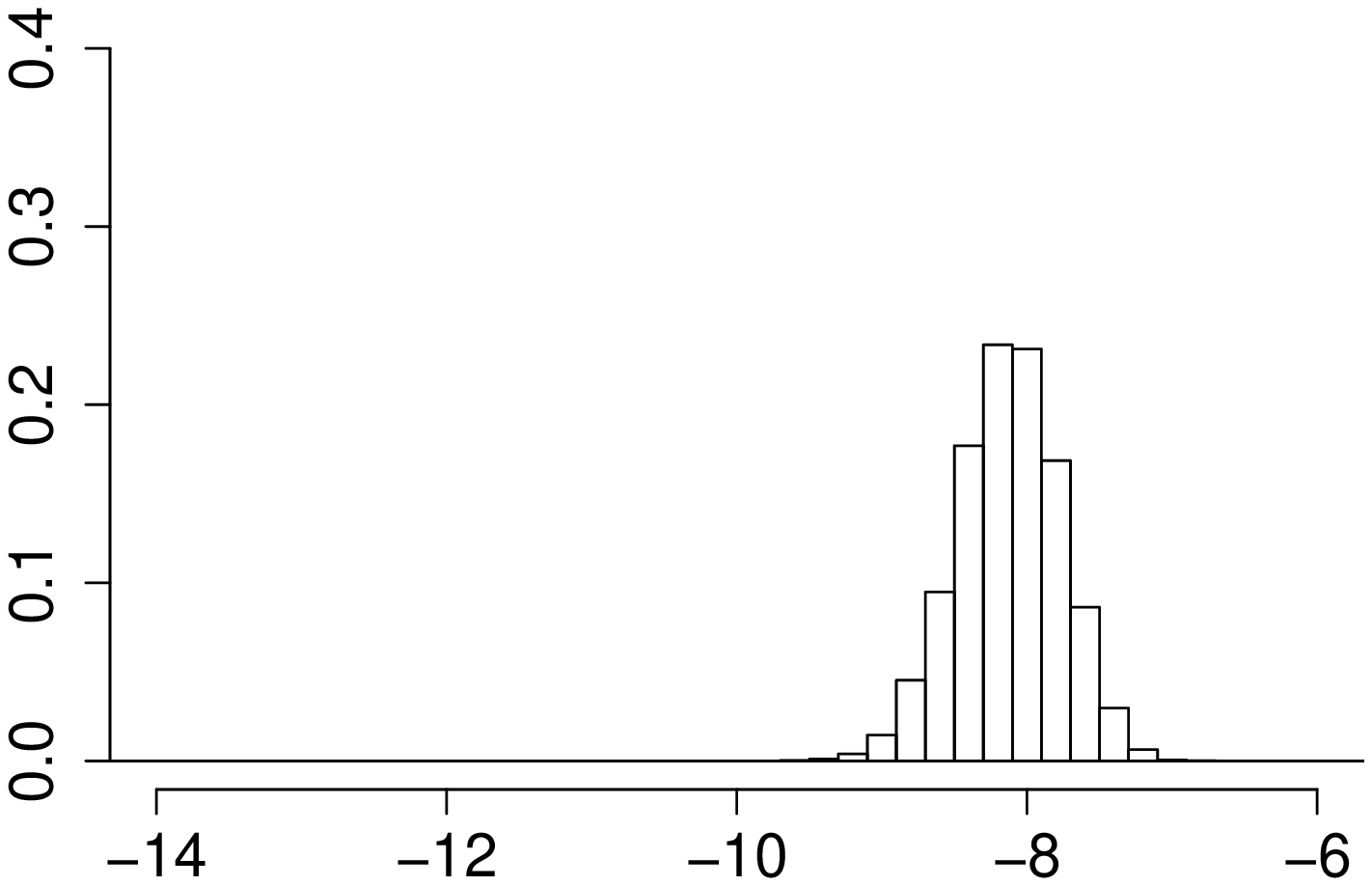}}
& \subfigure{\includegraphics[scale=0.28]{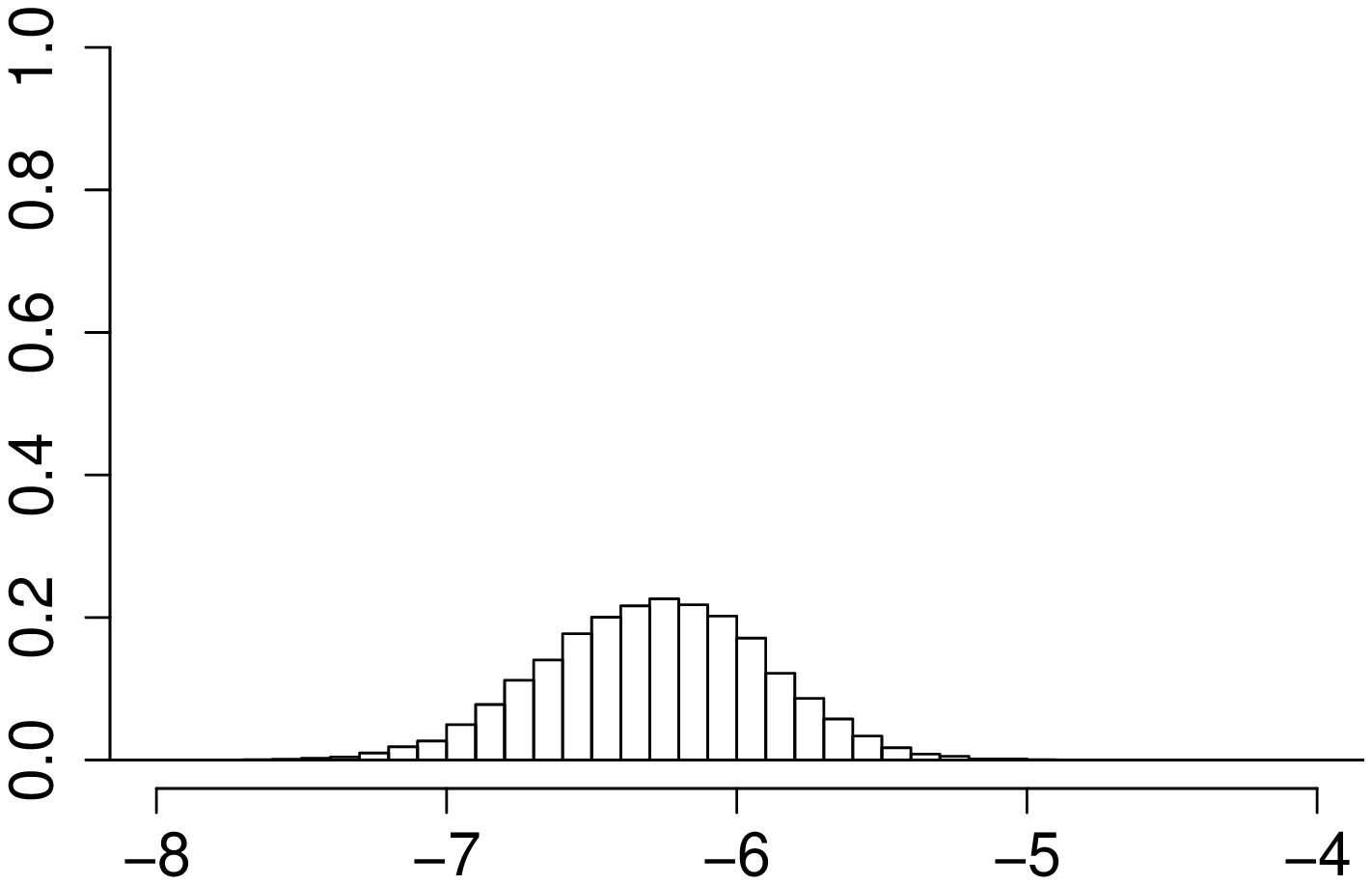}}
& \subfigure{\includegraphics[scale=0.28]{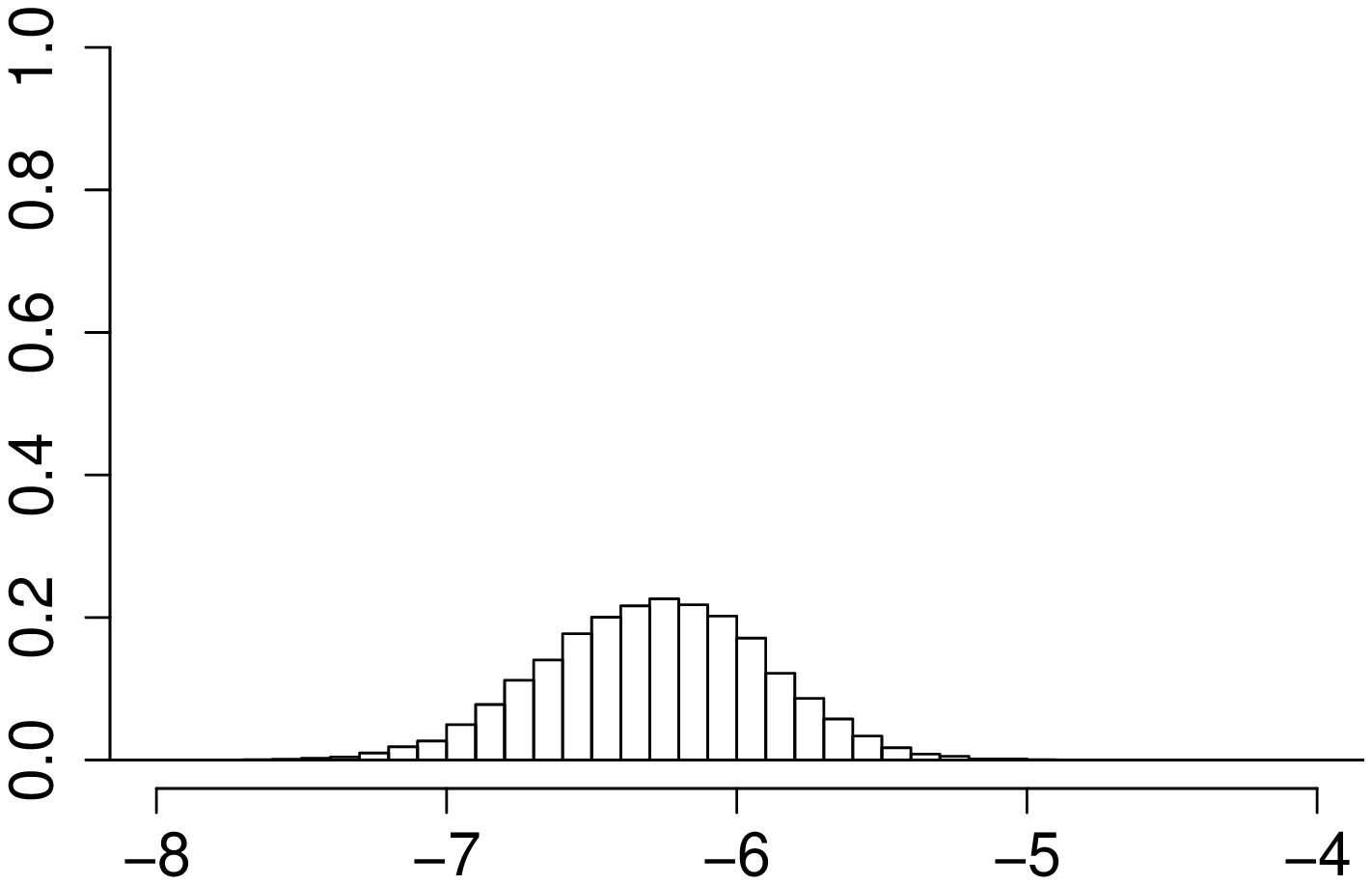}}\\
$M\neq M_{\max}$
& \subfigure{\includegraphics[scale=0.28]{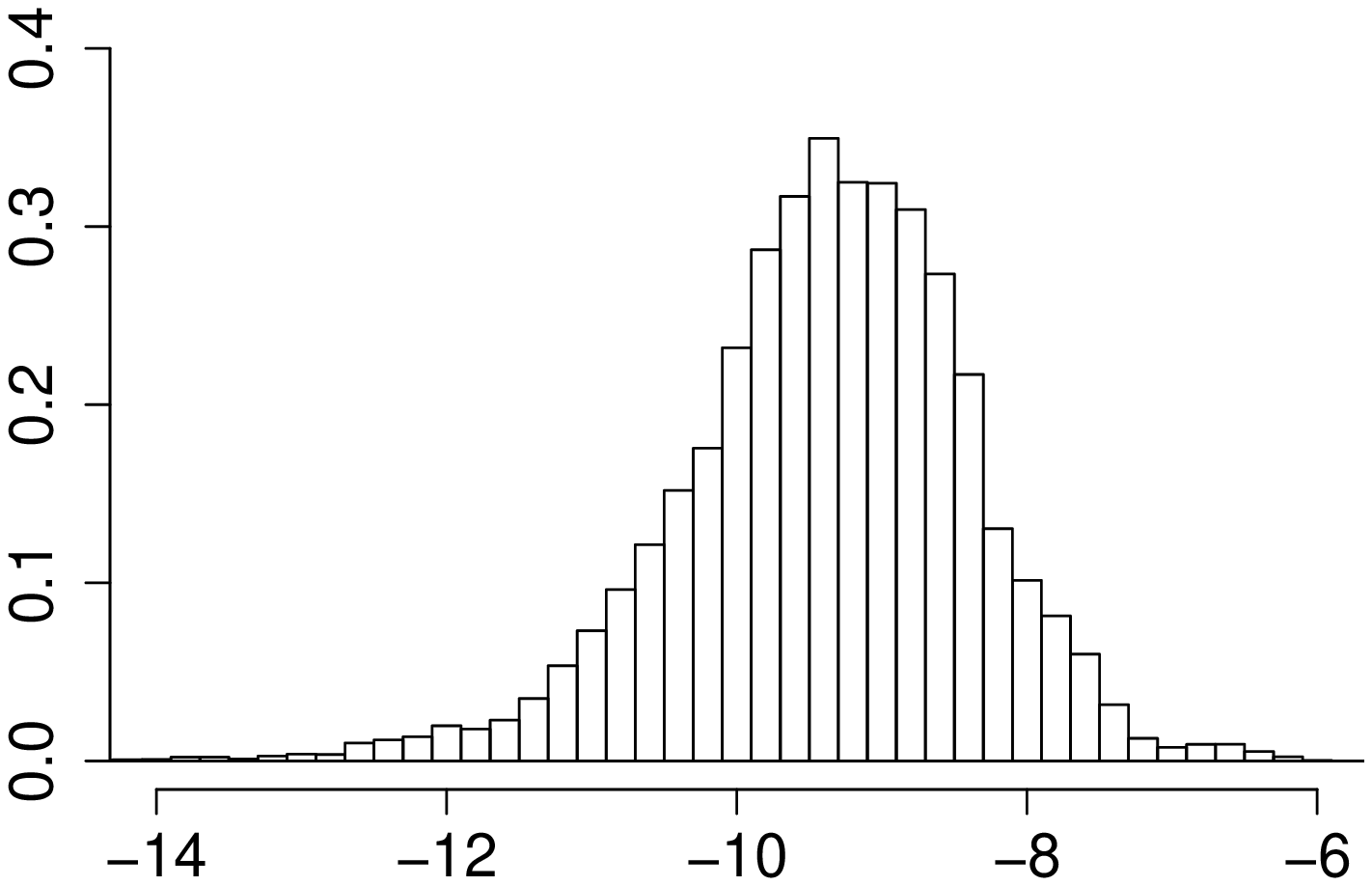}}
&\subfigure{\includegraphics[scale=0.28]{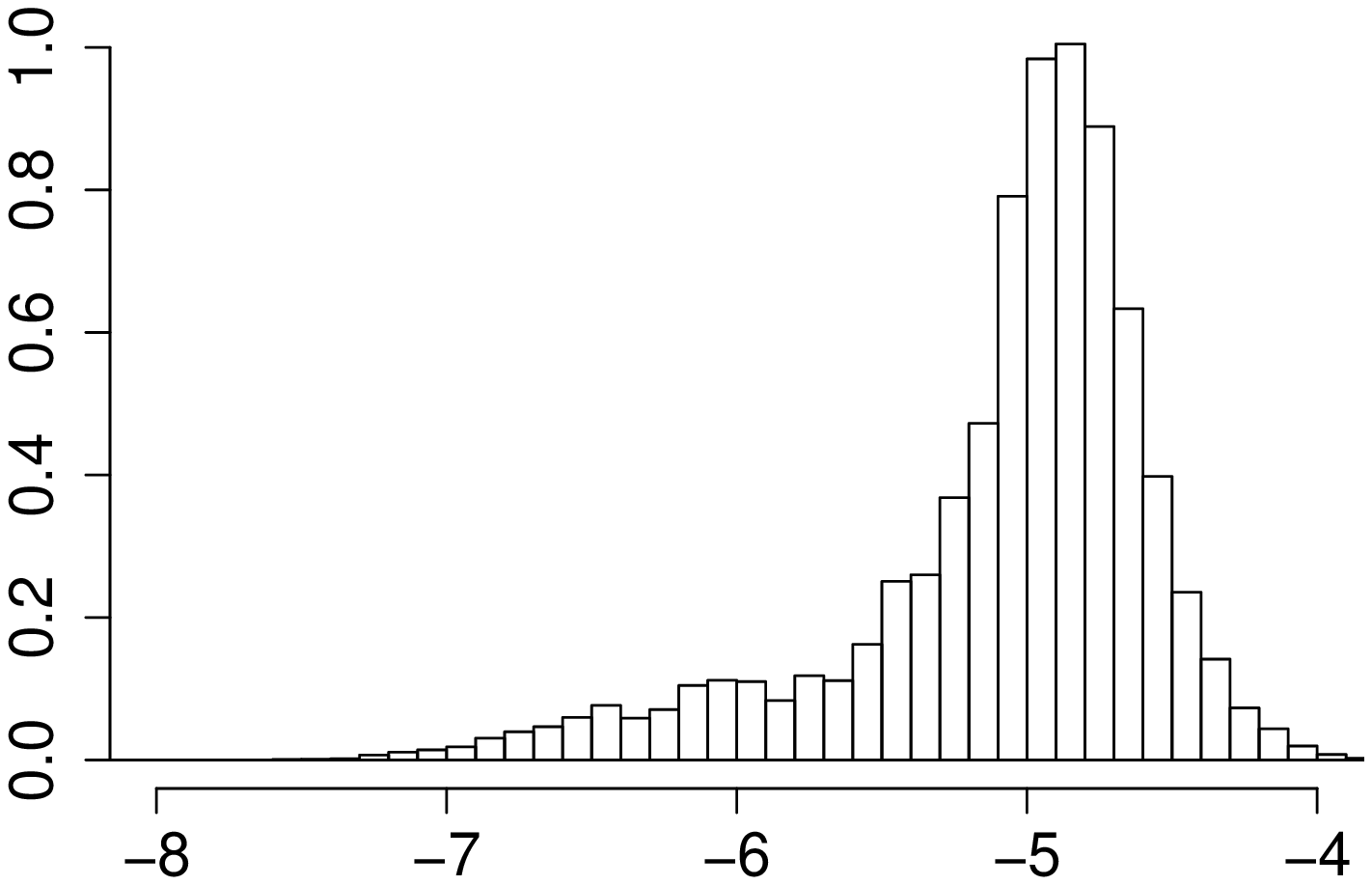}}
&\subfigure{\includegraphics[scale=0.28]{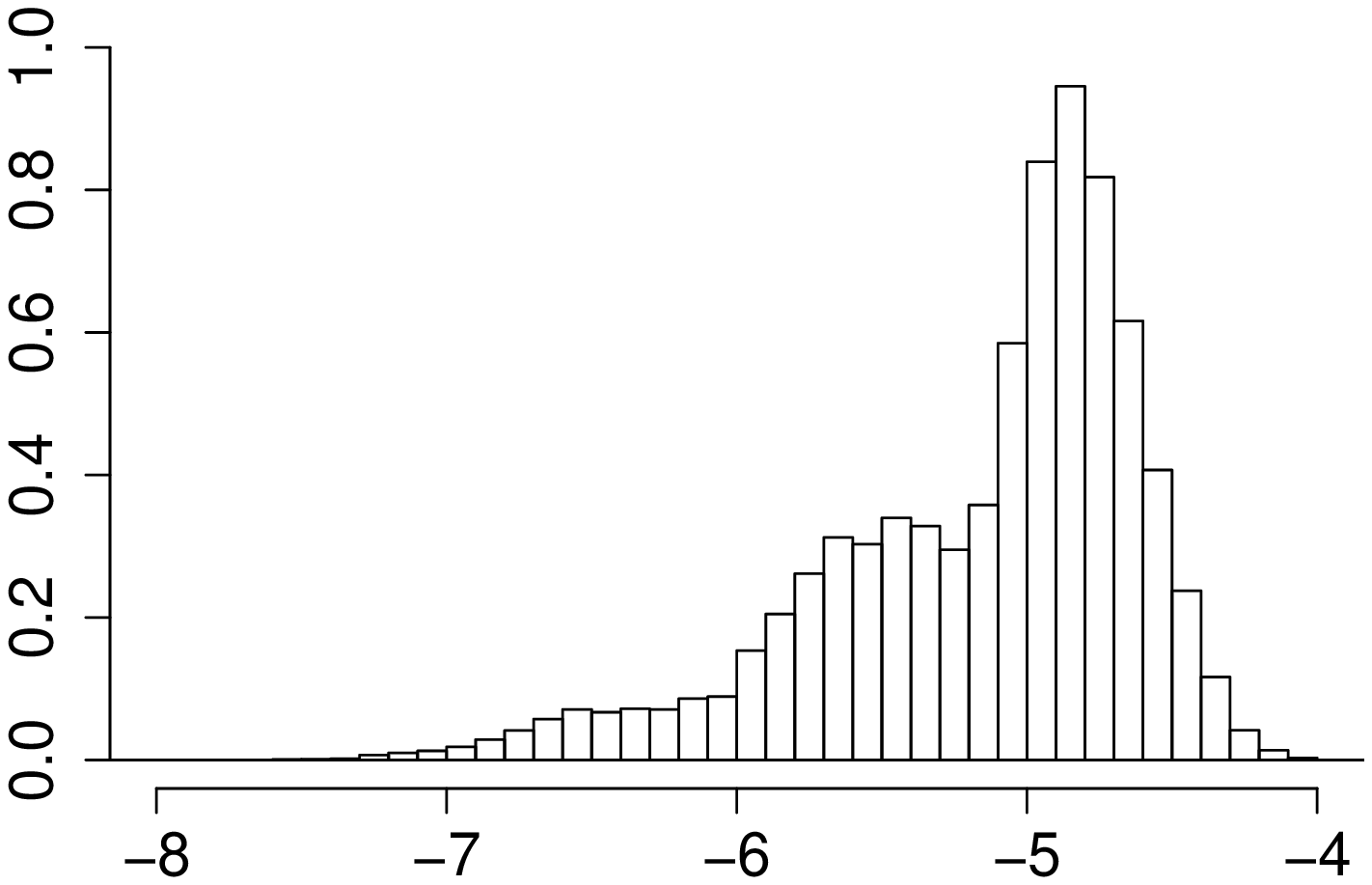}}\\
$M\neq M_{\max}$ and $\Lambda\in M$
&\hspace{1.8cm} NA %\subfigure{\includegraphics[scale=0.28]{Newpno0phi25.eps}}
& \subfigure{\includegraphics[scale=0.28]{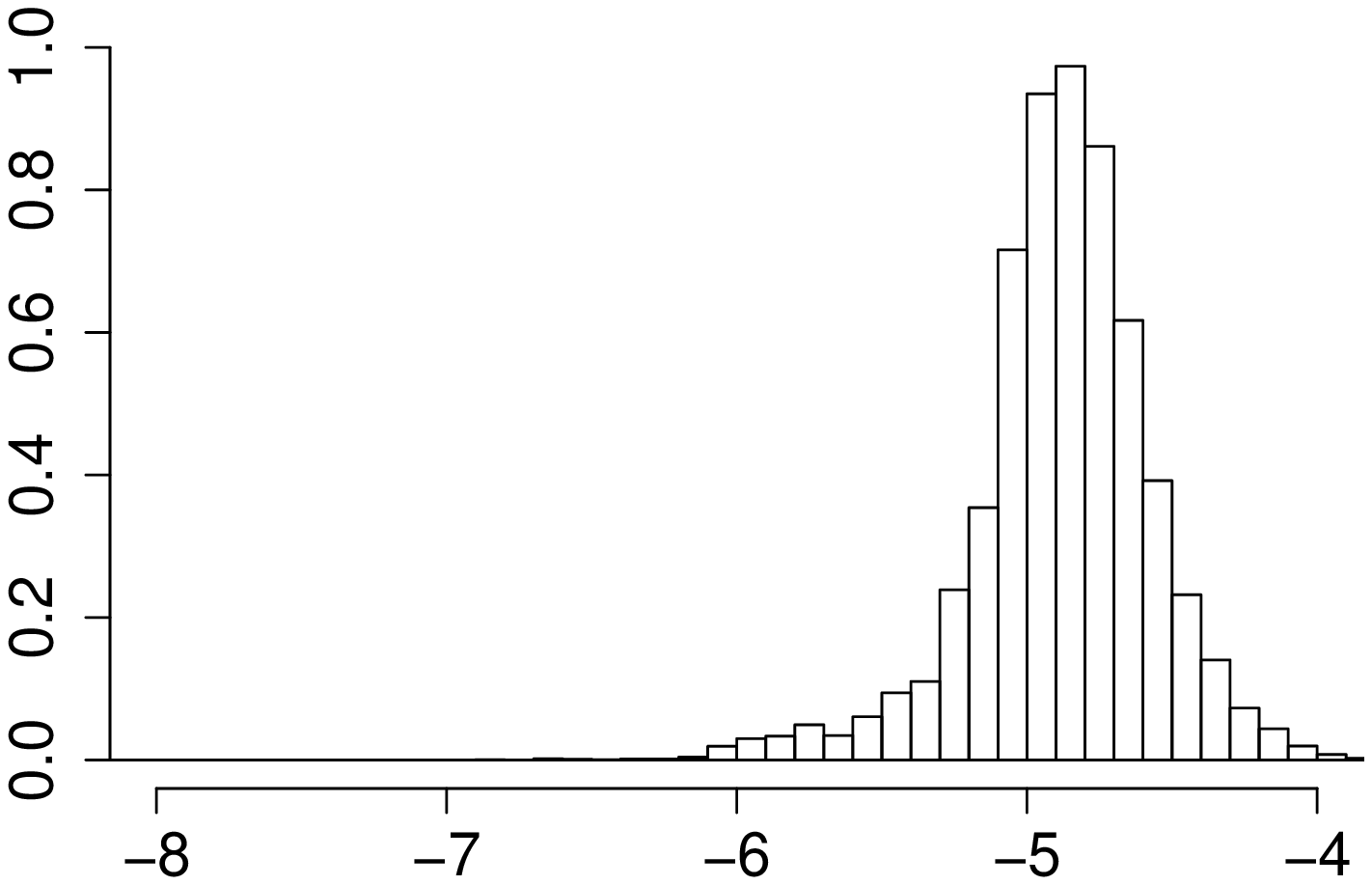}}
& \subfigure{\includegraphics[scale=0.28]{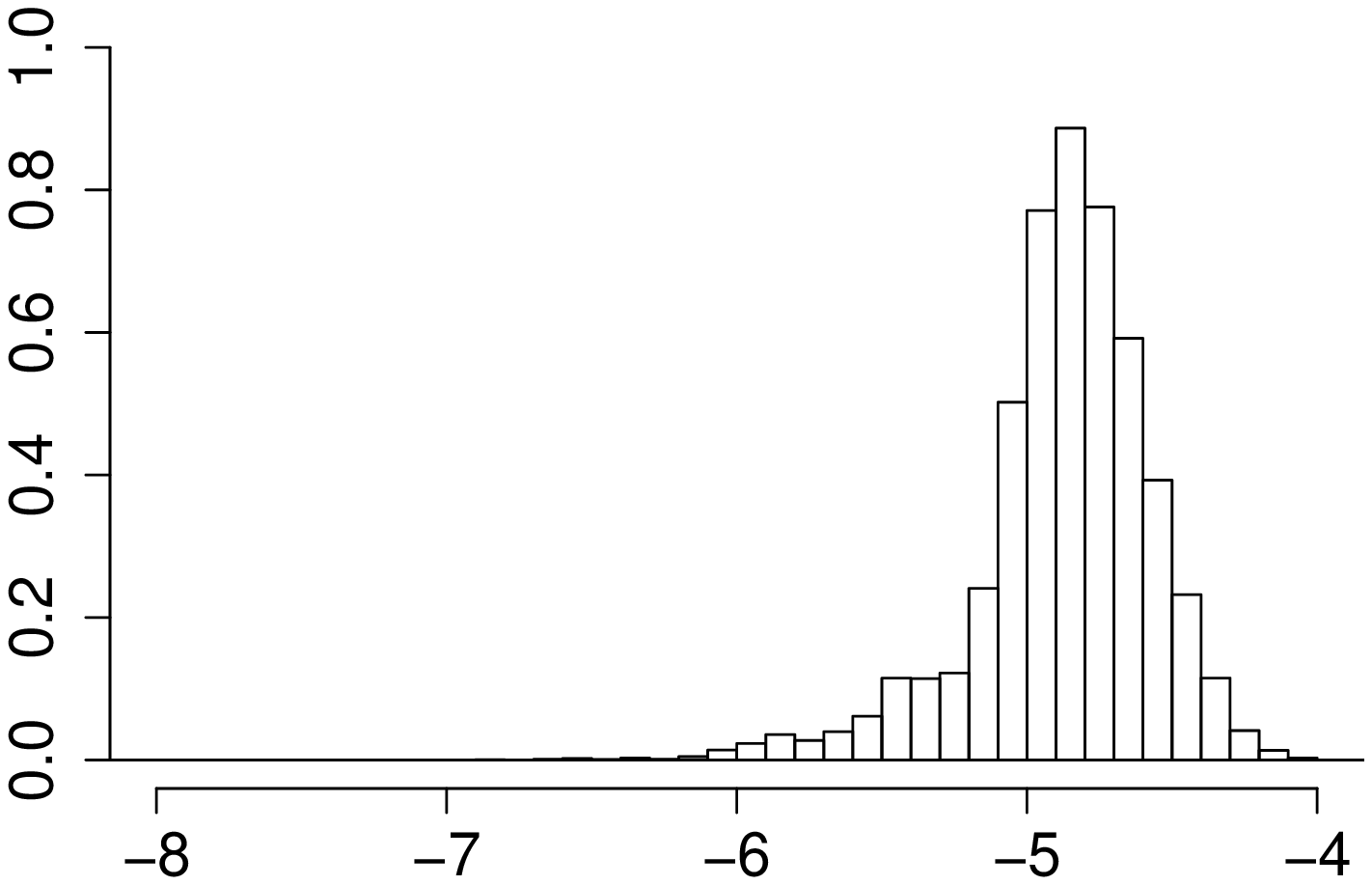}}\\
$M\neq M_{\max}$ and $\Lambda\not\in M$
& \subfigure{\includegraphics[scale=0.28]{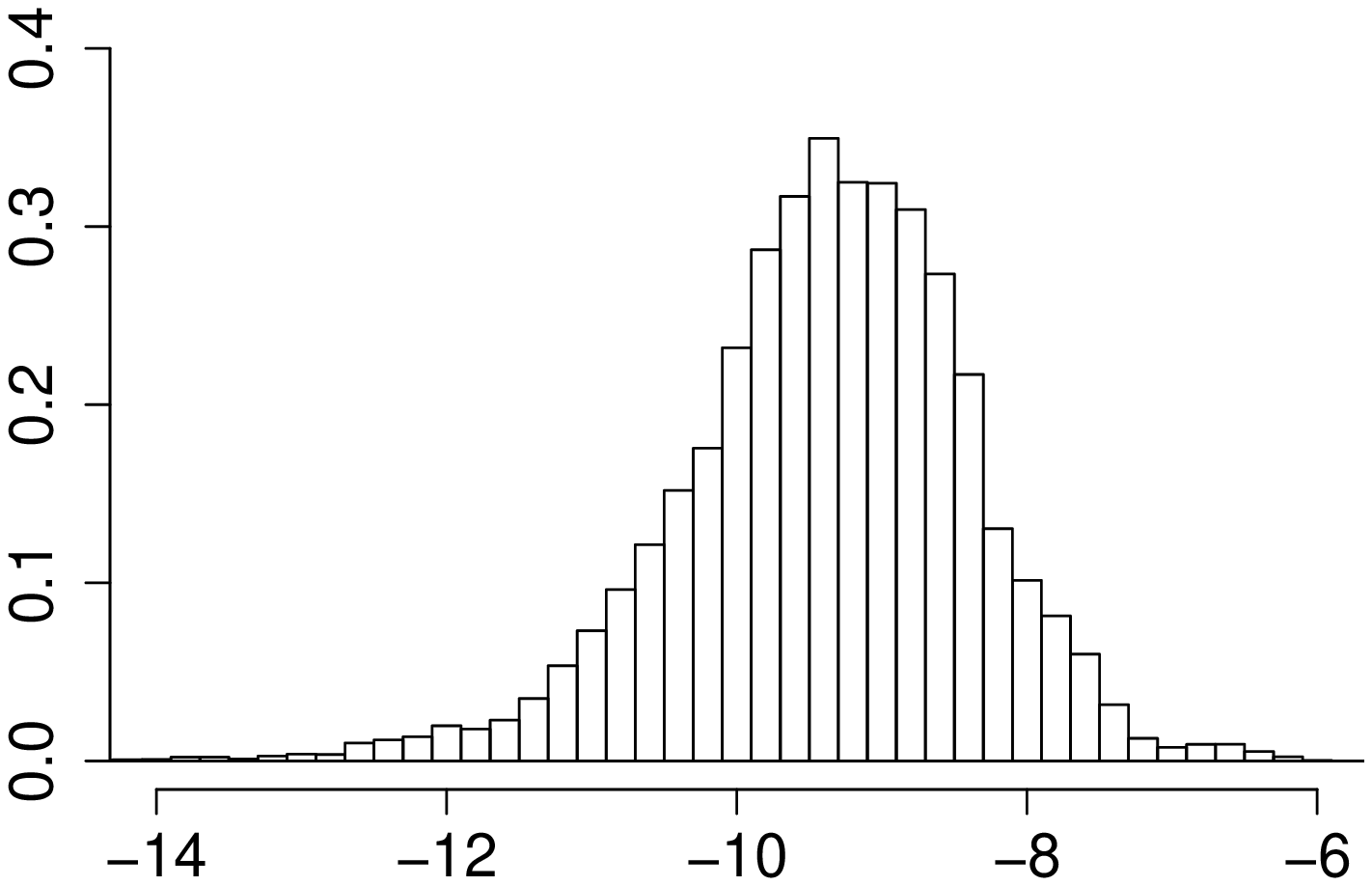}}
& \subfigure{\includegraphics[scale=0.28]{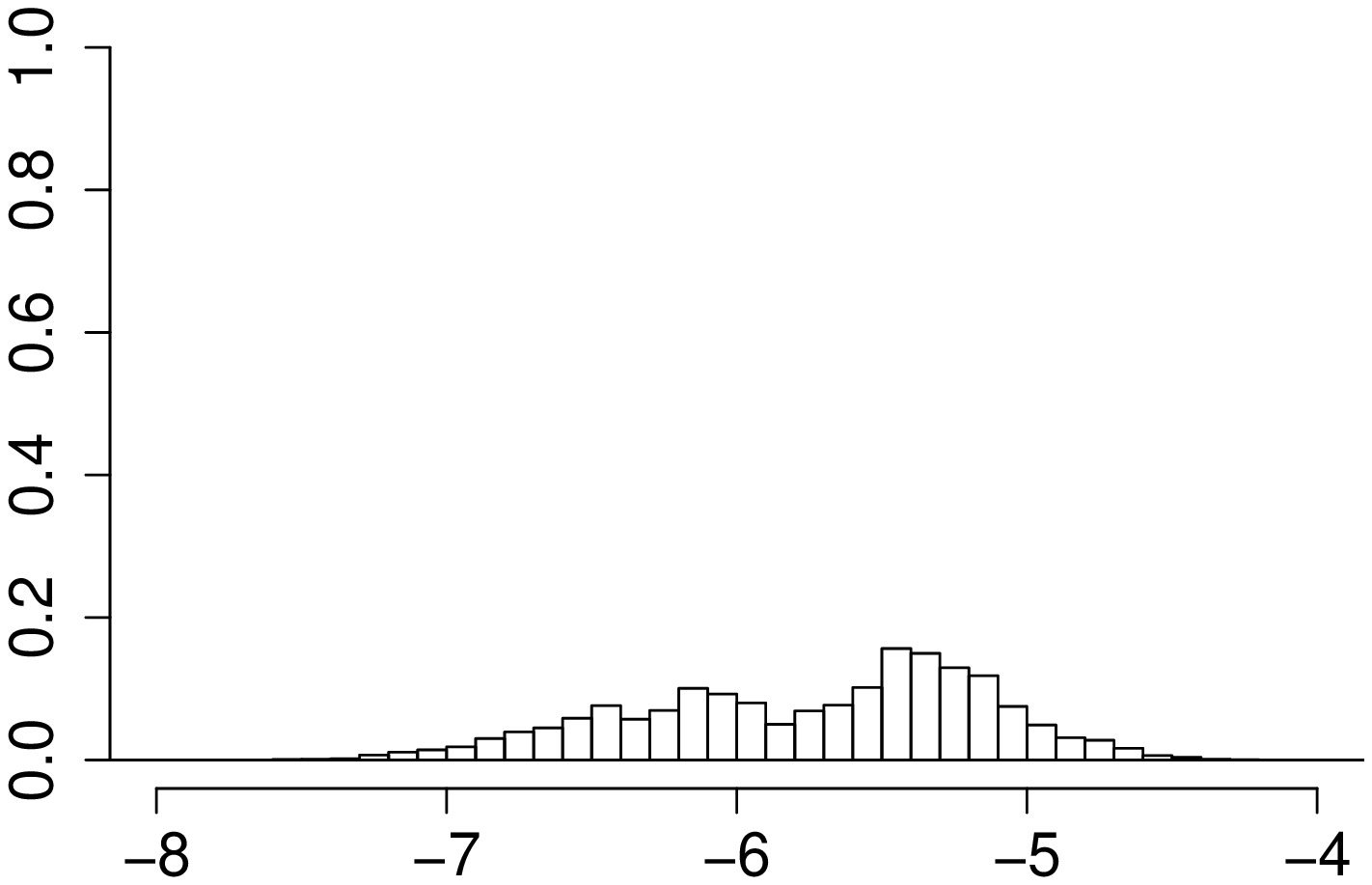}}
& \subfigure{\includegraphics[scale=0.28]{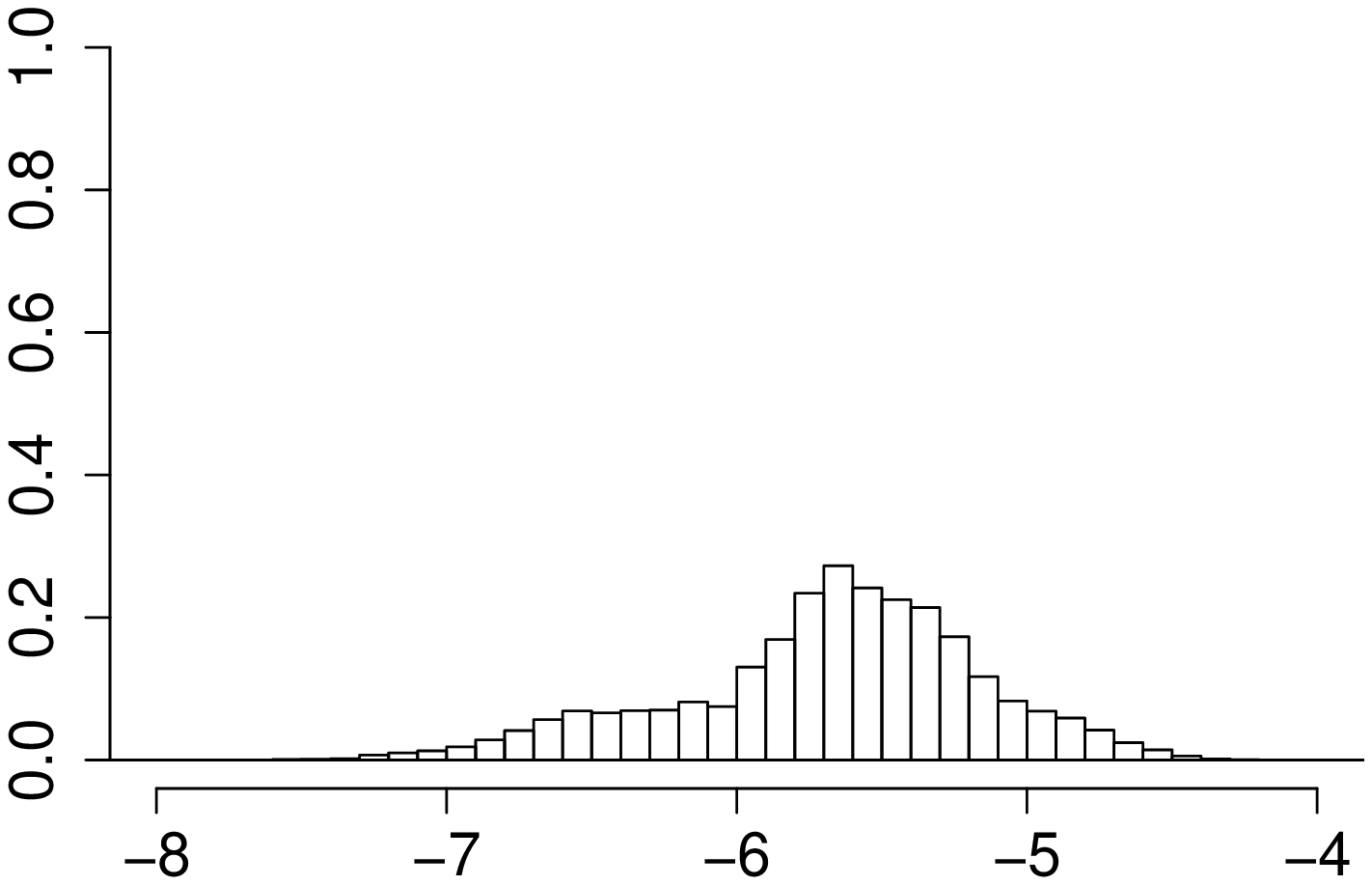}}\\
\end{tabular}
\caption{\label{fig:Sphi4}Red deer example: Histograms for 
$\phi^\Lambda$ for some $\Lambda\in L$ for different subgroups of the 
posterior samples. Results for different clique types $\Lambda$ in 
each column, and different subgroups in each row. The 
conditions given in the first column are defining the different subgroups.}
\end{figure}

In this section we provide a more detailed investigation of the multi-modality
in the posterior distributions 
for $\phi^\Lambda-\phi^{\{\emptyset\}}$ shown in Figure 12 in Section
7.3. As mentioned in the paper the modes can be explained
by first splitting the posterior samples into two groups, depending on 
the value of $M$. The first group is all samples where $M$ takes the 
aposteriori most probable value $M_{\max}=\{\{\emptyset\},\ABBB,\AABB,\ABAB,\ABBA,\AABA,\ABAA\}$, 
and the second group is the rest of the samples, i.e. samples where 
$M\neq M_{\max}$.
From Table 3 we see that the posterior probabilities
of these two groups are $0.284$ and $0.716$, respectively.
Making individual histograms for the $\phi$ values in
these two groups give the results shown in the second and third rows
in Figures \ref{fig:Sphi1}--\ref{fig:Sphi4}, which can be compared 
with the corresponding complete histogram in the first row.
As we can see the results for $M_{\max}$ are unimodal, whereas for $M\neq M_{\max}$
some of the histograms are still bimodal, see in particular the histograms
for $\AABB$, $\ABAB$, $\ABBA$, $\cc$ and $\cd$.
This bimodality can be explained by splitting the realisations in 
the group $M\neq M_{\max}$ into two subgroups, one for realisations of
$\phi^\Lambda$ where $\Lambda$ is on, and one for realisations of
$\phi^\Lambda$ where $\Lambda$ is off. The results for $\phi^\Lambda$ under this 
split is shown in the forth and fifth rows in Figures
\ref{fig:Sphi1}--\ref{fig:Sphi4}. Note however that this spilt results
in empty or almost empty sets of realisations of $\phi^{\Lambda}$ for
some $\Lambda$, in which case we do
not make a histogram (denoted NA in the figures).    
%\captionsetup[subfigure]{labelformat=empty}

